\documentclass[10pt,twoside,openright,b5paper]{book}
\usepackage{formatxxx}
\usepackage{textpos}
\usepackage{amsmath}
\usepackage[utf8]{inputenc}
\usepackage[page,toc,title]{appendix}
\usepackage[backend=biber]{biblatex}
\DeclareUnicodeCharacter{0301}{\'{e}}
\DeclareUnicodeCharacter{0302}{\'{e}}
\usepackage{imakeidx}
\makeindex[columns=2, title=Alphabetical Index, options= -s style.ist]

\begin{document}
\begin{titlepage}

\begin{center}
\large
	\textbf{Modeling and Suppressing Unwanted Parasitic Interactions in Superconducting Circuits}\\
			\vspace{2.0cm}
			\textbf{Xuexin Xu}\\
			\vspace{0.2in}
			\textbf{J\"ulich Forschungszentrum \& RWTH Aachen University}
\end{center}
		\vspace{2.0cm}


%
%
%
%
%
%
\end{titlepage}

\thispagestyle{empty}	

\frontmatter
\pagestyle{plain}
\pagenumbering{roman}
\chapter*{Abstract}
\addcontentsline{toc}{chapter}{Abstract}
\thispagestyle{myheadings}
Superconducting qubits are based on collective excitations in Josephson junctions. Their sufficiently high controllability and low-noise interaction with one another make them a viable candidate for implementing large-scale quantum computation. Although there are great improvements in the coherence of these qubits, the progression towards building a fault-tolerant quantum computer is still a major mission. This issue can be reflected in the imperfect gate fidelity. A source of such infidelity is the fundamental parasitic interaction between coupled qubits. This thesis deals with such spurious interaction in two- and three-qubit circuits. The parasitic interaction reflects a bending between computational and noncomputational levels. This bending creates a parasitic ZZ interaction. We first study the possibility of zeroing the ZZ interaction in two qubit combinations: a pair of interacting transmons, as well as a hybrid pair of transmon coupled to capacitively shunted flux qubit (CSFQ). We utilize our theory to accurately simulate experimental results taken by our collaborators from measuring a CSFQ-transmon pair in the absence and presence of cross-resonance (CR) gate.  The impressive agreements between our theory and experiment motivated us to study the characteristics of a CR gate that performs with 99.9\% fidelity in the absence of static ZZ interaction. Since the CR pulse produces an additional ZZ component on top of the static part, we propose a new strategy for zeroing total ZZ interaction, and name it dynamical ZZ freedom. This freedom can exist in all-transmon circuits and allow to make perfect entanglement. Based on our findings, we propose a new two-qubit gate, namely the parasitic-free (PF) gate. Moreover, we show that we can take advantage of the ZZ interaction and make it stronger in order to perform a controlled-Z gate.  Finally, we study the impact of a third qubit on the two-qubit gate performance and discuss several examples to illustrate the properties of two-body ZZ and three-body ZZZ interactions in circuits with more than two qubits.

\chapter*{Zusammenfassung}
\addcontentsline{toc}{chapter}{Zusammenfassung}
\thispagestyle{myheadings}
Supraleitende Qubits basieren auf kollektiven Anregungen in Josephson-Kontakten. Ihre ausreichend hohe Kontrollierbarkeit und rauscharme Wechselwirkung untereinander machen sie zu einem praktikablen Kandidaten für die Implementierung von Quantencomputern in großem Maßstab. Obwohl es große Verbesserungen in der Kohärenz dieser Qubits gibt, ist der Weg zum Bau eines fehlertoleranten Quantencomputers noch eine große Aufgabe. Dieses Problem kann sich in der unvollkommenen Gattertreue widerspiegeln. Eine Quelle solcher Untreue ist die fundamentale parasitäre Wechselwirkung zwischen wechselwirkenden Qubits. Diese Arbeit befasst sich mit dieser parasitären Wechselwirkung in Zwei- und Drei-Qubit-Schaltungen. Die parasitäre Wechselwirkung spiegelt eine Biegung zwischen rechnerischer und nicht rechnerischer Ebene wider. Diese Biegung erzeugt eine parasitäre ZZ-Wechselwirkung. Wir untersuchen zunächst die Möglichkeit, die ZZ-Wechselwirkung in zwei Qubit-Kombinationen auf Null zu setzen: ein Paar von wechselwirkenden Transmonen sowie ein hybrides Paar von Transmonen, die an ein kapazitiv geshuntetes Flux-Qubit (CSFQ) gekoppelt sind. Wir nutzen unsere Theorie, um die experimentellen Ergebnisse unserer Kollaborateure bei der Messung eines CSFQ-Transmon-Paares in Abwesenheit und Anwesenheit eines Cross-Resonance (CR) Gatters genau zu simulieren.  Die beeindruckenden Übereinstimmungen zwischen unserer Theorie und dem Experiment motivierten uns, die Eigenschaften eines CR-Gatters zu untersuchen, das mit 99,9\% Genauigkeit in Abwesenheit der statischen ZZ-Wechselwirkung arbeitet. Da der CR-Puls zusätzlich zum statischen Teil eine weitere ZZ-Komponente erzeugt, schlagen wir eine neue Strategie zur Aufhebung der gesamten ZZ-Wechselwirkung vor und bezeichnen diese als dynamische ZZ-Freiheit. Diese Freiheit kann in All-Transmon-Schaltkreisen existieren und erlaubt es perfekte Verschränkung zu erzeugen. Basierend auf unseren Erkenntnissen schlagen wir ein neues Zwei-Qubit-Gatter vor, nämlich das parasitenfreie (PF) Gatter. Außerdem zeigen wir, dass wir die ZZ-Wechselwirkung ausnutzen und verstärken können, um ein kontrolliertes Z-Gatter zu realisieren.  Schließlich untersuchen wir den Einfluss eines dritten Qubits auf das Verhalten des Zwei-Qubit-Gatters und diskutieren mehrere Beispiele, um die Eigenschaften der Zwei-Körper-ZZ- und Drei-Körper-ZZZ-Wechselwirkungen in Schaltungen mit mehr als zwei Qubits zu veranschaulichen.

\tableofcontents


\mainmatter
\pagestyle{scrheadings}

\chapter{Introduction}
\label{c1}
\vspace{-0.3in}
\section{Quantum Computation}
Quantum computers develop computation technology based on quantum mechanical phenomena, e.g. superposition, entanglement and interference. The field of quantum computing can be traced back to 1980's. Paul Benioff first put forward a theoretical demonstration of a quantum mechanical prototype of the Turing machine~\cite{benioff1980the-computer}, but it was not computationally powerful than a conventional classical one. Two years later, Richard Feynman proposed the first nontrivial application of quantum physics and believed that it can be only simulated on quantum computers~\cite{feynman1982simulating}.  Subsequently, David Deutsch developed this idea of applying quantum mechanics to solve computational problems in computer science~\cite{deutsch1985quantum}.  However, these nice ideas destined to remain on paper until 1994 when Peter Shor found the first potentially useful way to use a hypothetical quantum computer, and invented the polynomial-time quantum computer algorithm for integer factorization~\cite{shor1994algorithms}. Another remarkable algorithm was discovered by Lov Grover for searching an unsorted database~\cite{grover1996fast}. Both of them provide a superior speed up over classical counterpart.

The laws of quantum mechanics enable to do more than encoding what can be done with bit. They help to solve problems which no classical computer can solve in any feasible amount of time. This ability is known as quantum supremacy or quantum advantage~\cite{boixo2018characterizing}. Then it comes to the question, how to perform quantum computing on a certain physical system? About twenty years ago, David DiVincenzo organized a set of characteristics for the physical realization of a quantum computer, known as DiVincenzo's criteria~\cite{divincenzo2000physical}, and they will be briefly reviewed here.

\subsection{What is a qubit?}
In contrast to classical computers which use bits, quantum computing merges two great scientific revolutions of the 20th century: computer science and quantum physics, and uses the quantum states of an object to produce what's called ``quantum bit'', namely qubit. A well-characterized qubit should have a distinguished two-level system from higher energy levels.
The two energy levels of a physical qubit are often defined as ground state $|0\rangle$  and excited state $|1\rangle$, which are two orthogonal vectors in a two-dimensional vector space. The energy difference between the two states is referred to as the {\it qubit frequency} by assuming the reduced Planck constant $\hbar\equiv1$. 

Similar to classical computers, physical qubits should have the ability to be scaled up to any numbers.  A logical qubit used for programming, consisting of one or more physical qubits, can be in a superposition, partly in $|0\rangle$ and partly in $|1\rangle$ state. There are hundreds of physical systems which have well-defined two level system, and are easily scalable. These quantum systems can be classified in terms of the qubit characteristics. For instance, qubits can be based on atomic physics such as trapped ions and neutral atoms. Trapped ions are confined and suspended in free space using electromagnetic fields~\cite{harty2014high,ballance2016high,bruzewicz2019trapped}, while neutral atoms are realized in optical or magnetic traps encoded in hyperfine-Zeeman ground substates~\cite{mandel2003controlled,bloch2008quantum,cirac2012goals,ohl2019defect}. 

Other popular qubits are based on solid-state systems that are integrated into a solid material with the nanofabrication techniques. Qubits in solid state systems take advantage of spin states or charge states such as semiconductor qubits~\cite{loss1998quantum,petta2004manipulation,,hanson2007spins,gywat2009spins,muhonen2014storing}, superconducting qubits~\cite{bouchiat1998quantum,nakamura1999coherent,orlando1999flux,mooij1999flux,martinis2002rabi,yu2002coherent} and topological qubits~\cite{nayak2008non,stern2013topological,hyart2013flux,ben2015detecting}. 
Beyond natural and artificial atomic systems, photonics is also a promising platform and can be manipulated to create a quantum processor in semiconductor chips, e.g. Silicon chip~\cite{politi2008silica,peruzzo2010quantum,carolan2015universal,silverstone2016silicon}, making possible photon-photon interactions in a compact chip-integrated device. 

\subsection{How are qubit states reset?}
Registers in classical computers should be initialized to a known value before starting computation. Equivalently in quantum computers qubit states should be initialized usually to ground state. During quantum operations qubits heat up. Therefore in order to be able to run quantum algorithms several times one after the other, one needs a quick cooling method, i.e. resetting. This can be realized by natural cooling, also called passive cooling, with the residual coupling to the environment. Although passive cooling benefits in the energy efficiency and lower financial cost, this protocol is inherently time consuming. The reason is that passive cooling transfers heat from qubit systems to the environment at the speed of relaxation rate, which is always minimized to lower the probability of errors in a coherent quantum computation. Another approach is active cooling which becomes attractive by relying on an external approach to enhance the heat flow. For example,  by applying laser-based Doppler-cooling the effective temperature of trapped ions  was rapidly reduced~\cite{eschner2003laser,warring2013techniques,chepurov2014laser}; driving qubits can induce subsequent relaxation into the ground state for in superconducting circuits~\cite{valenzuela2006microwave,grajcar2008sisyphus,jin2015thermal}. Other promising tools like quantum-circuit refrigerator (QCR)~\cite{tan2017quantum-circuit,sevriuk2019fast,hsu2020tunable} are also demonstrated to be able to speed up qubit initialization.

\subsection{How long does a qubit last?}
\begin{figure}[h!]
	\centering
	\includegraphics[width=1\textwidth]{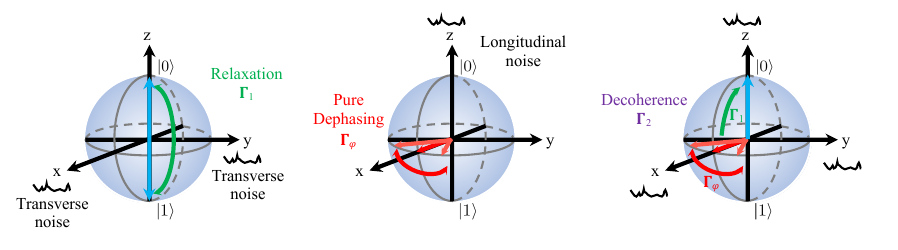}
	\put(-355,110){(a) Longitudinal relaxation}
	\put(-220,110){(b) Pure dephasing}
	\put(-110,110){(c) Transverse relaxation}
	\vspace{-0.2in}
	\caption{(a) Longitudinal relaxation (b) pure dephasing (c) transverse relaxation on the Bloch sphere.}
	\label{fig:bloch}
\end{figure}

Quantum systems cannot live long due to irreversible heat and quasiparticle exchange with environment, which is characterized by {\it coherence time}\index{coherence time}~\cite{burkard2004multilevel,catelani2011quasiparticle,ansari2013effect,bal2015dynamics,ansari2015rate}. Environment surrounding a qubit cannot be directly controlled or fully isolated, therefore the qubit-environment interaction can produce undesirable quantum state changes. Bloch-Redfield theory describes the time evolution of such qubit-environment coupling system with Markovian approximation and shows that decoherence can be represented on the Bloch sphere using longitudinal relaxation, transverse relaxation and pure dephasing\index{dephasing}~\cite{krantz2019quantum},  as shown in Fig.~\ref{fig:bloch}.
\vspace{-0.1in}
Longitudinal relaxation arises from the energy exchange between qubit and environment with transverse noise coupled to the qubit in the $x-y$ plane,  while pure dephasing results from longitudinal noise along the $z$ axis. Combination of these two losses leads to the transverse relaxation, which loses the orientation of the original trajectory. The two relaxations characterize coherence times $T_1$ and $T_2$ as
\begin{eqnarray}
\Gamma_1&\equiv&\frac{1}{T_1},\\
\Gamma_2&\equiv&\frac{1}{T_2}=\frac{\Gamma_1}{2}+\Gamma_{\varphi}.
\end{eqnarray}
$T_1$ is the time required by a qubit to relax from the first excited state to the ground state. $T_2$ is the average time in which the energy-level splitting remains unchanged. To perform high-fidelity quantum computing, coherence times of logical qubits must be longer than the operation time. Among physical qubits, trapped ions can last longer, e.g. up to a few hundred seconds or even a couple of minutes, by suppressing the limiting factors such as critical current noise~\cite{ansari2011noise}, magnetic-field fluctuations~\cite{simmonds2004decoherence}, frequency instability and leakage of the microwave reference-oscillator~\cite{schmidt2003coherence,wang2017single,wang2021single}.  In superconducting qubits  the lifetime $T_1$ in the past ten years for a few microseconds, has been improved to around 100 $\rm{\mu s}$~\cite{dunsworth2017characterization,nersisyan2019manufacturing}, and most recently exceeds 300 $\rm{\mu s}$ by replacing fabricating material niobium with tantalum~\cite{place2021new,wang2021transmon}. 
In semiconductor spin qubits, the relatively short coherence times of 0.1 to 1 $\rm{\mu s}$ ~\cite{nadj2010spin,de2011ultrafast,song2011manipulation}, has been improved by five orders of magnitude~\cite{kobayashi2021engineering}.

\subsection{What is a universal qubit gate set?}
Quantum gates are basic operations on a small number of qubits. They are the building blocks of quantum circuits, like XOR gates for conventional digital counterpart. Unlike many classical logic gates\index{logic gate}, quantum gates are reversible. Such reversible implementations can be expressed using a special class of logic gates, known as universal sets. In order to perform arbitrary unitary operations\index{unitary operator}, it is sufficient to have the ability of performing all members of a universal gate\index{universal quantum gates} set. The Solovay-Kitaev theorem~\cite{kitaev1997quantum} shows that it is possible to define a universal gate set with a discrete number of elements that can approximate any unitary with arbitrary precision, and the following gate sets can be  easily proved universal: the set of controlled NOT (CNOT) gate and all single qubit gates; the set of CNOT gate, Hadamard gate and suitable phase flips; the set of three-qubit Toffoli gate and Hadamard gate.

In reality, most of the physical systems proposed so far consider only two-qubit unitary transformations. Therefore realizing high-performance single qubit gates and CNOT gate is crucially important for quantum algorithms.

Single qubit gates correspond to rotations of a spin about the Cartesian axis, and can be implemented directly with external driving pulses. Usually quantum gates are written as Pauli matrices, e.g. several commonly used quantum gates are 
\begin{equation}
\rm X=\left[\begin{array}{c c}
0&1\\
1&0
\end{array}\right],
\rm Y=\left[\begin{array}{c c}
0&-i\\
i&0
\end{array}\right],
\rm Z=\left[\begin{array}{c c}
1&0\\
0&-1
\end{array}\right],
\rm S=\left[\begin{array}{c c}
1&0\\
0&-i
\end{array}\right],
\end{equation}
which change the amplitudes or phase of the states $|0\rangle$ and $|1\rangle$. Any single qubit operation can be decomposed into three Bloch sphere\index{Bloch sphere} rotations and a global phase factor, so it is easy to perform in experiments. To date single-qubit gate can be realized accurately enough for fault-tolerant quantum computing~\cite{brown2011single,chen2016measuring,yang2019achieving}.

A CNOT gate is the controlled version of the X gate and performs a selective negation of the target qubit as 
\begin{equation}
\rm CNOT=\left[\begin{array}{cccc}
1&0&0&0\\
0&1&0&0\\
0&0&0&1\\
0&0&1&0
\end{array}\right].
\end{equation}
Construction of CNOT gates varies in gate decomposition and qubits of physical properties. For instance, in trapped ions the CNOT gate can be realized by using a linear trap and laser beams~\cite{cirac1995quantum}; in silicon quantum dots, it can be completed by efficiently resonant drive~\cite{zajac2018resonantly}. The challenge to such  a gate is that two-qubit gates are not accurate enough for quantum error correction although these years some progresses have been made~\cite{nichol2017high-fidelity,wang2020high-fidelity,xu2020experimental}. More improvement is needed to suppress error rates such that high-fidelity quantum gates are realized. 

\subsection{Measurement}
Measuring quantum states determines computational results projected on the ``classical'' world. There are three required subsystems for such measurement\index{measurement}: (1) Quantum system whose properties are remaining unknown; (2) Measurement apparatus, behaves as a classical system that can interact with qubits; (3) Environment surrounding the apparatus whose presence supplies the decoherence.  One way to describe quantum measurements is to project the eigenstates of measurable quantities. A Hermitian measurement operator carries eigenvalues corresponding to the outcomes of measurement. For a mixed state, the unconditioned final state of the system is a classical mixture of the output states of all possible measurement outcomes. However, such projective measurements are too restrictive and can change or destroy the characteristics of quantum systems.  

More generalized quantum measurements are so called Positive Operator-Valued Measure (POVM)\index{POVM}, which can be viewed as a collection of positive Hermitian operators. The POVM process is a statistical measurement since each operator corresponds to an outcome of the measurement with a real and non-negative probability. This measurement is not repeatable, meaning that one can find different results when repeating the measurement which is independent of a particular realization. These properties make POVM the most general measurement one can perform on a quantum system. For instance, it can be used to measure phonon fluctuation in the ground state~\cite{hotta2009quantum} and hyperfine qubit levels~\cite{choudhary2013implementation} with trapped ions, perform state tomography of an exchange-only spin qubit~\cite{medford2013self}, entangle massive fermionic qubits~\cite{arvidsson2017protocol} and also study the characteristics of superconducting qubits~\cite{li2016robust,monroe2021weak}. 

\section{Superconducting Quantum Computing}
Superconducting qubits have become one of most promising candidates that satisfy  DiVincenzo's criteria and can implement large-scale quantum computing. This type of qubits is macroscopic in size and lithographically defined with microfabrication technology, making it easy to build and capacitively or inductively coupling to the others. Frequency domain of the superconducting qubits allows that the state of a qubit can be changed by existing commercial microwave devices and equipment. 

Superconducting quantum computing is an implementation of quantum computers based on superconducting circuits. Here an arbitrary single qubit gate is achievable by external microwave drive. Two-qubit gates such as CNOT gates are realized by combining cross-resonance gate~\cite{rigetti2010fully,chow2011simple} or microwave-activated phase (MAP) gate~\cite{noh2018construction} with single qubit rotations. The major milestone was the demonstration of the first quantum supremacy on a 53-qubit chip~\cite{arute2019quantum}. 

However, there are yet outstanding challenges against high-performance quantum computing~\cite{awschalom1992macroscopic,gottesman1998theory,kitaev2003fault-tolerant,raussendorf2007fault-tolerant}. Aside from the limited coherence times\index{coherence time}, the always-on parasitic interaction among qubits does affect the quality of multi-qubit gates. In such circuits, qubits are coupled to one another via shared couplers. Although the nonlinear circuit element Josephson junctions\index{Josephson junction} can approximate superconducting qubits as a two-level system, transitions between computational and noncomputational subspaces inevitably take place. In some cases, this interaction is wanted and should be strengthened to create desired entanglement. But most of time it is unwanted, and 
sets barriers to reach the error correction threshold. More precisely, the parasitic interaction accumulates phase error in the computational states, and eventually destroys the multi-qubit gates. Therefore, it must be carefully suppressed during the gate operations.

Particularly, the parasitic ZZ interaction between a pair of superconducting qubits is a limiting factor for two-qubit gates and quantum error correction. Although the ZZ interaction is always one or two orders of magnitude weaker than the coupling strength, it obviously degrades the performance of many quantum gates~\cite{gambetta2012characterization,caldwell2018parametrically,reagor2018demonstration}. Furthermore, driven two-qubit gates suffer from additional ZZ component produced by external pulses~\cite{chow2013microwave,mundada2019suppression,krinner2020benchmarking}. Therefore zeroing qubits from the undesirable ZZ interaction is highly demanded to boost up the gate quality, and is an important step toward bringing down the overhead of physical qubits. 
Specially, for a few gates such as controlled-Z (CZ) gates, the ZZ interaction is wanted, then we should take advantage of it and make it stronger~\cite{ghosh2013high-fidelity,li2019realisation,sung2020realization}. 

The main goal of this book is to improve two-qubit gate performance by either mitigating unwanted parasitic ZZ interaction, or improving wanted ZZ interaction, whenever the qubits are idle or driven by external pulses. We aim to develop a theory that can describe the characteristics of ZZ interaction in different circuits. Beyond that, we will also generalize our theory of parasitic interaction to a multi-qubit scheme, and find out the difference from the two-body interaction.

\section{Outline}
In chapter \ref{c2} we present generalized formalism of circuit quantization from corresponding Lagrangian and Hamiltonian. In particular, we introduce the basic elements of the superconducting circuit, e.g. the LC resonator and Josephson junctions, and describe several types of qubits in details such as transmon and capacitively shunted flux qubit (CSFQ), as well as their variations. After highlighting the Schrieffer-Wolff transformation, we move to the core part of the Hamiltonian model --- circuit quantum electrodynamics (circuit QED) by taking several examples.

In chapter \ref{c3}, we theoretically analyze the origin of static ZZ interaction using different theory models. We explore the possibility of static ZZ freedom in two types of devices:  CSFQ-transmon pair and transmon-transmon pair coupled via a harmonic bus resonator, and derive the condition of zeroing static ZZ interaction for both pairs. We apply our theory to simulate the experimental hybrid CSFQ-transmon circuit, which is fabricated at IBM and measured at Syracuse university by our collaborators. We calculate qubit and resonator frequencies and extract their coupling strengths as well as static ZZ interaction versus flux threading the CSFQ.

Following the experiment, in chapter \ref{c4} we explain what a cross-resonance (CR) gate is and how it is realized. We study the error source to the gate like classical crosstalk beyond the intrinsic noise, and show how the errors impact the CR gate and what we can do to improve the gate fidelity.  Our theory predicts the prerequisite to achieve high-performance CR gate with static ZZ freedom. Inspired by the experiment, we keep studying the inherent symmetry in a driven system, and demonstrate the presence of dynamical ZZ freedom.

Chapter~\ref{c5} introduces some novel gates based on the study of the tunable coupler. We first propose a parasitic free (PF) gate by cancelling unwanted ZZ interaction and suppressing external flux noise. Beyond that, we take advantage of the wanted ZZ interaction
to perform a CZ gate, and propose a scheme where the CZ gate can be fully turned off by tuning qubit frequency.

Chapter~\ref{c6} presents preliminary results of three-qubit interaction in two setups: triangle geometry and qubit chain. By calculating Pauli coefficients in the three-qubit computational subspace, we study characteristics of two-body ZZ and three-body ZZZ interactions, and take several examples to illustrate them.

\newpage
\thispagestyle{empty}

\chapter{Superconducting Qubits and Circuit QED}
\label{c2}
Superconducting qubits have been proved to be versatile on a macroscopic scale for the study of quantum technology, such as quantum computing~\cite{huang2020superconducting,stassi2020scalable,jurcevic2021demonstration}, quantum simulation~\cite{houck2012chip,paraoanu2014recent,barends2015digital}, quantum sensing~\cite{wolski2020dissipation,lachance2020entanglement,mcdonald2020exponentially}, and quantum information processing~\cite{billangeon2015circuit,wendin2017quantum,blais2020quantum}.
Due to the amazing properties of the nonlinear Josephson junctions\index{Josephson junction} at sufficiently low temperatures, superconducting qubits act as natural atoms with discrete level structures, and exhibit extraordinary properties on integration, readout and tunability. Over the past decades, this type of qubits has been become a leading candidate for realizing building blocks of a quantum computer. 

Generally speaking, there are three classes of superconducting qubits based on the relevant degrees of freedom: charge qubit \cite{bouchiat1998quantum,nakamura1999coherent}, flux qubit \cite{orlando1999flux,mooij1999flux} and phase  qubit\index{phase qubit}~\cite{martinis2002rabi,yu2002coherent}. However these qubits suffer from either charge or flux noise. To reduce the sensitivity to these noises, transmon \cite{koch2007charge-insensitive} and capacitively shunted flux qubit (CSFQ)~\cite{you2007csfq,steffen2010high-coherence} have been designed theoretically and tested experimentally by adding a large capacitance shunted to one Josephson junction, and nowadays widely used in the lab.

 Moreover, superconducting quantum circuits have led to circuit quantum electrodynamics (circuit QED\index{circuit QED}), which provides an approach to study the fundamental light-matter interaction in the solid-state platform. In contrast to cavity quantum electrodynamics (cavity QED\index{cavity QED}), quantized electromagnetic fields in the microwave frequency domain are stored in a on-chip resonator and superconducting qubits behave as quantum objects. 
The artificial atom–field coupling can be easily increased to hundreds of megahertz, and then stepped into the strong coupling regime~\cite{wallraff2004strong}. This results in faster operation since the transition dipole is mainly determined by the qubit geometry. Circuit QED now is a very active research direction as well as a leading architecture for quantum computation. Beyond superconducting qubits, circuit QED can also deal with hybrid quantum systems at its origin, losing their individual identity and combine into something new~\cite{xiang2013hybrid,clerk2020hybrid}.

This chapter starts with introduction of the basic concepts of superconducting qubits and circuit QED, including circuit quantization, Josephson effect and several types of superconducting qubits mainly focusing on transmon and CSFQ. By means of the  perturbative approach, we move to the regime where the qubit-resonator detuning is larger than coupling strength, namely dispersive regime\index{dispersive regime}, and study the models of one qubit coupled to a harmonic oscillator, and two qubits coupled via a bus resonator.

\section{Circuit Quantization}	
\begin{figure}[h!]
	\centering
	\includegraphics[width=0.6\textwidth]{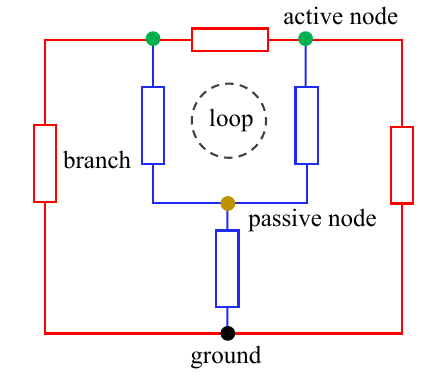}
	\vspace{-0.1in}
	\caption{An example of electrical circuit consists of several loops, same-species elements meet at active nodes while different species elements meet at passive modes.}
	\label{fig:network}
\end{figure}

In this section we briefly introduce how to quantize a non-dissipative electrical circuit, most of the discussion below is from Ref.~\cite{vool2017introduction}. The dimension of a physical circuit is small enough so that electromagnetic waves propagate across the circuit ``almost'' instantaneously, this is so called lumped approximation. Within this consideration, a circuit can be described as a network of elements meeting at nodes, and elements between two nodes construct a branch as shown in Fig.~\ref{fig:network}.
For each branch at time $t$ two independent variables voltage $v(t)$ and current $i(t)$ flow through it. Conventionally absorbed energy by an element $E=\int i(t)v(t)dt$ is assumed to be negative, in other words at $t=-\infty$ both voltage and current are zero. Then the two vectors should have opposite sign. To proceed Hamiltonian description of electrical circuits, we generalize the definition of fluxes and charges which are derived from branch voltages and currents by
\begin{eqnarray}\label{eq.fluxcharge}
&&\Phi(t)=\int_{-\infty}^{t}v(t')dt',\\
&&Q(t)=\int_{-\infty}^{t}i(t')dt'.
\end{eqnarray}

A dispersive element for which the voltage $v(t)$ is only a function of the charge $Q(t)$ and not directly of the time $t$ or any other variables, is said to be a capacitive element. Similarly, a dispersive element for which the current $i(t)$ is only a function of the flux $\Phi(t)$  is said to be an inductive element. For linear capacitance and inductance, the relation is given by $v(t)=Q(t)/C$ and $i(t)=\Phi(t)/L$, and energy stored in these basic elements can also  be calculated as  $Q^2(t)/2C$ and $\Phi^2(t)/2L$, respectively. Before finding the degrees of freedom of an arbitrary conservative circuit, one should take into account the build-in constitutive relations at each node and loop
\begin{eqnarray}\label{eq.kirchhoff}
&&\sum_k i_k(t)=0 \hspace{0.03in}\qquad\quad\qquad {\textrm {The Kirchhoff's current law}},\\
&&\sum_k v_k(t)=0 \qquad\quad\quad\quad {\textrm {The Kirchhoff's voltage law}},
\end{eqnarray} 
which is the famous Kirchhoff's law. The two equations above correspond to two standard methods to write the classical Hamiltonian: the node method and the loop method. In the next, we only show the node method which is applicable for most practical problems.

With the above background, we summarize the following  steps to write a Hamiltonian from the electrical circuit:

1. Separate the circuit into capacitive sub-network and inductive sub-network. Inductances and capacitances meet at {\it active} nodes, while only capacitances or only inductances converge at {\it passive} nodes.

2. Name a particular active node `‘ground'', and choose the set of branches that connect the ground via capacitances to every other active or passive node without forming any loops, this is so called spanning tree method.

3. Classify inductive energy into potential energy $U$ and capacitive energy into kinetic energy $T$, here one can rewrite the kinetic energy in terms of flux by noticing that $v=\dot\phi$ with $\phi$ being the node flux.

4. Write the Lagrangian of the network as a function of fluxes at nodes in the form of $\mathcal{L}(\cdots,\phi_i,{\dot\phi}_i,\cdots)=T-U$, one can check that by applying Lagrange’s equations.

5. Find the generalized momentum (node charge) which is canonical conjugate of node fluxes, defined as $q_i=\partial \mathcal{L}/\partial{\dot\phi}_i$;

6. Express the Hamiltonian in terms of the conjugate pair $q_i$ and $\phi_i$ by performing Legendre transform $H=\sum_iq_i\dot{\phi_i}-\mathcal{L}$.

Replace the variables with corresponding operators by straightforward adding a hat
$\phi\rightarrow\hat\phi$ and $q\rightarrow\hat q$ which satisfies the commutator relation $[\hat\phi,\hat q]=i\hbar$, description of electrical circuit is changed from classical to quantum regime. For simplicity, the hats on operators can be dropped. 

\section{LC Resonator}
\begin{figure}[h!]
	\centering
	\includegraphics[width=0.4\textwidth]{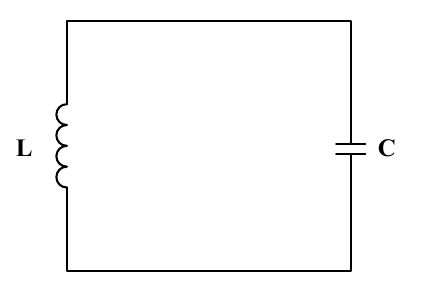}
	\vspace{-0.1in}
	\caption{Isolated LC resonator with kinetic energy $T$ stored in the inductor and potential $U$ stored in the capacitor.}
	\label{fig:LC}
\end{figure}

The transmission line resonator (TLR)~\cite{blais2004cavity} is a narrow one-dimensional device consisting of a full-wave section of superconducting coplanar waveguide. Each mode of the TLR is an independent harmonic oscillator. This 1D transmission line can be expressed by a chain of LC oscillators exhibiting standing wave resonances with less small loss, making it possible to readout and measure a circuit with high fidelity. In the vicinity of fundamental frequencies the line can be used as a LC circuit shown in Fig.~\ref{fig:LC}. 
The LC circuit is an idealized model of zero resistance since it assumes that currents can flow essentially without dissipation. The only one degree of freedom current, flows in the wire of the inductor and accumulates charge on the capacitor. A LC resonator only contains one inductor and one capacitor, so the Lagrangian can be easily obtained following steps described in the circuit quantization 
\begin{equation}
\mathcal{L}=T-U=\frac{1}{2}C{\dot\phi}^2-\frac{\phi^2}{2L},
\end{equation}
with $\phi$ being flux threading the inductor. Conjugate term charge $q$ can be derived from $q=\partial \mathcal{L}/\partial\dot\phi$.
By using Legendre transformation, the Hamiltonian of a LC resonator is written as
\begin{equation}\label{eq.LC}
H=\frac{q^2}{2C}+\frac{\phi^2}{2L}.
\end{equation}
Let us introduce standard creation and annihilation operators $r$ and $r^{\dagger}$ satisfying the commutation relation $[r,r^{\dagger}]=1$, and express them in the form of
 \begin{equation}\label{eq.Lca}
 \begin{split}
 \phi&=\phi_{\rm{ZPF}}(r+r^{\dagger}),\\
 q&=-i q_{\rm{ZPF}}(r-r^{\dagger}),
 \end{split}
 \end{equation} 
 where zero point fluctuation is expressed as $\phi_{\rm{ZPF}}=\sqrt{\hbar Z_0/2}$ and $q_{\rm{ZPF}}=\sqrt{\hbar/2Z_0}$ with $Z_0=\sqrt{L/C}$ being the characteristic impedance. The zero point energy is the lowest possible energy described by Heisenberg uncertainty principle such that $\phi_{\rm ZPF}q_{\rm ZPF}=\hbar/2$. Hamiltonian then is quantized as a harmonic oscillator
\begin{equation}\label{eq.QLC}
H=\hbar\omega_0\left(r^{\dagger}r+\frac{1}{2}\right),
\end{equation} 
whose eigenstates are the Fock states $|n\rangle$, satisfying $r=\sum_n\sqrt{n+1}|n\rangle\langle n+1|$, $r^{\dagger}=\sum_n\sqrt{n+1}|n+1\rangle\langle n|$ and $r^{\dagger}r\left|n\right>=n\left|n\right>$. Qubit frequency is written as $\omega_0=1/\sqrt{LC}$. In the rest of this manuscript, $1/2$ corresponding to zero-point energy is dropped.

\section{Josephson Junction}
\begin{figure}[h!]
	\centering
	\includegraphics[width=0.48\textwidth]{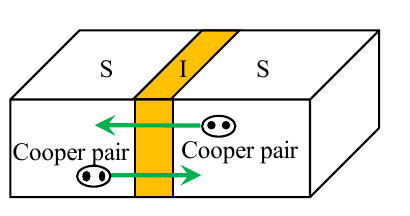}
	\vspace{-0.2in}
	\caption{Sketch of a Josephson junction.}
	\label{fig:JJ}
\end{figure}

A Josephson junction\index{Josephson junction} (JJ) is a quantum mechanical device which consists of two narrow superconductors separated by a thin insulator as shown in Fig.~\ref{fig:JJ}. 
In superconductors a pair of electrons are attractively bound and behave more or less like particles, namely Cooper pairs, and nearly all of the pairs will be locked down at the lowest energy in exactly the same state. However,
if the insulator is thin enough there is a probability for electron pairs to tunnel. This effect later became known as Josephson tunneling.

 Based on Ginzburg–Landau theory, in the superconducting state the wave function $\Psi_L$ is the amplitude to find all electrons on the left side, and $\Psi_R$ is the corresponding function to find it on the right side. By substituting two eigenvectors $\Psi_L=\sqrt{n_L}e^{i\varphi_L}$ and $\Psi_R=\sqrt{n_R}e^{i\varphi_R}$ into the Schr\"odinger equation across the junction, one can show
\begin{eqnarray}
i\hbar\frac{\partial}{\partial t}\sqrt{n_L}e^{i\varphi_L}&=&U_L\sqrt{n_L}e^{i\varphi_L}+K\sqrt{n_R}e^{i\varphi_R},\\  
i\hbar\frac{\partial}{\partial t}\sqrt{n_R}e^{i\varphi_R}&=&U_R\sqrt{n_R}e^{i\varphi_R}+K\sqrt{n_L}e^{i\varphi_L},
\end{eqnarray}
where $\varphi_L$ and $\varphi_R$ are the phases on the two sides of the junction and $n_L$, $n_R$ are the density of electrons at those two points. $K$ is a coupling constant for the wave functions across the barrier. If $K$ were zero, these two equations would describe the lowest energy state with energy $U$ of each superconductor.  The two equations can also be simplified as 
\begin{eqnarray}
\frac{\partial n_L}{\partial t}&=&-\frac{\partial n_R}{\partial t}=\frac{2K}{\hbar}\sqrt{n_L n_R}\sin\varphi,\\
\frac{\partial \varphi}{\partial t}&=&-\frac{U_R-U_L}{\hbar},
\end{eqnarray}
with $\varphi=\varphi_R-\varphi_L$ being the phase difference across the junction. The first equation says what $n_L$ and $n_R$ would be if there were no extra electric forces due to an unbalance between the electron fluid and the background of positive ions. It tells how the densities would start to change, and therefore describe the kind of current that would begin to flow. Let us suppose the electric potential difference across the junction is $V$, then the energy difference between the two superconductors is $U_R-U_L=-2eV$. If the two superconductors are made of the same materials, then charge carrier density becomes universal such that $n_L=n_R=n$. Note that supercurrent $I$ is proportional to the time derivative of charge carrier density $n$. Simplifying the solution of Schr\"odinger equation yields the following relations
\begin{eqnarray}\label{eq.JJ}
I_s&=&I_c\sin\varphi \qquad\qquad \textrm{The first~Josephson~relation,}\\
\frac{\partial\varphi}{\partial t}&=&\frac{2eV}{\hbar} \quad\qquad\qquad  \textrm{The second~Josephson~relation,}
\end{eqnarray}
where $I_s$ is the supercurrent, $I_c$ proportional to $n$ is the critical current.

 Magnetic flux quantum is defined as $\Phi_0=h/2e$, therefore the voltage across the junction can be defined as $V=d\Phi/dt$ which is analogous to Faraday's law with $\Phi=(\Phi_0/2\pi)\varphi$ being the flux. The energy stored in the JJ is
\begin{equation}\label{eq.Jenergy}
U=\int I_s V dt=\int I_c \sin\varphi d\Phi=-E_J\cos\varphi,
\end{equation}
with $E_J=I_c\Phi_0/2\pi$ being called the {\it Josephson energy}. 

\section{Transmons}
\begin{figure}[h!]
	\centering
	\includegraphics[width=0.5\textwidth]{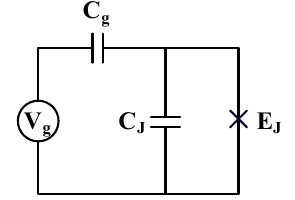}
	\caption{Circuit scheme of the charge qubit with external voltage source $V_g$.}
	\label{fig:charge}
\end{figure}

Connecting the JJ in series with a capacitor and a voltage source constitutes an artificial two-level electronic system: a single Cooper-pair box (CPB) also called charge qubit\index{charge qubit}~\cite{bouchiat1998quantum,nakamura1999coherent} as shown in Fig.~\ref{fig:charge}. 

One can take the node flux and write the circuit Lagrangian as
\begin{equation}
\mathcal{L}=\frac{C_J}{2}\dot\Phi^2+\frac{C_g}{2}(\dot\Phi-V_g)^2+E_J\cos\Phi/\Phi_0,
\end{equation}
with which conjugate variable of the flux is
\begin{equation}
Q=\frac{\partial\mathcal{L}}{\partial \dot\Phi}=(C_J+C_g)\dot\Phi-C_gV_g.
\end{equation}
In this case the Hamiltonian of the system is given by
\begin{equation}\label{eq.HtrQ}
H=\frac{({Q}-C_g V_g)^2}{2C_\Sigma}-E_J\cos{\Phi}/\Phi_0,
\end{equation} 
where $Q=2en$ is the charge with $n$ being the Cooper pair numbers and $C_\Sigma=C_J+C_g$. The Hamiltonian can also be rewritten in terms of dimensionless canonical variables $n=Q/2e$ and $\varphi=\Phi/\Phi_0$ with ${{\varphi}}\in [0,2\pi]$ 
\begin{equation}\label{eq.Htrn}
H=4E_C({n}-n_g)^2-E_J\cos{\varphi},
\end{equation} 
where $E_C=e^2/2C_\Sigma$ is the {\it charging energy} and  $n_g=C_g V_g/2e$ is the normalized gate induced charge. 
\subsection{Fixed-frequency transmons}
\begin{figure}[h]
	\centering
	\includegraphics[width=0.6\textwidth]{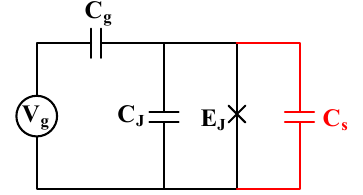}
	\caption{Circuit scheme of the transmon qubit with shunting capacitance $C_s$.}
	\label{fig:transmon}
\end{figure}

The most popular evolution of the charge qubit is transmon in which CPB is shunted by an additional large capacitance as shown in Fig.~\ref{fig:transmon}. 
This Hamiltonian is exactly the same as charge qubit with a new $C_{\Sigma}=C_s+C_J+C_g$. The difference is that large shunting capacitance can dramatically suppress charging energy. Namely  CPB operates in the limit $E_J/E_C\ll1$ whereas the transmon has $E_J/E_C\gg1$. Later we will show how the ratio makes the transmon exponentially less sensitive to charge noise.

To solve the eigenvalues of the Hamiltonian in Eq.~(\ref{eq.Htrn}), Schr\"odinger equation in the phase basis is written as \cite{koch2007charge-insensitive}
\begin{equation}\label{eq.schrn}
\left[4E_C(-i\frac{\partial}{\partial\varphi}-n_g)^2-E_J\cos{\varphi}\right]\psi_k(\varphi)=E_k\psi_k(\varphi),
\end{equation} 
where $\psi_k(\varphi)$ is the eigenstate which satisfies the periodic condition $\psi_k(\varphi)=\psi_k(\varphi+2\pi)$. By multiplying both sides of Eq.~(\ref{eq.schrn}) with $\exp{(-i n_g \varphi)}$ and assuming that $\Psi_k(\varphi)=\exp{(-i n_g \varphi)}\psi_k(\varphi)$, $n_g$ is absorbed into the new eigenstate, the equation above then is simplified to 
\begin{equation}\label{eq.ms}
\left[-4E_C \frac{\partial^2}{\partial\varphi^2}-E_J\cos\varphi\right]\Psi_k(\varphi)=E_k\Psi_k(\varphi).
\end{equation} 
This equation can be simply recast to the standard Mathieu equation, with the following solutions \cite{koch2007charge-insensitive}
\begin{align}\label{eq.mathieu}
E_k(n_g)&=E_C{\mathscr{M}}_A\left(2[m(k,n_g)-n_g],-E_J/2E_C\right),\\
m(k,n_g)&=\sum_{l=\pm1}\left[{\rm{int}}(2n_g+l/2){\rm{mod}}2\right]\left\{{\rm{int}}(n_g)+l(-1)^k[(k+1){\rm{div}}2]\right\},
\end{align} 
where ${\mathscr{M}}_A$ is Mathieu's characteristic value, int$(x)$ rounds to the integer closest to $x$, $a$ mod $b$ denotes the usual modulo operation, and $a$ div $b$ gives the integer quotient of $a$ and $b$. Note that eigenvalues only depend on the normalized gate charge $n_g$ and the ratio of $E_J$ and $E_C$.

\begin{figure}
	\centering
	\includegraphics[width=0.46\textwidth]{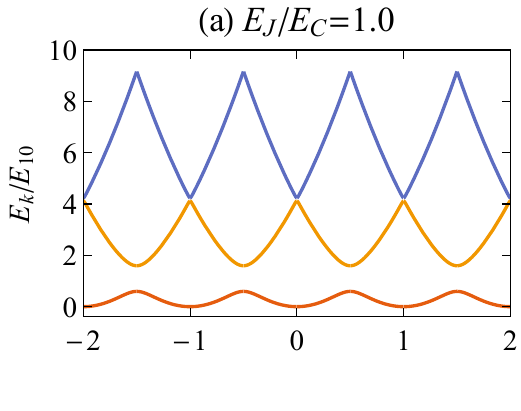}\vspace{-0.15in}\hspace{-0.1in}
	\includegraphics[width=0.46\textwidth]{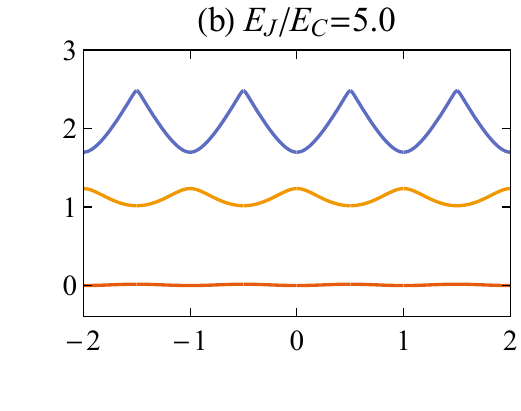}\\
	
	\includegraphics[width=0.46\textwidth]{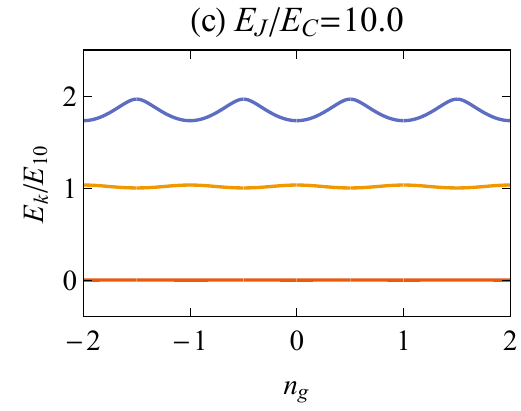}\hspace{-0.12in}
	\includegraphics[width=0.46\textwidth]{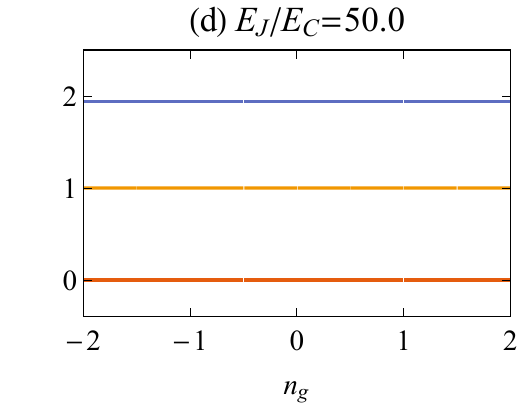}\vspace{-0.15in}
	\caption{The first three eigenvalues of the Hamiltonian Eq.~(\ref{eq.Htrn}) versus the offset $n_g$, at different $E_J/E_C$. Here $E_{10}$ is evaluated at the degeneracy point $n_g=0.5$.}
	\label{fig:eigvratio}
\end{figure}

The lowest three eigenvalues as a function of $n_g$ for several $E_J /E_C$ are shown in Fig.~\ref{fig:eigvratio}. One can see the anharmonic levels rely on the ratio of $E_J/E_C$ with anharmonicity defined by $\delta=E_{12}-E_{01}$.  Increasing $E_J /E_C$ definitely suppresses the sensitivity of the system with respect to offset charge $n_g$, finally resulting in a fixed frequency transmon as shown in Fig.~\ref{fig:eigvratio}(d). Large ratio and sufficient anharmonicity makes a transmon operationally a suitable qubit, since levels can be individually addressed with different frequencies. 

\begin{figure}
	\centering
	\includegraphics[width=0.48\textwidth]{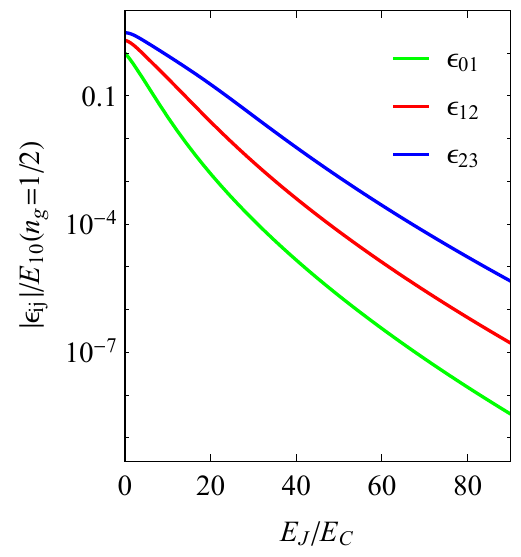}
	\includegraphics[width=0.48\textwidth]{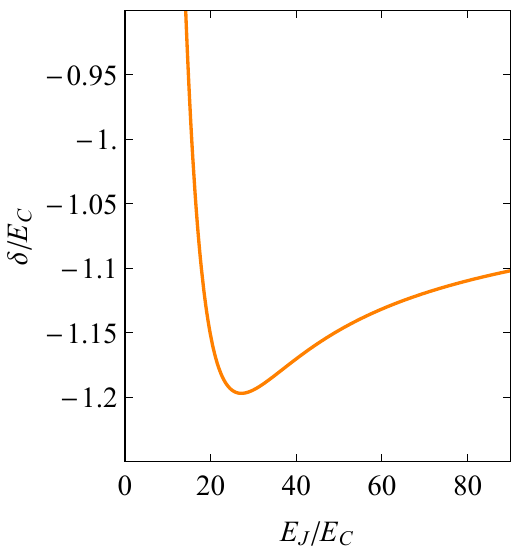}
	\put(-340,175){(a)}
	\put(-165,175){(b)}
	\vspace{-0.1in}
	\caption{(a) Charge dispersion (b) relative anharmonicity as a function of  $E_J/E_C$ at $n_g=1/2$.}
	\label{fig:anharmdispersion}
\end{figure}

In order to further quantify the dependency of energy levels on the ratio, we define the peak-to-peak charge dispersion with $\epsilon_{ij}=E_{ij}(n_g=1/2)-E_{ij}(n_g=0)$. Both anharmonicity and charge dispersion are plotted in Fig.~\ref{fig:anharmdispersion}. It shows that energy levels become less sensitive to the offset charge $n_g$ as $E_J/E_C$ increases, which is consistent with Fig.~\ref{fig:eigvratio}. In fig.~\ref{fig:anharmdispersion}(b) the relative anharmonicity $\delta/E_C$ is negative and can be approximated to $\delta\approx-E_C$, a result that we will derive via other approaches in the next section.

\subsection{Flux tunable transmons}
\begin{figure}[h!]
	\centering
	\includegraphics[width=0.75\textwidth]{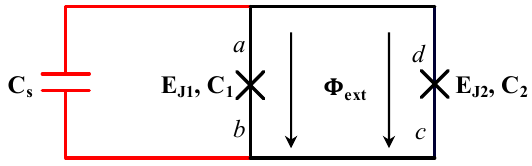}
	\vspace{-0.1in}
	\caption{Circuit scheme of the asymmetric transmon qubit.}
	\label{fig:astransmon}
\end{figure}

Due to charge-insensitive property, transmon has been employed as a central component of several scalable platforms~\cite{barends2013coherent,versluis2017scalable,ji2019scalable}, with applications to a wide range of quantum computation and quantum information processing. The transmon discussed in Fig.~\ref{fig:transmon} was frequency fixed, here we introduce another type of transmon whose frequency is tunable \cite{koch2007charge-insensitive,hutchings2017tunable}. The circuit scheme of the asymmetric transmon\index{asymmetric transmon} is shown in Fig.~\ref{fig:astransmon}. The JJ is replaced by the DC superconducting quantum-interference device (SQUID\index{SQUID}) consisting of two JJ's threading by external flux $\Phi_{\rm ext}$, with which the frequency is tuned within a specific domain. 

Circuit Hamiltonian with more than one junction cannot be directly recast to the Mathieu equation. Therefore it is useful to learn how to approximately simplify it. 
In phase basis, potential of the asymmetric transmon is written as $U=-E_{J1}\cos\varphi_1-E_{J2}\cos\varphi_2$ with $\varphi_1$ and $\varphi_2$ being the phases across two junctions. The change in macroscopic phase around a closed loop must be zero modulo $2\pi$, so we have 
\begin{equation}
\oint\nabla \varphi \cdot d\vec{l}=2\pi n=(\varphi_a-\varphi_b)+(\varphi_b-\varphi_c)+(\varphi_c-\varphi_d)+(\varphi_d-\varphi_a).
\end{equation}
From the definition of the gauge invariant phase, the phase difference is given by 
\begin{eqnarray}
\varphi_a-\varphi_b&=&\varphi_1+\frac{2\pi}{\Phi_0}\int_a^b \vec{A}\cdot d\vec{l},\\
\varphi_c-\varphi_d&=&-\varphi_2+\frac{2\pi}{\Phi_0}\int_c^d \vec{A}\cdot d\vec{l},
\end{eqnarray}
with $\vec{A}$ being the magnetic
vector potential. In superconductors the supercurrent equation gives
\begin{eqnarray}
\varphi_b-\varphi_c&=&\Lambda\int_b^c\vec{J}\cdot d\vec{l}+\frac{2\pi}{\Phi_0}\int_b^c \vec{A}\cdot d\vec{l},\\
\varphi_d-\varphi_a&=&\Lambda\int_d^a\vec{J}\cdot d\vec{l}+\frac{2\pi}{\Phi_0}\int_d^a \vec{A}\cdot d\vec{l}.
\end{eqnarray}
where $\vec{J}$ is the supercurrent density and $\Lambda$ depends on the number of Cooper pairs. Often the contour can be deep within the superconductor where $J = 0$, in this case flux quantization condition is
\begin{equation}
\varphi_1-\varphi_2=2\pi n+\frac{2\pi}{\Phi_0}\oint\vec{A}\cdot d\vec{l}=2\pi n+2\pi\frac{\Phi_{\rm ext}}{\Phi_0}.
\end{equation}
This equation reduces to $\varphi_1-\varphi_2=2\pi f$ at $n=0$ with $f=\Phi_{\rm{ext}}/\Phi_0$. 

The potential of the asymmetric transmon can be rewritten by defining $\varphi_p=(\varphi_1+\varphi_2)/2$ and $a=E_{J1}/E_{J2}$ as $U=-E_J(f)\cos\varphi$, here
\begin{eqnarray}\label{eq.asEJ}
\varphi&=&\varphi_p+\arctan[d\tan(\pi f)],\\
E_J(f)&=&E_{J\Sigma}\sqrt{\cos^2(\pi f)+d^2\sin^2(\pi f)},
\end{eqnarray}
with $E_{J\Sigma}=E_{J1}+E_{J2}$ and $d=(a-1)/(a+1)$, indicating that the effective Josephson energy can be tuned by threading a magnetic flux through the loop. Therefore the Hamiltonian of the asymmetric transmon is similar to Eq.~(\ref{eq.Htrn}).

Usually the asymmetric transmon have large enough $E_J/E_C$ such that $n_g$ can be safely dropped as shown in Fig.~\ref{fig:eigvratio}.
In this case the variance of variable $\varphi$ is small, so it is safe to expand the potential of a transmon up to fourth order in $\varphi$ as
\begin{equation}
H\approx 4E_Cn^2-E_J\left(1-\frac{1}{2}\varphi^2+\frac{1}{24}\varphi^4\right).
\end{equation}
Similar to LC resonator, a transmon canonical operators are defined as
\begin{equation}\label{eq.qac}
\begin{split}
{\varphi}&=\phiz (a +a^{\dagger}),\\
n&= i \nz (a -a^{\dagger}),
\end{split}
 \end{equation}
   with the zero-point fluctuations $\phiz$ and $\nz$, satisfying the minimum uncertainty $\phiz  \nz=1/2$. The fluctuations can be determined by the characteristic impedance $Z_t=\sqrt{E_C/E_J}$,  i.e. $\phiz = \sqrt{\hbar Z_t/2 }$. A transmon’s small anharmonicity allows to approximate the periodic potential energy in the vicinity of minimum with a 4th degree polynomial, making a Duffing oscillator ($\hbar\equiv1$)
\begin{equation}\label{eq.Hse}
H=\omega_0 a^{\dagger}a+\frac{\delta}{12}(a^{\dagger}+a)^4,
\end{equation}
with $\omega_0=\sqrt{8E_CE_J}$ being the {\it plasma frequency} and $\delta=-E_C$ being anharmonicity. Higher order corrections are  known in Ref.~\cite{didier2018analytical}. For $\omega\gg|\delta|$ the rotating wave approximation (RWA)\index{rotating wave approximation (RWA)} is valid, only excitation number conservation terms are kept since the other terms rapidly average out and then can be neglected. Then transmon Hamiltonian reduces to 
\begin{equation}
H_{\rm tr}=\omega a^{\dagger}a+\frac{\delta}{2}a^{\dagger}a(a^{\dagger}a-1),
\label{eq.Htrf}
\end{equation}
with $\omega=\omega_0+\delta$.
\section{Flux Qubits}
\begin{figure}[h!]
 	\centering
 	\includegraphics[width=0.42\textwidth]{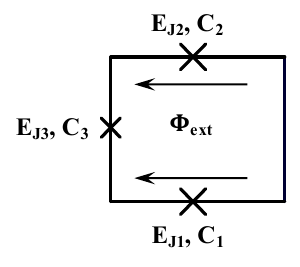}\vspace{-0.2in}
 	\vspace{-0.05in}
 	\caption{\label{fig:fluxqubit}Circuit scheme of three-junction flux qubit.}
 \end{figure}

Another type of superconducting qubit is the persistent current qubit, namely flux qubit~\cite{orlando1999flux,mooij1999flux}. The circuit scheme is depicted in Fig.~\ref{fig:fluxqubit}. 

Two in-series identical JJ's with $E_{J1}=E_{J2}=E_J$ and $C_1=C_2=C_J$ are in a closed loop with a third JJ that has critical current $\alpha  I_0$, capacitance  $C_3=\alpha C_J$  and Josephson energy $E_{J3}=\alpha E_J$ with $\alpha<1$.  In this qubit two supercurrents circulate with opposite directions. This type of qubit is insensitive to background charges and the states can be manipulated with magnetic fields. Based on the analysis of JJ, the potential of a flux qubit can be written as
 \begin{equation} U=-E_{J}(\cos\varphi_1-\cos\varphi_2+\alpha\cos\varphi_3),
 \end{equation}
with $\varphi_i$ being the phase across junction $i$ $(i=1, 2, 3)$. Similarly, integrating $\nabla \varphi$ around the loop gives the flux quantization condition $\varphi_1-\varphi_2+\varphi_3=2\pi f$. 

\begin{figure}[t!]
 	\centering
 	\includegraphics[width=0.49\textwidth]{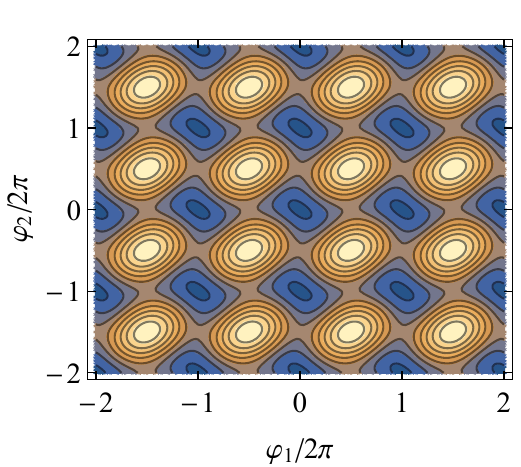}
 	\includegraphics[width=0.49\textwidth]{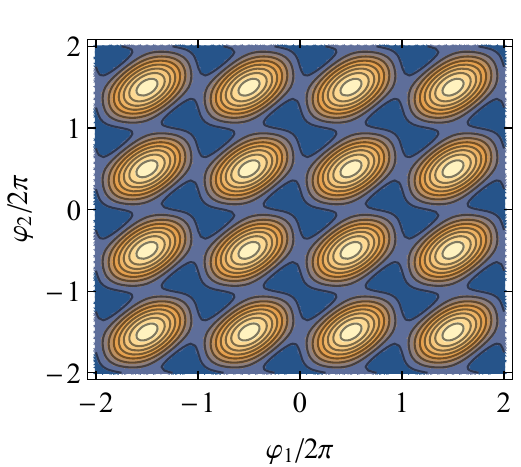}
 		\put(-350,145){(a)}
 	\put(-175,145){(b)}
 	\vspace{-0.1in}
 	\caption{Contour plot of potential for $f=0.5$ at (a) $\alpha=0.4$ and (b) $\alpha=0.8$. The nested nearly circular shapes mark the maxima in the potential, the near square and  hourglass-shaped contours enclose one and two minima, respectively.}
 	\label{fig:fluxqubitp}
 \end{figure}

To investigate how the flux influences the Josephson energy, we can take the flux derivative and get
\begin{equation}
\frac{\partial U}{\partial f}=-2\pi E_J \sin(\varphi_1-\varphi_2-2\pi f).
\end{equation}
One can see flux-noise-insensitive configurations take place at $f=n/2$ with $n$ being an integer.  These points are referred to as “sweet spots” (SS)\index{sweet spot (SS)}. At the first nontrivial SS $f=1/2$, we plot the potential for different $\alpha$ in Fig.~\ref{fig:fluxqubitp}. Qubit states are located inside the single and double well potential.

To find the minimum points of the potential, one can calculate  stable solutions of the following system of equations:
\begin{equation}
\begin{split}
\frac{\partial U}{\partial \varphi_1}&=\sin\varphi_1+\alpha\sin(2\pi f+\varphi_1-\varphi_2)=0,\\
\frac{\partial U}{\partial \varphi_2}&=\sin\varphi_2-\alpha\sin(2\pi f+\varphi_1-\varphi_2)=0.
\end{split}
\end{equation}
The solutions $(\varphi_1^*,\varphi_2^*)$ is given by combining the two equations
\begin{eqnarray}
\sin\varphi_1^*&=&-\sin\varphi_2^*=\sin\varphi^*,\\
\sin\varphi^*&=&-\alpha\sin(2\pi f+2\varphi^*).
\end{eqnarray}
At sweet spot when $\alpha<1/2$ there is only one minimum of the potential at $(0,0)$, while $\alpha\geq1/2$ there are two pairs of solutions $(\varphi^*,-\varphi^*)$ and $(-\varphi^*,\varphi^*)$, this is consistent with the trend of potential as shown in Fig.~\ref{fig:fluxqubitp}. For $\alpha>1/2$, the two minima in the eight-shaped unit cell of the double well can tunnel across the barrier between them as it is lower than that from the minimum in one unit cell to the neighboring unit cell. This makes energy levels anti-cross at sweet spot, which is not suitable to be considered as a qubit. More details can be found in Ref.~\cite{orlando1999flux}. 
\subsection{CSFQ}
\label{sec:csfq}
\begin{figure}[h!]
	\centering
	\includegraphics[width=0.6\textwidth]{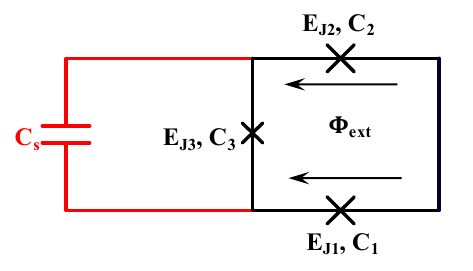}\vspace{-0.2in}
	\caption{Circuit scheme of CSFQ with shunting capacitance $C_s$.}
	\label{fig:CSFQ}
\end{figure}

In the last decade, a variation of flux qubit capacitively shunted flux qubit (CSFQ) has been developed by shunting a capacitance to the smaller JJ, and beneficial for the improvement in the coherence time\index{coherence time} to the microsecond magnitude \cite{you2007csfq,steffen2010high-coherence}. The circuit scheme is shown in Fig.~\ref{fig:CSFQ}.

\begin{figure}
	\centering
	\includegraphics[width=0.65\textwidth]{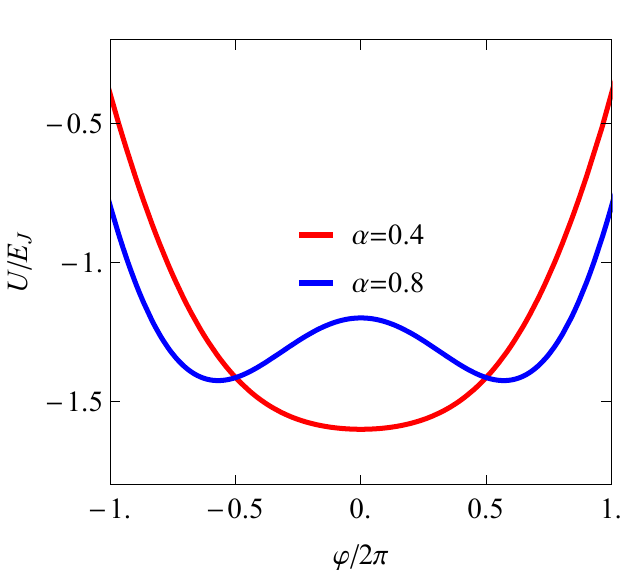}
	\caption{Potential of CSFQ as a function of phase at $\alpha=0.4$ and $\alpha=0.8$}
	\label{fig:fpotential}
	\end{figure}

Similar to the asymmetric transmon, we can reduce the degrees of freedom by defining $ \varphi_m=(\varphi_1-\varphi_2)/2$ and $ \varphi_p=(\varphi_1+\varphi_2)/2$, then Lagrangian of a CSFQ can be written as
 \begin{equation}
 \mathcal{L}=\frac{1}{2}\left(\frac{\Phi_0}{2\pi} \right)^2 C_J\left[\left( 2+4\beta\right)\dot\varphi_m^2+ 2\dot\varphi_p^2\right]+2E_J\cos\varphi_m\cos\varphi_p+\alpha E_J\cos(2\pi f-2\varphi_m),
 \end{equation}
where $\beta=\alpha+C_s/C_J$ with $C_s\gg C_J$.
Hamiltonian can be derived following Legendre transformation
\begin{equation}
\begin{split}
	H&=\sum_{i=m,p}\frac{\partial \mathcal{L}}{\partial \dot\varphi_i}\dot\varphi_i-\mathcal{L}\\&=4E_{Cm}n_m^2+4E_{Cp}n_p^2-2E_J\cos\varphi_m\cos\varphi_p-\alpha E_J\cos(2\pi f-2\varphi_m),
\end{split}
\end{equation}
where $E_{Cm}=e^2/2C_J(1+2\beta)$ and $E_{Cp}=e^2/2C_J$. The larger shunting capacitance permits the $\varphi_p$ direction to behave as a fast free oscillation and can be ignored, also confirmed using the Born-Oppenheimer analysis~\cite{divincenzo2006decoherence}. For simplicity we drop the subscript $m$ from variables, and the Hamiltonian reduces to \begin{equation}
H=4E_{C}n^2-2E_J\cos\varphi-\alpha E_J\cos(2\pi f-2\varphi),
\end{equation}
with potential reduced to $U=-2E_J\cos\varphi-\alpha E_J\cos(2\pi f-2\varphi)$. Similar to the three-junction flux qubit, corresponding minimum phase satisfies the relation $\sin\varphi_{\rm min}=-\alpha\sin2\varphi_{\rm min}$. At $\alpha>1/2$ the potential has two minima while at $\alpha\leq1/2$ it conveys one minimum. Due to the only one degree of freedom, the potential at SS is one-dimensional and plotted in Fig~\ref{fig:fpotential} for different $\alpha$.
 
In the following we only study the case that $\alpha<1/2$ in which the potential can be used as a qubit because each period conveys one minimum, not more. In this case energy barrier from the minimum in one unit cell to the neighboring unit cell is large enough to guarantee a discrete energy levels. If the external flux $f=1/2+\delta f$ is away from SS with $\delta f \ll 1/2-\alpha$, the minimum point, denoted by $\varphi_0$, can be found by setting the derivative of potential to zero. Since $\varphi_0$ and $\delta f$ are very small, we assume $\sin\varphi_0\approx\varphi_0$ and $\sin(2\pi\delta f-2\varphi_0)\approx\sin\left(2\pi\delta f\right)-2\cos\left(2\pi\delta f\right)\varphi_0$, these assumptions yield
\begin{equation}
\varphi_0=-\frac{\alpha\sin\left(2\pi\delta f\right)}{1-2\alpha\cos\left(2\pi\delta f\right)}\approx -2\pi \delta f\frac{\alpha}{1-2\alpha}.\\
\end{equation}

Similarly, Hamiltonian of CSFQ can also be rewritten as Eq.~(\ref{eq.Htrf}) in the vicinity of minimum. This potential has the simple form of a Duffing oscillator at SS, i.e. $\delta f=0$,  with the periodic frequency and anharmonicity
\begin{eqnarray}
&&\omega= \sqrt{8 E_J E_C (1/2-\alpha)}+\delta,\\
&&\delta=E_C (8\alpha-1)/(2-4\alpha),
\end{eqnarray}
In contrast to a transmon,  a CSFQ with  $1/8<\alpha<1/2$ has positive anharmonicity. When $f$ is away from SS the less symmetric potential can be approximated to 
\begin{equation}
U(f)=\left(1 -2 \alpha \right) E_{J}    \Delta \varphi ^2  + \left( 8\alpha -1 \right)  E_{J} \frac{\Delta \varphi ^4}{4!}   +  2 \pi (\delta f) \alpha E_{J}\left( 2\Delta \varphi  -\frac{ (2\Delta\varphi )^3}{3!}\right)+\cdots 
\end{equation}
with $\Delta \varphi=\varphi-\varphi_0$. However higher order terms cannot be simply ignored due to nonzero flux shift. More accurate calculation can be found in Appendix~\ref{app:CSFQ}.

\section{The Schrieffer-Wolff Transformation}
An open question in dealing with quantum many-body systems is how to decouple lower-energy dynamics from higher-energy degrees of freedom. The Schrieffer-Wolff (SW) transformation\index{Schrieffer-Wolff (SW) transformation} has been proved to effective and consistent
to address this problem~\cite{winkler2003,bravyi2011schriefferwolff}. In mathematical language, SW transformation is unitary operation\index{unitary operator} which can remove generalized off-diagonal terms and hence serves as a way of diagonalization to determine effective qubit Hamiltonian. Before stepping into the section of circuit QED, we briefly introduce SW transformation which can block or fully diagonalize a Hamiltonian in the dispersive regime.

In general, qubit-resonator interacting Hamiltonian can be decomposed into
\begin{equation}
H=H_0+H'=H_0+\epsilon V,
\end{equation}
with $H_0$ being the unperturbed component and $H'$ being the perturbation part. Unperturbed $H_0$ can be divided into two subspace A and B with assumption that A is separated from B of the spectrum by a gap $\Delta$. Let us consider sufficiently small $\epsilon$ so that weak perturbations are not strong enough to cover the gap, in other words the following limitation should be satisfied 
\begin{equation}
\epsilon\leq\frac{\Delta}{2||V||}.
\end{equation}

By translating the Hamiltonian into operator language, $H_0$ is diagonal while $H'=H_1+H_2$ is off-diagonal with $H_1$ being block diagonal and $H_2$ being block off-diagonal. The unitary operator $U=e^{-S}$ with anti-Hermitian $S$  is defined to diagonal or block-diagonal the Hamiltonian, therefore the choice of $S$ is critical to this transformation.
Effective Hamiltonian\index{effective Hamiltonian} is given by
\begin{equation}
H_{\rm eff}=e^{-S}H e^{S}=\sum_{j=0}^{\infty}\frac{1}{j!}[H,S]^{(j)},
\end{equation}
here Campbell-Baker-Hausdorff formula is applied and $[H,S]^{(j)}=[[H,S]^{(j-1)},S]$.  The commutation relation between block diagonal and block off-diagonal operators gives rise to a block off-diagonal matrix and the commutator of two block off-diagonal results in a block diagonal matrix, therefore the following relation is obtained 
\begin{equation}
H_{\rm eff}^{\rm off}=\sum_{m=0}^{\infty}\frac{1}{(2m+1)!}[H_0+H_1,S]^{(2m+1)}+\sum_{m=0}^{\infty}\frac{1}{2m!}[H_2,S]^{(2m)}{\overset{!}{=}}0.
\end{equation}
One can expand $S$ in terms of $\epsilon$ as 
\begin{equation}
S=\sum_{j=0}^{\infty}\epsilon^j S_j.
\end{equation}
Since $H_1$ and  $H_2$ are of the first order in $\epsilon$, then Hermitian operator $S$ satisfies
\begin{eqnarray}
&&[H_0,S_1]=-H_2\label{eq.h0s1},\\ 
&&[H_0,S_2]=-[H_1,S_1],\\
&&[H_0,S_3]=-[H_1,S_2]-\frac{1}{3}[[H_2,S_1],S_1].
\end{eqnarray}
Effective Hamiltonian is expressed as 
\begin{equation}
H_{\rm eff}=\sum_{j=0}^{\infty}\epsilon^j (f_j(H_0,\{S_n\}_{n=1}^j )+f_{j-1}(H',\{S_n\}_{n=1}^{j-1})),
\end{equation}
where $f_j(A,\{S_n\}_{n=1}^j)=\sum_n^j i^n/n!{\mathcal F}(A,\{S_n\}_{n=1}^j,j,n)$ with ${\mathcal F}$
being the function that takes all n-fold commutation relations between $A$ and $S$. For instance, we list the several orders of expansion as
\begin{equation}
\begin{split}
f_0(A,S_0)&=A,\\
f_1(A,S_1)&=[A,S_1],\\
f_2(A,S_1,S_2)&=[A,S_2]+\frac{1}{2}[[A,S_1],S_1],\\
f_3(A,S_1,S_2,S_3)&=[A,S_3]+\frac{1}{2}([[A,S_1],S_2]+[[A,S_2],S_1])+\frac{1}{3!}[[[A,S_1],S_1],S_1].\\
\end{split}
\end{equation}
Up to second order the effective Hamiltonian reduces to
\begin{equation}
\begin{split}
H_{\rm eff}&=H_0+[H_0,S_1]+[H_0,S_2]+\frac{1}{2}[[H_0,S_1],S_1]+H'+[H',S_1]\\
&=H_0+H_1+\frac{1}{2}[H_2,S_1].
\end{split}
\end{equation}
Next we take two examples to present how to utilize the SW transformation in the form of matrix calculation~\cite{magesan2020effective}. The first one is to diagonalize a Hamiltonian. Since $H_0$ is unperturbed and diagonal, one can separate the effective Hamiltonian into $H_{\rm}=H_0+H^{(m)}$ with higher order corrections 
\begin{equation}
H^{(m)}=\left[H_0,S^{(m)}\right]+H_x^{(m)},
\end{equation}
where $H_x^{(m)}$ includes all other commutations. The diagonalized effective Hamiltonian requires non-diagonal elements must be zero, namely
\begin{equation}
\begin{split}
\left<p\right|H^{(m)}\left|q\right>=0
&=\left<p\right|H_0S^{(m)}-S^{(m)}H_0\left|q\right>+\left<p\right|H_x^{(m)}\left|q\right>,\\
&=(E_p^{(0)}-E_q^{(0)})\left<p\right|S^{(m)}\left|q\right>+\left<p\right|H_x^{(m)}\left|q\right>,
\end{split}
\end{equation}
with $p\neq q$ and $H_0\left|p\right>=E_p^{(0)}\left|p\right>$. $S^{(m)}$ then is solved as
\begin{equation}
\left<p\right|S^{(m)}\left|q\right>=-\frac{\left<p\right|H_x^{(m)}\left|q\right>}{E_p^{(0)}-E_q^{(0)}}.
\end{equation}
Each order of $S$ can be calculated in this way and then higher order corrections are obtained.

The second example is to block diagonalize a Hamiltonian, similar approach is applied. Considering $H^{(0)}=H_0$ is block diagonal, higher order terms can be expressed as~\cite{magesan2020effective} 
\begin{equation}
H^{(m)}=H_1^{(m)}\oplus\cdots \oplus H_k^{(m)}\cdots,
\end{equation}
with $\oplus$ being the direct matrix sum. Since $S$ is block off-diagonal, similarly $S_{j,k}^{(m)}$ is defined as 
\begin{equation}
\left<p\right|S_{j,k}^{(m)}\left|q\right>=-\frac{\left<p\right|H_{x_{j,k}}^{(m)}\left|q\right>}{\left<p\right|H_j^{(0)}\left|p\right>-\left<q\right|H_k^{(0)}\left|q\right>}.
\end{equation}

An alternative approach is black box quantization in which the circuit is more complex and might even be a ‘black box’ whose properties are unknown. One can first consider the black box as a linearized problem with respect to the coupling between qubit and resonators, and then account for the weak anharmonicity of the Josephson effect. This method can determine the qubit parameters numerically much easier and more accurately compared to perturbation theory \cite{ansari2019superconducting}.

\section{Circuit QED}
Now let us discuss the circuit QED. Analog to cavity QED which describes the interaction between light and matter at the quantum level, Circuit QED can describe the interaction between superconducting circuits behaving as artificial atoms and resonators. This section will build up the general model of qubit-resonator interaction based on circuit quantization, and revisit some ideas borrowed from cavity QED.
\subsection{Single transmon coupled to a resonator}
\begin{figure}[h!]
	\centering
\includegraphics[width=0.7\textwidth]{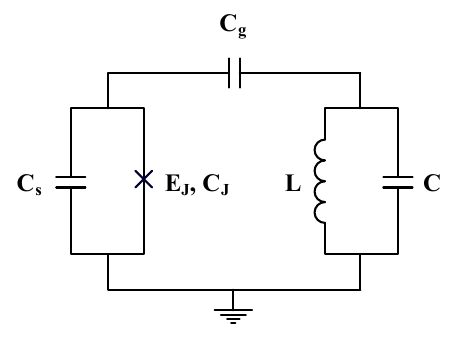}
\vspace{-0.2in}
\caption{One transmon couple to one resonator.}
\label{fig:cqed1}
	\end{figure}

First we consider  the interaction between a harmonic oscillator and one transmon as shown in Fig.~\ref{fig:cqed1}.
Compared to Fig.~\ref{fig:transmon}, the voltage source is replaced by a resonator. Lagrangian of the circuit is written as 
\begin{equation}
 \mathcal{L}=\left(\frac{\Phi_0}{2\pi} \right)^2\frac{1}{2}\left[C\dot\varphi_r^2+(C_s+C_J)\dot\varphi^2+ C_g(\dot\varphi_r-\dot\varphi)^2\right]-E_J\cos\varphi-\frac{\varphi_r^2}{2L},
\end{equation}
with $\varphi$ and $\varphi_r$ being the phase difference across the junction and the harmonic resonator, respectively. In the case that $C\gg C_g$, the corresponding conjugate terms $n$ and $n_r$ can be approximated as
\begin{equation}
\left(\begin{array}{c}
\dot\varphi \\ \dot\varphi_r
\end{array}\right)=\left(\begin{array}{cc}
\frac{1}{C}&\frac{C_g}{CC_{\Sigma}}\\\frac{C_g}{CC_{\Sigma}}&\frac{1}{C_{\Sigma}}\\
\end{array}\right)\left(\begin{array}{c}
n\\n_r
\end{array}\right).
\end{equation}
with $C_{\Sigma}=C_s+C_J+C_g$. Legendre transformation gives the Hamiltonian
\begin{equation}\label{eq.Hcqed1}
H=4E_rn_r^2+4E_Cn^2+\frac{C_g}{CC_{\Sigma}}nn_r-E_J\cos\varphi+\frac{\varphi_r^2}{2L},
\end{equation} 
where $E_C=e^2/2C_{\Sigma}$ and $E_r=e^2/2C$. By expanding Josephson energy to the 4th order and utilizing the creation and annihilation operators defined in Eq.~(\ref{eq.Lca}) and Eq.~(\ref{eq.qac}), the Hamiltonian can be quantized as
\begin{equation}\label{eq.Hcqedq}
H=\omega_r r^{\dagger}r+\omega_q a^{\dagger}a+\frac{\delta_q}{2}a^{\dagger}a(a^{\dagger}a-1)-g(a-a^{\dagger})(r-r^{\dagger}),
\end{equation}
where $\omega_r=1/\sqrt{LC}$ is the resonator frequency, $\omega_q\approx\sqrt{8E_CE_J}+\delta$ is the qubit frequency with anharmonicity $\delta\approx-E_C$ and coupling strength is 
\begin{equation} g=\frac{1}{2}\frac{C_g}{\sqrt{CC_\Sigma}}\sqrt{\omega_q\omega_r}
\label{eq.g}
\end{equation}
Fast oscillation terms, also called counter-rotating terms can be ignored based on the rotating-wave approximation (RWA), then the Hamiltonian simplifies to
\begin{equation}
H=\omega_r r^{\dagger}r+\omega_q a^{\dagger}a+\frac{\delta_q}{2}a^{\dagger}a(a^{\dagger}a-1)+g(ar^{\dagger}+a^{\dagger}r).
\label{eq.gJC}
\end{equation}
By restricting the transmon to the first two levels and replacing creation and annihilation operators with Pauli matrix such that  $a^\dagger\rightarrow\sigma^+=\left|e\right>\left<g\right|$ and $a\rightarrow\sigma^-=\left|g\right>\left<e\right|$, the famous Jaynes-Cummings Hamiltonian\index{Jaynes-Cummings Hamiltonian} is obtained
\begin{equation}
H=\omega_r r^{\dagger}r+\frac{\omega_q} {2}\sigma^z+g(\sigma^-r^{\dagger}+\sigma^+r),
\label{eq.JC}
\end{equation}
with the convention $\sigma^z=\left|e\right>\left<e\right|-\left|g\right>\left<g\right|$. 

Previously, we perturbatively expand the cosine potential to fourth order to quantize the qubit Hamiltonian. However, to proceed accurate calculation, multilevel form of transmon Hamiltonian is needed,  Eq.~(\ref{eq.gJC}) is thus rewritten as 
\begin{equation}
H=\omega_r r^{\dagger}r+\sum_j\omega_j \left|j\right>\left<j\right|+\sum_j \sqrt{j+1}g_j\left(r^{\dagger}\left|j\right>\left<j+1\right|+r\left|j+1\right>\left<j\right|\right),
\label{eq.Hfull}
\end{equation}
where $\left|j\right>$ is the bare eigenvector of the transmon and $g_j$ is the coupling strength, and $\omega_j$ is the corresponding bare energy by assuming $\hbar\equiv1$. With respect to the dispersive regime  $|g_j/(\omega_r-\omega_j)|\ll1$, the Hamiltonian can be full-diagonalized using SW transformation. Given that the first order in $S$ has the form of 
\begin{equation}
S_1=\alpha r^{\dagger} \sum_j\sqrt{j+1}\left|j\right>\left<j+1\right|-\alpha^* r\sum_j\sqrt{j+1}\left|j+1\right>\left<j\right|.
\end{equation}
Substituting this ansatz into Eq.~(\ref{eq.h0s1}) yields
\begin{equation}
S_1=- \sum_j \frac{\sqrt{j+1}g_j}{\omega_r-\omega_j}\left(r^{\dagger}\left|j\right>\left<j+1\right|- r\left|j+1\right>\left<j\right|\right).
\end{equation}
Calculating the first order correction and ignoring the two-photon processes leaves 
\begin{equation}
\frac{1}{2}[H_2,S_1]=\sum_j[r^{\dagger}r(g_{j+1}^2\mu_{j+1}-g_j^2\mu_j)-g_j^2\mu_j]\left|j\right>\left<j\right|,
\end{equation}
where $\mu_j=j/(\omega_r-\omega_j+\omega_{j-1})$. Summing it together with the unperturbed part $H_0$ gives the effective Hamiltonian
\begin{equation}
H_{\rm eff}=\left( \omega_r+\sum_j\chi_j\left|j\right>\left<j\right|\right) r^{\dagger}r+\sum_j\tilde{\omega}_j\left|j\right>\left<j\right|,
\end{equation}
with $\chi_j=g_{j+1}^2\mu_{j+1}-g_j^2\mu_j$ and $\tilde{\omega}_j=\omega_j-g_j^2\mu_j$. The shift in resonator frequency is called AC Stark shift\index{AC-Stark shift}, as the nonlinear shift is qubit-state dependent, it is also known as cross-Kerr\index{cross-Kerr} effect; While the shift in qubit frequency is called Lamb shift, as it arises from nonlinearity of JJ's, it is also known as self-Kerr\index{self-Kerr} effect~\cite{ansari2019superconducting}.

Generally speaking, the coupling strength $g_j$ cannot be readout and measured directly, but it is worth noting that dressed resonator frequency in the dispersive regime relies on the state of transmon. Therefore  if the resonator is measured with a swapping frequency, there should be two peaks in the spectrum, with which the coupling strength can be estimated. More accurate derivation can be found in Ref.~\cite{ansari2019superconducting,gely2018nature}.

\subsection{Two transmons coupled via a resonator}
\begin{figure}[h!]
	\centering
	\includegraphics[width=0.85\textwidth]{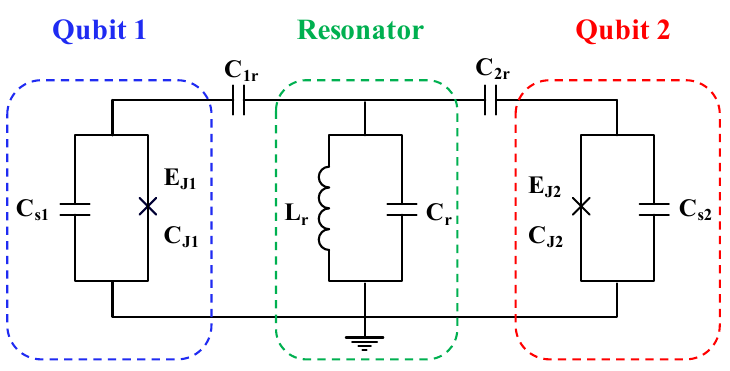}
	\vspace{-0.05in}
	\caption{Two transmon qubits couple to a LC resonator.}
	\label{fig:cqed2}
\end{figure}

A more complicated example is that two transmons are coupled to each other via a bus resonator as shown in Fig.~\ref{fig:cqed2}.
Similarly, by using standard circuit quantization, circuit Lagrangian can be written as
\begin{equation}
\begin{split}
\mathcal{L}&=\frac{1}{2}(C_{J1}+C_{s1})\dot{\phi}_1^2+\frac{1}{2}(C_{J2}+C_{s2})\dot{\phi}_2^2+\frac{1}{2}C_r\dot{\phi}_r^2+\frac{1}{2}C_{1r}(\dot{\phi_1}-\dot{\phi_r})^2\\
&+\frac{1}{2}C_{2r}(\dot{\phi_r}-\dot{\phi_2})^2+E_{J1}\cos\left( {\frac{2\pi}{\Phi_0}\phi_1}\right)+E_{J2}\cos\left( {\frac{2\pi}{\Phi_0}\phi_2}\right)-\frac{\phi_r^2}{2L_r},
\end{split}
\label{eq.cqed}
\end{equation}
where $\phi_i =(\Phi_0/2\pi)\varphi_i$ with $i=1,2,r$ is the flux across each island node. The kinetic part can also be expressed in matrix form as
\begin{equation}
T=\frac{1}{2}\left(\frac{\Phi_{0}}{2\pi}\right)^{2}\vec{\dot{\varphi}}^{T}\mathbf{C}\vec{\dot{\varphi}},
\label{eq.tcqed}
\end{equation}
with $\vec{\varphi}=(\varphi_1,\varphi_r,\varphi_2)$ being the phase at each node and capacitance matrix given by
\begin{equation}
\mathbf{C}=\begin{pmatrix}C_{J1}+C_{s1}+C_{1r} & -C_{1r} & 0\\
-C_{1r} & C_r+C_{1r}+C_{2r}& -C_{2r} \\
0& -C_{2r} & C_{J2}+C_{s2}+C_{2r}
\end{pmatrix}.
\end{equation}
One can rewrite the Hamiltonian in terms of the conjugate term of phase $\varphi$ as
\begin{equation}\label{eq.Hmatrix}
H=4\overrightarrow{n}^{T}\frac{e^{2}}{2\mathbf{C}}\overrightarrow{n}-E_{J1}\cos\varphi_1-E_{J2}\cos\varphi_2+\left(\frac{\Phi_{0}}{2\pi}\right)^{2}\frac{\varphi_r^2}{2L_r},
\end{equation}
with Cooper pair number operator $\overrightarrow{n}=(n_1,n_r,n_2)$. By defining $C_1=C_{J1}+C_{s1}$, $C_2=C_{J2}+C_{s2}$, and assuming $C_{1r},C_{2r}\ll C_{1},C_2,C_r$, the inverse of the matrix can be simplified to 
\begin{equation}
\mathbf{C}^{-1}\approx\begin{pmatrix}\frac{1}{C_1} & \frac{C_{1r}}{C_1C_r} &\frac{C_{1r} C_{2r}}{C_1C_2C_r} \\[6pt]
\frac{C_{1r}}{C_1C_r}& \frac{1}{C_r}&\frac{C_{2r}}{C_2C_r}\\[6pt]
\frac{C_{1r} C_{2r}}{C_1C_2C_r}&\frac{C_{2r}}{C_2C_r}&\frac{1} {C_2}
\end{pmatrix}.
\end{equation}
Expanding the Hamiltonian in Eq.~(\ref{eq.Hmatrix}) to the general form gives 
\begin{equation}
\begin{split}
H&=4E_{C1}n_1^2-E_{J1}\cos\varphi_1+4E_{C2}n_2^2-E_{J2}\cos\varphi_2+4E_{Cr}n_r^2+E_{Lr}\varphi_r^2\\
&+\frac{8C_{1r}}{\sqrt{C_1C_r}}\sqrt{E_{C1}E_{Cr}}n_1n_r+\frac{8C_{2r}}{\sqrt{C_2C_r}}\sqrt{E_{C2}E_{Cr}}n_2n_r+\frac{8C_{1r}C_{2r}}{\sqrt{C_1C_2}C_r}\sqrt{E_{C1}E_{C2}}n_1n_2,
\end{split}
\end{equation}
where $E_{Ci}=e^2/2C_i$ with $i=   1,2,r$. Second quantization of the Hamiltonian by introducing creation and annihilation operators yields
\begin{equation}
\begin{split}
H&=\omega_1a_1^{\dagger}a_1+\frac{\delta_1}{2}a_1^{\dagger}a_1(a_1^{\dagger}a_1-1)+\omega_2a_2^{\dagger}a_2+\frac{\delta_2}{2}a_2^{\dagger}a_2(a_2^{\dagger}a_2-1)+\omega_r r^{\dagger}r\\
&-g_{1c}(a_1-a_1^{\dagger})(r-r^{\dagger})-g_{2c}(a_2-a_2^{\dagger})(r-r^{\dagger})-g_{12}(a_1-a_1^{\dagger})(a_2-a_2^{\dagger}),
\end{split}
\label{eq.fullH1}
\end{equation}
where the $a_q$ ($a_q^{\dagger}$) and $r$ ($r^{\dagger}$) are annihilation (creation) operators of qubits and the resonator, respectively. Frequency, anharmonicity and coupling strength are given by
\begin{eqnarray}
\omega_q&=&8\sqrt{E_{Cq} E_{Jq}}+\delta_q,\\ 
\delta_q&=&-E_{Cq},\\ 
\omega_r&=&4\sqrt{E_{Cr} E_{Lr}},\\
g_{q r}&=&\frac{1}{2}\frac{C_{q r}}{\sqrt{C_r C_q}}\sqrt{\omega_q\omega_r}\label{eq.gir},\\
g_{12}&=&\frac{1}{2}\frac{C_{1r}C_{2r}}{\sqrt{C_1C_2}C_r}\sqrt{\omega_1\omega_2}.
 \label{eq.g12}
\end{eqnarray}
To decouple the resonator from the system, the SW transformation should be applied. Unitary operator can be solved using Eq.~(\ref{eq.h0s1}) as
\begin{equation}
U=\exp\left\{\sum_{q=1,2}\left[\frac{g_{q r}}{\Delta_q}(a_q^{\dagger}r-a_qr^{\dagger})+\frac{g_{q r}}{\Sigma_q}(a_q^{\dagger}r^{\dagger}-a_qr)\right]\right\},
\end{equation}
with $\Delta_q=\omega_r-\omega_q$ and $\sum_q=\omega_r+\omega_q$. Effective Hamiltonian then can be written as 
\begin{equation}
H_{\rm eff}=\tilde{\omega}_1a_1^{\dagger}a_1+\frac{\tilde{\delta}_1}{2}a_1^{\dagger}a_1(a_1^{\dagger}a_1-1)+\tilde{\omega}_2a_2^{\dagger}a_2+\frac{\tilde{\delta}_2}{2}a_2^{\dagger}a_2(a_2^{\dagger}a_2-1)+J(a_1^{\dagger}a_2+a_2^{\dagger}a_1),
\label{eq.Heff1}
\end{equation}
where parameters in the dressed frame are
\begin{eqnarray}
\tilde{\omega}_q&=&\omega_q-g_{q r}^2\left( \frac{1}{\Delta_q}+\frac{1}{\sum_q}\right), \\
\tilde{\delta}_q&\approx&\delta_q,\\
J&=&g_{12}-\frac{g_{1r}g_{2r}}{2}\left(\frac{1}{\Delta_1}+\frac{1}{\Delta_2}+\frac{1}{\sum_1}+\frac{1}{\sum_2} \right). 
\end{eqnarray}
Note that the above expression for $J$ is only valid for single-mode oscillators~\cite{solgun2019simple}. In the following chapters, we turn to applications of the circuit QED on two- and three-qubit interactions.

\newpage
\thispagestyle{empty}

\chapter{Static ZZ Freedom}
\label{c3}
An arbitrary unitary operation\index{unitary operator} can be decomposed into a set of single-qubit and two-qubit gates. In today's quantum processors made up of superconducting qubits, high-performance two-qubit gates are challenging to realize compared to fast single qubit rotations~\cite{mckay2017efficient}. State-of-the-art two-qubit gate error rates remain $0.4\%-1\%$,  which is still above the threshold for error correction~\cite{mckay2016universal,walter2017rapid}, despite tremendous improvements in coherence and optimal control~\cite{negirneac2020high,foxen2020demonstrating,ganzhorn2020benchmarking}. One limiting factor is the intrinsic parasitic interaction between a pair of interacting qubits. Due to the multilevel system of the physical qubits,  computational and noncomputational energy levels can interact with each other.  This coupling shifts the levels up or down, resulting in the always-on ZZ interaction. Such an interaction can accumulate quantum phase in states with Z being the Pauli operator $\sigma_{\rm z}$. On one hand, the ZZ interaction is unwanted for most gates and degrades the performance of desired entanglement~\cite{mundada2019suppression,mckay2019three-qubit,krinner2020benchmarking}. Therefore eliminating it is crucially important to achieve a high-contrast on/off operation modes. On the other hand, we can utilize  ZZ interaction and strengthen it to implement high-speed controlled-phase (CPHASE) gates such as controlled-Z (CZ) gates, which will be introduced in chapter~\ref{c5}.

In this chapter we propose several strategies for freeing two idle qubits from the ZZ interaction, more distinctly, static ZZ without driving. We demonstrate that in a circuit consisting of the same-sign-anharmonicity superconducting qubits, e.g. transmon qubits, the static ZZ interaction vanishes only by switching off qubit-qubit coupling strength. However applying two-qubit gates on such a noninteracting pair does not let them entangle, therefore static ZZ freedom in transmon-transmon pair is not applicable.  We also present a more interesting case, in which we let qubits eliminate their static ZZ in the presence of qubit-qubit coupling, therefore can produce parasitic free entanglement. For this aim, We combine qubits with different-sign anharmonicity,  and show that ZZ freedom can take place under certain circumstance. 

This chapter is based on some parts of our two publications Ref.~\cite{ku2020suppression} published in Physical Review Letters and Ref.~\cite{xu2021zz-freedom} published in Physical Review Applied. Other parts will be discussed in next chapter. First we introduce circuit models used to study the qubit-qubit coupling problem, both perturbatively and nonperturbatively.  Using circuit QED theory we then evaluate all interactions between qubits, including the parasitic interaction. We use delicated theory that static ZZ is considered to be 2 or 3 orders of magnitude weaker than normal coupling between qubits. We search for conditions
 that warrant the elimination of static ZZ interaction between idle qubits. We separate our analysis into transmon-transmon pair with the same sign anharmonicity, and CSFQ-transmon pair with opposite sign anharmonicity. Lastly, we use our theory model to simulate an experimental hybrid CSFQ-transmon circuit  performed by our collaborators at IBM and Syracuse University.

\section{General Model} \label{sec:general model}  
\begin{figure}[h!]
	\centering
	\includegraphics[width=0.75\textwidth]{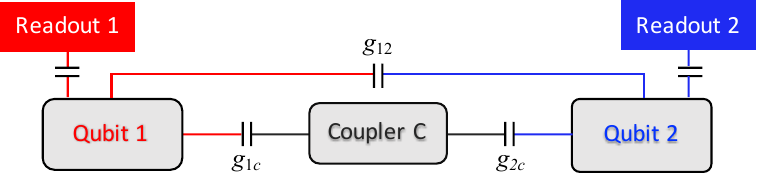}

	\caption{Two qubits coupled via a coupler C with indirect coupling  $g_{1c}$ and $g_{2c}$, and also capacitively coupled to each other with direct coupling $g_{12}$. Each qubit is measured by a readout resonator.}
	\label{fig:CQED}
\end{figure}

Two qubits can interact with each other in two ways: (i) direct capacitive coupling\index{direct coupling}, (ii) indirect coupling\index{indirect coupling} via a coupler. Figure~\ref{fig:CQED} shows a circuit with both couplings combined together. Qubits 1 and 2 are coupled via a coupler C with coupling strength $g_{1c}$ and $g_{2c}$, and also capacitively coupled to each other with coupling strength $g_{12}$. The two qubits can be the same species e.g. transmon-transmon pair or different species e.g. CSFQ/transmon pair. 

To study the circuit Hamiltonian shown in Fig.~\ref{fig:CQED} , we can either keep the coupler throughout our analysis
 or decouple it and then work in the two-qubit basis.  This in fact refers to two approaches of studying the circuit QED: full Hamiltonian model and effective Hamiltonian\index{effective Hamiltonian} model. The full Hamiltonian model evaluates higher excitations contribution in the coupler, which is more eivdent if the coupler is tunable. While the  effective Hamiltonian is usually applied if the coupler is harmonic and is far detuned from qubits. To make our analysis more precise, we consider both co-rotating terms and counter-rotating terms in our calculation. 

\subsection{Full Hamiltonian Model}
 By considering both qubits and the coupler as  multilevel systems, the circuit Hamiltonian can be quantized as 
\begin{align}
H =&\sum_{i, n_i} \omega_i(n_i) |n_i+1\rangle \langle n_i+1| - \sum_{j (\neq i)}{g}_{ij}(a_i-a^{\dagger}_i)(a_j-a^{\dagger}_j), 
\label{eq.ham}  
\end{align}
where $i=1,2$ is the qubit labels and $i=c$  the coupler label, $a_i^{\dagger} =\sum_{n_i} \sqrt{n_i+1} |n_i+1\rangle \langle n_i |$ and $a_i = \sum_{n_i} \sqrt{n_i} |n_i\rangle \langle n_i +1 |$ are the creation and annihilation operators with $|n\rangle$ being the Fock states of each element in the circuit, separately. The eigenstates of the noninteracting Hamiltonian compose of the bare basis. Here we denote bare energy of each level by $E_1({n_{1}})$ and $E_2({n_2})$ for qubits  and $E_c({n_c})$ for the coupler.  $\omega_i(n_i)$ is the energy gap between the two energy levels $E_i({n_i+1})$ and $E_i({n_i})$ of subsystem $i$. A qubit bare frequency is the difference between the lowest two levels i.e. $\omega_{1/2}(0)$ or simply $\omega_{1/2}$. The full Hamiltonian of the circuit contains all possible interactions even the fast oscillating terms, also called co-rotating terms between levels in the bare basis of all circuit elements.  In the presence of the non-zero coupling strength, bare energy levels will be shifted up or down, and this causes modification of qubit frequencies and interaction strengths, sometime appearance of new unwanted interactions. The new values for such quantities are labeled as ``dressed'' values to make them distinct from bare ones.

\begin{figure}[h!]
	\begin{center}
		\includegraphics[width=0.75\textwidth]{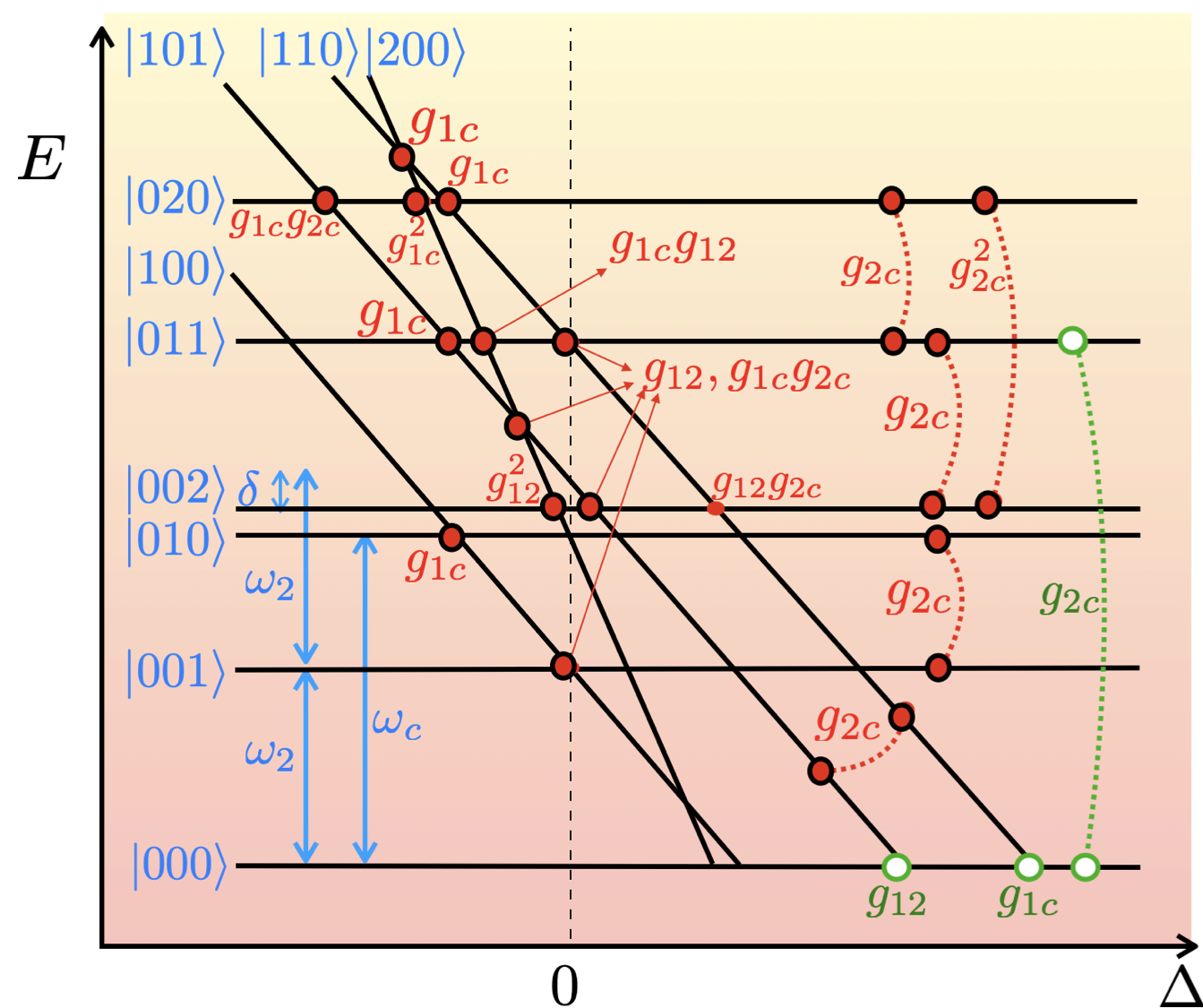}
		\caption{Energy diagram of the circuit in Fig.~\ref{fig:CQED}. Red heads indicate avoided crossing by co-rotating terms with one or two particles exchange, while green marks with white heads indicate counter-rotating interactions.} \vspace{-0.27in}
		\label{fig:diag}
	\end{center}
\end{figure}

Once these dressed states are found, and their energies are known, the dynamics of the system is simple: the eigenstate is a superposition of these bare states with the amplitudes being constant. To understand how the coupling terms change the bare basis, let us apply the full Hamiltonian in Eq.~(\ref{eq.ham})  on the bare states $\left|n_1,n_c,n_2\right>$ for Q1, coupler and Q2, respectively. It is worth noticing that  in Eq.~(\ref{eq.ham}) only one excitation is allowed in each qubit, but in some cases multi-photon processes can also take place. Generally speaking, one can consider the first order of $g_{ij}$ with $i\neq j$ and $i,j=1,2,c$, and expand Eq.~(\ref{eq.ham}) in terms of single particle exchange terms:
\begin{equation}
\begin{split}
 \hat{H}{|n_1,n_{c},n_2\rangle} &=  [\omega_1(n_1)+\omega_{c}(n_{c}) +\omega_2(n_2) ]  {|n_1,n_{c},n_2\rangle}  \\  
 &+ g_{12}  \left[   \sqrt{(n_1+1) n_2} |n_1+1,n_{c},n_2-1\rangle +  \sqrt{n_1 (n_2+1)}  |n_1-1,n_{c},n_2+1\rangle \right.   \\  
 &\left.+\sqrt{n_1 n_2} |n_1-1,n_{c},n_2-1\rangle 
+ \sqrt{(n_1+1) (n_2+1)} |n_1+1,n_{c},n_2+1\rangle \right]\nonumber
\end{split}
\end{equation}

\begin{equation}
\begin{split}
 & + g_{1c}   \left[    \sqrt{(n_1+1) n_{c}} |n_1+1,n_{c}-1,n_2\rangle +  \sqrt{n_1 (n_{c}+1)} \ |n_1-1,n_{c}+1,n_2\rangle \right.   \\ 
  &\left.+\sqrt{n_1 n_{c}} |n_1-1,n_{c}-1,n_2\rangle  
+ \sqrt{(n_1+1) (n_{c}+1)} |n_1+1,n_{c}+1,n_2\rangle \right]\\
&   + g_{2c}   \left[  \sqrt{(n_2+1) n_{c}} |n_1,n_{c}-1,n_2+1\rangle     +  \sqrt{n_2 (n_{c}+1)} |n_1,n_{c}+1,n_2-1\rangle \right. \\ 
& \left.+ \sqrt{n_2 n_{c}} |n_1,n_{c}-1,n_2-1\rangle + \sqrt{(n_2+1) (n_{c}+1)} |n_1,n_{c}+1,n_2+1\rangle\right].\label{eq.fullH}
\end{split}
\end{equation}
The equation above shows different ways of energy transfer between two circuit elements. We can show energy level-crossing diagram as a function of qubit-qubit detuning, see Fig. \ref{fig:diag}.

Let us assume a fixed frequency Q2 is detuned from Q1 by $\Delta$, i.e. $\omega_1=\omega_2-\Delta$. Now we consider more than one particle exchange cases. Given that an eigenstate of unperturbed Hamiltonian is $|n_1,n_{c},n_2\rangle$, interacting Hamiltonian provides transition to $ |n_1\pm k,n_{c}\mp k,n_2\rangle$ via the coupling strength $(g_{1c})^k$, to $|n_1,n_{c}\mp k,n_2\pm k\rangle$ via the coupling strength  $(g_{2c})^k$ , to $ |n_1\pm k,n_{c},n_2\mp k\rangle$ via the coupling strength $(g_{12})^k$ , to $|n_1\pm 1,n_{c}\mp2,n_2\pm1\rangle$ via the strength  $g_{1c}g_{2c}$, also to $|n_1\pm1,n_{c}\pm1,n_2\rangle$ via counter rotating terms with the strength $g_{1c}$, to $|n_1,n_{c}\pm1,n_2\pm1\rangle$ via counter rotating terms with the strength $g_{2c}$, to $|n_1\pm1,n_{c},n_2\pm1\rangle$ via counter rotating terms with the strength $g_{12}$,  and so on.   Figure \ref{fig:diag}  marks these transitions with red and green heads either at intersections or on dashed-coupled lines, and label them by maximum two photon couplings. Green marks with white heads indicate counter rotating interactions such as $a_ia_j$, and red heads represent the co-rotating terms like $a_i^{\dagger}a_j$. One example is that the transition between  $|200\rangle$ and  $|101\rangle$ takes place either by direct coupling $g_{12}$ or via the intermediate state $| 110\rangle$ with coupling $g_{1c}$ and $g_{2c}$. In absence of direct coupling the avoided crossing between the two levels has been found to be $\sqrt{2} g_{1c}g_{2c}/\Delta_{2} (\Delta_{1}+\delta)$ in Ref~\cite{reed2012realization}. The seemingly non-interacting levels may stay so, or may interact via $n$ photons for $n>2$ or under external gates although the interaction may be way smaller than the first order coupling. 

The full Hamiltonian model is complicated and usually performed numerically. By writing the Hamiltonian in the bare basis, namely the matrix form, and fully diagonalizing it, one can find all dressed energies. In this way, the parasitic ZZ interaction is defined as
\begin{equation}
\zeta={\tilde E}_{101}+{\tilde E}_{000}-{\tilde E}_{100}-{\tilde E}_{001},
\label{eq.fzz}
\end{equation}
with ${\tilde E}_{n_1n_cn_2}$ being the dressed energy and subscripts indicating the state of $\left|n_1,n_c,n_2\right>$. This is the original definition of static ZZ interaction without any approximation.
\subsection{Effective Hamiltonian Model}
The coupler frequency $\omega_{c}$ is usually assumed to be far detuned from qubits in order to warrant no backaction from itself. Therefore one can treat the Hamiltonian perturbatively in the dispersive regime\index{dispersive regime} if  $g_{ij}\ll\Delta_{ij}$ with $\Delta_{ij} \equiv \omega_i-\omega_j$ being frequency detuning~\cite{koch2007charge-insensitive}. Within this regime, the coupler can become decoupled from the system via Schrieffer-Wolff (SW) transformation. Then the circuit Hamiltonian $H$ is block diagonalized into the qubit subspaces associated with different photon number of the coupler, the same as Eq.~(\ref{eq.Heff1}).  In the multilevel two-qubit basis, the qubit-qubit Hamiltonian reduces to 
\beqr
H&=& \sum_q\sum_{n_{q}} \bar{\omega}_q({n_q}) |n_q+1\rangle \langle n_q+1| + \sum_{n_1,n_2} \sqrt{(n_1+1)(n_2+1)} \nonumber \\  &\times&J_{n_1,n_2}   \left(| n_1, n_2+1\rangle \langle n_1+1,n_2| +h.c. \right),
\label{eq.effH} 
\eeqr
where dressed frequency $\bar{\omega}_q({n_q})$ is the difference between $\bar{E}_q({n_q+1})$ and $\bar{E}_q({n_q})$ in dressed basis associated to interacting Hamiltonian, and expressed as
\begin{equation}
\bar{\omega}_q({n_q}) = \omega_q({n_q}) -g_{qc}^{2}(n_q+1)/\Delta_{q}({n_q}),
\end{equation}
with $g_{qc}$ being the coupling strength between qubit $q$ and the coupler $c$, $\Delta_q(n_q)=\omega_c-\omega_q(n_q)$ being the qubit-coupler frequency detuning. We can generalize the definition of anharmonicity to 
\begin{equation}
\bar{\delta}_q=\bar{E_q}(2)+\bar{E_q}(1)-\bar{E_q}(0)=\delta_q[1-2g_{qc}^2/\Delta_{q}(\Delta_{q}-\delta_q)],
\end{equation}
and the qubit-qubit coupling strength within two photons limit is
\beq
J_{n_1n_2}\equiv g_{12} -\frac{g_{1c}g_{2c}}{2} \sum_{q=1,2}\left[ \frac{1}{{\Delta_{q}(n_{q})}}+\frac{1}{{\Sigma_{q}(n_{q})}}\right],
\label{eq.J}  \vspace{-0.1in}
\eeq 
with $\Sigma_q(n_q) \equiv  \omega_{c} + \omega_q(n_q)$. For simplicity, $n_q=1$ will be dropped from all notations, e.g.  $\bar{\omega}_q(n_q=1)=\bar{\omega}_q$, $\Delta_q(n_q=1)=\Delta_q$ and $\Sigma_q(n_q=1)=\Sigma_q$.

Second round of SW diagonalization can separate the computational subspace from higher levels. Within the dispersive regime qubit-qubit Hamiltonian in the the computational form turns out to have the following operator structure:
\beq
\begin{split}
H_{\textup{eff}}&=\tilde{\omega}_1\left|10\right>\left<10\right|+\tilde{\omega}_2\left|01\right>\left<01\right|+(\tilde{\omega}_1+\tilde{\omega}_2+\zeta)\left|11\right>\left<11\right|\\
&=-\frac{\tilde{\omega}_1+\zeta/2}{2} {\rm ZI}-\frac{\tilde{\omega}_2+\zeta/2}{2} {\rm IZ}+ \frac{\zeta}{4}{\rm ZZ},
\label{eq.qqH}
\end{split}
\eeq   
where static ZZ interaction in Eq.~(\ref{eq.fzz}) reduces to
\beq 
\zeta= \tilde{E}_{11}-\tilde{E}_{10}-\tilde{E}_{01}+\tilde{E}_{00}.
\label{eq.ZZoriginal}
\eeq

Let us emphasize that Eq. (\ref{eq.ZZoriginal}) is a variant definition of static ZZ interaction in two-qubit basis compared to Eq.~(\ref{eq.fzz}). SW transformation can evaluate the following $\zeta$ and doubly-dressed frequencies within the dispersive regime
\beqr
&& \zeta={2J_{10}^2}/({\bar\Delta-\bar{\delta}_1})-{2J_{01}^2}/({\bar\Delta + \bar{\delta}_2}), \label{eq.zeta}\\
&& \tilde{\omega}_1=\bar{\omega}_1-J_{00}^2/\bar\Delta, \ \ \ \text{and} \ \ \tilde{\omega}_2=\bar{\omega}_2+J_{00}^2/\bar\Delta, 
\label{eq.dw}
\eeqr
with $\bar\Delta\equiv \bar{\omega}_2-\bar{\omega}_1$. One should notice that perturbation theory predicts there are a number of divergences in Eq. (\ref{eq.zeta}) and Eq. (\ref{eq.dw}) at $\bar{\Delta}=-\bar{\delta}_2,0,\bar{\delta}_1,\cdots$. However the original definition Eq. (\ref{eq.ZZoriginal}) is free of such divergence due to finiteness of energy levels and the divergences are mathematical artefacts of perturbation theory. Therefore ZZ interaction from such theory in the vicinity of these divergences is inaccurate.  

\begin{figure}[h!]
	\centering
	\includegraphics[width=0.48\textwidth]{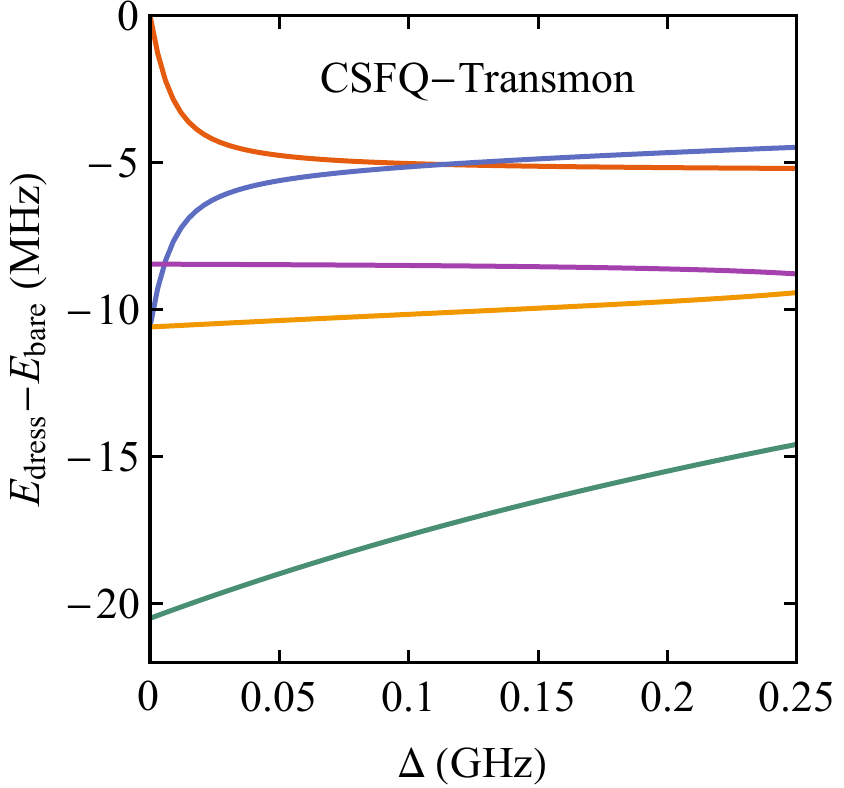}
	\put(-175,160){(a)}
	\includegraphics[width=0.48\textwidth]{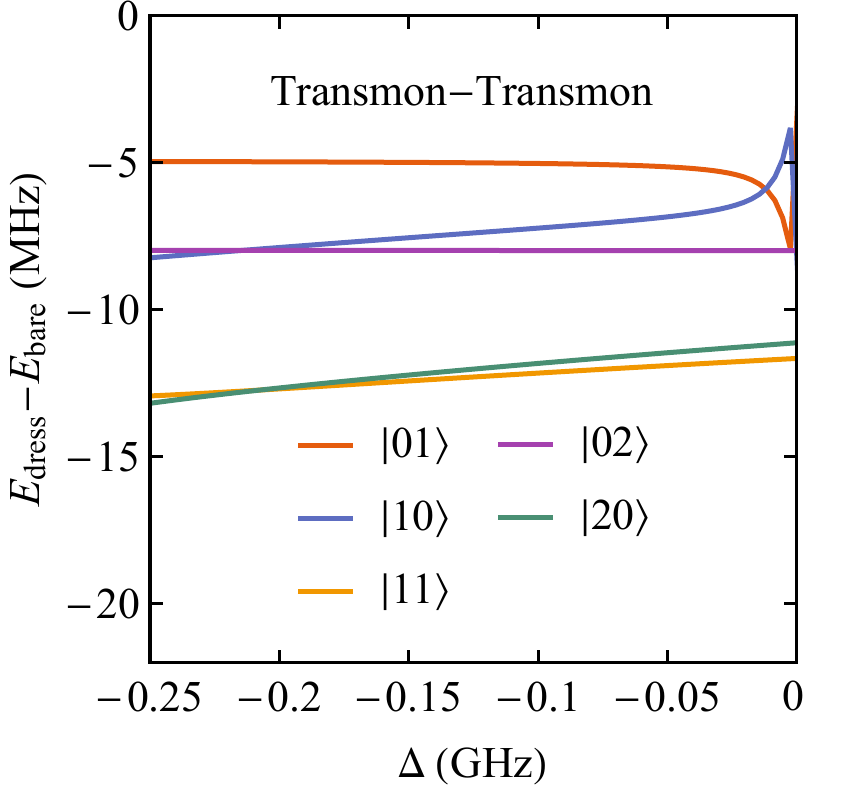}
	\put(-175,160){(b)}
	\vspace{-0.1in}
	\caption{Dispersion between free and interacting energy levels obtained from Hamiltonian (\ref{eq.ham}) vs.  qubit-qubit detuning in (a) a CSFQ-transmon device and (b) a transmon-transmon device.  Circuit parameters are the same as those used in Fig. \ref{fig:ctZZ} and Fig. \ref{fig:ttZZ}.}
	\label{fig:dress}
\end{figure}

To illustrate the absence of these fake divergences, let us numerically evaluate the first five nontrivial eigenvalues of free and interacting Hamiltonian by diagonalizing Eq.~(\ref{eq.effH}). Figure~\ref{fig:dress} shows the energy dispersion ($E_{\rm dress}-E_{\rm bare}$) versus qubit-qubit detuning in a CSFQ-transmon and a transmon-transmon circuit. The detuning is tuned by changing Q1 frequency while keeping Q2 and coupler fixed. At the divergence point $\Delta=0$ two qubits have the same bare frequency, however the computational dressed energy levels are changed by small shift, which corresponds to a finite ZZ interaction. 

Usually the frequency shift due to the interaction is much smaller than the  energy difference of the adjacent levels, so there will be no change in the relative position between computational and noncomputational subspaces. Intuitively, in a CSFQ-transmon pair with coupler frequency far detuned from both qubits, as well as within the limit of $|\Delta|<|\delta_{1/2}|$, $E_{11}$ falls in between noncomputational levels $E_{02}$ and $E_{20}$. $E_{11}$ can interact with the two levels at  the effective coupling strength $J_{01}$ and $J_{10}$, respectively.  The lower frequency decreases and the higher increases, this is called level repulsion. In this device, it is possible to make the two repulsions cancel each other at certain frequency detuning. However, in a transmon-transmon pair, the two noncomputational levels $E_{02}$ and $E_{20}$ are in one side of $E_{11}$, which makes the two repulsions to sum and not cancelled.


\section{Zeroing Static ZZ Interaction}
\label{sec:statZZfree}
Perturbation theory shows that parasitic ZZ interaction relies on the effective $J$ coupling defined in Eq.~(\ref{eq.J}). In particular, the frequency shifts depend on the effective coupling strength $J_{00}$, which causes $|01\rangle \leftrightarrow \langle10|$ transition with excitation from $|0\rangle$ to $|1\rangle$ in one qubit and absorption in the other qubit from $|1\rangle$ to $|0\rangle$. As one can see in Eq. (\ref{eq.dw}), the two dressed levels are further separated by the additional gap $2J_{00}^2/\bar{\Delta}$ with $\bar{\Delta}$ being the dressed qubit-qubit detuning. $J_{00}^2/\bar{\Delta}$ is added on the dressed frequency of one qubit, and the same amount is subtracted from the other qubit.
Therefore based on the original definition Eq. (\ref{eq.ZZoriginal}), $J_{00}$  has no contribution to the static ZZ coupling strength. However in higher levels the scenario is different; $J_{01}$ couples $|02\rangle \leftrightarrow \langle11|$, and $J_{10}$ does $|02\rangle \leftrightarrow |11\rangle$ transition.  The two interaction can cancel one another and free ZZ interaction of $E_{11}$, only  if $J_{01}$ and $J_{10}$ become zero simultaneously or the two noncomputational levels are on both sides of $E_{11}$. 

\subsection{Perturbative ZZ Freedom Condition}
Now let us explore the evolution of the computational basis $|00\rangle$, $|01\rangle$, $|10\rangle$ and $|11\rangle$ by applying time-evolution operator in the form of $U=e^{-iH_{\rm eff}t}$. In contrast to the free Hamiltonian, static ZZ interaction can accumulate additional phase error relying on the states. Evolving the idle quantum states $|00\rangle$ and  $|11\rangle$ after time $t$ results in the phase error $\exp(+i\zeta t/4)$, while the states $|01\rangle$ and  $|10\rangle$ return a opposite-sign phase $\exp(- i\zeta t/4)$. This error will be transmitted across a circuit and finally lower the two-qubit gate fidelity. 

Based on the expression of ZZ in Eq. (\ref{eq.zeta}), we can search for the possibility of eliminating the phase error for interacting qubits. One easy solution is to set $J_{01}\approx J_{10}\approx0$. A more interesting solution, as discussed before, is to enable the elimination by finding ways to cancel level repulsions between computational and noncomputational levels while keeping two qubits coupled. To find out such a nontrivial solution, we solve the equation $\zeta=0$ and obtain the following condition:  
\beq
\bar{\Delta}=\frac{\bar{\delta}_1+\bar{\delta}_2 \gamma^2}{1- \gamma^2},
\label{eq.ZZfeq}
\eeq
where $\gamma \equiv J_{10}/J_{01}$ with $J_{01}$ and $J_{10}$ defined in Eq.~(\ref{eq.J}). Counter-rotating terms and direct coupling are one order smaller than the co-rotating terms and indirect coupling, separately, therefore can be ignored. This simplifies $\gamma$  to the following expression
\beq
\gamma\approx \frac{1-\delta_1/ (2\Delta_{2} +\Delta)}{1-\delta_2/(2 \Delta_{2} +\Delta)} \frac{1-\delta_2/\Delta_{2}}{1-\delta_1/(\Delta_{2}+\Delta)},
\label{eq.Jratio}
\eeq
with $\Delta_2=\omega_c-\omega_2$ being the frequency difference between qubit 2 and coupler. The condition of the static ZZ freedom\index{static ZZ freedom} in Eq. (\ref{eq.ZZfeq}) makes it possible to explore such a  possibility for certain circuit parameters.  
\subsection{Examples of Static ZZ Freedom}
Now we study the elimination of ZZ interaction in two types of circuits: (a) a CSFQ-transmon pair with the opposite-sign anharmonicity coupled via a coupler; (b) a transmon-transmon pair with same-sign anharmonicity coupled via a coupler.
\subsubsection*{(a) CSFQ-transmon pair} 
This is an example of opposite-sign anharmonicity pair. Combining a positive and a negative anharmonic qubits makes it possible to realize static ZZ freedom. This freedom can take place in the presence of finite $J$ interaction between qubits that let them entangle once they are driven by gates. Here we search for circuit characteristics to achieve static ZZ freedom. 

\begin{figure}[h!]
	\centering
	\includegraphics[width=0.9\textwidth]{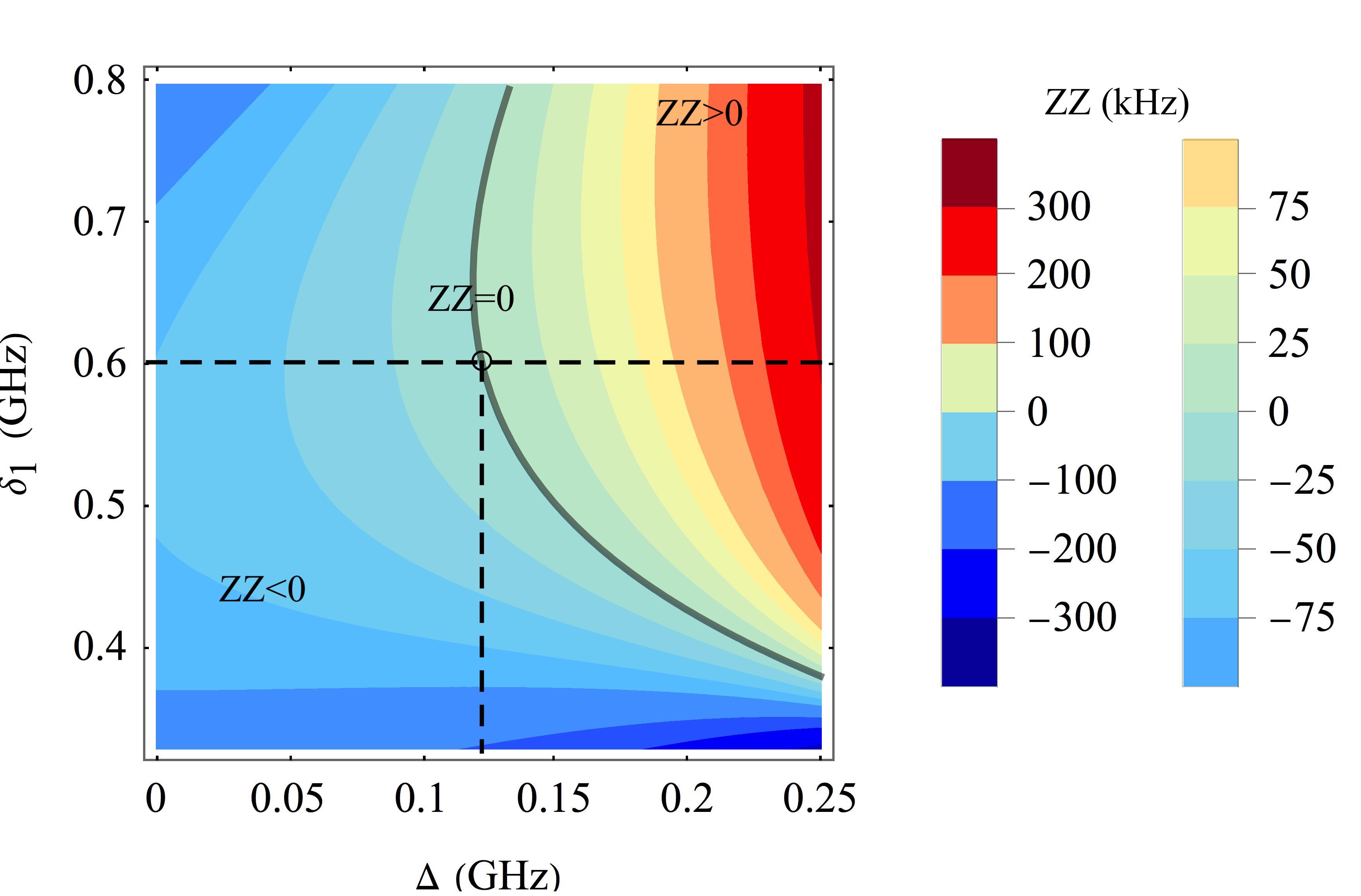} 
	\caption{Static ZZ interaction in a CSFQ-transmon device. Q1 has the anharmonicity $\delta_1$ and is detuned from a transmon Q2 by $\Delta$. Static ZZ interaction dependence on CSFQ anharmonicity and qubit qubit detuning. The two qubits are uncoupled directly, but they are indirectly coupled via a bus resonator with frequency $\omega_{c}=6.492$ GHz and coupling strengths $g_{1c}=g_{2c}=g=80$ MHz with Q2 frequency and anharmonicity $\omega_2=5.292$ GHz, $\delta_2=-0.33$ GHz. The effective ZZ interaction is positive (negative) in red (blue) areas and it vanishes on the solid lines.}
	\label{fig:ctZZ}
\end{figure}

Let us consider Q2 in Fig.~\ref{fig:CQED} is a transmon with  fixed frequency $\omega_2$  coupled to Q1, a CSFQ with $\Delta$ detuned frequency, i.e. $\omega_1=\omega_2-\Delta$.  We write the Hamiltonian Eq.~(\ref{eq.effH}) in matrix form with maximum excitation number being four, i.e. all levels from $E_{00}$ up to $E_{04}$ and $E_{40}$. We block diagonalize the matrix into two blocks: computational subspace and noncomputational subspace. Using Eq.~(\ref{eq.zeta}) we can evaluate the repulsion between computational and noncomputational levels, namely the static ZZ interaction. Figure~\ref{fig:ctZZ} shows ZZ  over a wide range of the frequency detuning $\Delta$ in unit of gigahertz on $x$ axis, and CSFQ anharmonicity $\delta_1$ on $y$ axis. The ZZ interaction can be negative or positive with a nontrivial borderline marked by solid black line between the two regions on which it vanishes. The marked circle indicates similar  parameters of the CSFQ-transmon circuit experimented which will be introduced in section~\ref{sec:exp}.

\begin{figure}[t!]
	\centering
	\includegraphics[width=0.49\textwidth]{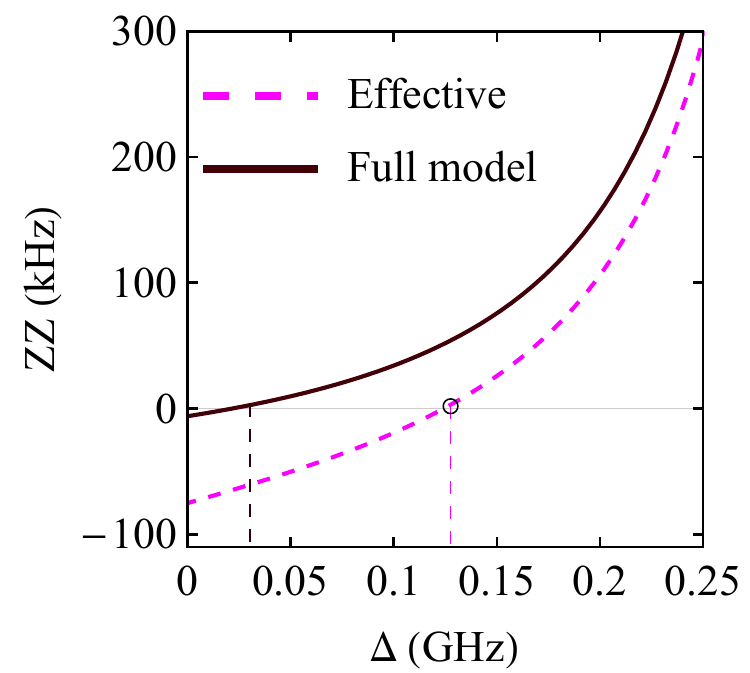}
	\includegraphics[width=0.49\textwidth]{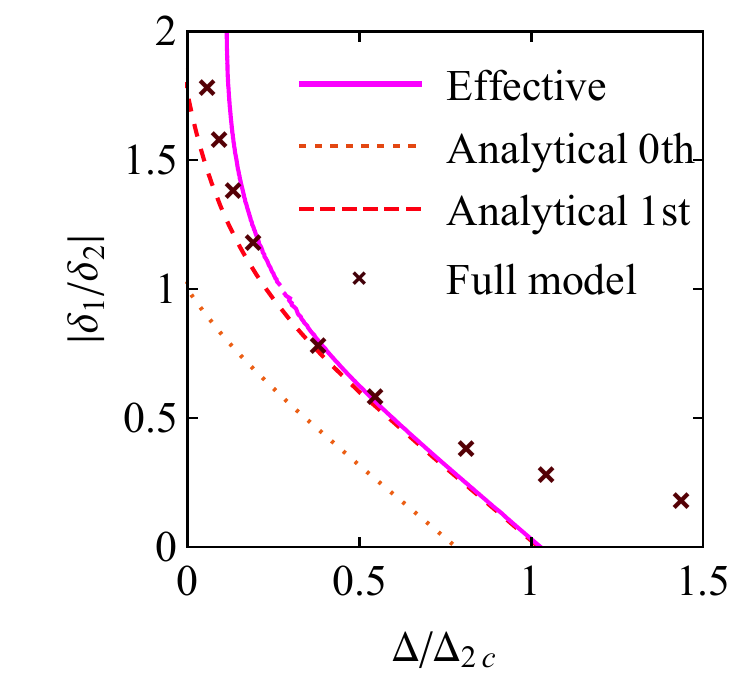} \put(-350,150){(a)}
	\put(-165,150){(b)}
	\caption{(a) Static ZZ interaction with $\delta_1=0.6$ GHz using the effective (dashed magenta) and full model (solid brown) individually. (b) Static ZZ freedom criteria via the effective (solid), analytical zeroth-order (dotted), first order (dashed) and full Hamiltonian model (cross). }
	\label{fig:fullsw}
\end{figure}

Figure \ref{fig:fullsw}(a) compares two different approaches to determine the static ZZ for CSFQ-transmon pair as a function of different  frequency detuning: dashed line calculating ZZ  from effective Hamiltonian\index{effective Hamiltonian} in Eq. (\ref{eq.zeta}), solid line obtained from numerical analysis by fully diagonalizing the Hamiltonian in Eq.~(\ref{eq.ham}). 
Comparing the two methods reveals that: both effective and full Hamiltonian models show the consistent  presence of ZZ freedom, and perturbation theory becomes more accurate as detuning $\Delta$ increases.  Figure \ref{fig:fullsw}(b) shows the parameters criteria at which the static ZZ is zero. The parameter landscape explored here are the normalized frequency detuning $\Delta$ by transmon-coupler detuning $\Delta_{2}$ and the magnitude of anharmonicity ratio by keeping the others unchanged. Crossed points show exact results from full Hamiltonian model, and the solid line is obtained from Eq. (\ref{eq.ZZfeq}). The two dashed lines are analytical solutions of Eq.~(\ref{eq.ZZfeq}) in the zeroth and first order in $|\delta_2/\Delta_{2}|$.  For obtaining these analytical solutions we rewrite ZZ interaction in terms of  $\delta_1=  k \delta$,  $\delta_2=-\delta$ with  $k, \delta>0$. We consider no direct coupling between qubits and the universal qubit-coupler coupling strength $g_1=g_2=g$, and $\Delta=b\Delta_{2}$. By substituting these parameters in Eq.  (\ref{eq.Jratio}), $\gamma$ is simplified to
\beq 
\gamma= \frac{(1+a)(1+b)(ak - b -2)}{(2+b+a) (ak -b -1)},
\eeq
with $a\equiv \delta/\Delta_{2}$. In the absence of $a$, static ZZ cancellation condition in Eq. (\ref{eq.ZZfeq}) then reduces to
\beq
 k=\frac{2+b-3b^2-2b^3}{2+5b+b^2},
 \eeq
 i.e. zeroth order solution.  This result gives similar trend as the other two numerical methods, but the precision is quite low. Adding the first order of $a$ increases the solution accuracy. One can see for increasing absolute ratio of CSFQ and transmon anharmonicity the analytical approximation is more trustable.

\subsubsection*{(b) transmon-transmon pair}  
This is an example of same-sign anharmonicity pair. Typically there are many choices of pairs of qubits coupled to each other, but here we only study the most popular one, transmon. In such a pair both anharmonicities are negative. Let us start with the trivial solution that $J_{01}\approx J_{10}\approx0$, and revisit the definition of $J$ coupling in Eq. (\ref{eq.J}). One can find that zeroing it requires that direct coupling $g_{12}$ cancels out the indirect couplings. This freedom has been realized recently in several experiments \cite{zhao2020high-contrast,li2020tunable,malekakhlagh2020first-principles}.  

However non-interacting qubits cannot be operated to achieve entanglement at such an operating point and therefore are not useful for quantum computation. To perform gates on such a transmon-transmon pair with high fidelity, one can suppress ZZ interaction by tuning circuit parameters. For this purpose we consider a transmon-transmon pair with almost the same anharmonicity and different frequencies. 

We simulate this device and extract static ZZ and $J$ coupling by diagonalizing the full Hamiltonian.
\begin{figure}[h!]
	\centering
	\includegraphics[width=0.48\textwidth]{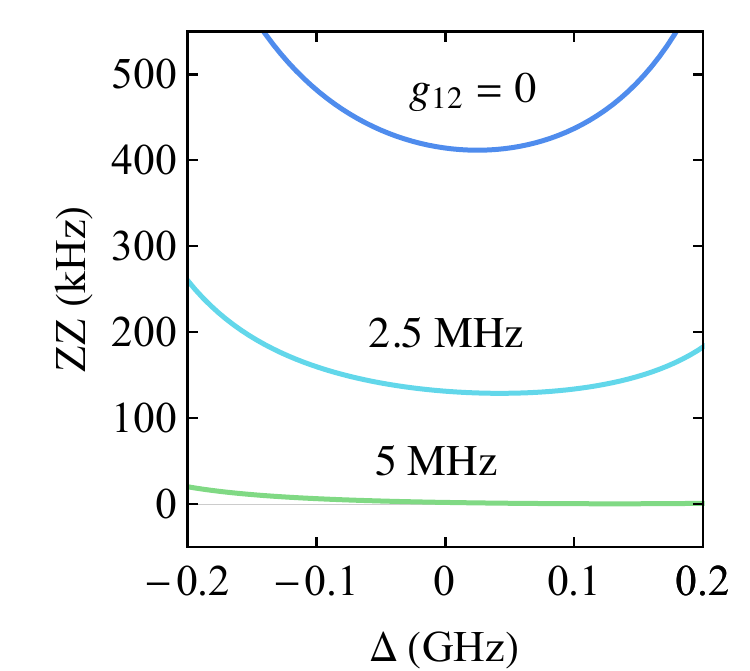}
	\includegraphics[width=0.48\textwidth]{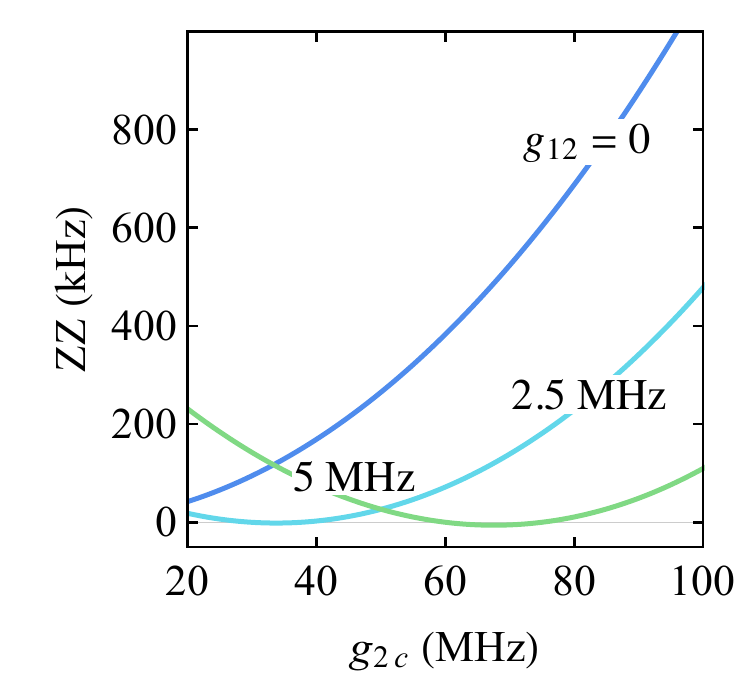}
	\put(-345,145){(a)}
	\put(-170,145){(b)}\\
	\includegraphics[width=0.48\textwidth]{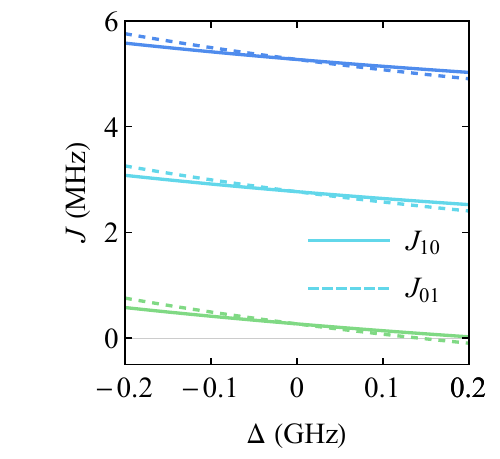}
	\includegraphics[width=0.48\textwidth]{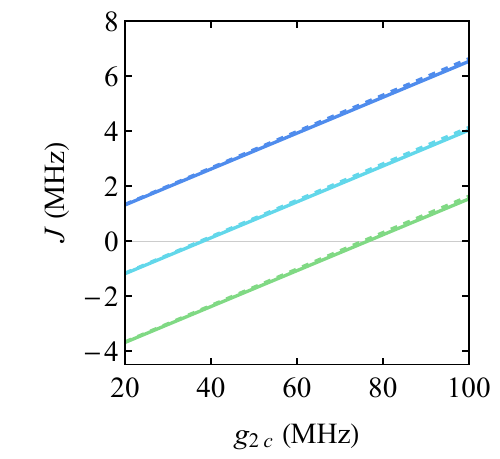}
	\put(-345,150){(c)}
	\put(-170,150){(d)}
	\vspace{-0.1in}
	\caption{Static ZZ interaction in a transmon-transmon device with $\omega_2=4.914$ GHz, $\omega_c=6.31$ GHz, $\delta_1=\delta_2=-0.33$ GHz and direct coupling $g_{12}=0, 2.5, 5$ MHz as a function of (a) qubit-qubit detuning with $g_1=98$ MHz, $g_2= 83$ MHz, (b) transmon-resonator coupling strength $g_{2c}$ with $g_{1c}=98$ MHz and $\Delta=-0.1$ GHz, separately. Corresponding $J_{01}$ and $J_{10}$ are plotted in (c) and (d).}
	\label{fig:ttZZ}
\end{figure}
Figure \ref{fig:ttZZ}(a) shows static ZZ interaction is near-symmetric and decreases by lowering the magnitude of detuning frequency $\Delta$ without direct coupling. On top of this we also study for the other two nonzero direct couplings $g_{12}$. At $g_{12}=5$ MHz the ZZ interaction can be dramatically suppressed to a few tens of kilohertz, and at specific detuning static ZZ freedom is realized with almost zero $J$ coupling as shown in Fig.~\ref{fig:ttZZ}(c). Apart from the static ZZ freedom\index{static ZZ freedom}, parasitic interaction increases slowly while $J$ coupling can go higher and become non-zero. 

Interestingly the weaker $g_{12}=2.5$ MHz case shows that suppression of  ZZ interaction can be achieved within a rather large domain of detuning frequency  $\Delta$, making it possible to design a transmon-transmon circuit with suppressed ZZ interaction within the dispersive regime $\Delta \gg J$.  Figure \ref{fig:ttZZ}(b) shows how ZZ coupling changes by varying transmon-resonator coupling $g_{2c}$.  Although the zero ZZ points belong to $J$-interaction-free domain, however in their vicinity one can see a large class of transmons that not only interact but also their unwanted ZZ interaction is suppressed. To strengthen $J$ around this region, one can vary the coupler frequency.  

\begin{figure}[h!]
	\centering
	\includegraphics[width=0.82\textwidth]{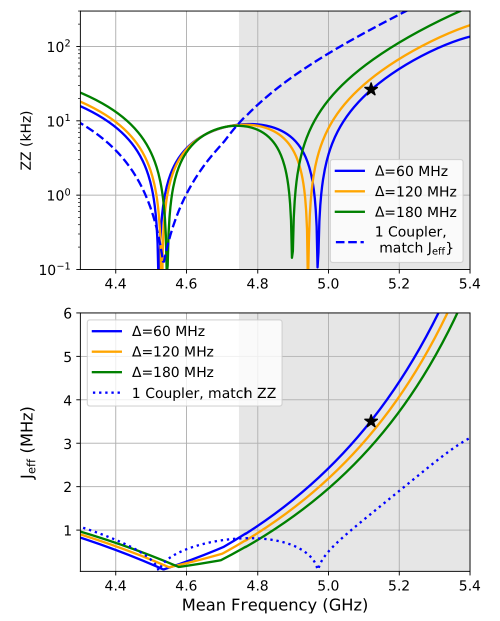}
	\put(-310,360){(a)}
	\put(-310,185){(b)}
	\vspace{-0.1in}
	\caption{(a) ZZ vs the mean qubit frequency at different qubit-qubit detunings. The experimental data is highlighted by the star. The dashed line is the ZZ for a pair of qubits with $\Delta = 60$ MHz coupled via  a direct coupler. (b) The effective $J$ at different qubit-qubit detunings, the experiment value is the star. The dotted line is the effective $J$ for a $\Delta = 60$ MHz direct coupler device. The shaded region represents the frequency region where the multi-path coupler shows an improvement in $J_{\rm eff}$~\cite{kandala2020demonstration}.}
	\label{fig:07050}
\end{figure}

A few months after we post this work on arXiv~\cite{xu2021zz-freedom}, the idea of suppressing static ZZ interaction in transmon-transmon devices was tested and verified in a IBM paper~\cite{kandala2020demonstration}. In that paper, they strengthened the direct coupling by adding a direct coupler to make static ZZ freedom possible in the all frequency-fixed transmons circuit. In the vicinity of static ZZ freedom shaded in Fig.~\ref{fig:07050}. They improved $J_{\rm eff}$ and performed an experiment at the frequency marked by the star. They achieved an effective $J$ of 3.5 MHz with 26 kHz ZZ interaction, and demonstrated a CNOT gate\index{CNOT gate} fidelity of 99.77(2)\%.

 Searching for other possibilities of ZZ freedom in the presence of $J$ coupling gives that qubit-coupler coupling  $g_{qc}$ is equal or larger than the qubit-resonator frequency detuning $\Delta_{q}$, and this goes beyond the dispersive regime and therefore  perturbation theory fails. Exact numerical result from full Hamiltonian presents consistent non-divergent solution and does not show any possibility for ZZ freedom in the transmon-transmon pair within the domain of quantum computation. Another approach is to replace frequency-fixed harmonic resonator with a tunable coupler\index{tunable coupler} i.e. an asymmetric transmon, making it possible
to vary static ZZ interaction close to zero \cite{krinner2020demonstration, collodo2020implementation, xu2020high-fidelity}. Furthermore, a ZZ-free iSWAP gate is implemented experimentally with the fidelity close to the coherence limit via optimal control~\cite{sung2020realization}.

\section{Experimental Setup}
\label{sec:exp}
In this section, we present the detailed theoretical model of the first experimented hybrid CSFQ-transmon device. The device was fabricated at IBM, and consists of one fixed-frequency transmon, one bus resonator, one CSFQ, and readout resonators for each qubit. The measurement was taken by our collaborators at Syracuse University. More details can be found in our publication Ref.~\cite{ku2020suppression}.
\subsection{Impact of Measurement}
Let us briefly discuss the impact of measurement\index{measurement} of qubits. Readout measurement usually takes place by means of weakly coupling a qubit to a resonator with the interaction $H_{qR}=g_{qR}(\hat{a}_q+\hat{a}_q^\dagger) (\hat{a}_R+\hat{a}^\dagger_R)$.  In a circuit where qubits are both coupled to the coupler and readout resonator, the process of eliminating the readout resonator\index{readout cavity} can take place before or after eliminating the coupler. Since both of the couplings are within the dispersive regime, SW are insensitive to the decoupling orders. However in reality the two orders sometime show inconsistent results~\cite{pommerening2020what}. To study the difference of the two decoupling orders, we derive corresponding dressed frequency and anharmonicity.  Decoupling first readout and then bus resonator in the dispersive regime gives rise to
\begin{equation}
\tilde\omega_{1/2}=\breve\omega_1\pm\frac{\left[g_{12}-\frac{g_{1r}g_{2r}}{2}\left(\frac{1}{\omega_r-\bar\omega_1}+\frac{1}{\omega_r-\bar\omega_2}\right)\right]^2}{\breve\omega_1-\breve\omega_2},
\end{equation}
defining
\begin{equation}
\begin{split}
\breve\omega_q&\equiv\bar\omega_q-\frac{g_{qr}^2}{\omega_r-\bar\omega_q},\\
\bar\omega_q&\equiv\omega_q-\frac{g_{qR}^2}{\omega_R-\omega_q}.
\end{split}
\end{equation}
Similarly, first decoupling bus and then readout gives 
\begin{equation}
\tilde\omega'_{1/2}=\check\omega_1\pm\frac{\left[g_{12}-\frac{g_{1r}g_{2r}}{2}\left(\frac{1}{\omega_r-\omega_1}+\frac{1}{\omega_r-\omega_2}\right)\right]^2}{\check\omega_1-\check\omega_2},\\
\end{equation}
defining
\begin{equation}
\check\omega_q\equiv\omega_q-\frac{g_{qr}^2}{\omega_r-\omega_q}-\frac{g_{qR}^2}{\omega_R-\omega_q-\frac{g_{qr}^2}{\omega_r-\omega_q}}.
\end{equation}
By assuming zero direct coupling $g_{12}=0$ and universal indirect coupling $g_{qR}=g_{qr}=g$, we derive the inconsistency in dressed qubit frequency $\textup{\dj} \omega\equiv\tilde\omega_1-\tilde\omega_1'=\tilde\omega_2-\tilde\omega_2'$ and  anharmonicity $\textup{\dj} \delta$ as
\beqr 
&& \frac{\textup{\dj} \tilde{\omega}_2}{g^6}=-\frac{\textup{\dj} \tilde{\omega}_1}{g^6}= \frac{(\Delta+2\Delta_{2})^2(\Delta^2+\Delta\Delta_{2}+\Delta_{2}^2)}{2\Delta \Delta_{2}^4 (\Delta+\Delta_{2})^4},\ \  \\   
&& \frac{\textup{\dj} \tilde{\delta}_{1/2}}{2 g^6}=-  \frac{(\Delta_{2}-\delta_{1/2})^3+(\delta_{1/2}+\Delta_{2})\Delta_{2}^2}{(\Delta_{2}^3 \Delta\mp\delta_{1/2})(\Delta_{2}-\delta_{1/2})^4} \pm 
\frac{2}{\Delta\Delta_{2}^4}, \ \ \ \ 
\label{eq.order}
\eeqr
with universal qubit-resonator detuning $\Delta_2=\omega_{r/R}-\omega_2$, two qubit detuning $\Delta=\omega_2-\omega_1$, and $\delta_{1},\delta_2,\Delta\ll\Delta_{2}$. In the dispersive regime, the difference is in the order of $g^6/\Delta_2^4\Delta$ and can be ignored. In the following discussion, we decouple readout resonator first then bus resonator to reduce the Hamiltonian.

\subsection{Modeling the Experimental Circuit}
\begin{figure}[h!]
	\labellist
	\bfseries
	\pinlabel Readout at 40 80
	\pinlabel Cavity at 40 65
	\pinlabel Transmon at 140 190
	\pinlabel {Bus cavity} at 240 190
	\pinlabel CSFQ at 360 190
	\pinlabel Readout at 485 80
	\pinlabel Cavity at 485 65
	\endlabellist
	\centering
	\includegraphics[width=1\textwidth]{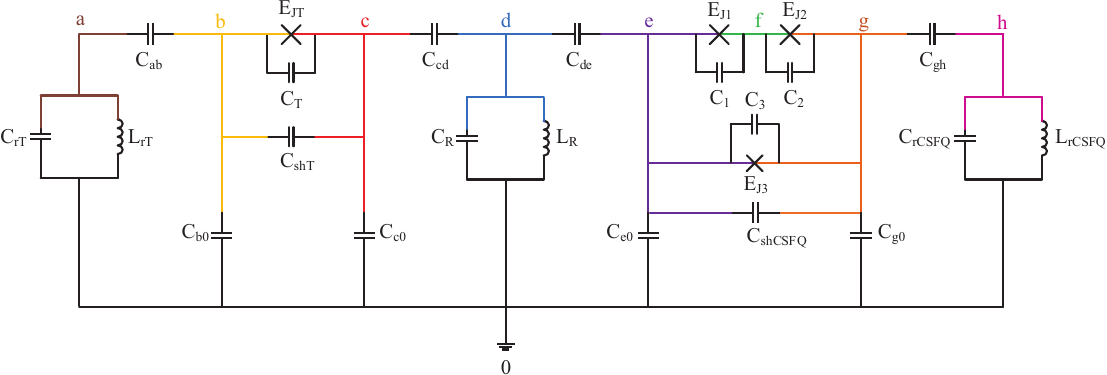} 
	\vspace{-0.1 in}
	\caption{Circuit model for CSFQ-transmon device. Each qubit has its own readout resonator and they are coupled via a bus resonator.}
	\label{fig:csfq-transmon}
\end{figure}
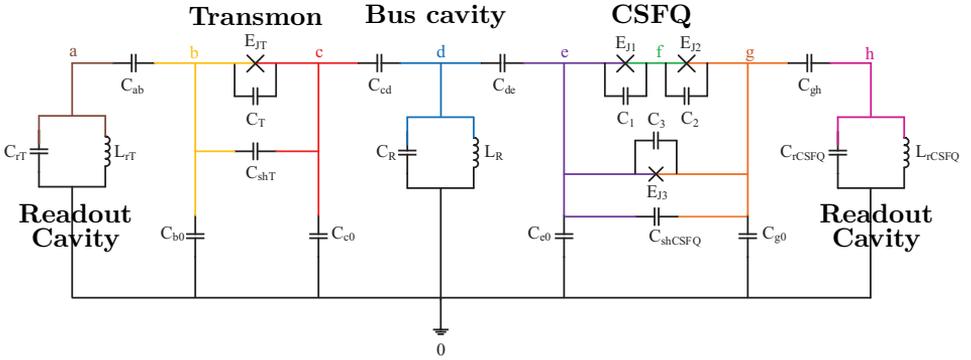

First we build a full-circuit Hamiltonian from a lumped-element circuit model for our CSFQ-transmon device in Fig.~\ref{fig:csfq-transmon}, and the corresponding design parameters are listed in Tab.~\ref{table:para}. The circuit contains three subsystems: transmon and its readout cavity, bus cavity, CSFQ and its readout cavity. By following standard circuit quantization, Lagrangian of the total system can be obtained. Note that we label all seven nodes from $a$ to $h$, and kinetic energy should include all charging energy stored in the capacitors. 
Similarly, we define $\varphi_T\equiv\varphi_b-\varphi_c$, $\varphi_m\equiv(\varphi_e-\varphi_g)/2$ and  $\varphi_p\equiv(\varphi_e+\varphi_g)/2-\varphi_f$. The superscript and subscript $T$, $m/p$ represent transmon and CSFQ, respectively. Particularly, $m$ means the phase difference between two large junctions in CSFQ while $p$ is the phase sum. The Lagrangian is then  simplified into the following form:
\begin{equation}
\begin{split}
\mathcal{L}&=\frac{1}{2}\left(\frac{\Phi_{0}}{2\pi}\right)^{2}\left[C_{rT}\dot{\varphi_{a}}^2+C_{ab}(\dot{\varphi_{a}}-\dot{\varphi_{b}})^{2}+C_{b0}\dot{\varphi_{b}}^{2}+C_{cd}(\dot{\varphi_{b}}-\dot{\varphi_{T}}-\dot{\varphi_{d}})^{2}\right.\\
&+(C_{shT}+C_{T})\dot{\varphi_{T}}^{2}+C_{c0}\left(\dot{\varphi_{b}}-\dot{\varphi_{T}}\right)^{2}+C_{R}\dot{\varphi_{d}}^{2}+C_{de}(\dot{\varphi_{e}}-\dot{\varphi_{d}})^{2}+C_{e0}\dot{\varphi_{e}}^{2}\\
&+C_{gh}\left(\dot{\varphi_{e}}-2\dot{\varphi_{m}}-\dot{\varphi_{h}}\right)^{2}+2C_J\left(\dot{\varphi_{m}}^{2}+\dot{\varphi_{p}}^{2}\right)+4(C_{3}+C_{shCSFQ})\dot{\varphi_{m}}^{2}\\
&+\left.C_{g0}\left(\dot{\varphi_{e}}-2\dot{\varphi_{m}}\right)^{2}+C_{rCSFQ}\dot{\varphi_{h}}^{2}\right]+E_{JT}\cos\varphi_{T}+2E_{J}\cos{\varphi_{p}}\cos{\varphi_{m}}\\
&+\alpha E_{J}\cos\left(2\pi f-2\varphi_{m}\right)-\left(\frac{\Phi_{0}}{2\pi}\right)^{2}\left(\frac{\varphi_{a}^{2}}{2L_{rT}}+\frac{\varphi_{d}^{2}}{2L_{R}}+\frac{\varphi_{h}^{2}}{2L_{rCSFQ}}\right),
\end{split}
\end{equation}
where $f=\Phi/\Phi_0$ is the normalized magnetic flux, $C_1=C_2\equiv C_J$, and $ E_{J1}=E_{J2}\equiv E_J$. The Hamiltonian is calculated from the Legendre transformation of the Lagrangian
\begin{eqnarray}
H&=&\sum_{i}\dot{\varphi_{i}}\frac{\partial \mathcal{L}}{\partial\dot{\varphi_{i}}}-\mathcal{L}=T+U,\\
T&=&\frac{1}{2}\left(\frac{\Phi_{0}}{2\pi}\right)^{2}\vec{\dot{\varphi}}^{T}\mathbf{C}\vec{\dot{\varphi}},\\
U&=&E_{La}\varphi_{a}^{2}+E_{Ld}\varphi_{d}^{2}+E_{Lh}\varphi_{h}^{2}+E_{JT}\cos\varphi_{T} \\ \nonumber
&-&2E_{J}\cos{\varphi_{p}}\cos{\varphi_{m}}-\alpha E_{J}\cos\left(2\pi f-2\varphi_{m}\right),
\end{eqnarray}
where the phase vector in the circuit is defined as $\vec{{\varphi}}^{T}=\left({\varphi_{b}},{\varphi_{e}},{\varphi_{a}},{\varphi_{T}},{\varphi_{d}},{\varphi_{m}},{\varphi_{p}},{\varphi_{h}}\right)$, and the energies stored in the readout resonators for the transmon, the CSFQ, and the bus resonator are $E_{La}=\Phi_0^2/8\pi^2 L_{rT} $, $E_{LCSFQ}=\Phi_0^2/8\pi^2 L_{rCSFQ}$, and $E_{LR}=\Phi_0^2/8\pi^2 L_{R} $, respectively.

\begin{table}[h!]
	\begin{center}
		\begin{tabular}{|c|c|c|c|c|c|}
			\hline 
			\multicolumn{4}{|c|}{Capacitance (fF)} & \multicolumn{2}{c|}{Josephson energy (GHz)}\tabularnewline
			\hline 
			$C_{rT}$& $452^*$& $C_{rCSFQ}$ &  $439^*$ & $E_{J1}=E_{J2}$ & $109^*$\tabularnewline
			\hline 
			$C_{ab}$&$6.6$  &$C_{gh}$  &$6.6$ & $E_{J3}=\alpha E_{J1}$ & $46.8^*$\tabularnewline
			\hline 
			$C_{b0}$&$66$  &$C_{g0}$  & $66$  & $E_{JT}$ & $13.7^*$\tabularnewline
			\hline 
			$C_{shT}$&$26$ &  $C_{shCSFQ}$&$26$& $\alpha$ & $0.43^*$\tabularnewline
			\hline 
			$C_{T}$&$0.5-1^*$ &  $C_1=C_2$&$1.2-2.3^*$ & \multicolumn{2}{c|}{Inductance (nH)}\tabularnewline
			\hline 
			$C_{c0}$&$63$ & $C_{e0}$ &  $63$ & $L_{R}$ & $1.3^*$\tabularnewline
			\hline 
			$C_{cd}$ &$16 $ & $C_{de}$ & $16$& $L_{rT}$ & $1.2^*$\tabularnewline
			\hline 
			$C_R$ &$469^*$ & $C_3=\alpha C_1$ & $0.5-1^*$& $L_{rCSFQ}$ & $1.2^*$\tabularnewline
			\hline 
		\end{tabular}
	\end{center}
\vspace{-0.2in}
	\caption{Circuit parameters corresponding to each circuit element in Fig.~\hyperref[fig:csfq-transmon]{\ref{fig:csfq-transmon}}. The parameters with $*$ symbol are calculated based on experimental data, while the parameters without $*$ are design values that are extracted using ANSYS Q3D Extractor simulation of the qubit layout.}
	\label{table:para}
\end{table}

We can decouple the first two redundant phase degrees of freedom since they are not included in the potential, and ignore the phase sum of CSFQ as  discussed in section~\ref{sec:csfq}. More details can be found in Appendix~\ref{app:cirH}. Now let us rewrite the Hamiltonian in terms of canonical conjugate $n$ and $\varphi$ as 
\begin{equation}
\begin{split}
H&=4E_{Ca}n_{a}^{2}+E_{La}\varphi_{a}^{2}+4E_{Cd}n_{d}^{2}+E_{Ld}\varphi_{d}^{2}+4E_{Ch}n_{h}^{2}+E_{Lh}\varphi_{h}^{2}\\&+4E_{CT}n_{T}^{2}-E_{JT}\cos\varphi_{T}+4E_{Cm}n_{m}^{2}-2E_{J}\cos\varphi_{m}-\alpha E_{J}\cos\left(2\pi f-2\varphi_{m}\right)\\&+\lambda_{dm}n_{d}n_{m}+\lambda_{hm}n_{h}n_{m}+\lambda_{aT}n_{a}n_{T}+\lambda_{dT}n_{d}n_{T}+\lambda_{mT}n_{m}n_{T},
\end{split}
\label{eq.ge}
\end{equation}
with charging and Josephson energy listed in Tab.~\ref{table:junction_params}.
\begin{table}[h!]
\centering
\begin{tabular}{| l | l | l |}
	\hline
	Description & Symbol & Value \\ \hline \hline
	Transmon Josephson energy & $E_{JT}$ & 13.7 GHz  \\ \hline
	Transmon charging energy &  $E_{CT}$ & 0.286 GHz  \\ \hline
	\hline
	CSFQ Josephson energy & $E_{Jm}$ & 109  GHz \\ \hline
	CSFQ charging energy & $E_{Cm}$ & 0.292  GHz \\ \hline
	CSFQ critical current ratio & $\alpha$ & 0.43 \\ \hline
\end{tabular}
\caption{\label{table:junction_params}Junction parameters and charging energy of the CSFQ and transmon. CSFQ Josephson energy is for the larger junctions. The two transmon parameters were calculated using the measured dressed qubit frequency. Meanwhile, the three CSFQ parameters were obtained by fitting spectroscopy data of the dressed qubit frequencies, $\tilde{\omega}_1(0)/2\pi$ and $\tilde{\omega}_1(1)/2\pi$ vs. flux with 1D potential approximation~\cite{steffen2010high-coherence}.}
\end{table}

This Hamiltonian can be decomposed into two parts: free part including readout resonators, qubits and bus cavity; interacting parts containing all the interaction. As discussed before, accurate frequency and anharmonicity of transmon and CSFQ can be derived from Ref.~\cite{gely2018nature} and Appendix \ref{app:CSFQ}, separately. By introducing several pairs of creation and annihilation operators, the circuit Hamiltonian can be quantized as
\begin{equation}
\begin{split}
H_{\rm circuit}&=\omega_{a}a^{\dagger}a-g_{aT}\left(a-a^{\dagger})(T-T^{\dagger}\right)+\omega_{h}h^{\dagger}h-g_{hm}\left(h-h^{\dagger})(m-m^{\dagger}\right)\\
&+\omega_{r}r^{\dagger}r+\sum_j\omega_T^j\left|j\right>\left<j\right|+\sum_k\omega_m^k\left| k \right>\left< k\right|-g_{rm}\left(r-r^{\dagger})(m-m^{\dagger}\right)\\
&-g_{rT}\left(r-r^{\dagger})(T-T^{\dagger}\right)-g_{mT}\left(m-m^{\dagger})(T-T^{\dagger}\right),
\end{split}
\label{eq.Heff}
\end{equation}
where $a$ and $h$ represent the readout resonators, $r$ is the bus resonator, $\omega_T^j$ is the transmon bare energy at level $j$, and $\omega_m^k$  is the transmon bare energy at level $k$. Corresponding annihilation operators are defined as $m=\sum_k\sqrt{k+1}\left|k\right>\left<k+1\right|$  for CSFQ and  $T=\sum_j\sqrt{j+1}\left|j\right>\left<j+1\right|$ for transmon.

\begin{figure}[h!]

	\centering
	\includegraphics[width=0.9\columnwidth]{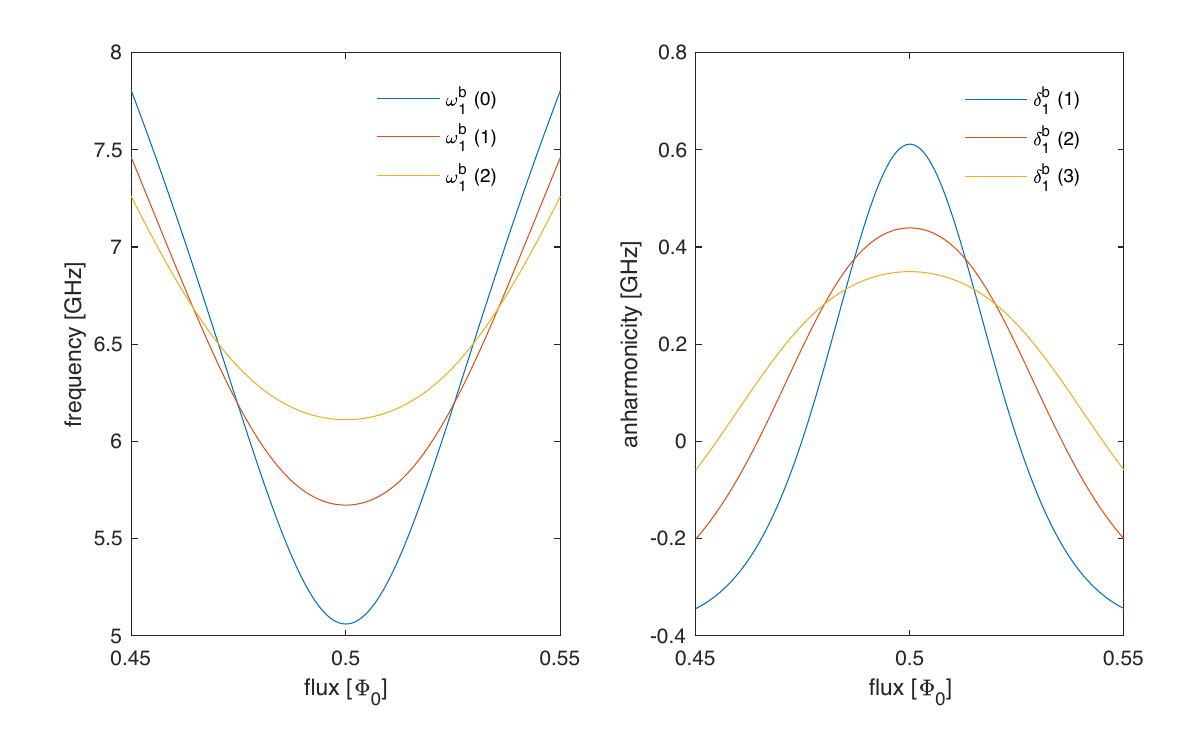}
	\put(-320,185){(a)}
	\put(-165,185){(b)}
		\vspace{-0.1in}
	\caption{Bare CSFQ frequency and anharmonicity versus flux. (a) The first three bare transition frequency $\omega_1^b(n)$ between levels $n$ and $n+1$. (b) The first three bare CSFQ anharmonicity $\delta_1^b(n)=\omega_1^b(n+1)-\omega_1^b(n)$.}
	\label{fig:freandanharm}
\end{figure}

We use the notation $\omega_1^b(j)$ to denote the bare transition frequency between the energy levels, $j+1$ and $j$ in the transmon; similarly, $\omega_2^b(k)$, the bare transition frequency between the energy levels, $k+1$ and $k$ in the CSFQ.  Figure~\ref{fig:freandanharm} shows the theoretical flux dependence of the bare CSFQ frequency and anharmonicity.

In the dispersive regime we first eliminate the readout resonators and then the bus, the Hamiltonian in Eq.~(\ref{eq.Heff}) reduces to qubit-qubit Hamiltonian. For simplicity, we replace the labels $j$, $k$ with $n_1$ for transmon and $n_2$ for CSFQ. The Hamiltonian in the form of multilevel systems is then written as 
\begin{equation}\label{eq:twoQ_Hamil}
\begin{split}
H_{\rm circuit}&=H_r+H_q\\
&=\tilde{\omega}_{r}r^{\dagger}r+\sum_{q=1,2}\sum_{n_q}{\omega}_q(n_q)\left|n_q\right>\left<n_q\right|+\sqrt{(n_1+1)(n_2+1)}\\
&\times J_{n_1,n_2}\left(\left|n_1+1,n_2\right>\left<n_1,n_2+1\right|+\left|n_1,n_2+1\right>\left<n_1+1,n_2\right|\right),
\end{split}
\end{equation}
where the dressed bus frequency is $\tilde{\omega}_r=\omega_r+\Sigma_q \chi_{n_q}^q\left|n_q\right\rangle \left\langle n_q\right|$ with $\chi$ being the dispersive AC-Stark shift\index{AC-Stark shift} of the resonator frequency. $\omega_q(n_q)$ is the energy difference between levels $n_q+1$ and $n_q$ for qubit $q$. $J_{n_1, n_2}$ is the virtual photon exchange rate calculated from $J_{n_1,n_2}=J^{\text{dir}}+J_{n_1,n_2}^{\text{indir}}$ with the direct coupling being $J^{\text{dir}}=g_{mT}$, and the indirect coupling $J_{n_1,n_2}^{\text{indir}}$ being~\cite{gambetta2013control}
\begin{equation}
\begin{split}
J_{n_1,n_2}^{\text{indir}}&=-\frac{g_{rm}g_{rT}}{2}\left(\frac{1}{\Delta_1(n_1)}+\frac{1}{\Delta_2(n_2)}+\frac{1}{\Sigma_1(n_1)}+\frac{1}{\Sigma_2(n_2)}\right) \\ 
\Delta_1(n_1)&=\omega_r-\omega_1(n_1), \\
 \Delta_2(n_2)&=\omega_r-\omega_2(n_2),\\
\Sigma_1(n_1)&=\omega_r+\omega_1(n_1), \\ 
\Sigma_2(n_2)&=\omega_r+\omega_2(n_2),
\end{split}
\end{equation}
where counter-rotating terms have already been included in the indirect coupling.

In the dispersive regime that $|\Delta|\gg J$, the Hamiltonian in Eq.~(\ref{eq:twoQ_Hamil}) can be fully diagonalized via SW transformation\index{Schrieffer-Wolff (SW) transformation} using a unitary operator\index{unitary operator} $U$ as:
\begin{equation}
\tilde{H_q}=U^{\dagger}H_q U=\sum_{q=1,2}\sum_{n_q}\tilde{{\omega}}_q(n_q)\left|n_q\right>\left<n_q\right|.
\label{eq.diagH}
\end{equation}
where $U$ can be derived from Eq.~(\ref{eq.h0s1}). The dressed qubit frequencies, anharmonicity, bare bus frequency, coupling strength, and two-photon exchange rate are presented in Tab.~\ref{table:freq}. Note that we define $\tilde{\omega}_q\equiv\tilde{\omega}_q(0)$.
\begin{table}[h!]
	 \centering
	\begin{tabular}{| l | l | l |}
		\hline
		Description & Symbol & Frequency  \\
		\hline \hline
		CSFQ bare freq.& $\omega_1^b(0)/2\pi$ & 5.0616 GHz \\ \hline
		CSFQ dressed  freq.& $\tilde{\omega}_1(0)/2\pi$ & 5.0511 GHz     \\ \hline
		CSFQ anharmonicity & $\delta_1/2\pi$ & +592.7 MHz \\ \hline
		CSFQ bare readout  freq. & $\omega_{h}/2\pi$ & 6.9065  GHz  \\ \hline
		CSFQ dressed readout  freq. & $\tilde{\omega}_{h}/2\pi$ & 6.9074 GHz   \\ \hline
		CSFQ-readout coupling & $g_{hm}/2\pi$ & 34 MHz  \\ \hline
		CSFQ-readout freq. shift & $\chi_{hm}/2\pi$ & 550 kHz  \\ \hline
		\hline
		Trans. bare  freq. & $\omega_2^b(0)/2\pi$ & 5.2920 GHz   \\ \hline
		Trans. dressed freq. & $\tilde{\omega}_2(0)/2\pi$ & 5.2855 GHz \\ \hline
		Trans. anharmonicity & $\delta_2/2\pi$ & -326.6 MHz \\ \hline
		Trans. bare readout freq. & $\omega_a/2\pi$ & 6.8050 GHz \\ \hline
		Trans. dressed readout freq.& $\tilde{\omega}_{a}/2\pi$ & 6.8059 GHz\\ \hline
		Trans.-readout coupling & $g_{aT}/2\pi$ & 36.2 MHz  \\ \hline
		Trans.-readout freq. shift & $\chi_{aT}/2\pi$ & 200 kHz  \\ \hline
		\hline
		Bus bare  freq.& $\omega_r/2\pi$ & 6.3062 GHz  \\ \hline
		Bus dressed  freq. & $\tilde{\omega}_r/2\pi$ & 6.3226 GHz \\ \hline
		Bus-Trans. freq. shift & $\chi_{rT}/2\pi$ & -2.2 MHz  \\ \hline
		Bus-CSFQ freq. shift & $\chi_{rm}/2\pi$ & 5.9 MHz \\ \hline
		Bus-CSFQ coupling & $g_{rm}/2\pi$ & 111.7 MHz  \\ \hline
		Bus-Trans. coupling & $g_{rT}/2\pi$ & 76.4 MHz   \\  \hline
		\hline
		virtual exchange coupling & $J_{00}/2\pi$ & 6.3  MHz \& fit\\ \hline
		direct coupling & $g_{mT}/2\pi$ & -2.7  MHz  \\ \hline
	\end{tabular}
\caption{\label{table:freq}Frequency scales on device with CSFQ at the sweet spot. Dressed qubit frequencies are measured from qubit spectroscopy, readout resonator frequencies and qubit-readout coupling strengths are from resonator measurement. Bare frequencies are calculated using the same method in Ref.~\cite{gely2018nature}. Bus dressed frequencies and qubit-bus coupling strengths are from bus spectroscopy~\cite{sheldon2017characterization}. Virtual exchange rate is extracted from CSFQ spectroscopy as well as theory fit, and direct coupling is estimated using Eq.~(\ref{eq.g12}). }
\end{table}

\begin{figure}[h!]
	\centering
	\labellist
	\bfseries
	\pinlabel (a) at 15 465
	\pinlabel (b) at 335 465
	\endlabellist
	\includegraphics[width=0.85\textwidth]{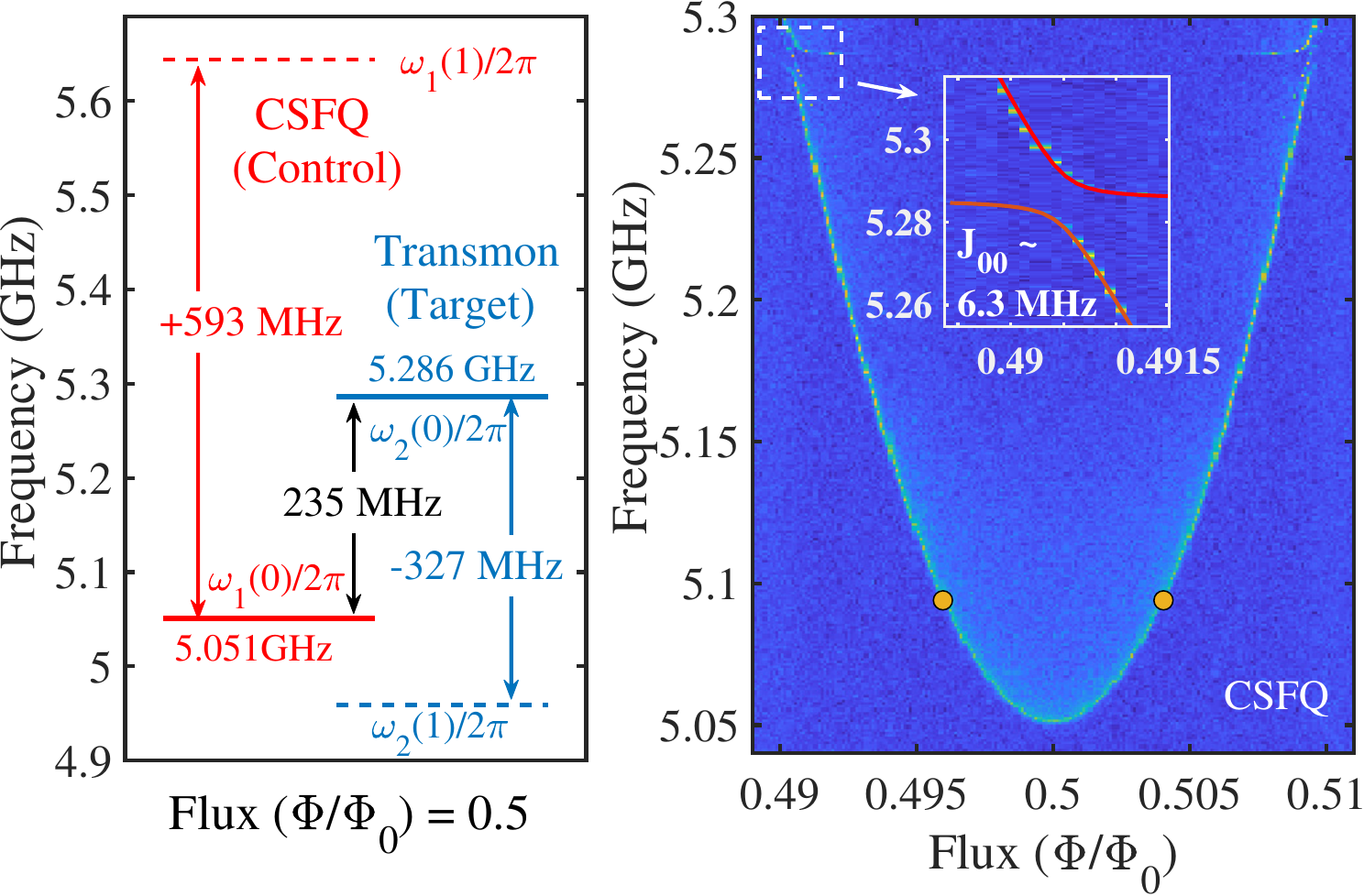}	
\caption{\label{fig:schematic} (a) Frequency diagram of transmon and CSFQ at flux sweet spot in the hybrid circuit. (b) CSFQ qubit frequency spectrum vs. external magnetic flux. Orange dots indicate flux points where static ZZ becomes zero. (Inset) Anti-crossing of transmon and CSFQ with fit (red solid line).}
\end{figure}

To illustrate two-qubit levels in the computational subspace, we plot the frequency diagram in Fig.~\ref{fig:schematic}(a). The measured primary qubit frequency and anharmonicity  are 5.051~GHz and $+593$~MHz for the CSFQ at SS\index{sweet spot (SS)}, 5.286~GHz and $-327$~MHz for the transmon, and thus the qubit-qubit detuning is 235~MHz. The flux dependence of the CSFQ spectrum in Fig.~\ref{fig:schematic}(b) allows us to explore a range of qubit-qubit detuning in the following experiments. By tuning CSFQ frequency to be exactly the same as transmon, we can extract $J_{00}$ coupling within the computational subspace. This is easy to be proved by truncating the Hamiltonian in Eq.~(\ref{eq:twoQ_Hamil}) to a $4\times4$ matrix  
\begin{equation}
H_q=\left(\begin{array}{cccc}
	0 & 0 & 0 &0\\
	0 & \omega_1 & J_{00} & 0\\
	0 & J_{00} & \omega_2 & 0\\
	0 & 0 & 0 & \omega_1+\omega_2
	\end{array}\right).
\end{equation}
Please note that this truncation does not influence the eigenvalues of $|01\rangle$ and $|10\rangle$ since all interaction has been considered. In the dressed basis eigenvalues can easily be found as 
\begin{equation}\label{eq.Jde}
\tilde{\omega}_{01/10}=\frac{1}{2}\left( \omega_1+\omega_2\pm\sqrt{(\omega_1-\omega_2)^2+4J_{00}^2}\right). 
\end{equation}
When $\omega_1=\omega_2$ the difference between the two levels is $2J_{00}$. We fit the anti-crossing between CSFQ and transmon in \ref{fig:schematic}(b) (inset) to obtain the zeroth-order exchange coupling strength $J_{00}= 6.3$~MHz. 

\begin{figure}[h!]
	\centering
	\includegraphics[width=0.49\columnwidth]{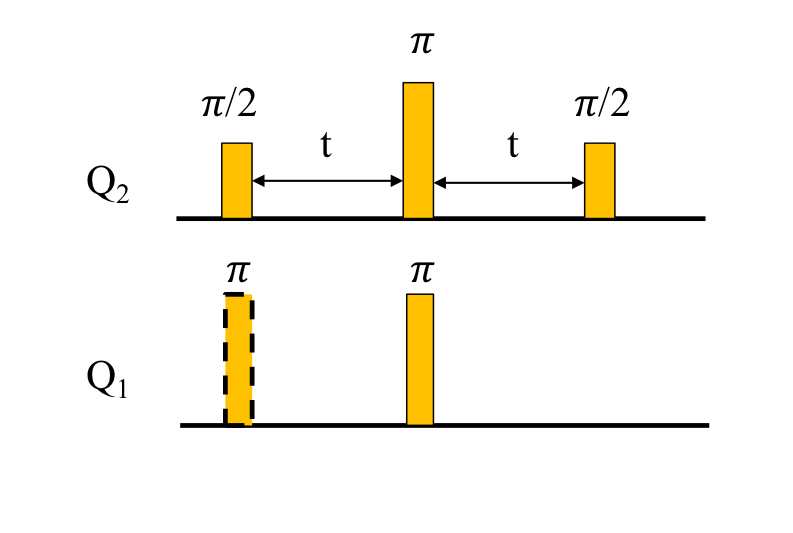}
	\includegraphics[width=0.49\columnwidth]{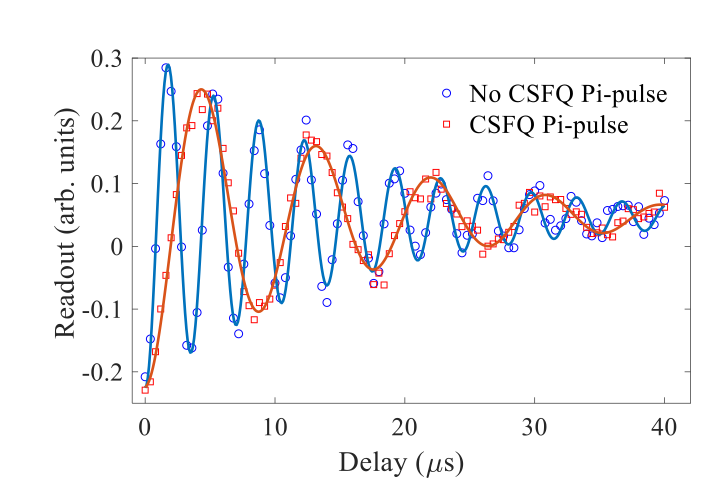}
	\put(-350,105){(a)}
	\put(-190,105){(b)}
	\vspace{-0.1in}
	\caption{\label{fig:JAZZ} (a) Joint Amplification of ZZ (JAZZ)~\cite{gambetta2012characterization} pulse sequences: Ramsey on one qubit with echo pulses on both qubits. (b) Measured frequencies with (without) $\pi$ rotation on control qubit CSFQ at the sweet spot.}
\end{figure}

In a CSFQ-transmon circuit the second term of static ZZ interaction in Eq.~(\ref{eq.zeta}) can be negative, due to the large and positive anharmonicity of the CSFQ. Note that for a hybrid CSFQ-transmon device, $J_{01}$ is quite different from $J_{10}$, in contrast to a transmon-transmon pair where they are almost the same. This allows the hybrid CSFQ-transmon pair to be free of ZZ interaction. This interaction is measured using a Joint Amplification of ZZ (JAZZ) experiment~\cite{gambetta2012characterization}, which is a modified Bilinear Rotational Decoupling (BIRD) sequence~\cite{garbow1982bilinear}. The standard BIRD sequence used in nuclear magnetic resonance (NMR) is a Ramsey\index{Ramsey interferometry} experiment on one qubit, with echo pulses on both the measured qubit and the coupled qubit. In the JAZZ experiment, this sequence is performed twice, for each initial state of the control qubit~\cite{takita2017experimental}. Static ZZ $\zeta$ is equal to the frequency difference found between the two experiments. The JAZZ experiment is shown in the Fig.~\ref{fig:JAZZ}(a) and measured two frequencies are shown in Fig.~\ref{fig:JAZZ}(b).

\begin{figure}[h!]
	\centering
	\includegraphics[width=0.85\columnwidth]{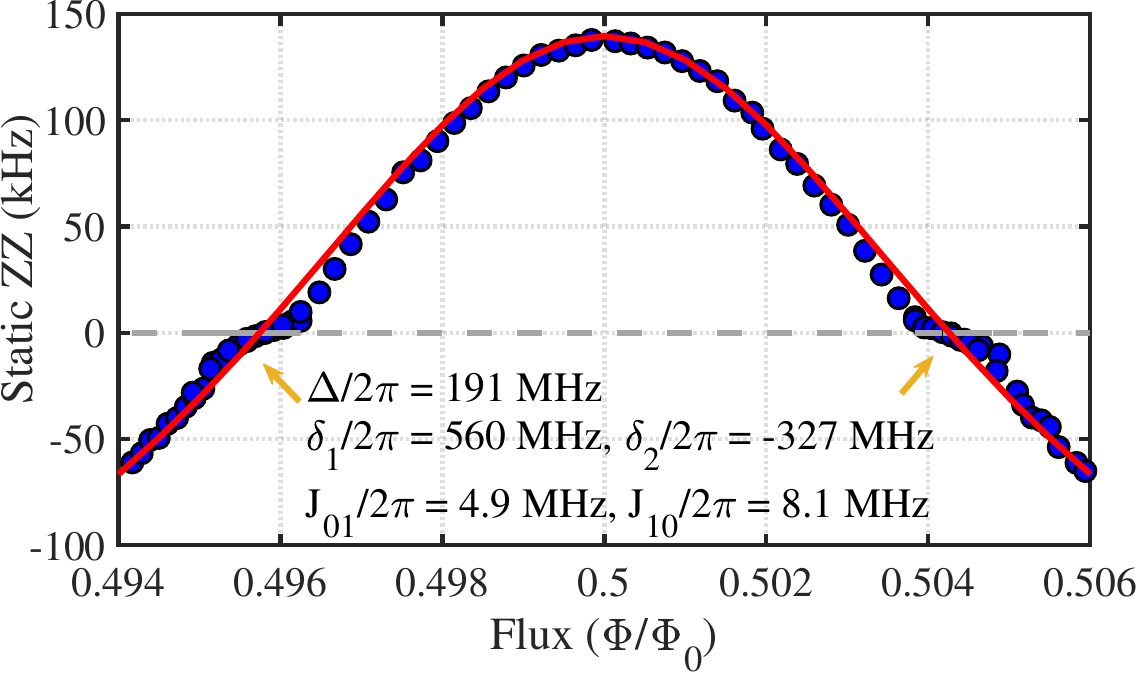}
	\caption{\label{fig:ZZ_flux} Static ZZ measured as a function of flux. Error bars are comparable with or smaller than the size of the data symbols. Red solid line represents a theory calculation using Eq.~(\ref{eq.zeta}). Static ZZ becomes zero at two flux points (0.496, 0.504), and the corresponding device parameters are shown in the plot.}
\end{figure}

Based on the measured and estimated parameters, static ZZ interaction has been extracted and fitted in Fig.~\ref{fig:ZZ_flux}. The maximum ZZ is around 140~kHz at the SS, away from this point it decreases and eventually crosses zero symmetrically near $\Phi/\Phi_0$ = 0.496 and 0.504. Corresponding circuit parameters are listed in the plot. It is worth noting that $J_{10}$ is much larger than $J_{01}$ in the hybrid circuit. Our theory agrees well with the experiment except the vicinity of zero-crossing points, where experimental data exhibit a kink. We speculate the kink could be caused by the breakdown of perturbation theory used in Eq.~(\ref{eq.zeta}). CSFQ frequency goes higher when the flux is away from the SS, in other words, qubit-qubit detuning decreases while $J_{10}$ increases, thus pushing the ratio $J/\Delta$ beyond the dispersive regime. A framework for treating such situation is discussed in Ref.~\cite{ansari2019superconducting}.

So far, we have studied the origin of parasitic static ZZ interaction and the possibility of freeing qubits from it. We show that the static ZZ freedom can take place in two-types of circuits: 1) In transmon-transmon pair zero ZZ is found with no entanglement by canceling indirect coupling\index{indirect coupling} via increasing direct coupling\index{direct coupling}; 2) In CSFQ-transmon pair, eliminating ZZ interaction is realized by combining opposite sign anharmonicity meanwhile keeping two qubits entangled. We also use our theory to model the experimental hybrid  CSFQ-transmon circuit and demonstrate the presence of static ZZ freedom. In the next chapter, we will study how the ZZ interaction impacts quantum gates when qubits are driven.

\chapter{Cross-Resonance Gate and Dynamical ZZ Freedom}
\label{c4}
Static ZZ interaction is one of the major problems to prevent fast and  high-performance two-qubit entanglement such as MAP gate~\cite{chow2013microwave}, iSWAP gate~\cite{mckay2016universal} and cross-resonance gate~\cite{sheldon2016procedure} in superconducting circuits. In last chapter we discussed the cancellation of static ZZ in several devices. This chapter we focus on the cross-resonance gate and show how the ZZ freedom can impact the gate fidelity. The main content is from the other parts of our two publications Ref.~\cite{ku2020suppression} published in Physical Review Letters and  Ref.~\cite{xu2021zz-freedom} published in Physical Review Applied.
\section{Cross Resonance Gate}
A cross-resonance (CR) gate\index{CR gate} is enabled by driving the control qubit at the frequency of the target qubit, and this CR drive causes the target qubit state to rotate in a direction depending on the control qubit state, and thus corresponds to a ZX gate~\cite{chow2011simple}. This is a general method to implement CNOT gate\index{CNOT gate} by $\rm CNOT=ZX_{\pi/2}IX_{-\pi/2}ZI_{-\pi/2}$ with additional single-qubit rotations. The CR driving Hamiltonian needs to be transferred to the same regime as the the qubit Hamiltonian in Eq.~(\ref{eq.diagH}). In the dressed frame it is written as
\begin{equation}
\tilde{H_d}=U^{\dagger}H_{d}U=\Omega\cos(\omega_d t) \sum_{n_1} U^{\dagger}(\left|n_1\right\rangle \left\langle n_1+1\right|+\left|n_1+1\right\rangle \left\langle n_1\right|)U,
\end{equation}
where $U$ is a unitary operator that diagonalizes the circuit Hamiltonian to Eq.~(\ref{eq.diagH}), $\Omega$ is the driving amplitude at driving  frequency $\omega_d$. Now let us move into the rotating frame at the frequency  $\omega_d$ on both qubits and ignore the counter rotating terms, total Hamiltonian then becomes
\begin{equation}
H_r=W^{\dagger}(\tilde{H_q}+\tilde{H_{ d}}) W-i W^{\dagger}\dot{W},
\label{eq.Hr}
\end{equation}
where $W=\sum_n\exp(-i\omega_{ d} t \hat{n})\left|n\right\rangle \left\langle n\right|$ with $\omega_d={\tilde
\omega}_2$. External pulses sometime are strong enough such that $0.1<\Omega/\Delta<1$ or even $\Omega/\Delta>1$ with $\Delta$ being the qubit-qubit detuning.  Since effective Hamiltonian approach is valid only in the dispersive regime, as shown in Ref. \cite{ansari2019superconducting}, and cannot be performed on qubits in the limit of strong drives.  Therefore we should treat block diagonalization of driving Hamiltonian with a nonperturbative method, other than SW.

\subsection{Principle of Least Action}
One approach to explore strong driving impacts is the so called ``least action method'' described in Ref. \cite{cederbaum1989block,magesan2020effective}.  This method aims to find a unitary operator $T$ that is closest to the identity operation and can perform a block diagonalization. The least action unitary operator $T$ that satisfies $H_{BD}=T^{\dagger} H_r T$ is given by
\begin{equation}
T=S S_{BD}^{\dagger}S_P^{-\frac{1}{2}},
\label{eq.leastaction}
\end{equation}
where $S$ is the nonsingular eigenvector matrix of $H_r$, $S_{BD}$ is the block-diagonal matrix of $S$, and $S_P=S_{BD}S_{BD}^{\dagger}$.  However, only part of the full matrix is required when block diagonalizing a Hamiltonian, and $T$ can be simplified as follows:  given a general $d\times d$ Hermitian Hamiltonian, it can be transformed into a block diagonal matrix $\mathscr{H}$ using Eq. (\ref{eq.leastaction})  with  two blocks $\mathscr{H}_{nn}$ and $\mathscr{H}_{mm}$, where $n$ and $m$ are the dimensions of the blocks and satisfy $d=n+m$ and $n\leq m$. The corresponding matrix of all eigenvectors $S$ can also be divided into 4 blocks, namely
\beq
S =
\begin{pmatrix}
	S_{nn} &S_{nm}\\
	S_{mn}& S_{mm}
\end{pmatrix}.
\eeq
As described in Ref. \cite{cederbaum1989block}, the unitary transformation $T$ can be simplified as $T=U(U^{\dagger}U)^{-1/2}$ where
\beq
U=
\begin{pmatrix}
	1 &X\\
	-X^{\dagger}& 1
\end{pmatrix},  U^{\dagger}U=\begin{pmatrix}
	1 +X X^{\dagger}&0\\
	0& 1+X^{\dagger}X
\end{pmatrix},
\eeq
with  $X=-(S_{mn}S_{nn}^{-1})^{\dagger}=S_{nm}S_{mm}^{-1}$. Therefore, only part of the eigenvector matrix $S$ is needed, for example, the first $n$ eigenvectors. 

\subsection{Effective CR Hamiltonian}
To study the impact of the driving pulse on the ZX gate, we block-diagonalize the total Hamiltonian in the rotating frame into two individual qubit blocks within the computational subspace and one block for all higher excited levels  to decouple them. In the case that driving amplitude is large enough to go beyond the dispersive regime, the block diagonalization should be applied under the principle of least action: First two blocks $4\times4$ and $11\times11$, and then repeat this approach in the $4\times4$ matrix to divide it into two 2$\times$2 blocks, each associated to a qubit. Finally, the effective CR Hamiltonian in the computational subspace can be written as 
\begin{equation}
H_{\rm CR}=\alpha_{\rm ZI}\frac{\rm ZI}{2}+\alpha_{\rm IX}\frac{\rm IX}{2}+\alpha_{\rm ZX}\frac{\rm ZX}{2}+\alpha_{\rm ZZ}\frac{\rm ZZ}{4}.
\label{eq.Hcr1}
\end{equation}
For a CR gate, only ZX term in the Hamiltonian is desired, the others such as ZZ, IX, and ZI are unwanted interactions. ZI and IX can be cancelled out by echoed CR sequences~\cite{corcoles2013process}, leaving ZX term accompanied with accumulated phase from ZZ interaction, resulting in the rotation of the target qubit with errors on the Bloch sphere.
On top of the static ZZ interaction $\zeta$, which solely comes from the contribution of higher excitation in the qubit-qubit interaction,  the CR drive with the amplitude $\Omega$ introduces an additional ZZ interaction that depends quadratically on $\Omega$. The two together produce the total ZZ interaction as follows
\begin{equation}
\alpha_{\rm ZZ}=\zeta + \eta\Omega^2+O(\Omega^3),
\label{eq.totZZ}
\end{equation}
where $\eta\Omega^2$ is what we refer to as the {\it dynamical ZZ interaction}. 

\subsection{Classical Crosstalk and Active Cancellation}
\label{sec:ccac}
When applying microwave on one qubit, some additional classical changes happen to another qubit, which is undesirable and should be cancelled, this is called {\it classical crosstalk}. In a CR gate, classical crosstalk is assumed to produce new unwanted interactions due to targeted drives on non-local elements. The new unwanted terms have been demonstrated to be existing in experiments~\cite{sheldon2016procedure}. In the presence of classical crosstalk\index{classical crosstalk}, the normal driving Hamiltonian is modified by adding another term as
\begin{equation}\label{eq:cc_driving_Hamil}
\begin{split}
H_{d}^{\rm ct}&=\Omega\cos(\omega_{ d} t+\phi_0)\sum_{n_1}(\left|n_1\right\rangle \left\langle n_1+1\right|+\left|n_1+1\right\rangle \left\langle n_1\right|)\\
&+R\Omega\cos(\omega_{ d} t+\phi_0+\phi_R)
\sum_{n_2}(\left| n_2 \right\rangle \left\langle n_2+1\right|+\left|n_2+1\right\rangle \left\langle n_2\right|),
\end{split}
\end{equation}
where $R$ is a scaling factor for classical crosstalk amplitude, and depends on both two-qubit gate length and flux. $\phi_0$ is the new phase of the CR drive to the control qubit and $\phi_R$ is the phase to the target qubit. When the Hamiltonian with classical crosstalk is taken to the dressed frame and block diagonalized, one can find the new terms IY and ZY in the effective CR Hamiltonian:
\begin{equation}
H_{\rm CR}^{\rm{ct}}=\beta_{\rm ZI}\frac{\rm ZI}{2}+\beta_{\rm ZX}\frac{\rm ZX}{2}+\beta_{\rm ZY}\frac{\rm ZY}{2}+\beta_{\rm IX}\frac{\rm IX}{2}+\beta_{\rm IY}\frac{\rm IY}{2}+\beta_{\rm ZZ} \frac{\rm ZZ}{4}.
\label{eq.ctp}
\end{equation}

\begin{figure}[h!]
	\centering
	\includegraphics[width=0.8\textwidth]{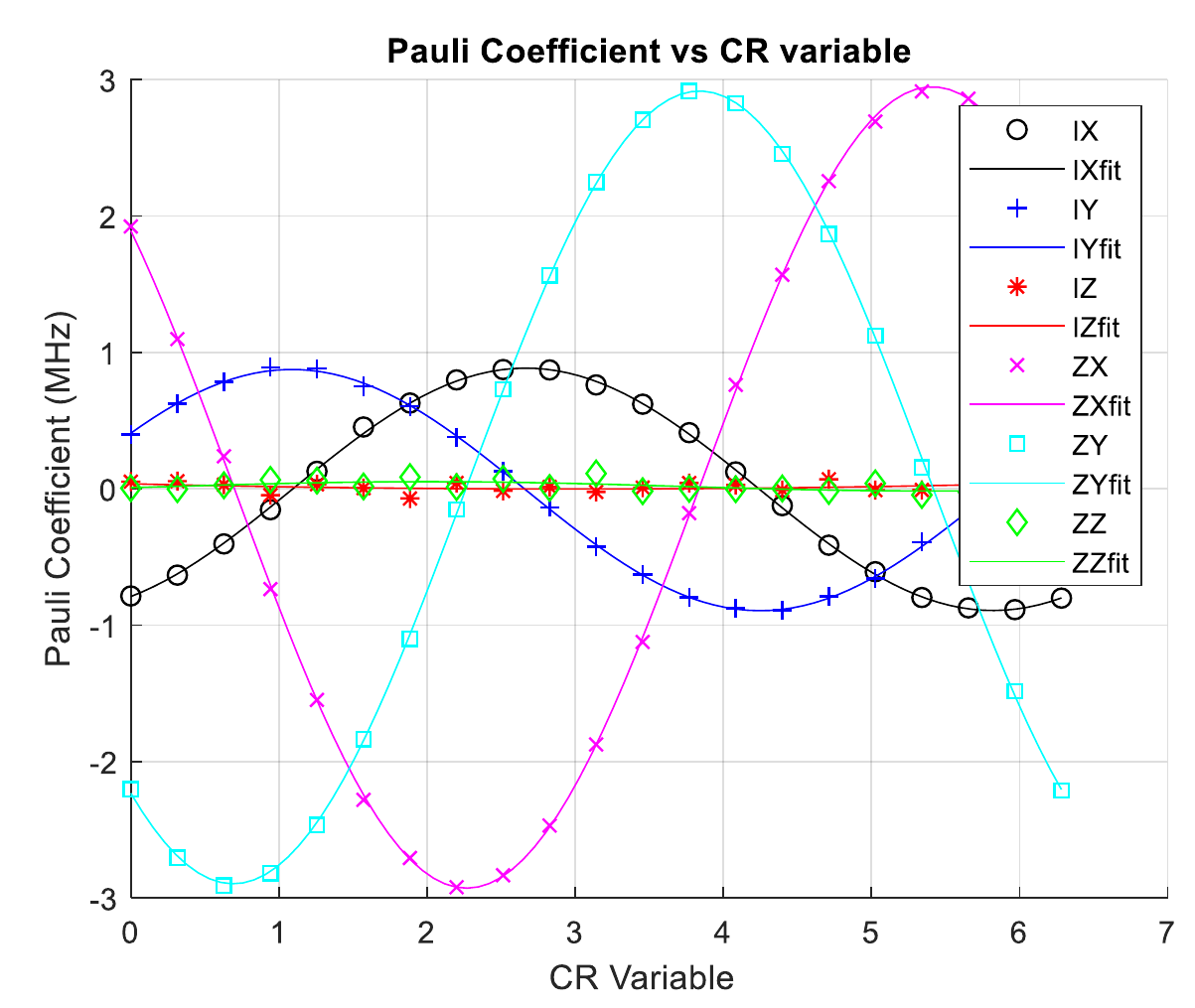}
	\vspace{-0.1in}
	\caption{Pauli coefficients as a function of CR variable extracted from quantum state tomography (dotted) and fitted using effective Hamiltonian (solid) for the hybrid circuit.}
	\label{fig:CRphase}
\end{figure}

Our collaborators also applied CR drive on the CSFQ to perform the CR gate and investigate the characteristic behavior of the CR effect. Here we present the theory analysis of the same chip under CR drive: CSFQ is the control qubit while transmon is the target qubit. On the experimental chip,  an external flux threads CSFQ and sometime is tuned away from the SS. By ignoring higher order perturbative results in $\Omega$, e.g. $O(\Omega^4)$, Pauli coefficients can be approximated as 
\begin{equation}\label{eq:Pauli_coeff}
\begin{split}
\beta_{\rm ZX}&\approx(B_f\Omega+C_f\Omega^3)\cos\phi_0,\\
\beta_{\rm ZY}&\approx(B_f\Omega+C_f\Omega^3)\sin\phi_0,\\
\beta_{\rm IX}&\approx(D_f\Omega+E_f\Omega^3)\cos\phi_0+RK_f\Omega\cos(\phi_0+\phi_R)\\
\beta_{\rm IY}&\approx(D_f\Omega+E_f\Omega^3)\sin\phi_0+RK_f\Omega\sin(\phi_0+\phi_R)\\
\beta_{\rm ZI}&\approx\alpha_{\rm ZI},\\
\beta_{\rm ZZ}&\approx\alpha_{\rm ZZ},
\end{split}
\end{equation}
where $B_f$, $C_f$, $D_f$, $E_f$  are flux-dependent quantities that can be evaluated numerically. The phase shift can be extracted using quantum state tomography. One intermediate result during the experiment is shown in Fig.~\ref{fig:CRphase}. 

All Pauli coefficients at a certain driving amplitude are extracted by varying the CR phase $\phi_x$ added on top of $\phi_0$. Note that the phase difference between IX and IY, ZX and ZY remains $\pi/2$ which is consistent with Eq.~(\ref{eq:Pauli_coeff}).  $\phi_0$ can be find at which the ZX component is maximized and ZY is zero. When IY component is zero, the phase $\phi_R$ is also extracted by setting $\beta_{\rm IY}=0$. These terms except desired ZX will accumulate errors, therefore must be eliminated in the process. $\phi_0$ can be calibrated to $0$ or $\pi$ for achieving the maximum. IX and IY can be removed by either applying an active cancellation pulse on the target qubit~\cite{sheldon2016procedure} or optimized rotary echoes~\cite{sundaresan2020reducing}. Here we introduce the theory behind the active cancellation. 

Active cancellation is accomplished by introducing a second microwave drive tone on the target qubit. The new pulse has the similar form as classical crosstalk, i.e. $A\cos(\omega_d t+\phi)\sum_{n_2}(\left| n_2 \right\rangle \left\langle n_2+1\right|+\left|n_2+1\right\rangle \left\langle n_2\right|)$, and also adds new components on top of Pauli coefficients. By assuming $\phi_0$ is calibrated to $\pi$, then ZY term vanishes and ZX is maximum negative. The additional pulse on the target qubit is required to cancel the existing IX and IY terms, namely the following conditions should be satisfied
\begin{equation}
\begin{split}
(D_f\Omega+E_f\Omega^3)+RK_f\Omega\cos\phi_R&=A K_f\Omega\cos\phi,\\
RK_f\Omega\sin\phi_R&=AK_f\Omega\sin\phi.
\end{split}
\end{equation}
The next step is to set proper amplitude of the cancellation pulse to remove both IX and IY simultaneously. In experiment, this is realized by sweeping the cancellation pulse amplitude. If IX and IY are not zero at the same cancellation amplitude  then the cancellation phase is incorrect and has to be calibrated again~\cite{sheldon2016procedure}. 

\begin{figure}[h!]
	\centering
	\includegraphics[width=0.8\textwidth]{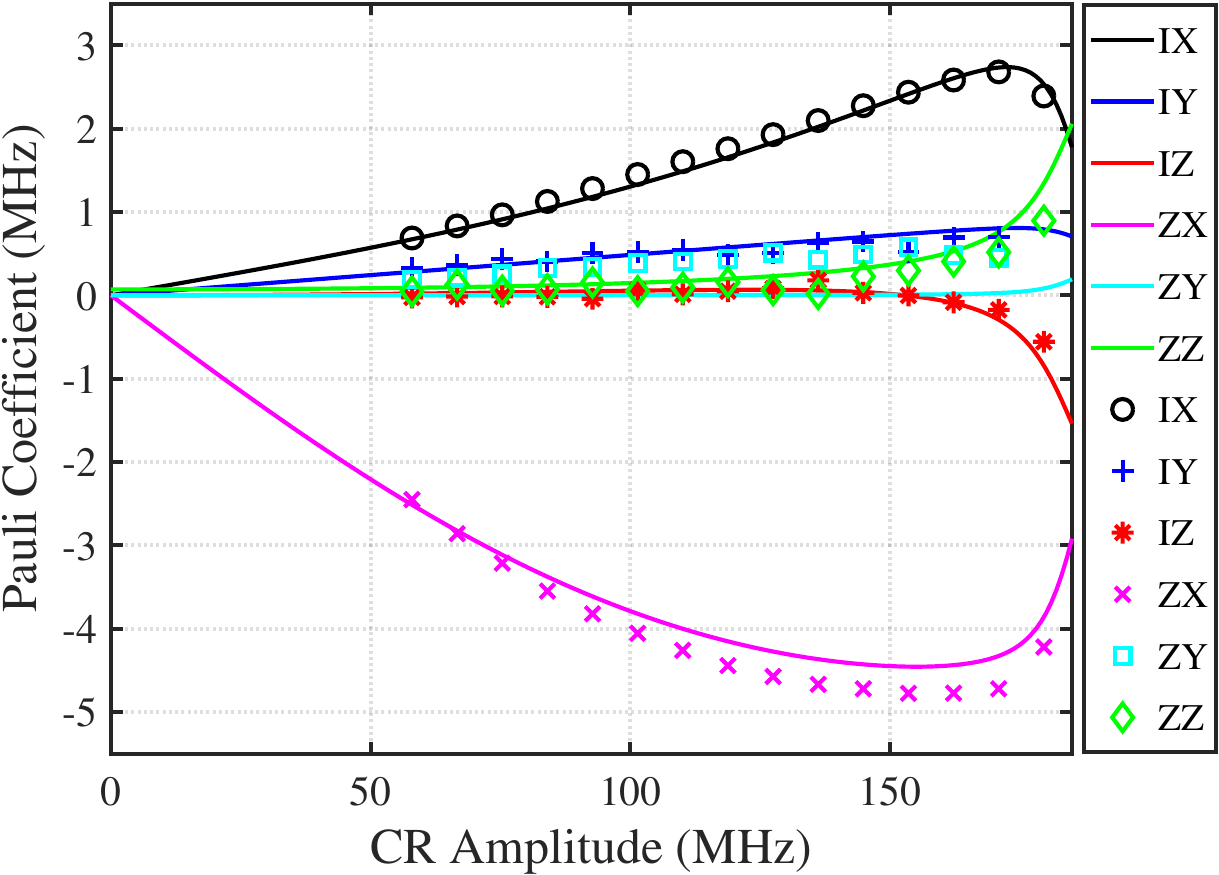}
	\vspace{-0.1in}
	\caption{Pauli coefficients vs. CR amplitude at the sweet spot: experimental tomographic measurements (points) and theoretical curves (solid lines). The parameters in Eq.~(\ref{eq:cc_driving_Hamil}) used for simulation are $R=0.0125$, $\phi_0=\pi$ and $\phi_R=\pi+0.4$.}
	\label{fig:paulicoeff}
\end{figure}

From Fig.~\ref{fig:CRphase} the phase of CR drive to the control qubit can be calibrated as $\phi_0=\pi$ and $\phi_R$ is measured as $\phi_R=\pi+0.4$. Due to the limit of experimental apparatus, active cancellation pulse was not performed in the final experiment, therefore IX and IY will be kept in the Hamiltonian. Using quantum state tomography experimental Pauli coefficients at the sweet spot\index{sweet spot (SS)} are plotted as a function of driving amplitude in Fig.~\ref{fig:paulicoeff}, together with our theory prediction. One can see that the unwanted ZY vanishes in the device as expected. It is also clear to see that components IZ and ZZ are small and insensitive to weak CR amplitude. The measured Pauli coefficients ZX and ZZ are consistent with the results predicted by the effective Hamiltonian~(\ref{eq.ctp}).
 
In brief, the active cancellation process includes two parts: phase calibration can effectively eliminate ZY term, while active cancellation pulse applied on the target qubit can totally remove both IX and IY terms, leaving the effective CR Hamiltonian only with Pauli components ZX and ZZ. 

\subsection{Echoed CR Rate}
\begin{figure}[h!]
	\centering
	\includegraphics[width=0.65\textwidth]{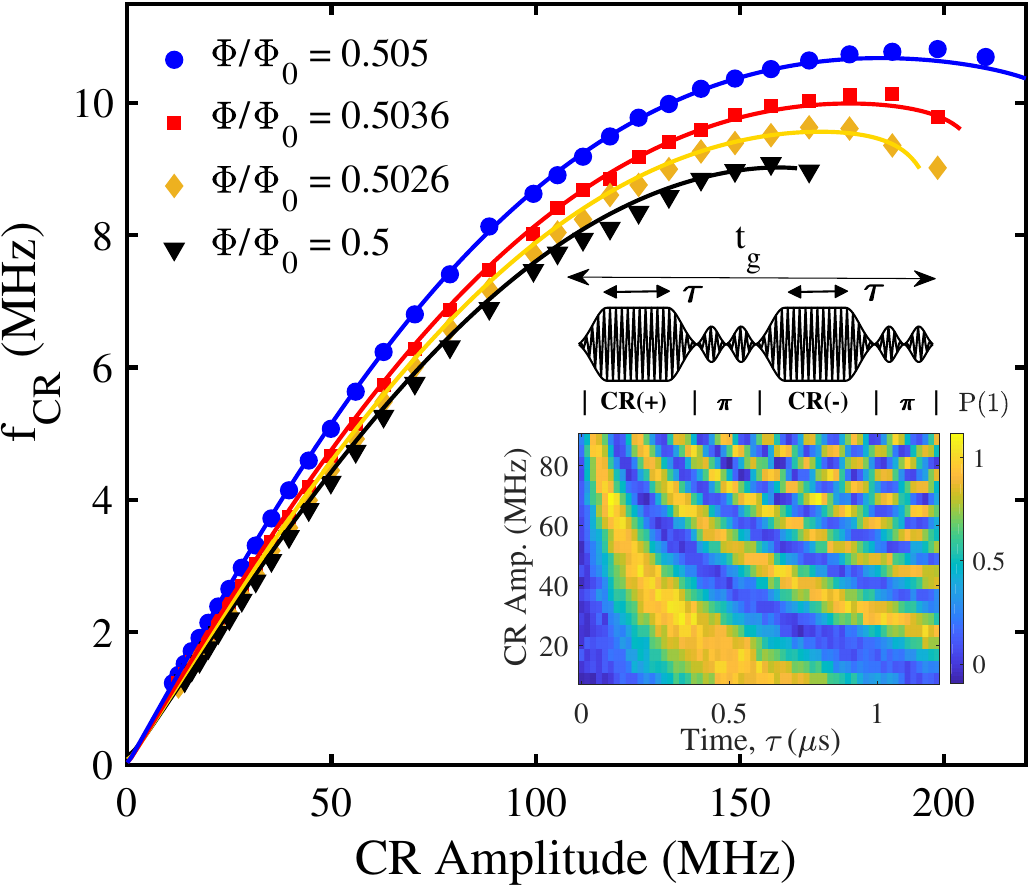}
	\vspace{-0.1in}
	\caption{Echoed CR rate vs. CR amplitude at four representative flux points. Device parameters are shown in Tab.~\ref{table:freq}. The corresponding qubit-qubit detunings are (234, 217, 166) in MHz. Solid lines correspond to theoretical model.  (Inset) Color density plot of the oscillation of target qubit driven with various CR amplitudes at sweet spot with CSFQ (control) in ground state. Colorbar represents first excited state probability of target qubit. Echoed CR pulse sequence illustrated above inset plots.}
	\label{fig:CR_flux}
\end{figure}

Although active cancellation can remove ZY, IX and IY terms, ZI remains large in the Hamiltonian~(\ref{eq.ctp}). The standard way to eliminate such a term is to use echoed CR pulse~\cite{corcoles2013process}, which involves two Gaussian flat-top CR pulses\index{Gaussian pulse}  with $\pi$ phase difference, and a $\pi$-pulse on the control qubit after each CR pulse as shown inset of Fig.~\ref{fig:CR_flux}. If the ZZ interaction can also be eliminated, then an ideal CR gate will effectively behave like a two-qubit gate corresponding to $\rm ZX_{\theta}$~\cite{gambetta2013control}:
\begin{equation}
{\rm ZX}_{\theta}=\exp{\left[-i \theta \left(\rm ZX/2\right)\right]}=\left(\begin{array}{cccc}
\cos(\theta/2)&-i\sin(\theta/2)&0&0\\
-i\sin(\theta/2)&\cos(\theta/2)&0&0\\
0&0&\cos(\theta/2)&i\sin(\theta/2)\\
0&0&i\sin(\theta/2)&\cos(\theta/2)\\
\end{array}\right).
\end{equation}
By choosing $\theta=\pi/2$ together with single qubit rotations, the CNOT gate can be realized.  However in reality the ZZ interaction is always on, the frequency of the echoed CR oscillation, i.e. $f_{\rm ECR}$, thus can be determined from the following relation:
\begin{equation}
\begin{split}
f_{\rm ECR}&=\sqrt{\left(\beta_{\rm ZX}+\beta_{\rm IX}\right)^2+\left(\beta_{\rm ZY}+\beta_{\rm IY}\right)^2+\left(\beta_{\rm ZZ}/2\right)^2}\\
&+\sqrt{\left(\beta_{\rm ZX}-\beta_{\rm IX}\right)^2+\left(\beta_{\rm ZY}-\beta_{\rm IY}\right)^2+\left(\beta_{\rm ZZ}/2\right)^2}.
\end{split}
\label{eq.fCR}
\end{equation}
If all unwanted terms are eliminated, Eq.~(\ref{eq.fCR}) reduces to 2$\rm \beta_{ZX}$. Figure~\ref{fig:CR_flux} shows the experiential oscillation frequency $f_{\rm ECR}$ of the transmon as a function of the CR amplitude at different flux bias $\Phi/\Phi_0$'s.

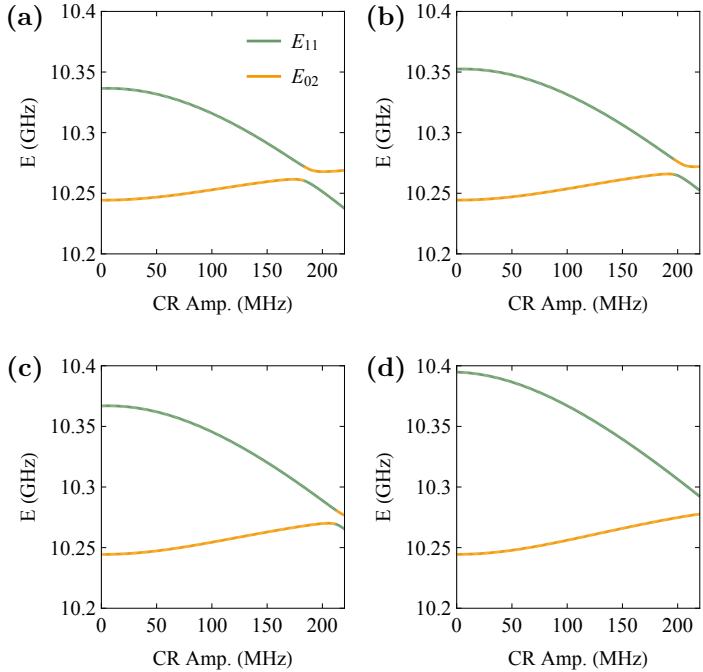
\begin{figure}[h!]
	\labellist
	\bfseries
	\pinlabel (a) at 5 380
	\pinlabel (b) at 215 380
	\pinlabel (c) at 5 175
	\pinlabel (d) at 215 175
	\endlabellist
	\centering
	\includegraphics[width=0.72\textwidth]{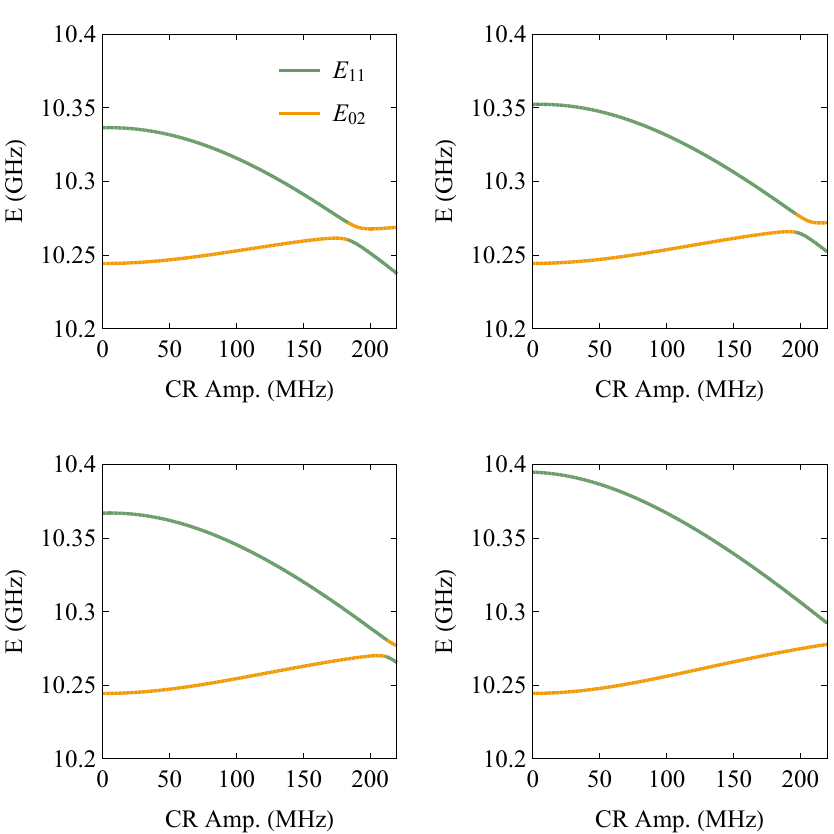}
	\vspace{-0.1in}
	\caption{Dressed energy levels $E_{11}$ and $E_{02}$ vs. CR amplitude for flux points: (a) $f=0.5$, (b) $f=0.5026$, (c) $f=0.5036$, and (d) $f=0.505$. Device parameters are shown in Tab.~\ref{table:freq}.}
	\label{fig:energy1102}
\end{figure}

\vspace{-0.05in}
At weak driving amplitude, $f_{\rm ECR}$ increases almost linearly, which corresponds to the first order perturbation theory. And by increasing $\Omega$, $f_{\rm ECR}$ deviates from linearity due to the off-resonant drive on CSFQ~\cite{chow2011simple}. As CR amplitude becomes stronger,  increase of $f_{\rm ECR}$ slows down, eventually changes the sign of slope to the opposite. This is because the energy levels $E_{11}$ and $E_{02}$ get closer to each other and at certain driving amplitude the anti-crossing takes place as shown in Fig.~\ref{fig:energy1102}. The resulting theoretical curves for $f_{\rm ECR}$ vs. CR amplitude under the principle of least action are consistent with the experimental points.

The $\rm ZX_{90}$ gate produces a Rabi oscillation occurred on the target, where the axis of oscillation will depend on the state of the control. The time length of CR pulse and the oscillation frequency of the target qubit satisfy
\begin{equation}
(2\pi f_{\rm ECR})\tau = \pi/2,
\label{eq.timelength}
\end{equation}
where $\tau$ is the total time length of the square echoed CR pulse. In experiment, one can use Gaussian flat-top CR pulses\index{Gaussian pulse} with rising and falling edges, where $\tau_0$ is defined to be the length of the flat-top part of each CR pulse. This $\pi$ rotation includes 20 ns derivative removal via adiabatic gate (DRAG)\index{DRAG} pulse, and is buffered by two 10-ns delays. The CR pulses are rounded square with equivalent rise and fall times of 20 ns. Thus we define the two-qubit gate length to be $t_g=2\tau_0+160$~ns, where $\tau_0$ is the average flat-top length of each CR pulse. Due to the finite rise and fall time, we have $\tau>\tau_0$, e.g., for $\tau_0=0$, $\tau$ is non-zero. In the weak driving regime, the linear oscillation frequency can be expressed as 
\begin{equation}
f_{\rm ECR} \approx \gamma(f)\Omega,
\end{equation}
with a flux-dependent coefficient $\gamma(f)$, e.g., $\gamma(0.5) \approx 0.1$. The exact flux-dependent $\gamma(f)$ can be found from Eq.~(\ref{eq.fCR}). By substituting this approximation into Eq.~(\ref{eq.timelength}), we obtain the following expression for the CR amplitude to make a $\rm ZX_{90}$ gate:
\begin{equation}\label{eq:Omega}
\Omega(f, \tau)=1/[4\gamma(f)\tau].
\end{equation}

\section{Two-qubit Gate Error}
To qualify the gate performance, we compute two-qubit error per gate in an echoed CR pulse sequence for implementing a $\rm ZX_{90}$ gate by considering the density matrix starting in the ground state in the Pauli basis.  CR gate errors can come from several aspects: (1) Limited coherence times compared to operation time. (2)  Classical crosstalk. (3) Parasitic ZZ interaction. Here, ZZ interaction is assumed to be a global error, and for each time step we apply actual gate unitary transformation as well as decoherence terms. The total map following the gate sequences is
\begin{equation}
\rho_{f}=\Lambda_{T1,T2,Q1}\circ\Lambda_{T1,T2,Q2} \circ\Lambda_{\rm ZZ}\circ \Lambda_{\rm XI} \circ \Lambda_{\rm CR-}\circ \Lambda_{\rm XI}\circ \Lambda_{\rm CR+}[\rho_i],
\end{equation}
and each map is defined by~\cite{mckay2019three-qubit}
\begin{equation}
	\begin{split}
		\Lambda_{\rm ZZ}[\rho]&=U_{\rm ZZ} \cdot \rho \cdot U_{\rm ZZ}^{\dagger},\\
		\Lambda_{\rm XI}[\rho]&=\rm XI \cdot \rho \cdot \rm XI ,\\
		\Lambda_{\rm CR\pm}[\rho]&=U_{\rm CR\pm} \cdot \rho \cdot U_{\rm CR\pm}^{\dagger},\\
		\Lambda_{T_1,T_2}[\rho]&=\frac{1-e^{-t/T_2}}{2} {\rm{Z}} \cdot \rho \cdot {\rm{Z}} +\frac{1+e^{-t/T_2}}{2} \rho \\
		&+\frac{1-e^{-t/T_1}}{2}\left|0\right>\left<1\right| \cdot \rho \cdot \left|1\right>\left<0\right|-\frac{1-e^{-t/T_1}}{2}\left|1\right>\left<1\right| \cdot \rho \cdot \left|1\right>\left<1\right|,
	\end{split}
\end{equation}
where the unitary operators $U_{\rm ZZ}$ and $U_{\rm CR\pm}$ are defined as $U_{\rm ZZ}=e^{-i 2 \pi \alpha_{\rm ZZ}t_g \rm ZZ/4}$ and $U_{\rm CR\pm}=e^{-i 2 \pi \tau H_{\rm CR}^{\rm{ct}}(\pm\Omega)}$ with $\alpha_{\rm ZZ}$ defined in Eq.~(\ref{eq.totZZ}) and $H_{\rm CR}^{\rm ct}$ defined in Eq.~(\ref{eq.ctp}).

\begin{figure}[h!]
	\centering
	\includegraphics[width=0.72\textwidth]{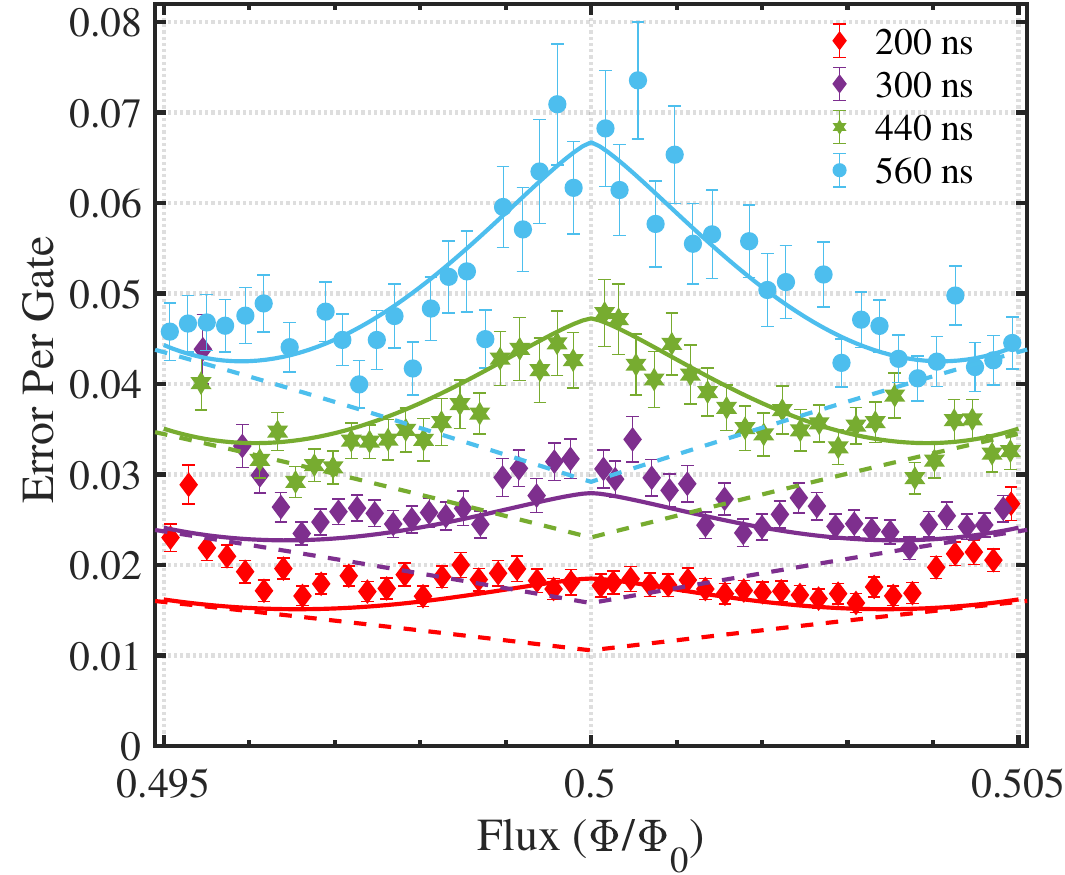}
	\vspace{-0.1in}
	\caption{\label{fig:twoQ_fid_flux} Average error per two-qubit gate plotted vs. flux for four representative gate lengths. Device parameters of the hybrid CSFQ-transmon circuit are shown in Tab.~\ref{table:freq} Dashed lines indicate theoretical coherence-limited two-qubit gate errors with no ZZ interactions; full theory simulations shown by solid lines.
	}
\end{figure}

To calculate the CR gate fidelity\index{fidelity}, we can use a different but efficient visual representation of process maps, the Pauli transfer matrix\index{Pauli transfer matrix} $\mathcal{R}$~\cite{chow2012universal}, which maps the input density matrix components to output density matrix components in the basis of Pauli operators. In contrast to process matrix representation~\cite{mohseni2008quantum-process}, $\mathcal{R}$ has
several useful properties. It consists of only real numbers and so contains exactly the same number of parameters as the gate, with no redundancy due to Hermiticity. Besides, since it represents the action of a process of density matrix evolution, it can help establish a number of properties of the underlying process which are otherwise hidden in the standard representation. Given a Hilbert space of $d$ dimensions $\mathcal{H}_d$ and space of linear operators $\mathcal{L}(H_d)$, any map $\Lambda$: $\mathcal{L}(H_d)\rightarrow\mathcal{L}(H_d)$ can be represented by the Choi matrix~\cite{choi1975completely,chow2012universal}
\begin{equation}
\rho_{\Lambda}=\frac{1}{d}\sum_{i,j}E_{ij}\otimes \Lambda(E_{ij}),
\label{eq.choi}
\end{equation}
with $E_{ij}$ being a matrix with 1 in the $ij^{\rm th}$ entry and 0's elsewhere. Now the density matrix is changed to the space $\mathcal{H}_{d}^{(A)}\otimes\mathcal{H}_{d}^{(B)}$. In the following, we assume $A=B$. In case of the Pauli transfer matrix, the qubit state basis is fixed to Pauli matrices $P_i\in\{\rm I, X,Y,Z\}$, and then $\mathcal{R}$ matrix elements are defined as 
\begin{equation}
\mathcal{R}_{ij}=\frac{1}{d}{\rm Tr}[P_i\Lambda(P_j)].
\end{equation}
Combining with Eq.~(\ref{eq.choi}) changes the expression to
\begin{equation}
\mathcal{R}_{ij}={\rm Tr}[\rho_{\Lambda}P_j^T\otimes P_i],
\end{equation}
with 
\begin{equation}
\rho_{\Lambda}=\frac{1}{d^2}\sum_{i,j}\mathcal{R}P_j^T\otimes P_i.
\end{equation}	
The average gate fidelity thus can be found as~\cite{pedersen2007fidelity}
\begin{equation}
F=\frac{d\rm Tr[\rho_{\rm ideal}\rho]+1}{d+1}=\frac{{\rm Tr}[\mathcal{R}_{\rm ideal}^T\mathcal{R}]/d+1}{d+1}.
\label{eq.fidelity}
\end{equation}
Since no active cancellation is performed on the final experiment, we need to model the classical crosstalk with amplitude $R(f, \tau)\Omega(f,\tau)$, where $R(f,\tau)$ is a flux and gate-length dependent scaling factor. The experimental data on the CSFQ-transmon device revealed that the classical crosstalk on the target qubit is much more significant with increasing gate length. This implies that $R(f, \tau)$ is an increasing function of the gate time length, which is consistent with the assumption that more power is pumped in the system if the gate length is longer.
For simplicity, we consider that flux and gate-length dependence are separable, and fit $R(f, \tau)\approx\alpha(f)\tau^{2/3}$ which agrees well with the experimental data. The nonlinearity with respect to $\tau$ was introduced otherwise the IY Pauli coefficient is consistently larger than the result from measured CR tomography, for example, in Fig.~\ref{fig:paulicoeff}. The average two-qubit error per gate for the experimental CR gate was measured via standard randomized benchmarking\index{randomized benchmarking} (RB)~\cite{magesan2011scalable} by varying the flux $f$ and gate length $t_g$ shown in Fig.~\ref{fig:twoQ_fid_flux}.  

The flux-dependence of $\alpha(f)$ was extracted by performing separate CR tomography measurements~\cite{sheldon2016procedure}. Our collaborators  performed such an active cancellation experiment for a fixed gate length at three different flux points during previous measurement, and found that away from the sweet spot, the classical crosstalk amplitude $R$ followed a nearly linear decrease. To be consistent with previous measurement results, we assume the fitting function of the classical crosstalk to be $R=(0.07-40|f-0.5|^{1.2})\tau^{2/3}$. For the simulation of two-qubit gate error at the flux sweet spot\index{sweet spot (SS)}, we calculate $R=(0.0123,0.0220,0.0322,0.0383)$ and $\Omega/2\pi=\{70,30,17,12\}$~MHz for the gate length $t_g=(200, 300, 440, 560)$~ns, respectively. 

Figure~\ref{fig:twoQ_fid_flux} shows that with increasing gate length, a characteristic ``W''-shaped pattern develops with flux, both from theory and experiment, showing the largest errors at sweet spot and two minima at the two flux points $f = 0.496, 0.504$, where static ZZ interaction becomes zero.
The fastest gate measured in this experiment with smallest error rate $1.6\times10^{-2}$ is realized at $t_g = 200$~ns. This behavior can be described by the interplay between fidelity loss from the ZZ interaction and classical crosstalk on the one hand, and fidelity gain from longer coherence times near the sweet spot on the other hand. Away from the sweet spot, the ZZ interaction and classical crosstalk decrease and the gate fidelity approaches the coherence limit.

One of the most prominent advantages of a CSFQ-transmon pair over a transmon-transmon pair is that the static ZZ interaction can be cancelled by carefully choosing qubit parameters.  To make a comparison between hybrid CSFQ-transmon circuits and transmon-transmon circuits, we calculate the gate error in three samples: current CSFQ-transmon device (1), a state-of-the-art tested transmon-transmon device (2)~\cite{sheldon2016procedure}, and an ideal CSFQ-transmon device (3) with flux threading CSFQ at sweet spot. On CSFQ-transmon sample (1), we compare four scenarios: in presence of classical crosstalk and nonzero ZZ, only ZZ interaction, static ZZ freedom, only coherence limit. On transmon-transmon sample (2), we compare experimental results and theoretical prediction with ZZ interaction, as well as coherence limit with experimental coherence times\index{coherence time} and ideal coherence times.
On ideal CSFQ-transmon sample (3), we assume both CSFQ and transmon have very long coherence times, and compare the results of static ZZ freedom with coherence limit. In the ideal circuit, static ZZ freedom at the sweet spot can be realized by only changing Josephson energy $E_J$. In practice, such a device could be made by potentially keeping CSFQ at sweet spot, while making the transmon slightly tunable~\cite{hutchings2017tunable}.  Corresponding device parameters are listed in Tab.~\ref{table:coherence limit}.
\begin{table}[h!]
	\centering
	\begin{tabular}{|c|c|c|c|c|c|c|c|c|c|}
		\hline 
		Device & $T_{1}^{(1)}$& $T_{2}^{(1)}$& $T_{1}^{(2)}$ &$T_{2}^{(2)}$&$\tilde{\omega}_1$ &$\tilde{\omega}_2$ &$\delta_1$&$\delta_2$&$\eta$ \tabularnewline
		& $\mu$s&$\mu$s& $\mu$s &$\mu$s& GHz& GHz& MHz& MHz&1/MHz\tabularnewline
		\hline
		(1) &18&15&40&45&5.051&5.286&$+593$&$-327$&$6.0\times 10^{-5}$ \tabularnewline
		\hline 
		(2)&40&54&43&67&5.114&4.914&$-330$&$-330$ & $1.6\times 10^{-5}$\tabularnewline
		\hline 
		(3) &200&200&200&200&5.094&5.286&$+593$&$-327$&$8 \times 10^{-6}$\tabularnewline
		\hline 
	\end{tabular}
	\caption{Coherence time, dressed frequency, anharmonicity and nonlinear ZZ interaction rate in Eq.~(\ref{eq.totZZ}) for the current device (1), a transmon-transmon device (2)~\cite{sheldon2016procedure}, and an ideal CSFQ-transmon device (3), respectively. $\eta$ is fitted using the data from Fig.~\ref{fig:paulicoeff}. }
	\label{table:coherence limit}
\end{table}

\begin{figure}[h!]
	\centering
	\vspace{-0.1in}
	\includegraphics[width=0.8\textwidth]{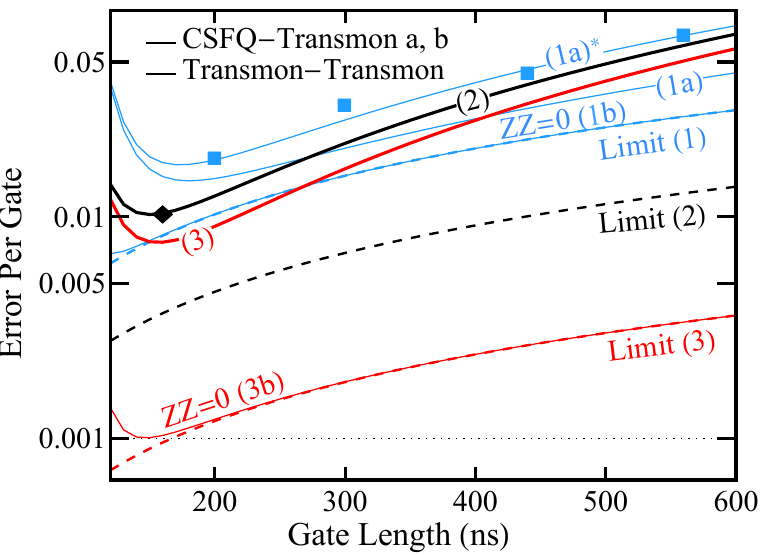}
	\vspace{-0.1in}
	\caption{\label{fig:twoQ_fid_predic}Experimental data and theory simulation for two-qubit gate error vs. gate length for our present CSFQ-transmon (a), static ZZ-free CSFQ-transmon (b), and a transmon-transmon device with non-zero ZZ (thick lines). The CSFQ was placed at sweet spot. Square CR pulses were used in theory simulation. Three sets of coherence times used in simulation were color-coded in blue, black, and red, and numbered by $n=\{1,2,3\}$. ``Limit (n)" represents coherence-limited gate error. Classical crosstalk is not included except $(\rm 1a)^*$. Blue squares and black diamond are experimental data points from present device and Ref.~\cite{sheldon2016procedure}, respectively.
	}
\end{figure}

By varying the gate length, we plot the gate error in Fig.~\ref{fig:twoQ_fid_predic}. Here (a) represents the device with nonzero static ZZ interaction, while (b) represents the device with static ZZ freedom. Classical crosstalk is only considered in the device marked with $*$. Figure~\ref{fig:twoQ_fid_predic} shows that the CSFQ-transmon device should achieve coherence times at least as long as this experimental device, while maximizing $T_2$ of the CSFQ thus enabling a gate error as the device (1b) comparable to the simulated transmon-transmon results on device (2). For the projected longer coherence times 200 $\mu$s~\cite{serniak2019direct}, the gate error on device (3b) subject to elimination of classical crosstalk can reach $1\times10^{-3}$. This level is inaccessible for a transmon-transmon device, even with the projected longer coherence times of device (3). Separate comparison of each device can be found in Appendix~\ref{app:CRerror}.

While coherence-limited gate errors denoted by dashed lines in Fig.~\ref{fig:twoQ_fid_predic} decrease monotonically with gate length, the total error reaches a minimum at an optimum gate length. This is a universal behavior, even in the absence of static ZZ or classical crosstalk e.g. device (3b). It can be explained by the dynamic ZZ that arises from strong CR drive. Since larger CR amplitude is required for shorter gate length and the dynamic ZZ scales quadratically with CR amplitude, at short gate times a large ZZ interaction can still exist, which thus limits the minimum gate error. In next section, we will focus on the additional ZZ component produced by CR drive, and added on top of static ZZ interaction. The new term is a device-dependent quantity, so we could find a way to change it, thus making the total ZZ interaction suppressed.

\section{Dynamical ZZ Freedom}
\label{sec. dynzz}

In previous section we study the CR gate error and predict that at larger driving amplitude, the dynamical ZZ induced by CR drive leads to a dramatic decrease in the gate fidelity\index{fidelity}. In this section, we take several samples of CSFQ-transmon and transmon-transmon pairs by varying the two-qubit detuning as well as anharmonicity of the two qubits to explore the possibility of zeroing total ZZ interaction at a large domain of CR pulse amplitudes. The general idea is to tune the circuit parameters and driving amplitude to make the dynamical part cancel out the static part, meaning that dynamical ZZ should at least has opposite sign with static part. We find that CR amplitude can control the magnitude of dynamic part and allows for vanishing total ZZ strength.  Moreover the freedom is persistent as long as CR gate is active and then improves the CR gate fidelity. 

As discussed, when qubits are driven, an additional component is added on top of the static ZZ interaction described in Eq.~(\ref{eq.totZZ}) with $\eta$ being the dynamical quadratic factor. To achieve dynamical ZZ freedom, the minimum requirement is that dynamical part should have opposite sign with static part. Static ZZ interaction can be easily calculated from either full Hamiltonian or effective Hamiltonian\index{effective Hamiltonian} in the dispersive regime in chapter~\ref{c3}. In order to evaluate $\eta$, we first derive it from perturbation theory. Although perturbation theory is not a proper approximation for such a driving Hamiltonian, especially at strong driving amplitude, it enables us to find the relation between the dynamical factor and circuit parameters. By block diagonalizing the Hamiltonian in Eq.~(\ref{eq.Hr}) using SW transformation, dynamical quadratic factor $\eta$ is given by 
\begin{equation} 
\eta=\frac{J_{01}^2}{2\Delta^2\delta_2(\Delta+\delta_2)^2(\Delta-r\delta_2)^3(2\Delta-r\delta_2)}\sum_{i=0}^6 A_i(r,\gamma)\Delta^i\delta_2^{(6-i)},
\label{eq.eta}
\end{equation}
with $\gamma$ being the ratio of $J_{10}$ and $J_{01}$ in Eq.~(\ref{eq.J}) and $ r\equiv \delta_1/\delta_2$. Detailed $A_i(r,\gamma)$ can be found in Appendix~\ref{app.eta}.

\begin{figure}[h!]
	\begin{center}
		\includegraphics[width=0.49\textwidth]{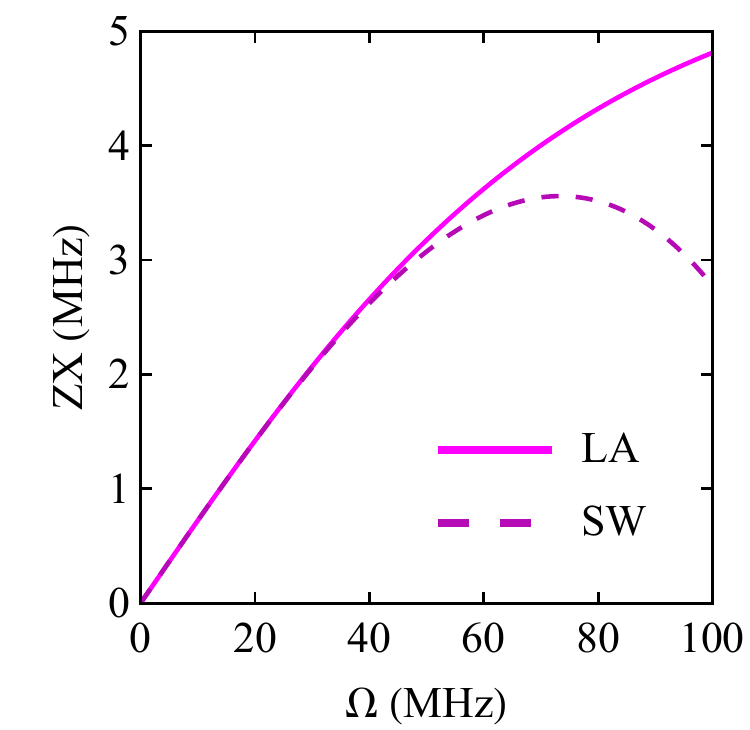}   
		 \put(-180,165){(a)}
		\includegraphics[width=0.49\textwidth]{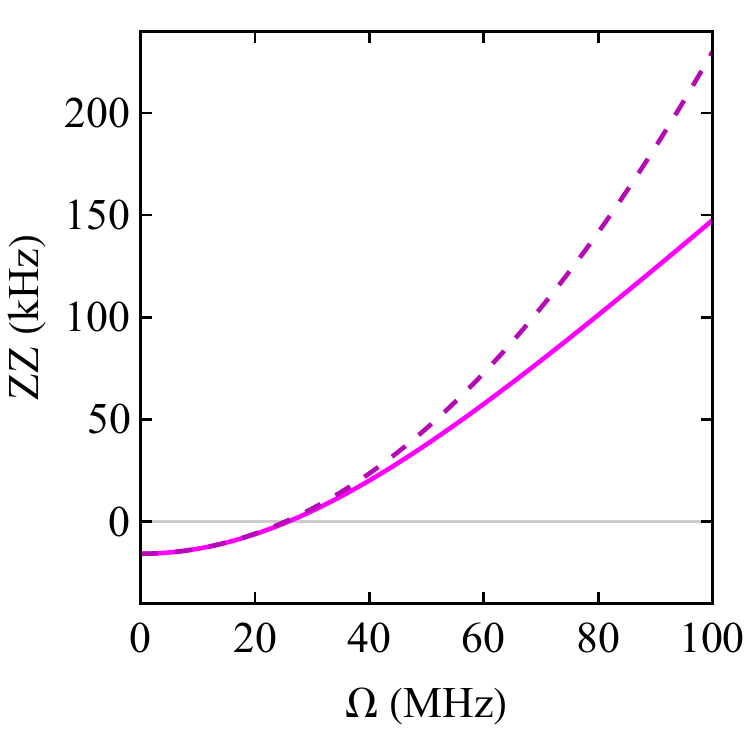}
		\put(-180,165){(b)} 
		\vspace{-0.1in}
		\caption{ZZ and ZX coupling strengths in a CSFQ-transmon pair versus CR  amplitude $\Omega$, using Schrieffer-Wolff transformation (dashed) and Least Action transformation (solid).  $\Delta=$0.1 GHz and the other parameters similar to Fig. \ref{fig:fullsw}(a).} 
		\vspace{-0.15in}
		\label{SW_LA}
	\end{center} 
\end{figure}

To see how accurate the perturbative result is, we make a comparison between perturbative approach SW transformation\index{Schrieffer-Wolff (SW) transformation} and non-perturbative approach least action (LA) principle in terms of Pauli coefficients of ${\rm ZZ}$ and ${\rm ZX}$. Given that the other unwanted terms IX, IY and ZY in the CR Hamiltonian Eq.~(\ref{eq.ctp}) are eliminated by applying active cancellation pulse on the target qubit and calibrating the global phase of CR pulse to $\pi$, the Hamiltonian is only left with $\rm ZX$ and $\rm ZZ$ terms. Here we take an example of CSFQ-transmon circuit and plot the corresponding Pauli coefficients in Fig. \ref{SW_LA}. One can see that for weak driving amplitude perturbative and non-perturbative methods give consistent results, however as expected they deviate from each other within strong driving domain. This is because in Eq.~(\ref{eq.totZZ}) higher order corrections denoted by $O(\Omega^3)$ starts to contribute in strong driving limit. In the next we keep a record of the both sets of coupling strengths and compare them in the results of $\eta$.   

\begin{figure}[h!]
	\begin{center}
		\includegraphics[width=0.49\textwidth]{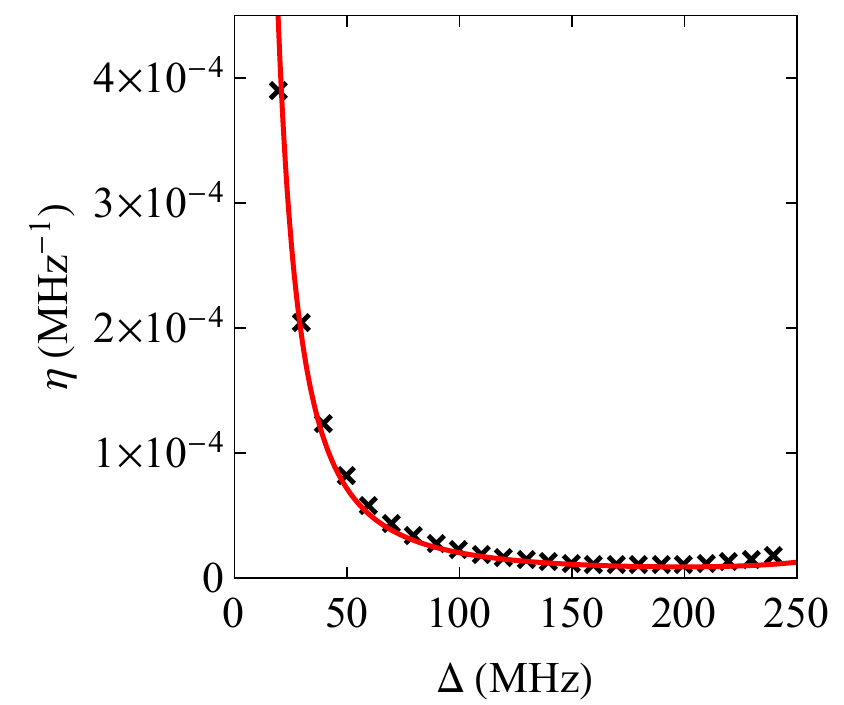}    
		\put(-180,145){(a)}
		\includegraphics[width=0.49\textwidth]{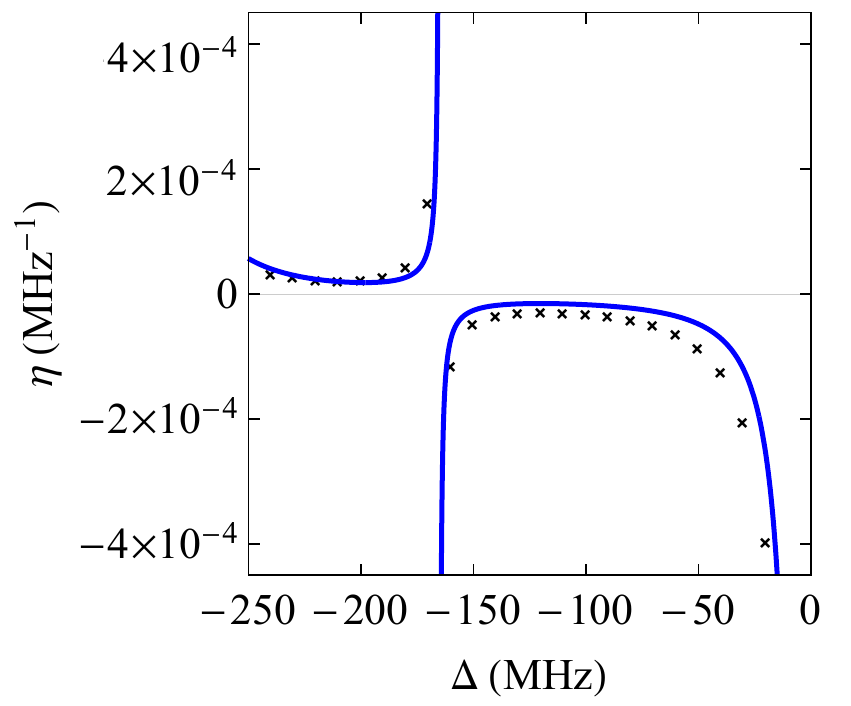}
		\put(-180,145){(b)}
		\vspace{-0.1in}
		\caption{$\eta$ in Eq.~(\ref{eq.eta}) as a function of qubit frequency detuning, (a)  CSFQ-transmon devices similar to  Fig. \ref{fig:fullsw}(a), and  (b)  transmon-transmon devices similar to Fig. \ref{fig:ttZZ}(a) with $g_{12}=2.5$~MHz. Cross points are numerical results from LA transformation and solid line from SW transformation.}
			\vspace{-0.1in}
		\label{fig:eta}
	\end{center}
\end{figure}

 In Fig. \ref{fig:eta} we plot the dynamical quadratic factor $\eta$ in two setups using perturbative approach up to the second order in solid lines as well as LA principle in cross points. Figure~\ref{fig:eta}(a) shows that in a CSFQ-transmon pair $\eta$ is always positive under current circuit parameters, which makes CR drive to add up positive dynamic ZZ component on top of the static part. This may result in suppression of total ZZ strength if the static part is negative. 
While in Fig.~\ref{fig:eta}(b) a transmon-transmon pair carries both positive and negative $\eta$ separated by a divergence limit at certain detuning. Perturbation theory shows that in Eq.~(\ref{eq.eta}) there exists poles at detuning values $\Delta=-\delta_2, 0, \delta_1/2, \delta_1$, which originates from unexpected higher-level resonance. However LA transformation finds that the divergence is nonphysical and that ZZ strength remains finite, similar to Fig.~\ref{fig:dress}.

\subsection{Dynamical ZZ cancellation}
\begin{figure}[h!]
	\centering
	\includegraphics[width=0.9\textwidth]{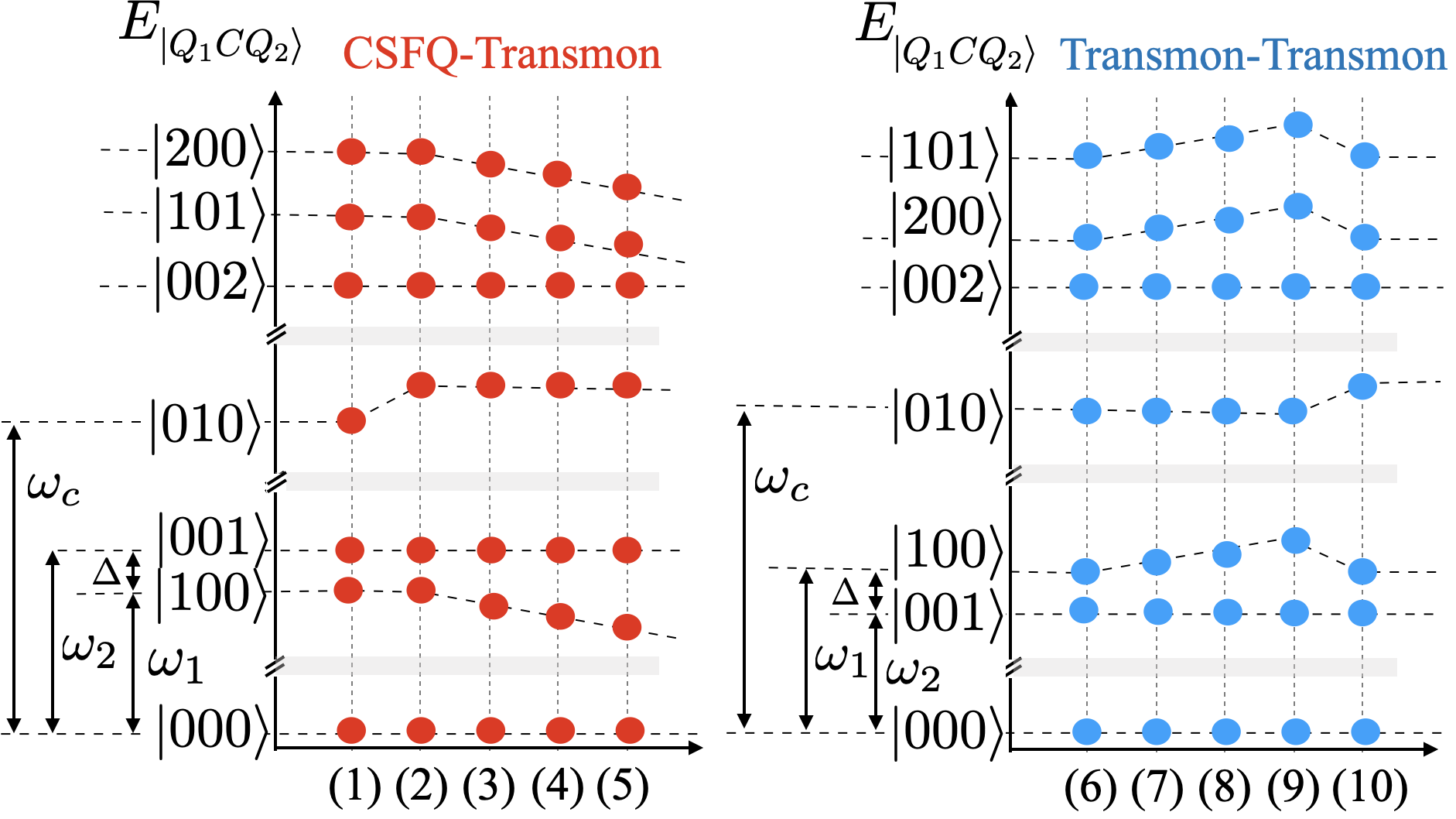}
	\put(-330,175){(a)}
	\put(-170,175){(b)}
	\vspace{-0.1in}
	\caption{ (a) Energy diagrams for  five CSFQ-transmon circuits 1-5. (b) Energy diagrams for five transmon-transmon circuits 6-10.} \label{fig:ene}
\end{figure}

The opposite sign of dynamical part and static part makes it possible to realize ZZ freedom when the two-qubit are driven. To further explore the possibility, we study five  CSFQ-transmon samples labeled from 1 to 5, and five transmon-transmon samples labeled from 6 to 10, the relative relations of the energy diagrams among the samples are plotted  in Fig. \ref{fig:ene}.

\begin{figure}[h!]
	\centering
	\includegraphics[width=0.49\textwidth]{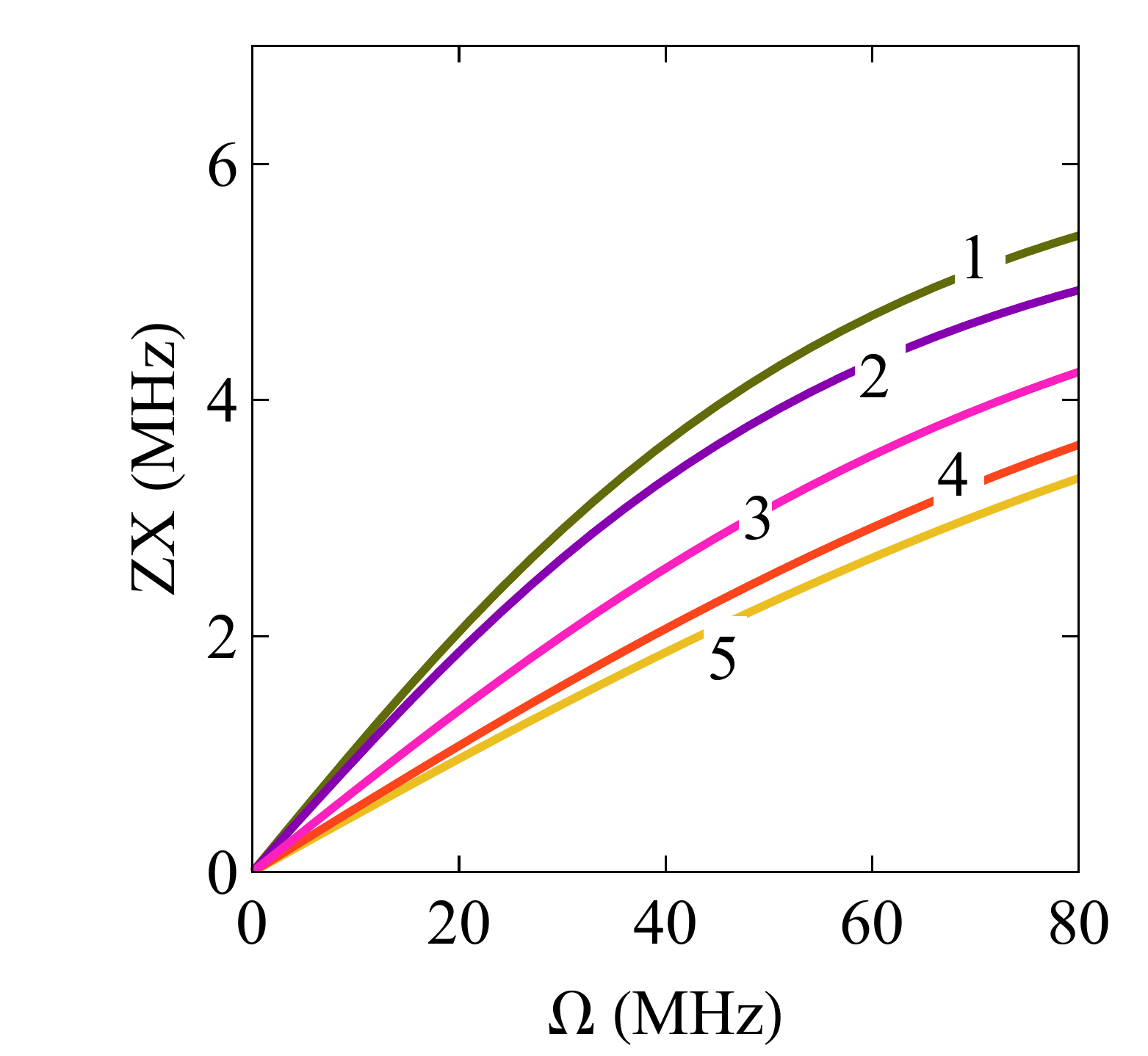}
	\includegraphics[width=0.49\textwidth]{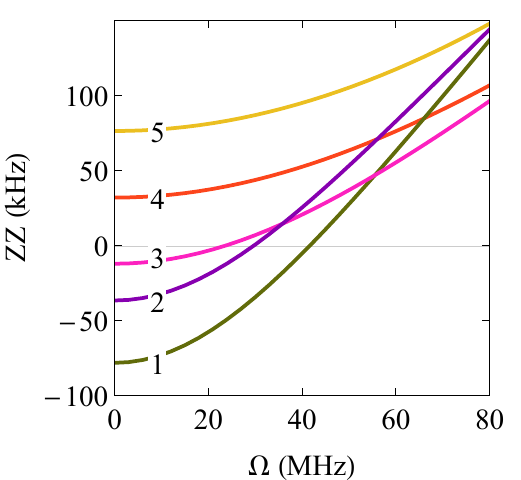}
	\put(-355,155){(a)}
	\put(-180,155){(b)}
	\caption{ The coupling strength under CR gate using non-perturbative least action transformation for: (a) ZX interaction in devices 1-5 and (b) ZZ interaction in 1-5. Common  parameters in 1-5 are $\delta_{1/2}=0.6,-0.33$ GHz, $g_{1c/2c}=80$ MHz, and $g_{12}=0$. In 1: $\Delta=70$ MHz and $\Delta_2=1.1$ GHz.  In 2-5: $\Delta=70, 105, 150, 180$ MHz  and $\Delta_2=1.2$ GHz.} \label{fig:dynamiccsfq}
\end{figure}

Let us first talk about CSFQ-transmon devices. In Fig.~\ref{fig:ene}(a) the energy levels of $|010\rangle$, $|100\rangle$, and $|001\rangle$ show their differences in the harmonic resonator and the two qubit frequencies, noncomputational levels $|002\rangle$ and $|200\rangle$ are in two sides of $|101\rangle$. Applying CR pulse produces desired ZX entanglement between the two qubits accompanied with unwanted ZZ term. To get precise prediction we determine dynamical ZZ interaction only from non-perturbative LA transformation. ZX strengths in CSFQ-transmon devices have been plotted In Fig. \ref{fig:dynamiccsfq}(a). For each device the strength of ZX coupling increases with the CR amplitude, but monotonously decreases with increasing two-qubit detuning, which can be seen by comparing devices 2 to 5. Furthermore, devices 2 has a larger bus resonator frequency, and shows that smaller qubit-resonator detuning can lead to a fast ZX rate in contrast to device 1. Total ZZ strengths in CSFQ-transmon devices are plotted in Fig. \ref{fig:dynamiccsfq}(b). In samples 1-3 the static ZZ, i.e. at $\Omega=0$, are negative and in 4 and  5 positive. The dynamical part added on top of static part is always positive under current circuit parameters, which is consistent with what has been shown in Fig.~\ref{fig:eta}(a). Adding the positive dynamic ZZ component can mitigate the negative static ZZ in samples 1-3, making the total ZZ to be zero at certain driving amplitude $\Omega$.

\begin{figure}[h!]
	\centering
	\includegraphics[width=0.49\textwidth]{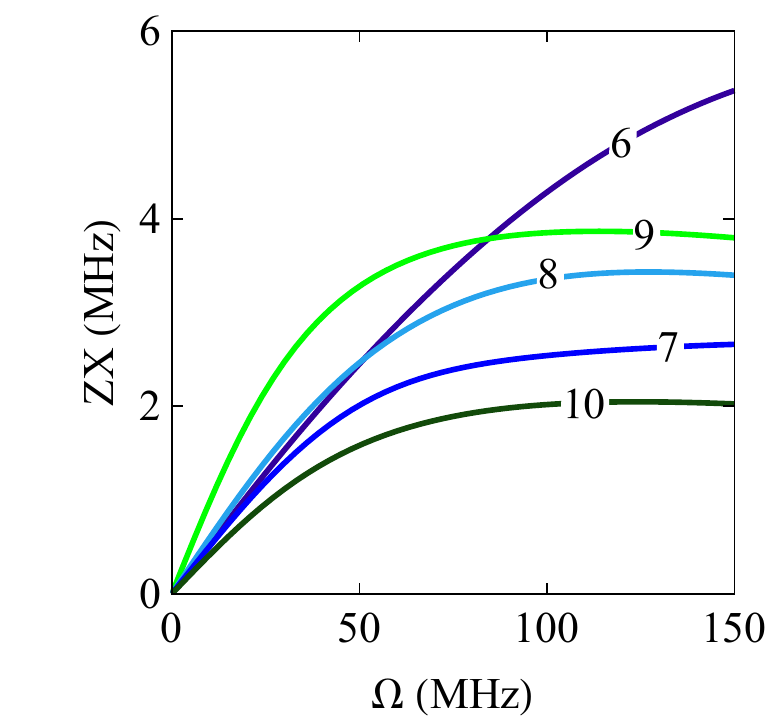}
	\includegraphics[width=0.49\textwidth]{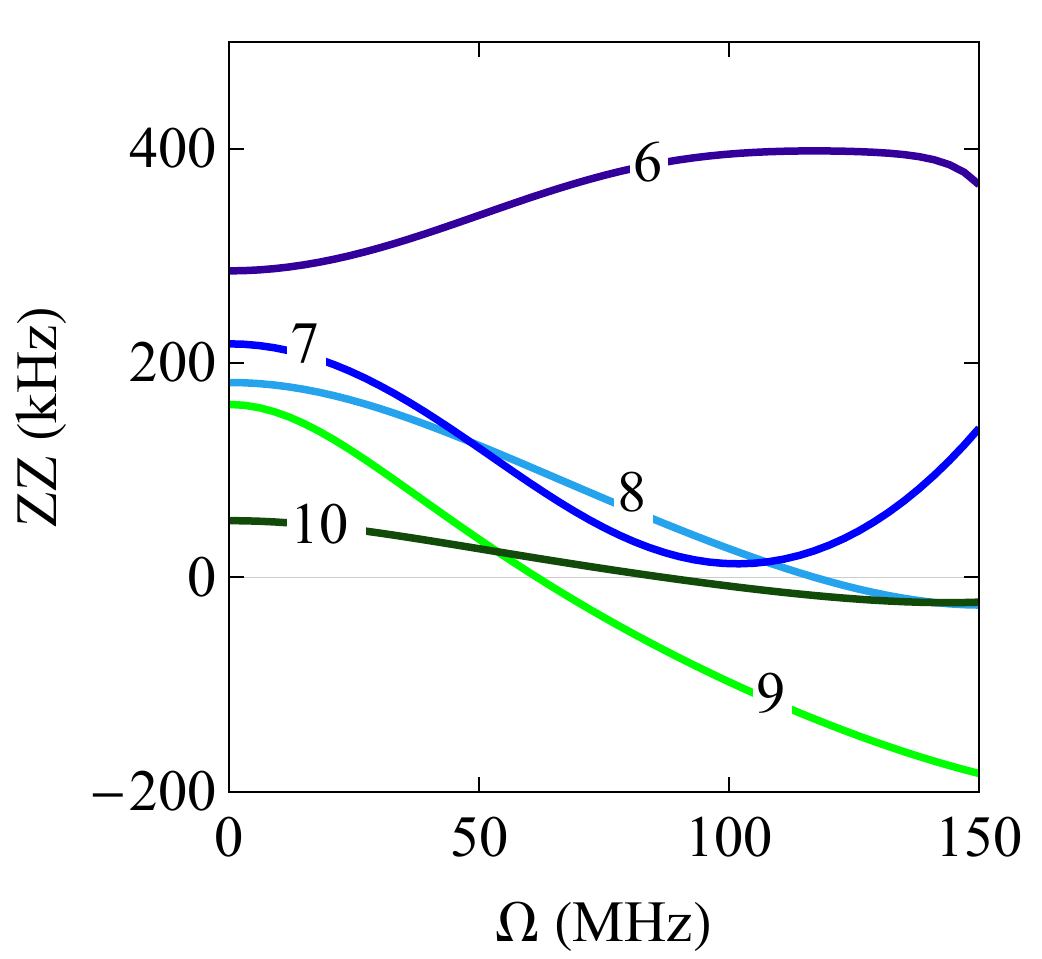}
\put(-355,160){(a)}
\put(-180,160){(b)}
	\caption{   The coupling strength under CR gate using nonperturbative least action transformation for:  (a) ZX interaction in 6-10 and (b) ZZ interaction in 6-10.  Common parameters in 6-10 are $\delta_{1/2}=-0.33$ GHz, $g_{1c/2c}=98, 83$ MHz, and $g_{12}=2.5$ MHz. In 6-9: $\Delta=-200, -150, -100, -50$ MHz and $\Delta_2=1.4$ GHz. In 10: $\Delta=-70$ MHz and $\Delta_2=2$ GHz.} \label{fig:dynamictr}
\end{figure}

Similar calculation is also performed in five transmon-transmon samples labeled from 6 to 10. The corresponding energy levels are depicted in Fig. \ref{fig:ene}(b). In contrast to CSFQ-transmon devices, the noncomputational states $|002\rangle$ and $|200\rangle$ in transmon-transmon circuits are both below $|101\rangle$. Figure~\ref{fig:dynamictr}(a) shows that ZX rate is disordered in terms of qubit-qubit detuning, this is because the existence of the divergence-like condition is satisfied at $\Delta=\delta/2=-165$~MHz where multilevel resonance can take place. Samples 6 and 7 are in the two sides of this particular detuning, and device 7 is closer to the point, then the ZX rate is further suppressed, this can be explained by working out the higher order contributions proportional to $\Omega^3$~\cite{malekakhlagh2020first-principles}. Since in these examples the coupler frequency is far detuned from qubits, the repulsions between $|101\rangle$ and noncomputational levels in transmon-transmon devices have the same sign and sum together. Dynamical ZZ freedom\index{dynamical ZZ freedom} takes place in transmon-transmon circuits as it can be seen in devices 8-10, and the cancellation driving amplitude decreases as the bus resonator frequency is farther away from the qubits by comparing devices 9 with 10. Note that device 6 is the IBM experimental circuit used in Ref. \cite{sheldon2016procedure} and we can see it does not show total ZZ freedom since static part has the same sign with dynamical part. One interesting device is 7 where the total ZZ can approach to zero but never cross it, behaving as a minimum at certain driving amplitude $\Omega$, this is because higher order corrections e.g. $O(\Omega^3)$ at larger amplitude becomes predominant and cannot be ignored.

To further explore the condition of dynamical ZZ cancellation, we can use the dynamical factor $\eta$ in Eq.~(\ref{eq.eta}) with sufficient accuracy for weak driving. By solving $\alpha_{\rm ZZ}=\zeta+\eta\Omega^2=0$, the condition for dynamical ZZ freedom\index{dynamical ZZ freedom} in the first order of $\Delta/\delta_2$ can be solved at the particular CR amplitude in the limit of $\Delta/\delta_2\ll 1$:
\beq
\Omega^* =|\Delta|\sqrt{ \frac{2  (r+\gamma^2)}{r+\gamma(2+\gamma)}} \sqrt{ 1 - C\frac{\Delta}{\delta_2} },
\label{eq.dyncanc} 
\eeq
with 
\begin{equation}
C\equiv \frac{ 1/2+2\gamma+ \gamma^2 + r^2  +  r \gamma (2+\gamma) + \gamma^2 (1+2 \gamma^2)/2r}{ (r +  \gamma^2) [r  + \gamma (2+\gamma) ]}.
\end{equation} 
Table \ref{tab} compares the ten samples including both CSFQ-transmon and transmon-transmon pairs, and corresponding cancellation CR amplitude $\Omega^*$ at which dynamical part cancels out the static ZZ interaction. The amplitude $\Omega^*$ is determined using three different methods: In the row labeled by LA we use non-perturbative least action method to determine total ZZ and find where it is zero; In $O(n)$ row we use the SW-evaluated  static ZZ coupling  $\zeta$ of Eq. (\ref{eq.zeta}) and the SW-evaluated $\eta$ in Eq.~(\ref{eq.eta}) and substitute them in Eq. (\ref{eq.totZZ}) to find the solution at which amplitude ZZ becomes zero; Below it we present the results from Eq. (\ref{eq.dyncanc}) and in the last row we evaluate the ratio of $\Delta/\delta_2$ in each device. One can see the results are better consistent in the limit of    $\Delta/\delta_2\ll 1$.

\begin{table}[h!]
	\label{tab.1}
	\centering
	\begin{tabular}{|c|c|c|c|c|c|c|c|c|c|c|}
		\cline{2-11} \cline{3-11} \cline{4-11} \cline{5-11} \cline{6-11} \cline{7-11} \cline{8-11} \cline{9-11} \cline{10-11} \cline{11-11} 
		\multicolumn{1}{c|}{} 
		& (1) & (2) & (3) & (4) & (5) & (6) & (7) & (8) & (9) & (10)\tabularnewline
		\hline 
		LA & 42 &  30 & 24 & No &  No & No & No & 115  & 61 & 82 \tabularnewline
		\hline 
		$O(n)$ & 41 & 31 & 24 & No & No & No & 71 & 83 & 46 & 62\tabularnewline
		\hline 
		Eq(\ref{eq.dyncanc}) & 41 & 34 & 40 & 20 & No & 110 & 104 & 81 & 46 & 61\tabularnewline
		\hline 
		\hline
		$\Delta/\delta_{2}$ & 0.11 & 0.11 & 0.17 & 0.25 & 0.3 & 0.61 & 0.45 & 0.3 & 0.15 & 0.21\tabularnewline
		\hline 
	\end{tabular}
\vspace{-0.1in}
	\caption{$\Omega^*$ from different methods in devices 1-10 in unit of MHz. `No' indicates devices with no dynamic ZZ freedom.}
	\label{tab}
\end{table}

\begin{figure}[h!]
	\centering
		\includegraphics[width=0.48\textwidth]{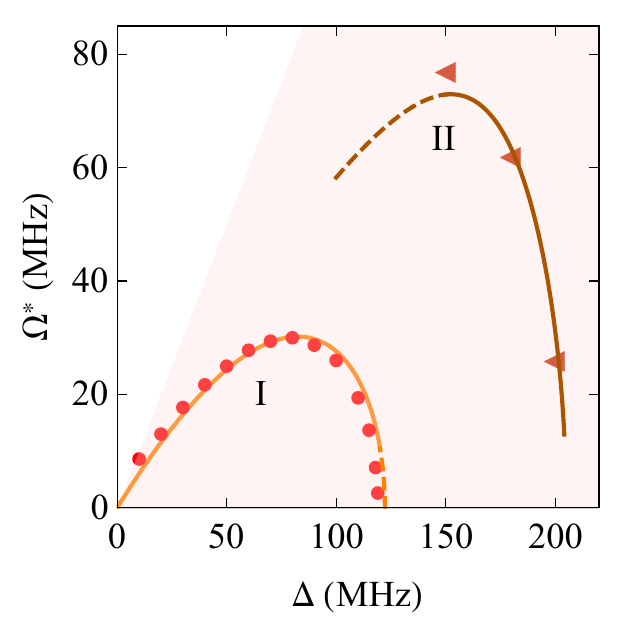}
		\put(-175,165){(a)}\hspace{2mm}
		\includegraphics[width=0.48\textwidth]{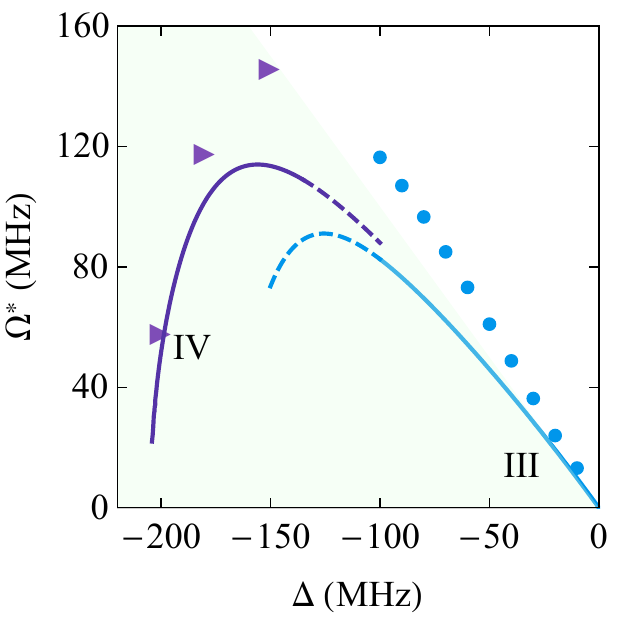}
		\put(-180,165){(b)}
		\vspace{-0.15in}
		\caption{Cancellation amplitude of dynamical ZZ freedom versus qubit detuning, (a) for CSFQ-transmon with parameters similar to circuit 2-5 devices, \rn{1} with qubit anharmonicity $\delta_{1/2}=0.6, -0.33$ GHz and \rn{2} with $\delta_{1/2}=0.41, -0.39$ GHz,  (b)  for transmon-transmon with parameters similar to 6-9 devices, \rn{3} with qubit anharmonicity $\delta_{1/2}=-0.33,-0.33$ GHz and \rn{4} with $\delta_{1/2}=-0.41,-0.33$ GHz. Lines are perturbative results, dots and triangles are from nonperturbative least action. Shaded area are the validity domain of perturbation theory.}
		\label{fig:ZZdy}	
\end{figure}

Let us further study the needed cancellation CR amplitudes at which dynamic freedom is achieved. In what comes next we work with the parameters from CSFQ-transmons devices 2-5  and transmon-transmons devices 6-9. We plot the numerical result  of dynamic ZZ freedom condition within a large range of qubit-qubit detuning $\Delta$ in Fig. \ref{fig:ZZdy}(a) and \ref{fig:ZZdy}(b).  Dots show non-perturbative results using LA transformation and lines are SW perturbative results.  Shaded area is the validity domain within dispersive regime in terms of $\Omega/\Delta$. 
Although the ZZ cancellation on samples takes place below 150 MHz, it does not mean that it is limited by the two-qubit detuning. To illustrate this, we add three more triangular points using LA transformation in each setup by changing CSFQ and transmon anharmonicity accompanied with perturbative fitting lines. In CSFQ-transmon pairs,  cancellation CR amplitude is small so perturbative and non-perturbative results agree well with each other.  However, in transmon-transmon pairs one can see perturbation theory is a crude approximation, this is mainly because the static ZZ strength in transmon-transmon pairs is usually large and then cancelling it requires strong driving amplitudes, which falls outside of the dispersive regime. One of the noticeable characteristics of dynamic ZZ freedom in CSFQ-transmon pairs as seen in Fig. \ref{fig:ZZdy}(a)  is that by increasing detuning frequency first cancellation amplitude increases, and then the amplitude squeezes. In CSFQ-transmon devices set $\rm I$, 
cancellation CR amplitude is small, but becomes larger by changing qubit anharmonicity, as shown in devices set $\rm II$. In transmon-transmon pairs of Fig. \ref{fig:ZZdy}(b) the amplitude monotonically changes in a wide domain and then starts to drop both in device sets $\rm III$ and $\rm IV$. These behaviors are consistent with what we found above for the way how $\zeta$ and $\eta$ scale with detuning $\Delta$. 
\subsection{CR gate error}
\label{sec error}
To quantify the performance of echoed-CR gate with respect to the total ZZ interaction, we numerically simulate the CR gate\index{CR gate} for several CSFQ-transmon and transmon-transmon devices in Fig.~\ref{fig:dynamiccsfq} and Fig.~\ref{fig:dynamictr}, respectively. Here we consider the CR pulse is round square with negligible rise and fall times, and keep $\pi$ pulse 40 ns long as in the experiment. When performing a CR gate, the flat-top of the single CR tone $\tau$ satisfies $\tau=1/8\alpha_{\rm ZX}$ for small CR driving amplitude to achieve $\rm ZX_{90}$ rotation, therefore the total gate duration is $t_g=(2\tau+80)$ ns. Assuming that active cancellation pulse eliminate all unwanted terms except ZZ, then the Hamiltonian is only left with desirable ZX and the error source ZZ. We compute the two-qubit error per gate by evaluating how the unitary evolution of the echoed CR gate evolves an initial state as shown in Eq.~(\ref{eq.fidelity}),  and project it into the Pauli basis using Pauli transfer matrix\index{Pauli transfer matrix}.  

\begin{figure}[ht]
	\centering
	\includegraphics[width=.75\textwidth]{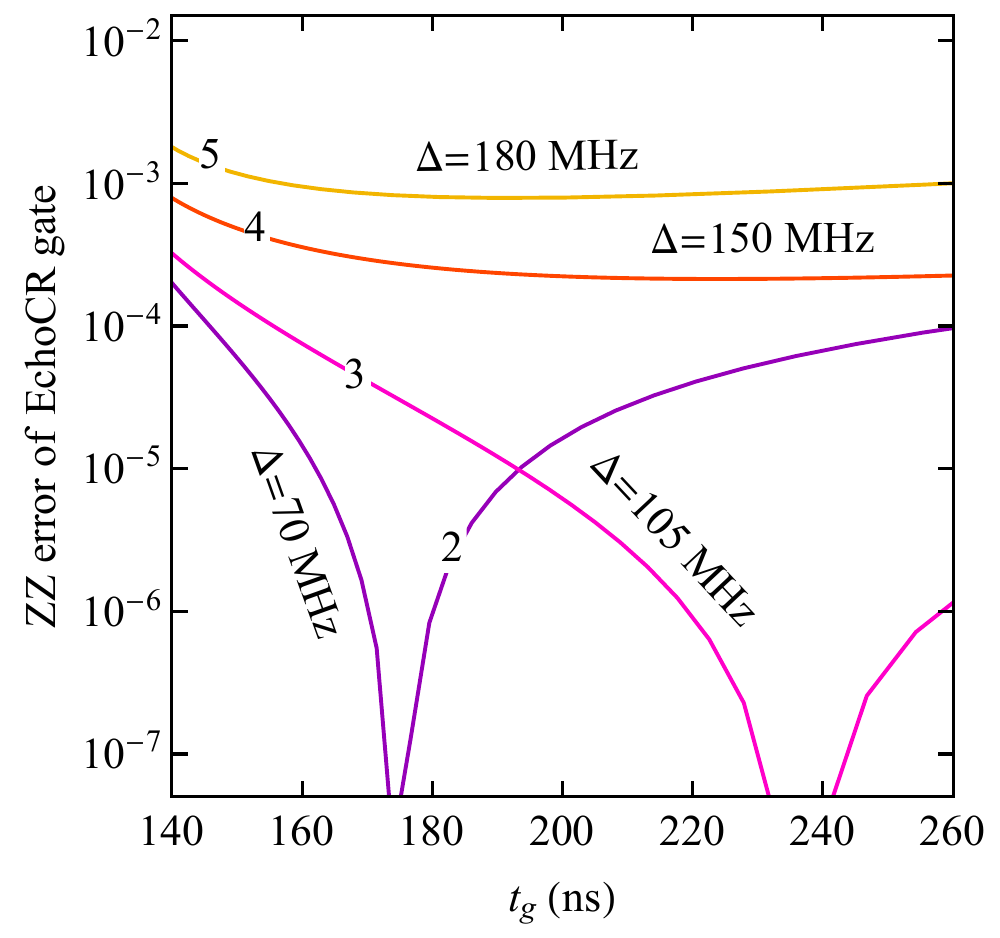}
	\put(-292,250){(a)}\\ \vspace{-0.1in}
	\includegraphics[width=.75\textwidth]{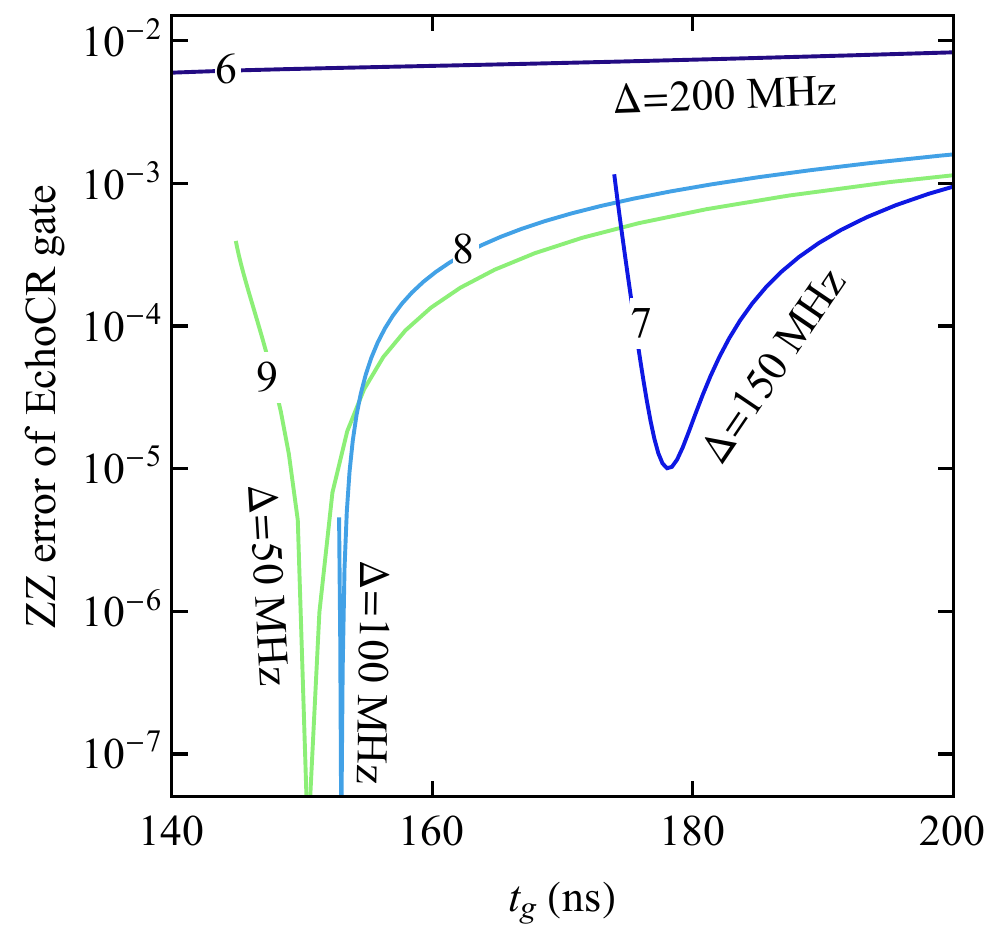}
	\put(-292,250){(b)}
	\vspace{-0.1in}
	\caption{ZZ error of the echo CR gate as a function of gate length and two-qubit detuning in (a) CSFQ-transmon devices (b) transmon-transmon devices.}
	\label{fig:dyfidelity}
\end{figure}

Figure \ref{fig:dyfidelity}(a) shows the CR gate error caused by ZZ interaction as a function of gate length in CSFQ-transmon pair with different qubit-qubit detuning. In these plots the decoherence effect on the gate is negligible as we assume that qubits can have desirably long coherence times $T_1$ and $T_2$. In CSFQ-transmon devices 2 and 3 the error drops to almost zero at certain gate times, this is exactly where total ZZ can be dynamically set to zero. Figure \ref{fig:dynamiccsfq}(a) and \ref{fig:dynamiccsfq}(b) show that in device 2 the dynamic freedom takes place at $\Omega\sim30$ MHz, where corresponding ZX rate is   $\alpha_{\rm ZX}\sim 2.7$ MHz.  Such a ZX rate requires the flap-top length to be $\tau\sim 46$ ns for each CR pulse to perform $\pi/2$ ZX gate.  The total echoed-CR pulse length then is around 172 ns, which is consistent with the plot. For device 3 zero ZZ interaction takes place at $\Omega\sim21$ MHz, where $\alpha_{\rm ZX}$
is smaller by a factor of 1/1.7 compared to device 2, this causes the prolongation of the gate length of the single CR pulse and makes the total gate time to be 235 ns. While the gate error in devices 4 and 5 monotonously decreases as the gate becomes longer, which corresponds to weak driving amplitude and then smaller total ZZ as the dynamical contribution has the same sign with static part. The gate error in absence of decoherence is in the scale of $10^{-3}$. 

Figure  \ref{fig:dyfidelity}(b) presents the gate error in transmon-transmon devices with expected similar behavior as the CSFQ-transmon pair. The time length of perfect ZX gate in devices 8 and 9 are shorter compared to CSFQ-transmon pairs. The reason for such improvement is that static ZZ is larger in these transmon-transmon devices, then the dynamic cancellation requires stronger amplitude where ZX rate is larger. For device 8 the dynamical ZZ freedom takes place at $\Omega\sim120$ MHz, where $\alpha_{\rm ZX}\sim 3.5$ MHz.  Such a ZX rate requires the flat-top length to be $\tau\sim 36$ ns for each CR pulse to perform $\pi/2$ ZX gate, making the total time to be 152 ns. While cancellation amplitude is smaller $\sim 60$ MHz in device 9, but the ZX rate increases a little faster to 3.7 MHz since the two-qubit detuning is far away from fake divergence point $-\delta/2$, resulting in a shorter gate at 148 ns. However, one can see that transmon-transmon devices 7, 8, and 9 show some cutoff in their minimum gate length for ${\rm ZX}_{\pi/2}$ gate. The answer can be found from the saturation of ZX in Fig. \ref{fig:dynamictr}(a), where $\alpha_{\rm ZX}$ rate starts to saturate after some amplitudes and cannot increase anymore. This saturation limits the flat-top length $\tau$ such that it cannot become shorter than a minimum $\tau_{\rm min} = 1/8\alpha_{\rm ZX}^{\rm max}$. More preciously, this cutoff can be calculated as $t_g^{\rm min}=(1/4\alpha_{\rm ZX}^{\rm max}+80)$ ns. For instance device 7 in Fig. \ref{fig:dynamictr}(b) reaches to a maximum at $\sim 2.5$~MHz and this introduces a gate time cutoff below $\sim 180$ ns as shown in Fig. \ref{fig:dyfidelity}(b). Moreover the device 7 shows a finite minimum at certain gate time where dynamic ZZ is suppressed to a minimum of $\sim$20 kHz, which results from the higher order contribution for stronger drive. Although th error cannot be eliminated, it can be reduced to $10^{-5}$ without decoherence error.  The worst scenario can be seen in device 6 where the gate error increases with gate time as total ZZ cannot be eliminated.

Following the experiment, we study what the CR gate is and how to constitute such a gate. We evaluate the gate performance with respect to ZZ interaction and classical crosstalk. Our theory also predicts that the CR gate with 99.9\% fidelity is achievable in the absence of static ZZ interaction with longer coherence times. The CR pulse produces new ZZ component adds on top of the static part, thus leading to a new strategy to cancel the total ZZ interaction, namely dynamical ZZ freedom. We show that this freedom is applicable in both transmon-transmon circuits and CSFQ-transmon circuits. In the next chapter, we will continue the discussion of parasitic ZZ interaction and propose a new gate at which both static and dynamical ZZ freedom is realized throughout the operation.
%
%


\chapter{Some Novel Gates}
\label{c5}
In chapter \ref{c4} we presented that static ZZ freedom is achievable experimentally on a superconducting circuit consisting of a pair of opposite sign anharmonicity at certain parameters. These parameters in the case of our experimental device~\cite{ku2020suppression} could be met by tuning external flux of CSFQ slightly away from the sweet spot. In that device CSFQ offers another degree of freedom, which is proved to be efficient for mitigating parasitic unwanted ZZ interaction. One can also achieve requirements for static ZZ freedom by changing coupling strength, which requires tuning coupler parameters. However, this tunability always challenges coherence characteristics of the coupler, as what has been seen in the CSFQ-transmon experiment that tuning flux away from sweet spot leads to a dramatic decrease in the decoherence time and then restricts the CR gate fidelity. Therefore suppressing flux noise while keeping tunability is the key factor to a coupler.

Alternatively,  there are several types of tunable couplers which have previously been both designed in theory and tested in experiment \cite{liu2006controllable,koch2007charge-insensitive,chen2014qubit,lu2017universal}, can also provide a new degree of freedom. One of the promising candidates is the asymmetric transmon~\cite{koch2007charge-insensitive,hutchings2017tunable}, which plays the same role as a tunable device, meanwhile the coherence time\index{coherence time} can slightly decrease to keep the balance of the two properties. Such an asymmetric transmon can either serve as a control/target qubit or a coupler instead of a harmonic oscillator. This type of qubit has been implemented in several devices and enables the improvement of entangling gates performance, by which desired entanglement can be enhanced or unwanted terms can be suppressed~\cite{sung2020realization,xu2020high,cai2021perturbation,stehlik2021tunable}. 

In this chapter we first introduce the properties of asymmetric transmons including frequency tunability and flux noise sensitivity. By including the flux-controlled tunable coupler with a continuous tunability in superconducting circuits, we take advantage of the cancellation for unwanted both static and dynamical ZZ interaction as described in chapter \ref{c3} and chapter \ref{c4},  then propose a new gate. Later we show how to take advantage of wanted ZZ and make it stronger to constitute a novel CZ gate\index{CZ gate}.

\section{Properties of an Asymmetric Transmon}
Let us revisit the discussion in chapter \ref{c2}, the asymmetric transmon\index{asymmetric transmon} consists of two Josephson junctions\index{Josephson junction} with external flux threading the loop, plasma frequency is then changed to 
\begin{equation}
\omega(f)\approx\sqrt{8E_CE_J(f)}=\sqrt{8E_CE_{J\Sigma}\sqrt{\cos^2(\pi f)+d^2\sin^2(\pi f)^2}},
\label{eq.plas}
\end{equation}
where $E_{J\Sigma}=E_{J1}+E_{J2}$ and $d=(a-1)/(a+1)$ with $a$ being the ratio of Josephson energy $E_{J1}$ and $E_{J2}$. Equation~(\ref{eq.plas}) is a periodic function in terms of flux, meaning that plasma frequency oscillates with $f
$. Here we consider external flux is limited to half a period such that $f\in[0,0.5]$ as it has covered the whole tunable window with two boundary frequencies calculated by
\begin{eqnarray}
\omega(f=0.5)&=&\sqrt{8dE_CE_{J\Sigma}},\\
\omega(f=0)&=&\sqrt{8E_CE_{J\Sigma}},
\end{eqnarray} 
and $\omega(f=0)>\omega(f=0.5)$ since $0<d<1$. Such two boundaries of the frequency window indicate that larger ratio $a$ leads to a narrow tunable domain. In the next we take an example to illustrate how the frequency is tuned. 

\begin{figure}[t!]
	\centering
	\includegraphics[width=0.75\textwidth]{{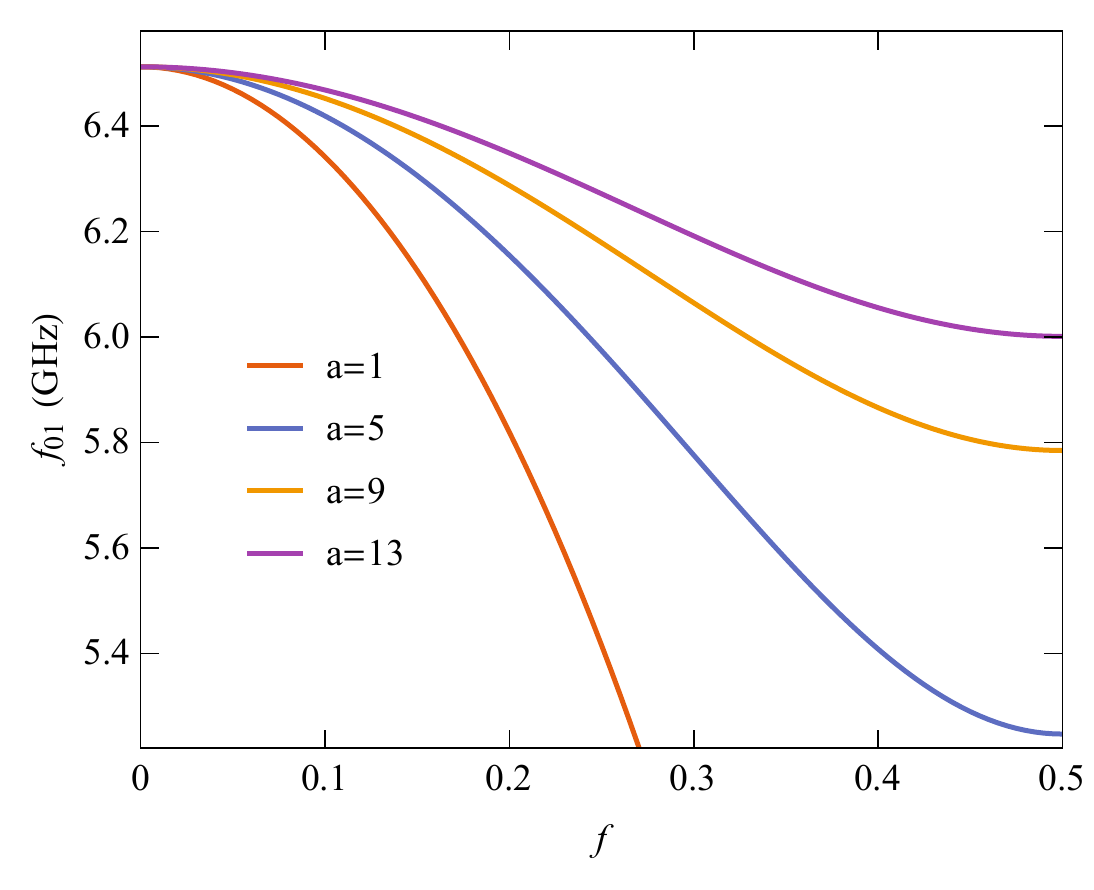}}		\put(-302,210){(a)}\\
	\vspace{-0.1in}
		\includegraphics[width=0.75\textwidth]{{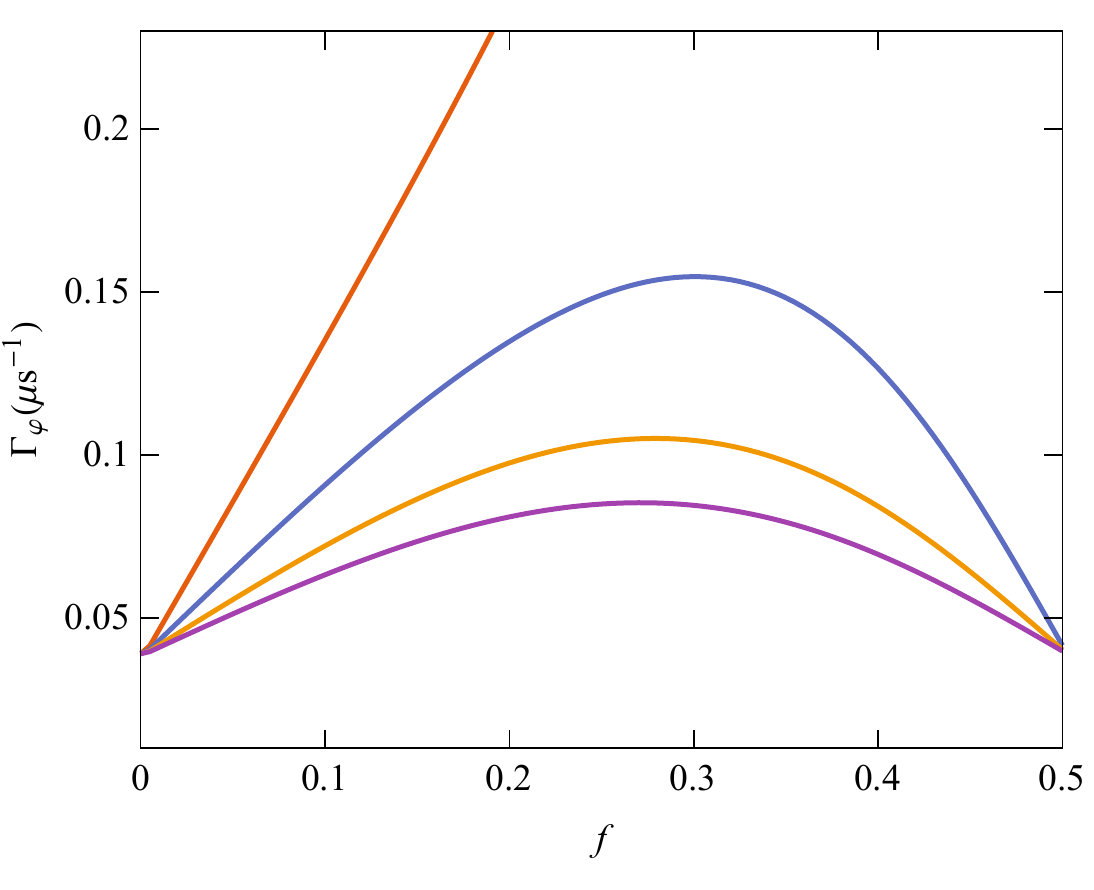}}
				\put(-302,210){(b)}
				\vspace{-0.15in}
	\caption{Asymmetric transmon properties (a) frequency $f_{01}$ versus flux (b) dephasing rate versus flux for different $a=E_{J1}/E_{J2}$, the ratio of two Josephson energies.}
	\label{fig:atransmonf}
\end{figure}

Given that the asymmetric transmon has $E_C=0.337$ GHz and $E_{J\Sigma}=17.5$ GHz, external flux $f$ is tuned between 0 and 0.5, we first plot the bare frequency $f_{01}$ as a function of the ratio $a=1,5,9,13$ in Fig.~\ref{fig:atransmonf}(a). One can see that the tunable frequency window becomes narrower with increasing ratio $a$, from more than 1.2 GHz at $a=1$ to around 500 MHz at $a=13$. 
However, the new degree of freedom also brings some troubles such as flux noise. 

Dephasing rate of a qubit is proportional to the frequency gradient as a function of flux $D_{\Phi}=|\partial f_{01}/\partial\Phi|$, details can be found in Appendix~\ref{app:T2}. This means that tuning the flux away from the sweet spot\index{sweet spot (SS)}  leads to decrease in coherence time. In other words, large tunability causes more loss. To assess the effect of flux noise on dephasing, we assume the flux noise power spectrum is identical for all different ratios, which allows background dephasing to be the same for each ratio. Another assumption is that the asymmetric transmon has the same modified expression of dephasing rate as the CSFQ with $\Gamma_{\Phi}=(0.00288{\rm m\Phi_0}) D_{\Phi}+0.039\mu s^{-1}$. By calculating $D_{\Phi}$ using data in Fig.~\ref{fig:atransmonf}(a) and substituting the derivative into $\Gamma_{\Phi}$, we obtain the dephasing rate and plot it for the four ratios in Fig.~\ref{fig:atransmonf}(b). It can be seen that at two ends dephasing rate is the lowest for all ratios, and the two points are called sweet spots of the asymmetric transmon. The dephasing rate can be approximated as parabolic function of flux, and the sensitivity to $1/f$ flux noise appears to be suppressed with increasing ratio $a$, i.e. $\Gamma_{\varphi}$ is comparably flat across the entire tuning range with a maximum value less than 0.1 $\mu s^{-1}$ when the ratio is 13. 

\section{The Parasitic Free (PF) Gate}
In chapter~\ref{c4} we introduce the dynamical ZZ freedom in driven circuits with a fixed-frequency bus resonator, which requires non-zero static ZZ interaction when qubits are idle. However, qubits are not always active or driven in a large quantum processor, therefore the crosstalk from surrounding qubits does affect the working qubits. 

One way to deal with this problem is replacing the harmonic bus cavity with a tunable coupler such as the asymmetric transmon. One can first tune the coupler frequency to turn off static parasitic interaction in idle status, and turn it on accordingly when qubits are driven to dynamically cancel out ZZ interaction. In this way we make a gate where ZZ freedom is achieved throughout the gate length, namely parasitic free (PF) gate\index{parasitic free (PF) gate}~\cite{xu23parasitic-free}. In the following, we discuss the case that two fixed-frequency transmons are coupled via the tunable coupler.  This circuit can produce perfect $\rm{ZX}_{90}$ entanglement when qubits are driven by the CR-type pulses, and is free of static ZZ interaction when qubits are idle.

\begin{figure}[h!]
	\centering
	\includegraphics[width=0.8\textwidth]{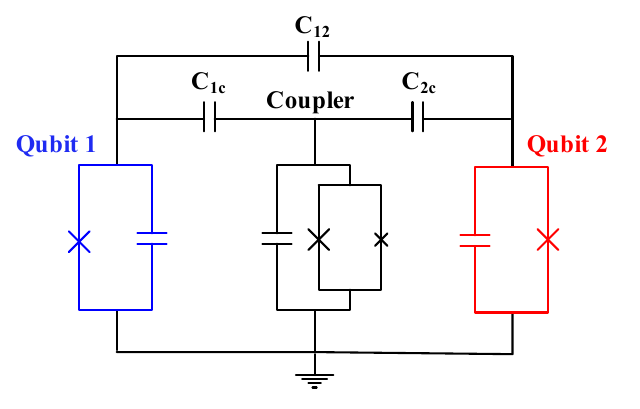}
	\vspace{-0.2in}
	\caption{Circuit schematic of two fixed-frequency transmon qubits directly coupled to a tunable coupler via $C_{1c}$ and $C_{2c}$, and capacitively couple to each other via $C_{12}$.}
	\label{fig:couplerzx}
\end{figure}

The circuit scheme is shown in Fig.~\ref{fig:couplerzx}. In the pairwise interacting quantum circuit, two fixed-frequency transmons are directly coupled via a capacitor $C_{12}$, and indirectly couple to each other mediated by the asymmetric transmon through $C_{1c}$ and $C_{2c}$, respectively. 

Usually $C_{12}$ is much smaller than $C_{1c}$ and $C_{2c}$, such that qubit-coupler interaction is much stronger than the qubit-qubit direct coupling\index{direct coupling}, i.e. $g_{1c}, g_{2c}\gg g_{12}$. The system Hamiltonian is written in the form of multilevel systems as 
\begin{equation}\label{eq.hcoupler}
\begin{split}
H&=\sum_{i=1,2,c}\sum_n\omega_i(n_i)|n_i+1\rangle\langle n_i+1|+\sum_{i<j}\sum_n \sqrt{(n_i+1)(n_j+1)}\\
&\times g_{ij}(|n_i\rangle\langle n_{i+1}|-|n_{i+1}\rangle\langle n_i|)\otimes(|n_j\rangle\langle n_{j+1}|-|n_{j+1}\rangle\langle n_j|),
\end{split}
\end{equation}
where $\omega_i(n_i)=E_i(n_i+1)-E_i(n_i)$ with $E_i(n_i)$ being the bare energy for subsystem $i~(i=1,2,c)$. There are two ways to describe this Hamiltonian: one is the sum of difference photon number blocks with interaction terms in the off-block part, the other is the sum of the same excitation number blocks with no interaction in between if counter-rotating terms are ignored~\cite{magesan2020effective}. Since the coupler is assumed to be far detuned from fixed-frequency transmons, here we choose the first approach and write the Hamiltonian in matrix form as 
\begin{equation}\label{eq.hmatrix}
H=\left( \begin{array}{cccccc}
0_p & G_{01} & G_{02} &&&\\
G_{01} & 1_p & G_{12} &&&\\
G_{02} & G_{12} & 2_p& && \\
&&&.&&\\
&&&&.&\\
& & & &&.\end{array}\right). \\
\end{equation}
where $i_p$ is the block containing $i$ photons and $G_{ij}$ includes all interaction between block $i$ and block $j$. 

To simplify the Hamiltonian, we can apply SW transformation\index{Schrieffer-Wolff (SW) transformation} to decouple $0_p$ block from higher photon excitation numbers, the system thus reduces to the two-qubit basis. As the coupler is an asymmetric transmon and the levels are anharmonic, so at least $2_p$ should be included in the full Hamiltonian before decoupling. After block diagonalizing Eq.~(\ref{eq.hmatrix}), the computational subspace can be fully diagonalized and then has the same form as Eq.~(\ref{eq.qqH}). In the dispersive regime\index{dispersive regime} that $g_{ij}/\Delta_{ij}\ll 1$ with $\Delta_{ij}=\omega_i-\omega_j$, we can calculate the eigenvalues of each level using perturbation theory up to fourth order correction, and then derive static ZZ interaction as
\begin{eqnarray}
\zeta&=&\zeta^{(2)}+\zeta^{(3)}+\zeta^{(4)}\label{eq.zetap},\label{eq.zetatot}\\
\zeta^{(2)}&=&2g_{12}^2\left(\frac{1}{\Delta_{12}-\delta_2}-\frac{1}{\Delta_{12}+\delta_1}\right)\label{eq.zeta2}, \\
\zeta^{(3)}&=&2g_{1c}g_{2c}g_{12}\left[ \frac{2}{(\Delta_{12}-\delta_2)\Delta_{1c}}-\frac{2}{(\Delta_{12}+\delta_1)\Delta_{2c}}\right.\nonumber\\
&&+\left.\frac{1}{\Delta_{1c}\Delta_{2c}}+\frac{1}{\Delta_{12}\Delta_{2c}}-\frac{1}{\Delta_{12}\Delta_{1c}}\right],
\end{eqnarray} 
\begin{eqnarray}
\zeta^{(4)}&=&g_{1c}^2g_{2c}^2\left[2\left(\frac{1}{\Delta_{1c}}+\frac{1}{\Delta_{2c}}\right)^2\frac{1}{\Delta_{1c}+\Delta_{2c}-\delta_c}  \right.\nonumber\\
&&-\left.\frac{2}{\Delta_{2c}^2(\Delta_{12}+\delta_1)} +\frac{2}{\Delta_{1c}^2(\Delta_{12}-\delta_2)}\right.\nonumber\\
&&-\left. \left(\frac{1}{\Delta_{2c}}+\frac{1}{\Delta_{12}} \right)\frac{1}{\Delta_{1c}^2}- \left(\frac{1}{\Delta_{1c}}-\frac{1}{\Delta_{12}} \right)\frac{1}{\Delta_{2c}^2}\right].\label{eq.zeta4}
\end{eqnarray} 

It is worth noting that higher excitation number in the coupler i.e. the second excited level with respect to the anharmonicity $\delta_c$ has already been included in the fourth order correction. 

\begin{figure}[h!]
	\centering
	\includegraphics[width=0.65\textwidth]{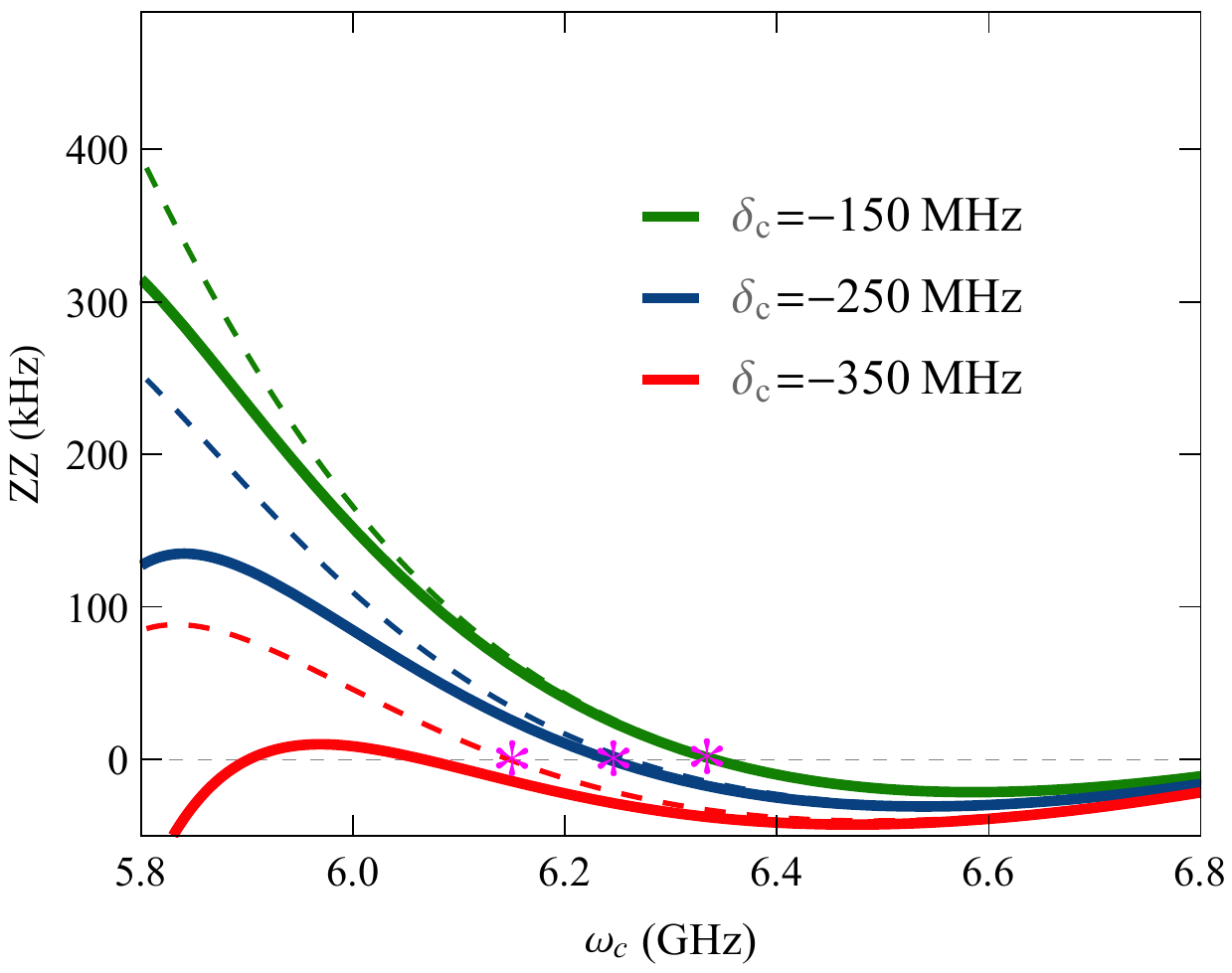}
	\vspace{-0.05in}
	\caption{Static ZZ interaction as a function of coupler frequency for three different coupler anharmonicity $\delta_c$. Solid line shows the results obtained from perturbation theory up to four order, dashed line are the results from full Hamiltonian model with off points marked by $\ast$.}
	\label{fig:coupler2zz}
\end{figure}

Circuit parameters are assumed as follows: the frequency and anharmonicity of qubit 1 and 2 are fixed at $\omega_1=5.12$ GHz, $\delta_1=-0.322$ GHz, and $\omega_2=5.0$ GHz, $\delta_2=-0.322$ GHz, separately, the tunable coupler frequency domain is $\omega_c=5.8-6.8$ GHz, qubit-coupler coupling strength $g_{1c}=g_{2c}=120$ MHz and direct coupling\index{direct coupling} strength $g_{12}=8$ MHz by considering a small direct capacitance. By substituting all parameters into Eqs~(\ref{eq.zetatot})$-$(\ref{eq.zeta4}), static ZZ interaction for three coupler anharmonicity is evaluated and plotted in Fig.~\ref{fig:coupler2zz}. We calculate static ZZ using two approaches: perturbation theory in Eq.~(\ref{eq.zetap}) with solid line and full Hamiltonian in Eq.~(\ref{eq.hcoupler}) with dashed line. 

One can see that the two approaches are consistent when coupler frequency is large while anharmonicity is small, but deviate from each other at lower coupler frequency with larger anharmonicity. This disagreement is due to the inaccuracy of perturbation theory on the edge of the dispersive limit, which is most evident at lower coupler frequency with anharmonicity $\delta_c=-350$ MHz. For instance, at $\omega_c=5.8$ GHz, $|020\rangle$ with bare energy 11.25 GHz, interacts with $|110\rangle$ with bare energy 10.92 GHz and $|011\rangle$ with bare energy 10.8 GHz, via the coupling strength $\sqrt{2}g_1=\sqrt{2}g_2\approx170$ MHz, leading to $g/\Delta\sim0.35$, which falls outside of the consistency domain~\cite{ansari2019superconducting}. 

{ \textbf{OFF status}:} Figure~\ref{fig:coupler2zz} shows that static ZZ freedom\index{static ZZ freedom} is achievable in all devices. As the anharmonicity is more negative, the required coupler frequency for zeroing ZZ interaction is lower. Such coupler frequency for ZZ freedom when qubits are in idle status corresponds to the ``off'' point of the flux sequence threading the asymmetric transmon.

\begin{figure}[h!]
	\centering
	\includegraphics[width=0.48\textwidth]{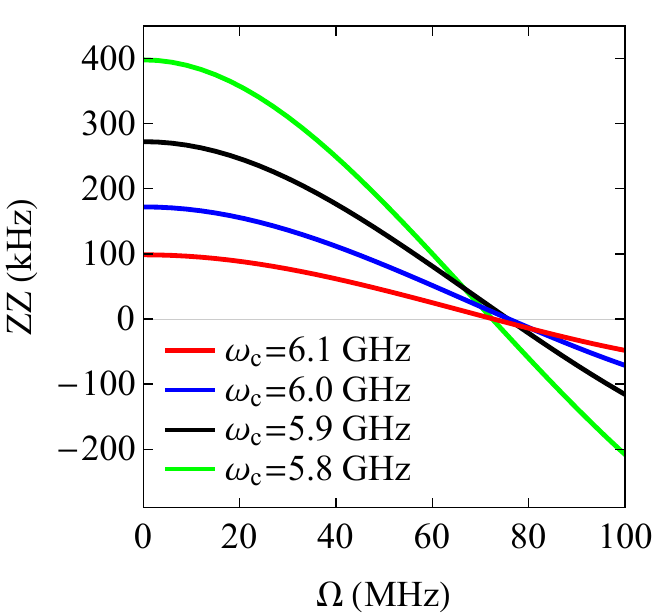}
	\includegraphics[width=0.48\textwidth]{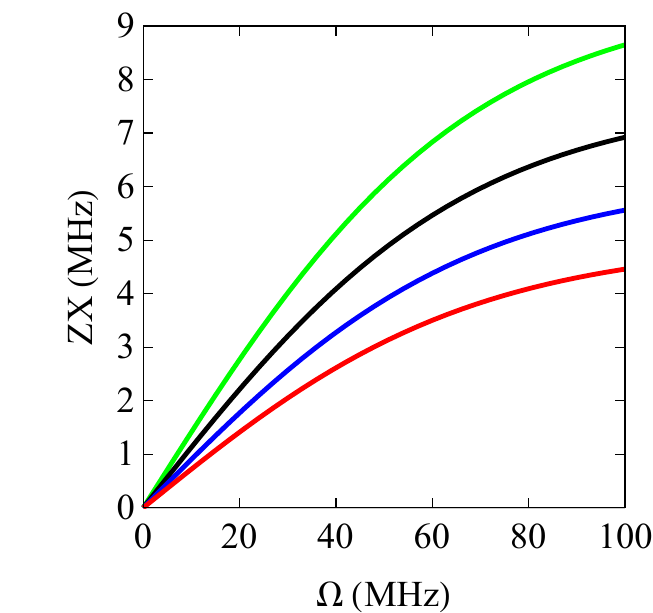}
		\put(-350,150){(a)}
	\put(-175,150){(b)}
	\caption{(a) Total ZZ interaction, the sum of static part and CR drive induced dynamical part.  (b) ZX rate versus CR driving amplitude $\Omega$ for four coupler frequencies with anharmonicity $\delta_c=-150$ MHz. Dynamical ZZ freedom takes place around $\Omega=75$ MHz.}
	\label{fig:coupler2dy}
\end{figure}

{ \textbf{ON status}:} Now let us move to driven regime. Based on the discussion in CR gates we know that fast ZX rate takes place at stronger driving amplitude, at which dynamical ZZ interaction is also larger. In such a device static ZZ should be designed large so that required cancellation driving amplitude is also strong. In the next we will focus on the device with anharmonicity $\delta_c=-150$ MHz as the static ZZ is the largest compared to the other two curves. The off frequency is about 6.345 GHz read from Fig.~\ref{fig:coupler2zz} when the qubits are in idle status. The system is driven by an echoed CR pulse, with which ZX gate is produced with angle $\theta=\alpha_{\rm ZX} t$ and other terms except ZZ can be eliminated via active cancellation pulse as described in section~\ref{sec:ccac}. External CR drive adds dynamical ZZ on top of the static component, and makes it possible to achieve dynamical ZZ freedom\index{dynamical ZZ freedom} if static and dynamical ZZ have the opposite sign, as shown in Fig.~\ref{fig:coupler2dy}.

Figure~\ref{fig:coupler2dy} shows corresponding ZX and ZZ strength for four different coupler frequencies using non-perturbative least action transformation. Total ZZ reduces to the static part at $\Omega=0$ which is consistent with Fig.~\ref{fig:coupler2zz}. Dynamical ZZ cancellation in all four devices can take place around driving amplitude $\Omega=80$ MHz, at which ZX component becomes larger with decreasing coupler frequency. For instance, if the ``on'' frequency is 5.8 GHz, the dynamic freedom takes place at $\Omega\sim75$ MHz, where corresponding ZX rate is   $\alpha_{\rm ZX}\sim 7.5$ MHz.  Such a ZX rate requires the flat-top length to be $\tau\sim 17$ ns for each CR pulse to perform $\pi/2$ ZX gate. By considering the each of the echo CR pulse is followed by a 40 ns $\pi$ rotation, the total pulse length is around 114 ns with respect to a square-shape CR pulse.

\begin{figure}[h!]
	\centering
	\includegraphics[width=0.65\textwidth]{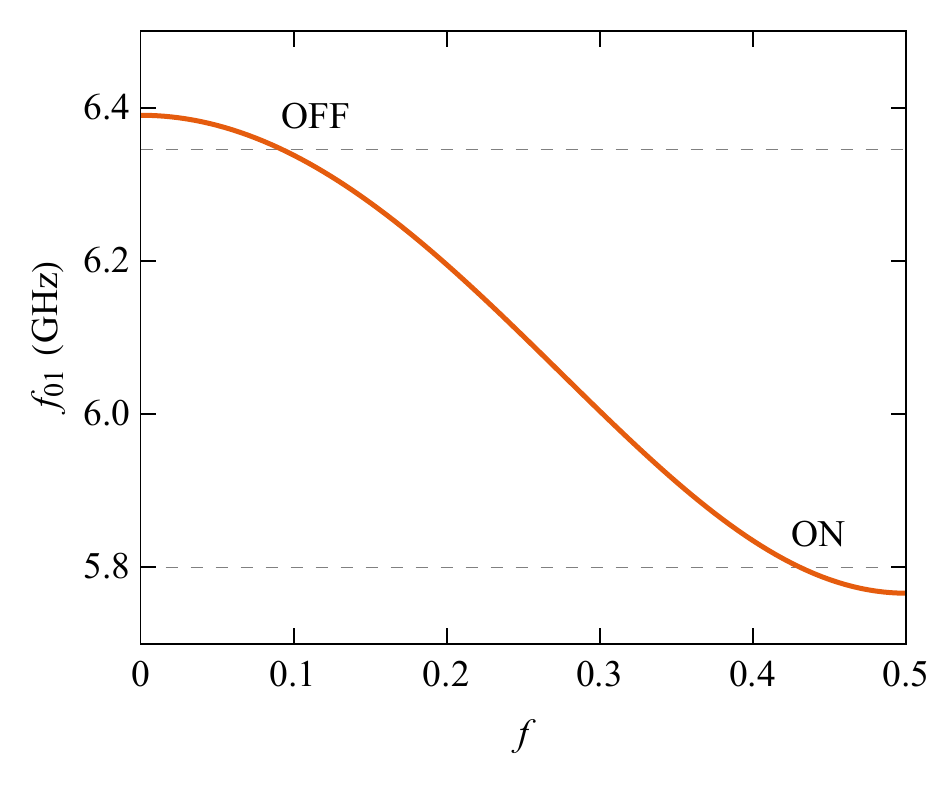}
	\vspace{-0.2in}
	\caption{Coupler frequency  versus external flux. On and off frequencies are designed to be close to sweet spot with a gap about 50 MHz.}
	\label{fig:acouplerpf}
\end{figure}

Another feature of the PF gate is that flux noise can be suppressed throughout the operation. As mentioned before, the asymmetric transmon suffers from $1/f$ noise when tuning the flux away from sweet spot. In order to reduce the error caused by flux noise, the circuit design and qubit control are needed to be optimized. As described in Fig.~\ref{fig:atransmonf}(b) the flux noise is suppressed most around the sweet spots $f=0$ and $f=0.5$. If we can find a device where the two on and off points are around sweet spots with  lower dephasing rate $\Gamma_\varphi$, the gate performance will undoubtedly be improved. Here we show one example of such an ideal device. The coupler can be designed as follows: $E_C=0.142$ GHz, $E_{J1}/E_{J2}=10$ and $E_{J\Sigma}=37.6$ GHz. Figure~\ref{fig:acouplerpf} plots the coupler frequency as a function of external flux. Dashed lines mark on and off frequency with about 50 MHz gap to the frequencies at sweet spots, which provides room for circuit parameters variations.

In such a device where the coupler frequency is tunable, we can realize the universal ZZ freedom. This means that in idle status tuning coupler frequency to the off point can decouple the ZZ interaction, with which the single-qubit operation accuracy can be improved; In driven status, first by tuning the coupler frequency to the on point, a non-zero static parasitic interaction is obtained, then applying external pulses especially driving control qubit with target  frequency adds exactly the same amount of ZZ but with an opposite sign. The total effect will make the dynamical ZZ cancellation possible. Compared to the devices described in chapter~\ref{c4}, the new circuit with the asymmetric transmon can produce a larger ZX rate, which can effectively reduce the gate length. 
\section{The Tunable CZ Gate}
In previous discussion we always aim to suppress or eliminate this parasitic ZZ interaction. In fact, without any control the free evolution of a system in presence of ZZ coupling can be used to constitute a controlled-Z (CZ) gate\index{CZ gate}~\cite{dicarlo2009demonstration}. In a large quantum processor, we want a pair of interacting qubits to be free of ZZ interaction in idle status but preserve strong ZZ when performing a CZ gate. Generally speaking, decoherence time of a superconducting qubit is a couple of microseconds. To realize high-performance gates the operation time should be shorter, i.e. several Hundreds of nanosecond, which limits the ZZ interaction to be in the magnitude of megahertz. Then it comes to the question how can we choose the frequency domain to fast tune the ZZ interaction from zero to several megahertz? A few experiments have been performed by replacing the harmonic resonator with the asymmetric transmon in transmon-transmon pairs~\cite{collodo2020implementation,xu2020high-fidelity,sung2020realization}, and successfully realized shorter CZ gates, but none of them can tune ZZ interaction to the exact zero.

 Here we introduce a device on which an asymmetric transmon coupled to a CSFQ via a harmonic bus resonator. As discussed, CSFQ-transmon pair is a good device where static ZZ freedom can be achieved without driving meanwhile entangling two qubits, although CSFQ suffers from shorter decoherence time as flux is away from sweet spot\index{sweet spot (SS)}. One way to deal with flux noise is fixing CSFQ at the sweet spot, and utilizing the asymmetric transmon instead to tune ZZ interaction. Thus we can totally turn off ZZ interaction in idle status, and turn it on to realize the CZ gate.
 
\begin{figure}[h!]
	\centering
	\includegraphics[width=1\textwidth]{{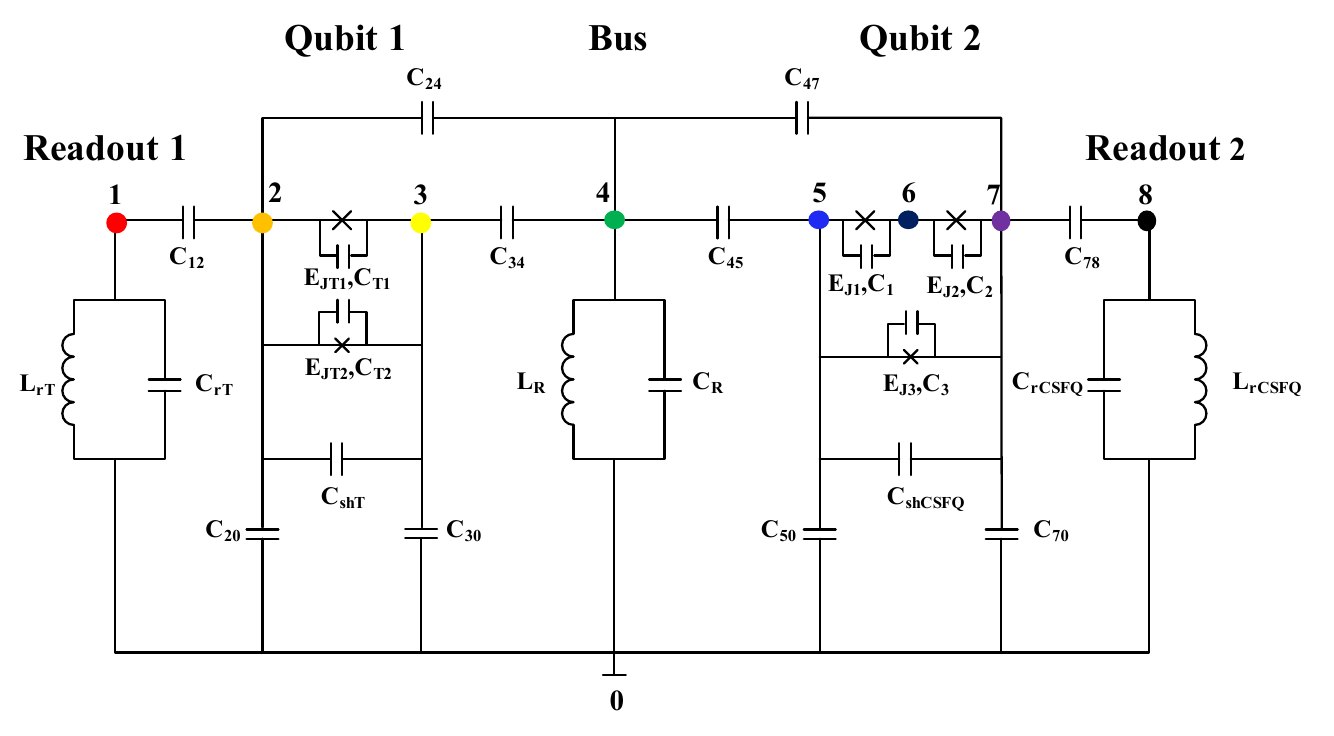}}
	\vspace{-0.4in}
	\caption{Circuit scheme of CSFQ coupled to an asymmetric transmon via a bus resonator.}
	\label{fig:acouplercir}
\end{figure}

The circuit is designed as follows: an asymmetric transmon couples to a CSFQ mediated by a bus resonator, each qubit is measured by a readout cavity\index{readout cavity}. The frequency is arranged that bus resonator is in between two qubits. Detailed circuit scheme is shown in Fig.~\ref{fig:acouplercir}. Compared to the experimented hybrid CSFQ-transmon circuit in chapter~\ref{c3} and~\ref{c4}, Josephson junction\index{Josephson junction} in transmon is replaced by a DC SQUID,  two more capacitors are included using ANSYS Q3D software, which efficiently performs the 3D and 2D quasi-static electromagnetic filed simulations required for the extraction of RLCG parameters from an interconnect structure. 

Before we go deeper into the analysis of the circuit and parameters, one question should be answered, that is how we can strengthen ZZ interaction and remain the zero point in the same device. When the bus resonator is far detuned between two qubits, maximally ZZ is less than 1 MHz as shown in Fig.~\ref{fig:ZZ_flux}. Perturbed derivation of ZZ interaction in Eq.~(\ref{eq.zeta}) indicates that increasing $J$ coupling can effectively enlarge it, and the most applicable method is to reduce the bus resonator frequency. Therefore the qubit-bus detuning should be changed, i.e. less than 1 GHz when the gate is on. Circuit Hamiltonian has the same form as Eq.~(\ref{eq.Heff}), with corresponding circuit parameters are listed in Tab.~\ref{tab:acoupler}.  

\begin{figure}[h!]
\centering
\includegraphics[width=0.48\textwidth]{{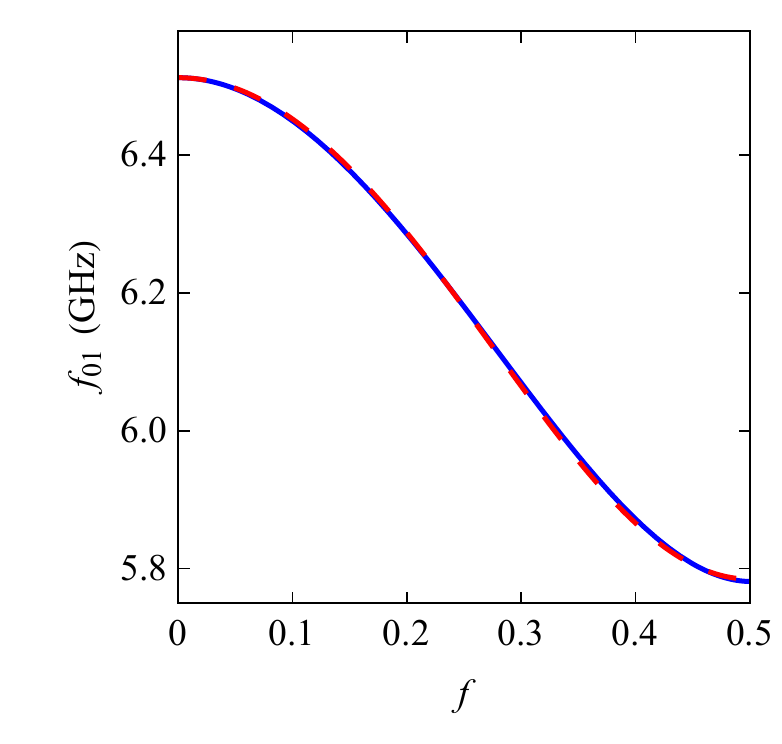}}\hspace{0.1in}
\includegraphics[width=0.48\textwidth]{{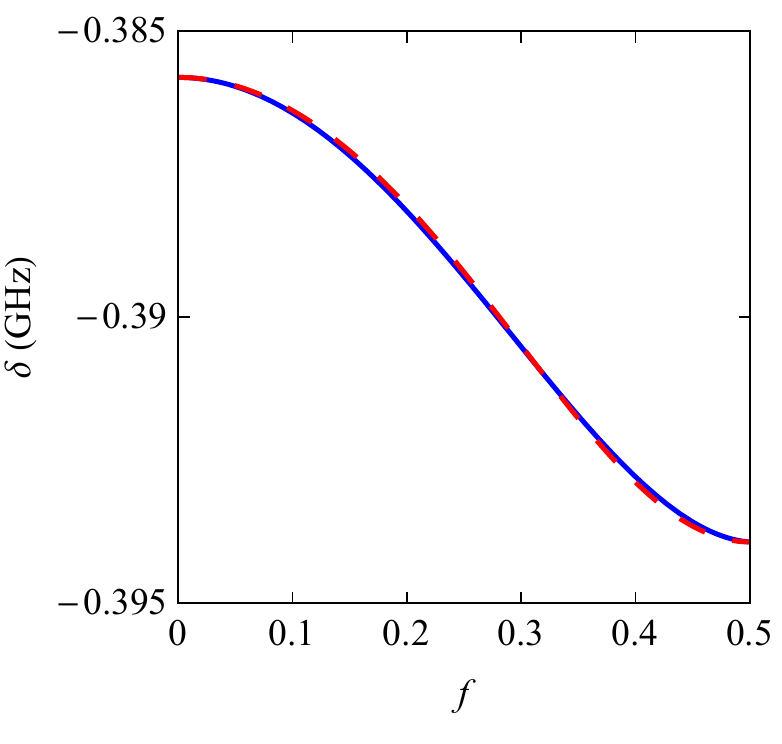}}
\put(-350,155){(a)}
\put(-185,155){(b)}
\vspace{-0.2in}
\caption{(a) Asymmetric transmon frequency $f_{01}$ as a function of flux $f$(b) Asymmetric transmon anharmonicity as a function of flux $f$. Solid line represents the result from numerical simulation, while dotted line represents corresponding fitting function.}
\label{fig:acoupler}
\end{figure}

Figure~\ref{fig:acoupler} shows frequency tuning window is about 700 MHz, anharmonicity changes slowly from $-385$ MHz to $-394$ MHz. Solid blue curve is the numerical derivation and dashed red is the result from fitting function expressed as
\begin{eqnarray}
\omega_1&=&6.51262 - 6.7 f^2 + 3.5 f^3 + 8.1 f^4,\label{eq.omega1}\\
\delta_1&=&-0.385807 - 0.0635 f^2 + 0.124 f^4.
\end{eqnarray}

\begin{table}[h!]
	\begin{center}
		\begin{tabular}{|c|c|c|c|c|c|}
			\hline 
			\multicolumn{4}{|c|}{Capacitance (fF)} & \multicolumn{2}{c|}{Josephson energy (GHz)}\tabularnewline
			\hline 
			$C_{12}$&$3.4$  &$C_{78}$  &$4.44$ & $E_{JT1}$ & $15.75$\tabularnewline\hline 
			$C_{20}$&$52.8$  &$C_{70}$  &$68$ & $E_{JT2}$  & $1.75$\tabularnewline
			\hline 
			$C_{30}$&$43.9$  &$C_{50}$  & $57.5$  & $a$ & $9$\tabularnewline
			\hline 
			$C_{ shT}$&$21$ &  $C_{shCSFQ}$&$27.5$& $E_{J1}=E_{J2}=E_J$ & $111.70$\tabularnewline
			\hline 
			$C_{T1}$&$4$ &  $C_1=C_{2}=C$&$4$ &$E_{J3}=\alpha E_J$ &$48.03$\tabularnewline
				\hline 
			$C_{T2}=C_{T1}$&$4/9$ & $C_{3}=\alpha C$ &  $1.72$ & $\alpha$& $0.43$\tabularnewline
			\hline 
			$C_{34}$ &$26.52 $ & $C_{45}$ & $32.6$&\multicolumn{2}{c|}{Inductance (nH)}\tabularnewline
			\hline 
			$C_{24}$ &$2.17$ & $C_{47}$ & $2.24$&$L_{rT}$&1.14\tabularnewline
			\hline 
				$C_{rT}$ &$454.73$ &$C_{rCSFQ}$  &$461.32$ & $L_{rCSFQ}$ & $1.15$\tabularnewline
			\hline 
				$C_{R}$ &$600.59$ &  & & $L_{R}$ & $1.50$\tabularnewline
			\hline 
		\end{tabular}
	\end{center}
\vspace{-0.15in}
\caption{\label{tab:acoupler}Circuit parameters design values that were extracted using ANSYS Q3D Extractor simulation of the qubit layout.}
\end{table}

Let us consider $C_{24}\ll C_{34}$ and $C_{47}\ll C_{C45}$, thus contribution from the two additional capacitors will not affect the expression in Eq.~(\ref{eq.ge}), and all $g$ couplings can be obtained by simply replacing $a-h$ with 1-8 correspondingly. By substituting circuit parameters into the expression in chapter~\ref{c3}, CSFQ bare frequencies, anharmonicity, bus frequency and coupling strengths are calculated in Tab.~\ref{tab:couplerfreq}. 

The variation of coupling strength due to tunability of the coupler is tiny and then ignored. Note that CSFQ-bus detuning is only 400 MHz while corresponding coupling strength is 167 MHz. Although the system is still within the dispersive regime, the accuracy of perturbation theory is not enough. In the following we will denote the asymmetric transmon as Q1, CSFQ as Q2 and deal with the system using full Hamiltonian model.
\begin{table}[h!]
	\centering
	\begin{tabular}{| l | l | }
		\hline
		CSFQ freq. &  4.9 GHz  \\ \hline
		CSFQ anharm. &  $577.6$ MHz \\ 
		\hline
		Bus  freq.&  5.3 GHz \\ \hline
		Bus-CSFQ coupling & 167 MHz  \\ \hline
		Bus-Trans. coupling &  161 MHz   \\ 
		\hline
		CSFQ readout & 6.9 GHz\\ \hline
		CSFQ readout g & 42 MHz \\ \hline
		trans. readout & 7.0 GHz\\ \hline
		trans. readout g & 40.6  MHz \\ \hline
		direct coupling & 10.1 MHz\\ \hline
	\end{tabular}
	\caption{\label{tab:couplerfreq}Frequency scales on device with CSFQ at the sweet spot.}
\end{table}

\begin{figure}[h!]
	\centering
	\includegraphics[width=0.85\textwidth]{{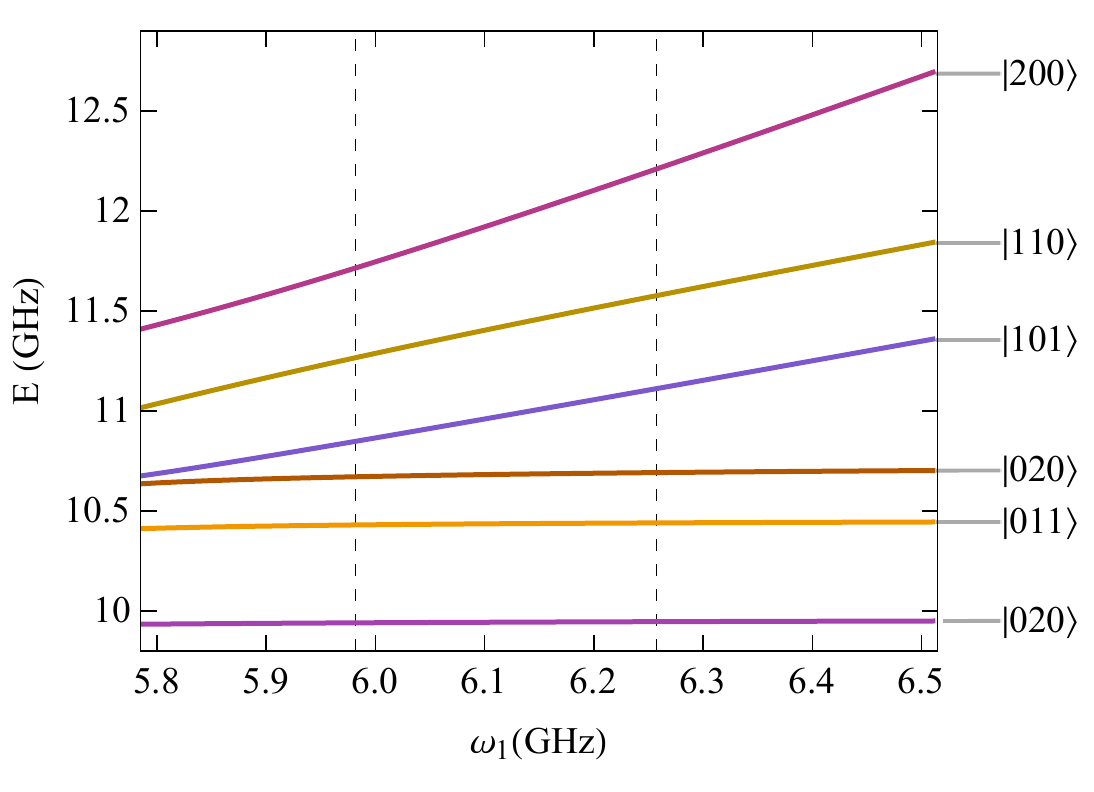}}
	\vspace{-0.2in}
	\caption{System eigenvalues as a function of the asymmetric transmon frequency. Only states in the two-excitation manifold are shown. }
	\label{fig:acouplerlevel}
\end{figure}

To easily illustrate how the CZ gate is operated, we use its energy-eigenstates $|Q1,R,Q2\rangle$ to describe the system. Figure~\ref{fig:acouplerlevel} shows system eigenvalues as a function of Q1 frequency, and only states in the two-excitation manifold are plotted. Although the anti-crossing is beyond tunable domain, it is also worth noticing that two-photon process between $|101\rangle$ and $|020\rangle$ is possible especially when the gap is small at lower transmon frequency.
For an ideal CZ gate, the computational state $|101\rangle$ should adiabatically evolve from the off point to the on point at which ZZ is strengthened, then back to the off point with no leakage\index{leakage} to non-computational states. This can be realized by tuning transmon frequency, during this process non-zero ZZ leads to a CZ gate if the total phase accumulation of Q1 is $\pi$. Here the adiabatic process\index{adiabatic process} means unwanted transitions should be avoided. Particularly, the minimum gap between |$101\rangle$ and other states is a key factor in determining adiabaticity.

In our device, the minimum gap occurs between $|101\rangle$ and $|020\rangle$ states, and can be evaluated through near-degenerate perturbation theory when the two states are near resonance~\cite{sakurai1994modern}. There is a two-fold degeneracy before interaction Hamiltonian is switched on, then we can separate the eigenstates into two sets: {$|m\rangle$} denotes the unperturbed eigenstates of the degenerate energy $E_d$ with $m\in d$, and {$|l\rangle$} is the set of unperturbed eigenstates with different energies. Note that the two sets do not coincide with each other. Define projectors $Q_d=\sum_{m\in d}|m\rangle\langle m|$ and $P_d=1-Q_d$, then second order perturbation theory in the degenerate subspace gives rise to~\cite{sakurai1994modern} 
\begin{equation}
H_0+Q_dVQ_d+Q_dVP_d(E_k-H_0)^{-1}P_dV-E_k=0,
\label{eq.degenerate}
\end{equation}
where $H_0$ is the unperturbed Hamiltonian, $V$ is the interaction Hamiltonian and $E_k$ is the eigenvalues. By substituting all parameters into the equation above, we find the minimum gap between two states is $\sim$ 20 MHz, corresponding to $2g_{101\leftrightarrow020}$, similar to what we found in Eq.~(\ref{eq.Jde}). When operating at transmon frequency away from the degenerate point, the condition $g_{101\leftrightarrow020}t>1$ should be satisfied, meaning that the CZ gate should be slow enough to avoid such a transition, i.e. $t>100$ ns.

\begin{figure}[h!]
	\centering
	\includegraphics[width=0.75\textwidth]{{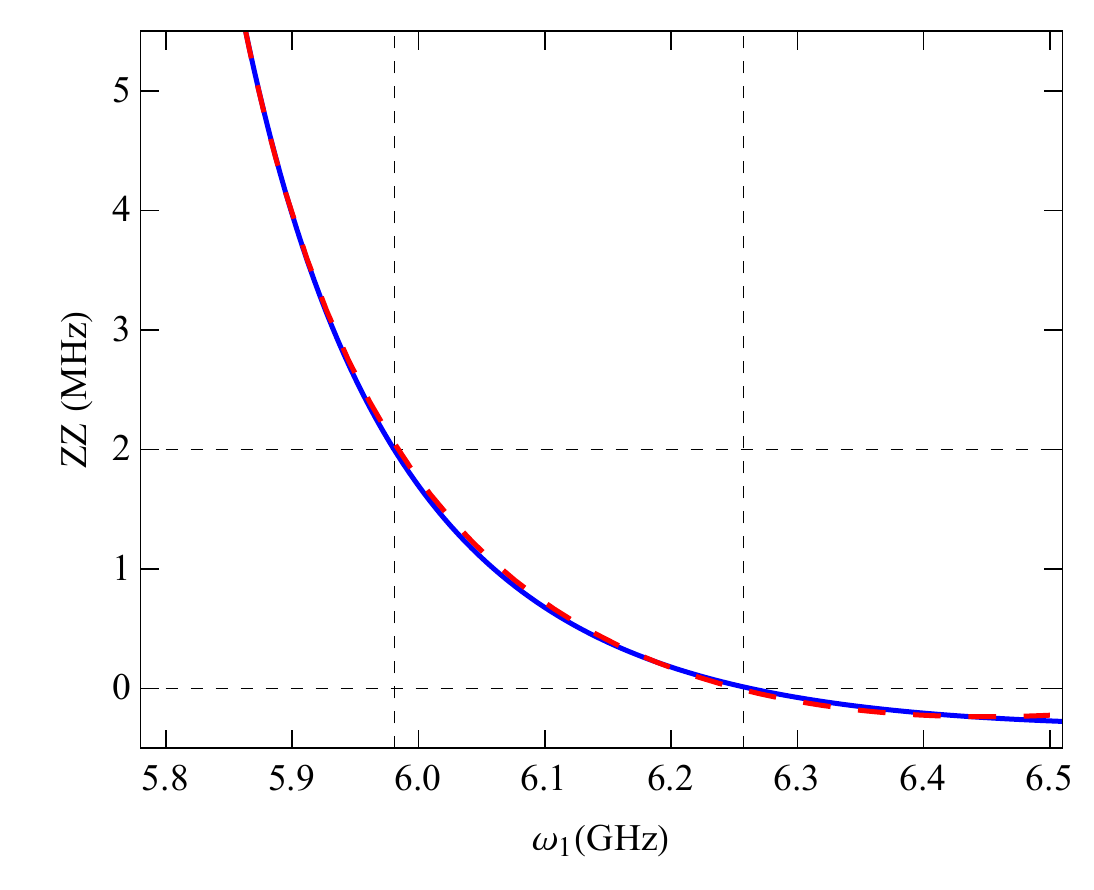}}
	\vspace{-0.2in}
	\caption{ZZ interaction as a function of asymmetric transmon frequency, on and off frequencies are 5.981 GHz and 6.257 GHz, respectively. Solid line represents the result from numerical simulation, while dotted line represents corresponding fitting function.}
	\label{fig:acouplerZZ}
\end{figure}

Similarly the full Hamiltonian can be diagonalized and rewritten as 
\begin{equation}
H=\tilde{\omega}_1|100\rangle\langle 100|+\tilde{\omega}_2|001\rangle\langle 001|+(\tilde{\omega}_1+\tilde{\omega}_2+\alpha_{\rm ZZ})|101\rangle\langle 101|,
\end{equation}
with $\alpha_{\rm ZZ}$ being plotted as a function of transmon frequency in Fig.~\ref{fig:acouplerZZ}. 

One can see that under current parameters ZZ coupling can be completely turned off at $\omega_1=6.257$  GHz. Corresponding fitting function marked by dashed line is 
\begin{equation}\label{eq.zfit}
\alpha_{\rm ZZ}(\omega_1)=\left[\frac{1}{0.73(\omega_1-5.7)}+1.75(\omega_1-5.75)^2-2.92\right] \rm MHz.
\end{equation}

To speed up CZ gate while maintaining no leakage to higher levels, we choose a relatively large ZZ coupling strength to be 2 MHz as the working point at $\omega_1=5.981$ GHz. Corresponding flux can be evaluated from fitting functions Eq.~(\ref{eq.omega1}) as
$f=0.214$ OFF and $f=0.338$ ON, separately.

\begin{figure}[h!]
	\centering
	\includegraphics[width=0.75\textwidth]{{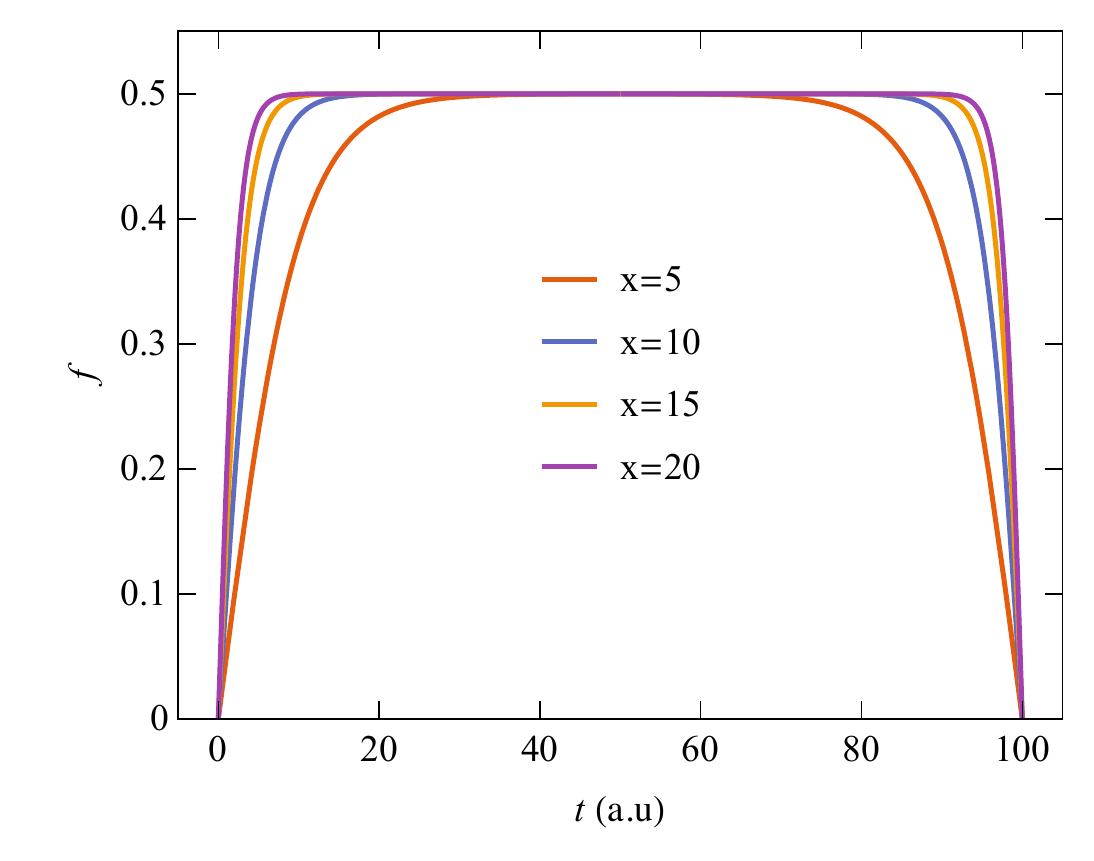}}
	\vspace{-0.2in}
	\caption{Control pulse with different flat top length versus time (arbitrary unit). $f$ is the flux threading the loop, and $x$ determines the flat top length as shown in Eq.~(\ref{eq.pulse}).}
	\label{fig:acouplertanh}
\end{figure}

To further suppress the leakage to noncomputational levels, an optimized control pulse is needed to be implemented such as flat-top Gaussian pulse~\cite{collodo2020implementation} and Slepian-based optimal control~\cite{sung2020realization}. In each case the pulse is not square, then accumulated conditional phase is expressed as
\begin{equation}\label{eq.zint}
\phi_{\rm ZZ}=\int\alpha_{\rm ZZ}(t)dt,
\end{equation}
and a CZ gate is realized when $\phi_{\rm ZZ}=\pi$. In the following we will take an example of tanh-shaped pulse to estimate the gate length. The pulse is expressed as 
\begin{equation}
\label{eq.pulse}
\begin{split}
f(f_0,f_{\rm end},x,s,t)&=[f_0+\tanh(xst)(f_{\rm end}-f_0)][\sign(1/s-t)+1]/2\\
&+\{f_0+\tanh[xs(2/s-t)](f_{\rm end}-f_0)\}[\sign(t-1/s)+1]/2,
\end{split}
\end{equation}
where $s=1/2t_g$ with $t_g$ being the gate length, $f_0=0.214$, $f_{\rm end}=0.338$ and $x$ determines the flat-top length. In arbitrary unit this pulse shape is plotted in Fig.~\ref{fig:acouplertanh}.

By combining all fitting functions and substituting all parameters, one can obtain the gate length. For instance, if the pulse take the shape of $x=5$, the pulse is only a function of gate length $t_g$. We first substitute Eq.~(\ref{eq.pulse}) into Eq.~(\ref{eq.omega1}), and then substitute Eq.~(\ref{eq.omega1}) into Eq.~(\ref{eq.zfit}) to get the gate-length dependent ZZ interaction, finally we can integrate ZZ interaction from $t=0$ to $t=t_g$ as shown in Eq.~(\ref{eq.zint}). By solving the integration $\phi_{\rm ZZ}=\pi$, a CZ gate is achieved at the gate length $t_g\sim419$ ns.

Compared to other schemes to perform CZ gates, this system can completely turn off parasitic ZZ interaction although the gate length is longer. Different from other protocols in which the asymmetric transmon serves as a coupler~\cite{collodo2020implementation,xu2020high-fidelity,sung2020realization}, the main leakage\index{leakage} source in this device is the transition between $|101\rangle$ and $|020\rangle$ since transmon frequency is far detuned from CSFQ and bus cavity. This transition needs to  be realized by two-photon process\index{two-photon process}, therefore is comparably weak. The CZ gate length can be further reduced by pushing transmon frequency closer to the bus cavity, meanwhile avoiding the unwanted transition by optimizing the control pulse.

We  have studied two novel gates based on the asymmetric transmon. The first one, PF gate, is realized in a circuit where two fixed-frequency transmons couple to each other via the asymmetric transmon. On such a gate we take advantage of the cancellation for unwanted ZZ interaction throughout the operation, namely static ZZ freedom in idle status and dynamical ZZ freedom in driven qubits. The second one, the tunable CZ gate, is performed in a circuit where the asymmetric transmon serves as a qubit, and couples to a CSFQ via a harmonic bus resonator. In this device wanted ZZ can be completely switched off when the gate is inactive, but wanted ZZ should be further strengthened to make the gate length shorter when it is active. 
\newpage
\thispagestyle{empty}

\chapter{Three-Qubit Interaction}
\label{c6}
In previous chapters we have analyzed the characteristics of wanted \& unwanted two-qubit interaction in superconducting qubits. However, as quantum processors with many qubits are developed, each qubit is inevitably exposed to interacting neighbors.  Interaction with nearby  idle  qubits will undoubtedly cause unwanted entanglement and accumulate errors across the system in the presence or absence of a two-qubit gate. Such interactions do not obviously affect the resting qubits, but can give rise to gate parameters change and lower gate performance. For instance, errors arising from coupling between control/target qubit of a CR gate and their surrounding idle qubits lead to reduction in the gate fidelity~\cite{malekakhlagh2020first-principles,sundaresan2020reducing,cai2021perturbation}. 

\begin{figure}[h!]
	\centering
	\includegraphics[width=0.95\textwidth]{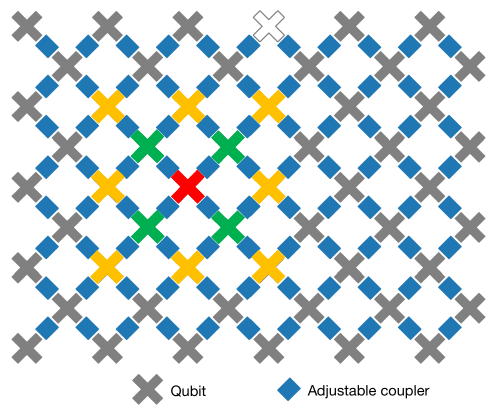}
	\vspace{-0.1in}
	\caption{\label{fig:sycamore}Layout of Sycamore processor~\cite{arute2019quantum}. Given a qubit marked by red, there are four qubits marked by green next to it, and each of which is surrounded by another three qubits marked by orange. }
\end{figure}

For a larger quantum processor, i.e. Sycamore~\cite{arute2019quantum} which uses 53 qubits and 86 couplers as shown in Fig.~\ref{fig:sycamore},  qubit at one vertex marked by red is connected with four neighboring qubits marked by green via adjustable couplers, thus four pairs of two-body interactions will change the qubit parameters and can accumulate errors to related quantum gates. Furthermore, each neighboring qubit is surrounded by another three idle qubits marked by yellow, thus there are 12 non-pairwise additive three-body interaction with respect to the qubit at red vertex. Although three-qubit interaction is usually weaker than two-body interaction~\cite{khoshnegar2014toward}, but in total the contribution to one qubit is nontrivial and can even destroy quantum gates.

This chapter we study the characteristics of such three-body interaction in two complicated circuit schemes: triangle and planar geometries of three qubits~\cite{xu2024lattice}. In triangle case there are three couplers among the three qubits and in planar one the number of coupler is two. These circuits can be seen as the continuation of previous two-qubit model by adding a third qubit. To study how the neighboring qubits affect two-qubit gates, we briefly describe a three-qubit model with respect to two-body interaction, and then propose a protocol for static three-qubit Hamiltonian tomography to find out the dependency of two- and three-body coupling strengths. Finally we present several examples with simulation results to illustrate the impacts and make a comparison with the two-qubit model. 
\section{Triangle Geometry\index{triangle geometry} of  Three-Qubit Coupling }
\begin{figure}[h!]
	\centering
	\includegraphics[width=0.8\textwidth]{{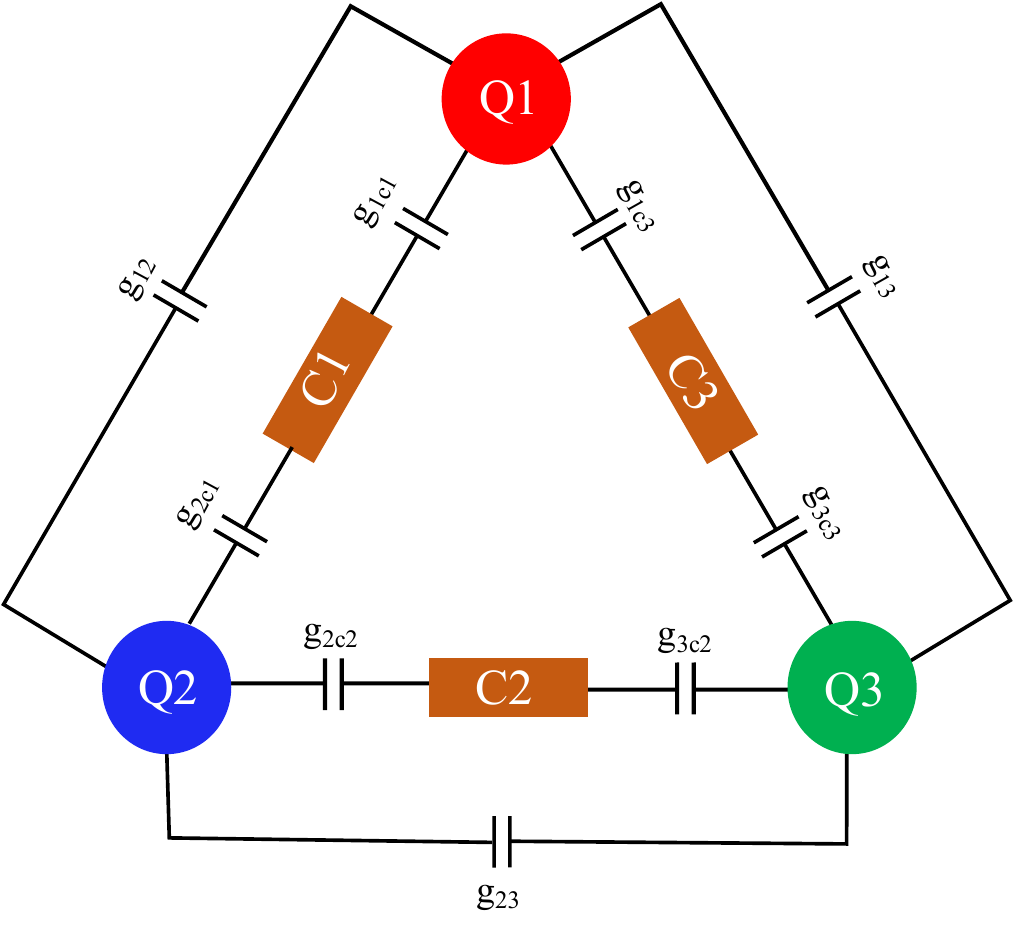}}
	\vspace{-0.1in}
	\caption{Triangle geometry of three qubit interaction. Each pair of qubits is indirectly coupled to each other via a coupler, and directly via a capacitor. }
	\label{fig:triangle}
\end{figure}

First, let us consider a triangle circuit which consists of three superconducting qubits and three couplers as shown in Fig.~\ref{fig:triangle}. Here circles represent qubits and rectangles represent couplers. Each pairwise qubits are connected to each other via a coupler in the coupling strength $g_{qc_i}$ between qubit $q$ and coupler $c_i$ with $q,i=1,2,3$, respectively. Meanwhile, they are also capacitively coupled to each other via direct coupling $g_{qq'}$ between qubit $q$ and $q'$. For simplicity, all couplers are assumed to be harmonic oscillators.

Following the convention used in chapter \ref{c3}, bare qubit $q$ frequency is $\omega_q(n_q)={E}_q(n_q+1)-{E}_q(n_q)$ with bare energy $E_q(n_q)$ at Fock state $|n_q\rangle$ by setting $\hbar\equiv1$, and coupler $c_i$ frequency is $\omega_{c_i}$. We can ignore the weak interaction between different resonators, and then write the circuit Hamiltonian $H$ in the basis of multilevels following the strategy used in the two-qubit model as
\begin{equation}
\begin{split}
H&=\sum_{i=1}^3\omega_{c_{i}}c_{i}^{\dagger}c_{i}+\sum_{q=1}^3\sum_{n}\omega_q(n_q)\left|n_q+1\right\rangle \left\langle n_q+1\right|\\
&+\sum_{i=1}^3\sum_{q=1}^3\sum_{n}\sqrt{n_q+1}g_{qc_{i}}\left(c_i-c_i^{\dagger}\right)\left(
\left|n_q+1\right\rangle \left\langle n_q\right|-\left|n_q\right\rangle \left\langle n_q+1\right|\right)\\
&+\sum_{q\neq q'}\sum_{n m}\sqrt{n_q+1}\sqrt{m_{q'}+1}g_{qq'}\\
&\times\left(|n_q\rangle\langle n_q+1|-|n_q+1\rangle\langle n_q|\right)\otimes\left(|m_{q'}\rangle\langle m_{q'}+1|-|m_{q'}\rangle\langle m_{q'}+1|\right).
\end{split}
\label{eq.threeH}
\end{equation}

The first line in Eq.~(\ref{eq.threeH}) is the unperturbed Hamiltonian of qubits and resonators, second line is qubit-coupler interacting Hamiltonian, and the rest is direct capacitive coupling part. In the case that coupler frequencies are far detuned from qubits, higher or lower, we can decouple all couplers from the circuit using SW transformation, then the qubit Hamiltonian $H_Q$ in the dispersive regime\index{dispersive regime} is reduced to 
\begin{equation}
\label{eq.threeQf}
\begin{split}
H_{Q}&=\sum_{q}\sum_{n}\sum_{i}\bar{\omega}_{q}(n_q)\left|n_q+1\right\rangle\left\langle n_q+1\right|-\sum_{q\neq q'}\sum_{nm}\sum_{i}\sqrt{n_q+1}\sqrt{m_{q'}+1}\\
&\times J_{c_{i}qq'}^{n_qm_{q'}}\left(|n_q\rangle\langle n_q+1|-|n_q+1\rangle\langle n_q|\right)\otimes\left(|m_{q'}\rangle\langle m_{q'}+1|-|m_{q'}\rangle_{q'}\langle m_{q'}+1|\right),
\end{split}
\end{equation}
where dressed frequency and effective virtual photon exchange rate are 
\begin{align}
\bar{\omega}_q(n_q)=&\omega_q(n_q)-\sum_i \frac{n_qg_{qc_{i}}^{2}}{\omega_{c_i}-\omega_{q}({n_q})},\label{eq.omegaq}\\
J_{c_{i}qq'}^{n_qm_{q'}}=&g_{qq'}-\frac{g_{qc_{i}}g_{q'c_{i}}}{2}\left[\frac{1}{\Delta_{qc_i}(n_q)}+\frac{1}{\Delta_{q'c_i}(m_{q'})}+\frac{1}{\Sigma_{qc_i}(n_q)}+\frac{1}{\Sigma_{q'c_i}(m_{q'})}\right],\label{eq.threeJ}
\end{align}
with $\Delta_{qc_i}=\omega_{c_i}-\omega_{q}(n_q)$ and $\Sigma_{qc_i}=\omega_{c_i}+\omega_{q}(n_q)$.  One can see that effective $J$ coupling depends on the transition states $n$ and $m$ of pairwise interacting qubits $q$ and $q'$. To illustrate how strong the two- and three-qubit interactions are, we calculate them using the following six approaches.

\textbf{RWA-PT:} The first approach is perturbation theory (PT) with RWA by ignoring fast-rotating terms. If the three qubits are the same species  with same sign anharmonicity, i.e. all transmons, the coupling strength $J_{c_iqq'}^{n_qm_{q'}}$ can be approximately state-independent for a pair of interaction as shown in Fig.~\ref{fig:ttZZ}(c) and Fig.~\ref{fig:ttZZ}(d) since the anharmonicity of a transmon is small compared to the frequency. The Hamiltonian then can be rewritten in terms of creation and annihilation operators as 
\begin{equation}
H_{Q}=\sum_{q=1}^3\left( \bar{\omega}_{q}a_{q}^{\dagger}a_{q}+\frac{\delta_{q}}{2}a_{q}^{\dagger}a_{q}^{\dagger}a_{q}a_{q}\right) +\sum_{q<q'}J_{qq'}(a_{q}^{\dagger}a_{q'}+a_{q'}^{\dagger}a_{q}),
\label{eq.threeQ}
\end{equation}
where $\bar{\omega}_q={\bar\omega}_q(0)$, $\bar{\delta}_q={\bar\omega}_q(1)-{\bar\omega}_q(0)\approx\delta_q$ and $J_{qq'}\approx J_{r_iqq'}^{01}$.  Note that here we drop all counter-rotating terms. The Hamiltonian in Eq.~(\ref{eq.threeQ}) describes an interacting three-body problem, but we can treat it as three pairs of two-body interaction, and tackle them using similar methods as introduced in previous chapters. 

The most straightforward way is the perturbation theory which gives the dressed energy of each level, then all Pauli coefficients can be derived.  The Hamiltonian in the computational subspace can thus be rewritten in terms of Pauli matrices as 
\begin{equation}
H_{\rm eff}=\rm \alpha_{ZII}\frac{ZII}{2}+ \alpha_{IZI}\frac{IZI}{2}+ \alpha_{IIZ}\frac{IIZ}{2}+ \alpha_{ZZI}\frac{ZZI}{4}+\alpha_{ZIZ}\frac{ZIZ}{4}+\alpha_{IZZ}\frac{IZZ}{4}+ \alpha_{ZZZ}\frac{ZZZ}{8},
\end{equation}
and Pauli coefficients from perturbation theory up to third order are

\begin{eqnarray}
{\rm \alpha_{ZII}}&=&\frac{1}{4}\left(E_{000}+E_{011}+E_{001}+E_{010}-E_{100}-E_{101}-E_{110}-E_{111}\right)\nonumber\\  
&=&-\omega_1-\frac{J_{12}^2}{\Delta_{12}}-\frac{J_{13}^2}{\Delta_{13}}-\frac{2J_{12}J_{13}J_{23}}{\Delta_{12}\Delta_{13}}-\frac{\alpha_{\rm ZZI}}{2}-\frac{\alpha_{\rm ZIZ}}{2}-\frac{\alpha_{\rm ZZZ}}{4},\\
{\rm \alpha_{IZI}}&=&\frac{1}{4}\left(E_{000}+E_{001}+E_{100}+E_{101}-E_{010}-E_{011}-E_{110}-E_{111}\right)\nonumber\\  &=&-\omega_2-\frac{J_{12}^2}{\Delta_{21}}-\frac{J_{23}^2}{\Delta_{23}}-\frac{2J_{12}J_{13}J_{23}}{\Delta_{21}\Delta_{23}}-\frac{\alpha_{\rm ZZI}}{2}-\frac{\alpha_{\rm IZZ}}{2}-\frac{\alpha_{\rm ZZZ}}{4},\\
{\rm \alpha_{IIZ}}&=&\frac{1}{4}\left(E_{000}+E_{100}+E_{010}+E_{110}-E_{001}-E_{011}-E_{101}-E_{111}\right)\nonumber\\
&=&-\omega_3-\frac{J_{13}^2}{\Delta_{31}}-\frac{2J_{23}^2}{\Delta_{32}}-\frac{2J_{12}J_{13}J_{23}}{\Delta_{31}\Delta_{32}}-\frac{\alpha_{\rm ZIZ}}{2}-\frac{\alpha_{\rm IZZ}}{2}-\frac{\alpha_{\rm ZZZ}}{4},\\
{\rm \alpha_{ZZI}}&=&\frac{1}{2}\left(E_{000}+E_{110}-E_{010}-E_{100}+E_{001}+E_{111}-E_{101}-E_{011}\right)\nonumber\\  &=&\frac{4J_{12}^2(\delta_1+\delta_2)}{(\Delta_{12}+\delta_1)(\Delta_{12}-\delta_2)}-\frac{4J_{12}J_{13}J_{23}(A_{123}+B_{123}+C_{123})}{(\Delta_{12}+\delta_1)(\Delta_{12}-\delta_2)},\label{eq.zzi}\\
{\rm \alpha_{ZIZ}}&=&\frac{1}{2}\left(E_{000}+E_{101}-E_{100}-E_{001}+E_{010}+E_{111}-E_{011}-E_{110}\right)\nonumber\\  &=&\frac{4J_{13}^2(\delta_1+\delta_3)}{(\Delta_{13}+\delta_1)(\Delta_{13}-\delta_3)}-\frac{4J_{12}J_{13}J_{23}(A_{132}+B_{132}+C_{132})}{(\Delta_{13}+\delta_1)(\Delta_{13}-\delta_3)},\label{eq.ziz}\\
{\rm \alpha_{IZZ}}&=&\frac{1}{2}\left(E_{000}+E_{011}-E_{001}-E_{010}+E_{100}+E_{111}-E_{101}-E_{110}\right)\nonumber\\  &=&\frac{4J_{23}^2(\delta_2+\delta_3)}{(\Delta_{23}+\delta_2)(\Delta_{23}-\delta_3)}-\frac{4J_{12}J_{13}J_{23}(A_{231}+B_{231}+C_{231})}{(\Delta_{23}+\delta_2)(\Delta_{23}-\delta_3)},\label{eq.izz}\\
{\rm \alpha_{ZZZ}}&=&E_{000}+E_{011}-E_{001}-E_{010}+E_{101}+E_{110}-E_{100}-E_{111}\nonumber\\  &=&\frac{8J_{12}J_{13}J_{23}(B_{123}+C_{123})}{(\Delta_{12}+\delta_1)(\Delta_{12}-\delta_2)},\label{eq.zzz}
\end{eqnarray} 
where $E_{ijk}$ is the eigenvalues of state $|ijk\rangle$, parameters $A$, $B$ and $C$ are defined as 
\begin{eqnarray}
A_{ijk}&=&\frac{\delta_i\delta_j-\delta_i\Delta_{ik}-\delta_j\Delta_{jk}}{\Delta_{ik}\Delta_{jk}},\\
B_{ijk}&=&\frac{\delta_i^2+\delta_j^2+\delta_i\delta_j+\delta_i\Delta_{ik}+\delta_j\Delta_{jk}}{(\Delta_{ik}+\delta_i)(\Delta_{jk}+\delta_j)},\\
C_{ijk}&=&\frac{\delta_i\delta_j+\delta_j\delta_k+\delta_i\delta_k-\delta_i\Delta_{ik}-\delta_j\Delta_{jk}}{(\Delta_{ik}-\delta_k)(\Delta_{jk}-\delta_k)},
\end{eqnarray}
with qubit-qubit detuning $\Delta_{ij}=\bar\omega_i-\bar\omega_j$ $(i,j=1,2,3)$. In contrast with two-qubit architecture, one can see that another term proportional to all three couplings is subtracted from  two-qubit ZZ interaction in Eq.~(\ref{eq.zeta}). Now the single-qubit flipping frequency is the combination of dressed qubit frequency with corresponding two- and three-qubit interactions. The new three-body interaction\index{three-body interaction} term is created in the form of ZZZ, corresponding coefficient is obviously smaller than two-body interaction as it is proportional to $J^3/\Delta^2$. In other words, the nonzero value  first shows up in the third order correction. 


It is worth mentioning that near-degenerate case, in which noncomputational levels could be in the vicinity of computational levels, is excluded in the calculation. But if conditions of detrimental multi-qubit frequency collisions are satisfied, e.g. $\omega_q(n)=\omega_{q'}(m)$, we can still deal with the problem using similar method in Eq.~(\ref{eq.degenerate}). 

\textbf{RWA-SSW:} The second approach is simplified SW (SSW) transformation\index{Schrieffer-Wolff (SW) transformation} by considering the Hamiltonian in Eq.~(\ref{eq.threeQ}), together with RWA\index{rotating wave approximation (RWA)} by ignoring fast-rotating terms. Since the Hamiltonian in matrix form is not in the ascending order of energies, so we need to reorder it as  
\begin{equation}
H=\left( \begin{array}{ccc|ccc}
E_{000} && &&& \\
& \ddots&  && {\makebox(0,0){\it{\huge J}}}&\\
&& E_{111}&  &&\\ \cline{1-6}
& & & E_{002}&& \\
& {\makebox(0,0){\it{\huge J}}}& & &\ddots& \\
& & & &&E_{222}\\
\end{array}\right)
\end{equation}

The upper block contains all eight computational levels which only allow transition from 0 to 1, the other higher levels are in the lower block with $J$ representing all couplings between the two blocks. To decouple higher levels, a second SW transformation is applied, therefore the Hamiltonian is reduced to a $8\times8$ matrix. By fully diagonalizing this matrix, all Pauli coefficients can be obtained.
  
\textbf{RWA-SW:} The third approach is SW transformation by considering the Hamiltonian in Eq.(\ref{eq.threeQf}), together with RWA by ignoring fast-rotating terms. Compared to RWA-SSW, here effective coupling discrepancy is included, the other diagonalization processes are the same.

\textbf{NRWA-PT:} The fourth approach is perturbation theory with NONRWA by including all counter-rotating terms in the calculation. Detailed formula of fast oscillation contribution can be found in Appendix~\ref{app:nonrwa}.  

\textbf{NRWA-SSW:} The fifth approach is simplified SW by assuming effective $J$ couplings are only qubit-dependent, and all counter-rotating terms are included in the calculation. Numerical simulation is the same as RWA-SSW.

\textbf{NRWA-SW:} The last approach is SW by including the state-dependent $J$ discrepancy, and all counter-rotating terms in the calculation. Numerical simulation is the same as RWA-SW. Note that NRWA-SW gives the most accurate Pauli coefficients among the six results since it includes both coupling discrepancy and counter-rotating terms.

Now let us take two examples  in a circuit with fixed-frequency transmons, and name them Device 1 and Device 2, to show what the difference is among the six approaches. Frequency, anharmonicity and coupling strength of each subsystem in Device 1 are listed in Tab.~\ref{tab:device1}.

\begin{table}[h!]
	\begin{center}
		\begin{tabular}{|c|c|c|c|c|c|c|c|c|c|}
				\hline 
			\multicolumn{5}{|l}{Device 1} 
		 &	\multicolumn{4}{r|}{unit: GHz} \tabularnewline
			\hline 
			$\omega_1$&$\omega_2$  &$\omega_3$&$\delta_1$&$\delta_2$&$\delta_3$ &$\omega_{c_1}$  & $\omega_{c_2}$&$\omega_{c_3}$  \tabularnewline\hline 
				4.9 &5.0  &5.1 &$-0.33$&$-0.33$ &$-0.33$ &6.0 & 6.3 &6.6  \tabularnewline
			\hline 
				$g_{1c_1}$&$g_{1c_3}$ &$g_{2c_1}$&$g_{2c_2}$&$g_{3c_2}$&$g_{3c_3}$ &$g_{12}$ & $g_{13}$&$g_{23}$  \tabularnewline
				\hline 
				0.08&0.08 &0.08&0.08&0.08&0.08 &0 &0&0\tabularnewline
					\hline 
		\end{tabular}
	\end{center}
	\vspace{-0.1in}
	\caption{\label{tab:device1}Device 1 parameters.}
\end{table}

\begin{figure}[ht]
	\centering
	\includegraphics[width=0.9\textwidth]{{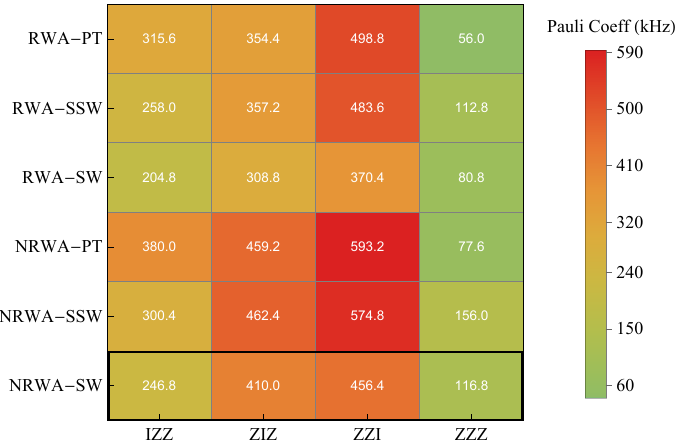}}
	\vspace{-0.in}
	\caption{Pauli coefficients of triangle geometry three qubit interaction in Device 1, with $\omega_{c_i}>\omega_j$ $(i, j=1, 2, 3)$ using six approaches. Perturbation theory (PT) and simplified SW (SSW) results are obtained by assuming effective qubit-qubit coupling $J$ is state-independent, while the most accurate result is for SW takes into account $J$ discrepancy. The most accurate result is from NRWA-SW, marked by black box.}
	\label{fig:example1}
\end{figure}

In Device 1, all coupler frequencies are way larger than qubits. With these numbers we evaluate Pauli coefficients both analytically and numerically using the six approaches.
Figure~\ref{fig:example1} shows that three-body ZZZ interaction is around 100 kHz while two-body ZZ interaction is at least double of that.  By comparing the RWA with NRWA results, we find that counter rotating terms have the same sign with co-rotating terms and make the total two- and three-body interaction stronger. State-independent $J$ for PT and SSW leads to a lager two-body interaction than the numerical SW by a few tens of kilohertz. In contrast to ZZI component, IZZ interaction is weaker although the two subsystems have same 100 MHz qubit-qubit detuning. This is because coupler C2 in between Q2 and Q3 has a higher frequency than C1 in between Q1 and Q2, which results in  $|J_{12}|>|J_{23}|$. This phenomenon is similar to what we found in Fig.~\ref{fig:ttZZ} for the two-qubit models.

Another scenario we explore is that one coupler frequency is below the qubit while keeping the other parameters of Device 1 unchanged. Frequency, anharmonicity and coupling strength of each subsystem are listed in Tab.~\ref{tab:device2}.

\begin{table}[h!]
	\begin{center}
		\begin{tabular}{|c|c|c|c|c|c|c|c|c|c|}
			\hline 
			\multicolumn{5}{|l}{Device 2} 
			&	\multicolumn{4}{r|}{unit: GHz} \tabularnewline
			\hline 
			$\omega_1$&$\omega_2$  &$\omega_3$&$\delta_1$&$\delta_2$&$\delta_3$ &$\omega_{c_1}$  & $\omega_{c_2}$&$\omega_{c_3}$  \tabularnewline\hline 
			4.9 &5.0  &5.1 &$-0.33$&$-0.33$ &$-0.33$ &6.0 & 6.3 &3.4  \tabularnewline
			\hline 
			$g_{1c_1}$&$g_{1c_3}$ &$g_{2c_1}$&$g_{2c_2}$&$g_{3c_2}$&$g_{3c_3}$ &$g_{12}$ & $g_{13}$&$g_{23}$  \tabularnewline
			\hline 
			0.08&0.08 &0.08&0.08&0.08&0.08 &0 &0&0\tabularnewline
			\hline 
		\end{tabular}
	\end{center}
	\vspace{-0.1in}
	\caption{\label{tab:device2}Device 2 parameters.}
\end{table}

\begin{figure}[ht]
	\centering
	\includegraphics[width=0.9\textwidth]{{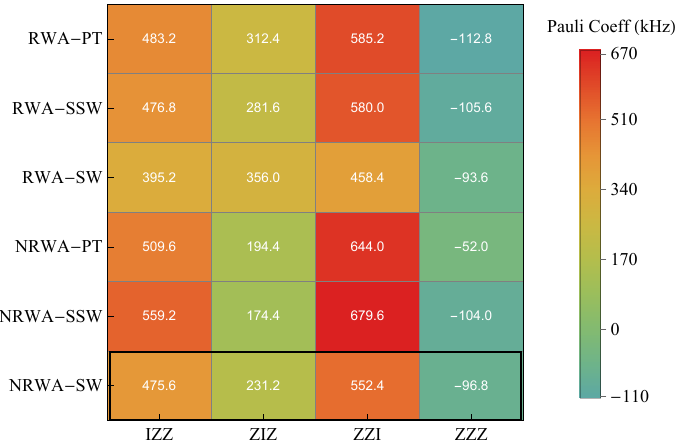}}
		\vspace{-0.in}
	\caption{Pauli coefficients of triangle geometry three qubit interaction in  Device 2 using six approaches in the condition that only C3 frequency is lower than qubits. Perturbation theory (PT) and simplified SW (SSW) results are obtained by assuming effective qubit-qubit coupling $J$ is state-independent, while SW takes into account $J$ discrepancy. The most accurate result is from NRWA-SW, marked by black box.}
	\label{fig:example2}
\end{figure}

Here  the third coupler frequency is changed from $6.6$ GHz to $3.4$ GHz, and corresponding Pauli coefficients are shown in Fig.~\ref{fig:example2}. One can see that now ZZZ coefficient is negative since  the sign of $J_{13}$ in Eq.~(\ref{eq.threeJ}) changes to positive under the condition that $\omega_{c_3}<\omega_1,\omega_3$. For RWA-PT and RWA-SSW, the decrease in coupler C3 frequency leads to increase in Q1 and Q3 dressed frequencies due to the sign change in Eq.~(\ref{eq.omegaq}). From the perturbation theory we find that third order correction has the same sign with second order under current condition that only $J_{13}$ is negative. ZZI and IZZ components then get strengthened in contrast to Fig.~\ref{fig:example1}.  As for ZIZ, increasing dressed frequency results in reduction of second order correction term, which is now added by a positive third order correction, the total effect is that we see a decrease in ZIZ interaction.

Now let us revisit the Sycamore layout in Fig.~\ref{fig:sycamore}. Given that three-qubit interaction has the same amplitude as what we show in Fig~\ref{fig:example1} and Fig.~\ref{fig:example2}, in total there are 12 non-pairwise three-body interactions accumulating phase errors on the state of the central qubit. Such total three-body interaction is even comparable to the total two-body ZZ interaction if they are linearly sum together, e.g. 1 MHz, and becomes more obvious in a complex quantum process, then should be cancelled out.


\section{Planar Geometry\index{planar geometry} of Three-Qubit  chain}
\begin{figure}[h!]
	\centering
	\vspace{-0.1in}
	\includegraphics[width=0.8\textwidth]{{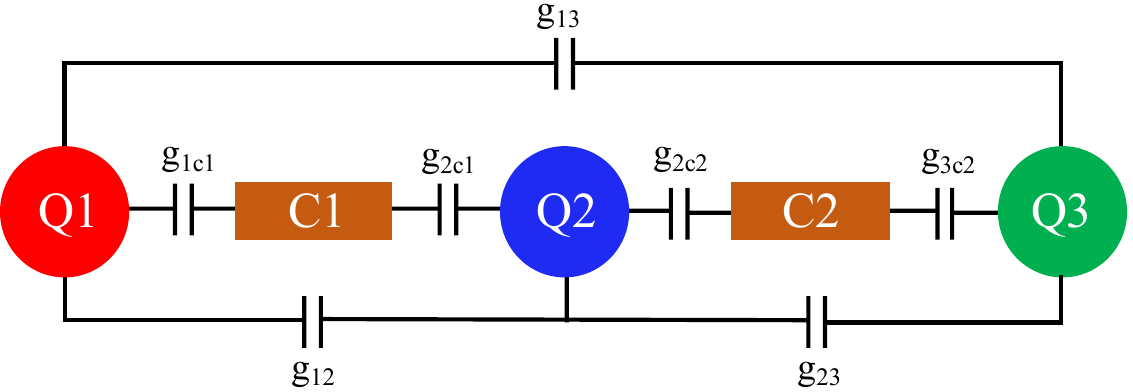}}
	\vspace{-0.1in}
	\caption{Geometry of three qubit chain Qubits Q1 and Q3 are coupled to Q2 via couplers C1 and C2, separately. Each pair of qubits are also capacitively coupled to each other. }
	\label{fig:plain}
\end{figure}

A simpler version of three-qubit coupling scheme is a qubit chain, as shown in Fig.~\ref{fig:plain}. This circuit is equivalent to stretch the triangle geometry in Fig.~\ref{fig:triangle} to a line by removing one coupler. 

\begin{figure}[h!]
	\centering
	\includegraphics[width=0.9\textwidth]{{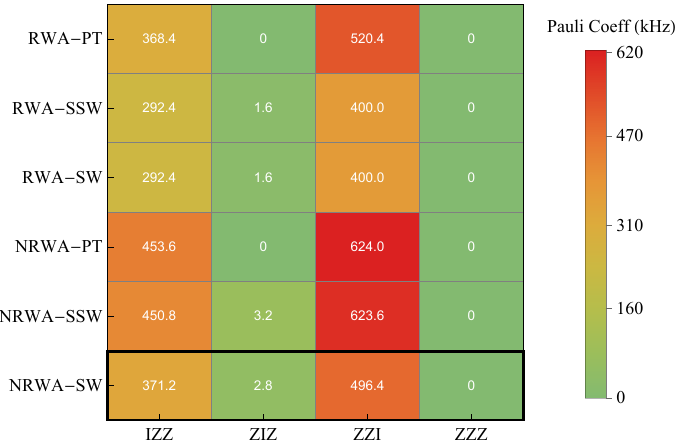}}
		\vspace{-0.in}
	\caption{Pauli coefficients of planar geometry three qubit interaction in Device 3. Perturbation theory (PT) and simplified SW (SSW) results are obtained by assuming effective qubit-qubit coupling $J$ is state-independent, while SW takes into account $J$ discrepancy. The most accurate result is from NRWA-SW, marked by black box.}
	\label{fig:example3}
\end{figure}

Here we also explore two examples, and name them Device 3 and Device 4. To make a comparison with triangle geometry, we use the same device parameter in Tab~\ref{tab:device1} for Device 3, and assume $\omega_{c_3}$, $g_{1c_3}$ and $g_{3c_3}$ to be zero. Corresponding Pauli coefficients are plotted in Fig.~\ref{fig:example3}.

Compared to the triangle geometry, now the effective coupling $J_{13}$ between Q1 and Q3 is zero since direct capacitive coupling $g_{13}=0$. Thus two-body  interaction obtained from perturbation theory reduces to the formula of static ZZ interaction that we have derived in Eq.~(\ref{eq.zeta}). Three-body ZZZ interaction in Eq.~(\ref{eq.zzz}) also turns out to be zero. PT gives exact zero ZIZ component, but RWA-SSW and RWA-SW show $\sim$ 1 kHz contribution. This is because we expand the SW transformation to the second order, in which cross terms like $J_{12}^2J_{23}^2/\Delta_{12}\Delta_{23}^2$ are included and lead to two orders weaker ZIZ interaction. In contrast to Fig.~\ref{fig:example1}, coefficients of IZZ and ZZI become larger since that the third order correction now is approximated to be zero.

\begin{figure}[h!]
	\centering
	\includegraphics[width=0.9\textwidth]{{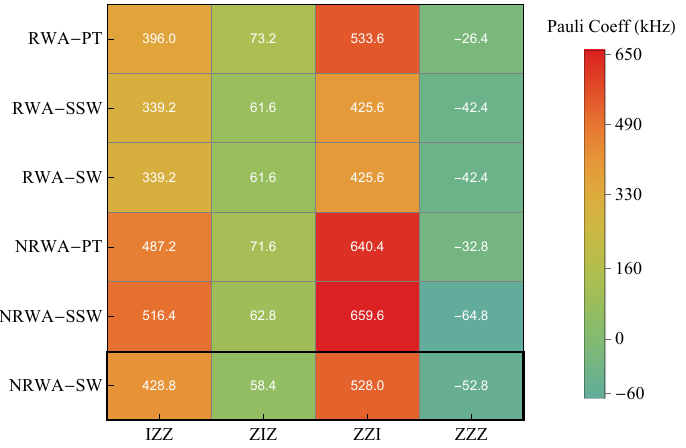}}
		\vspace{-0.in}
	\caption{Pauli coefficients of planar geometry three qubit interaction in Device 4 by assuming direct coupling $g_{13}=2$ MHz using six approaches. Perturbation theory (PT) and simplified SW (SSW) results are obtained by assuming effective qubit-qubit coupling $J$ is state-independent, while SW takes into account $J$ discrepancy. The most accurate result is from NRWA-SW, marked by black box. }
	\label{fig:example4}
\end{figure}

Note that in the discussion above we assume direct coupling $g_{13}$ to be zero. But in reality the two qubits can also capacitively interact with each other, leading to a nonzero direct coupling\index{direct coupling}. In this case we explore the Device 4 in which direct coupling $g_{13}=2$ MHz, other parameters are the same as Device 3.  Accordingly, Pauli coefficients are plotted in Fig.~\ref{fig:example4}.

One can see that both two-body interaction ZIZ and three-body interaction ZZZ are nonzero.
This is because nonzero direct coupling $g_{13}$ leads to the total effective coupling $J_{13}=2$ MHz as shown in Eq.~(\ref{eq.threeJ}). It is clear that direct coupling plays an important role in the two- and three-body interaction although it is small. ZIZ component increases from zero to 60 kHz while ZZZ is about $-50$ kHz. The sign of ZZZ interaction is changed to negative since now only $g_{13}$ is positive.

By comparing the triangle geometry and planar geometry of three-qubit interaction, we show that the sign of three-body ZZZ interaction can be changed to the opposite by either changing coupler frequency in triangle geometry or increasing direct coupling in planar geometry. This provides two potential ways to eliminate ZZZ interaction. In other words, it is possible to find out a boundary between the positive region and negative region by changing device parameters, on which ZZZ interaction is always zero, like what we can see in Fig.~\ref{fig:ctZZ}.

In a chain with more than three qubits, there are only 2 non-pairwise three-body interactions changing the state of a particular qubit. As shown Fig~\ref{fig:example4},  in the presence of nonzero direct coupling between the two ends of a three-qubit chain, the total three-body interaction is much weaker than two-body interaction, which can be safely ignored when performing quantum gates in this processor.

In brief, we explore three-qubit interaction in two setups: triangle geometry and a qubit chain. We derive the Pauli coefficients using both analytical and numerical approaches, and present several examples to illustrate how the two- and three-qubit interactions are affected by the circuit geometry and parameters. We show that the sign of ZZZ interaction can be changed to the opposite by either changing coupler frequency in triangle geometry or increasing direct coupling in the three-qubit chain. This provides two potential ways to eliminate ZZZ interaction.

\newpage
\thispagestyle{empty}

\chapter{Conclusion}
Superconducting qubits have become one of the most prominent candidates for the realization of scalable quantum computing. However, the performance of quantum logic gates still needs to be improved. One limiting factor for the two-qubit gate errors is parasitic interaction between qubits. In this book we mainly focus on ZZ interaction between pairwise coupled qubits, and develop several protocols to either eliminate or utilize it to perform high-fidelity quantum gates.

In chapter~\ref{c3}, we show the static ZZ interaction is originated from transitions between computational and noncomputational levels. We characterize such interaction in two circuits of qubit pairs: a hybrid CSFQ-transmon pair and a pair of interacting transmons. The hybrid pair with opposite anharmonicity between the qubits allows for the elimination of static ZZ interaction meanwhile keeps two qubits entangled. The same-specie pair can only achieve static ZZ freedom by almost decoupling two qubits. We also derive the condition of static ZZ freedom  from perturbation theory, and use this model to simulate the first hybrid CSFQ-transmon experiment performed by our collaborators, in which zero static ZZ interaction was observed. Our theoretical results are consistent with the experimental data.

In chapter~\ref{c4}, we follow the experiment and study the performance of a cross-resonance (CR) gate in the hybrid system. The  CR gate is realized by driving the control qubit CSFQ with the frequency of target qubit transmon. Our analysis shows that zeroing static ZZ interaction can effectively improve the CR gate fidelity, although the CR pulse produces additional dynamical component on top of it. We predict that two-qubit gate error rate of the order of 0.001 is feasible if the coherence times can be prolonged to 200 $\mu$s and static ZZ freedom is realized. Inspired by the experiment, we notice that static and dynamical ZZ interaction can cancel each other. Therefore we  derive the condition of dynamical ZZ freedom for CSFQ-transmon pair as well as transmon-transmon pair, and show that this type of freedom can lead to a large  increase in the CR gate fidelity.

Chapter~\ref{c5} describes our proposal for two novel two-qubit gates by including a tunable coupler: the parasitic free (PF) gate and the tunable controlled-Z (CZ) gate. We first study the parasitic free (PF) gate in a circuit of two fixed-frequency transmons coupled via a frequency-tunable coupler. In the circuit both static ZZ and dynamical ZZ freedom can be realized by tuning the flux threading the frequency-tunable coupler. By fine designing the pulse we achieved ZZ freedom at two sweet spots with extremely reduced flux noise. The second gate we propose is a tunable CZ gate in a circuit of a CSFQ coupled to a frequency-tunable transmon threading by external flux via a bus resonator. In this circuit ZZ interaction can be totally turned off at one flux point, and dramatically increased to several MHz at another flux point. 

In last chapter, we study three-qubit circuits in triangle  and chain geometry. In each setup we study and evaluate the two-body ZZ and three-body ZZZ interaction between qubits. We show that the sign of ZZZ interaction can be changed to the opposite by either changing coupler frequency in triangle circuit or increasing direct coupling in the chain. This provides two potential ways to eliminate such ZZZ interaction.

\begin{appendices}
\chapter{CSFQ Hamiltonian Quantization}
\label{app:CSFQ}
\begin{figure}
	\centering
	\includegraphics[width=.5\textwidth]{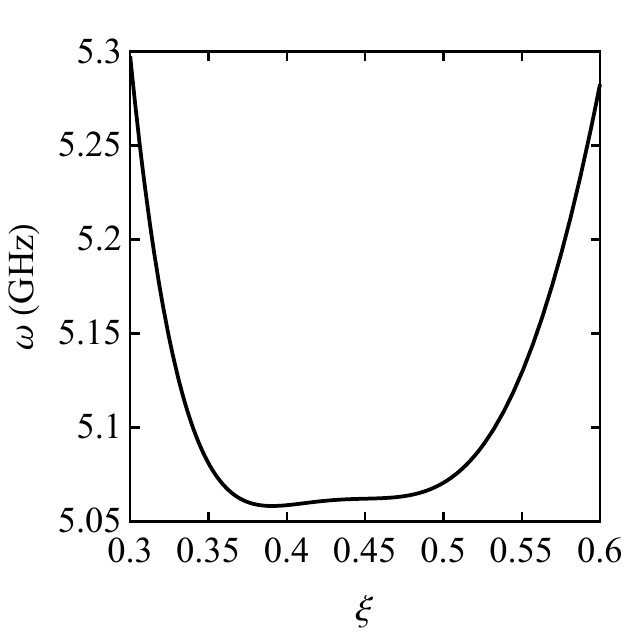}
	\caption{Frequency as a function of $\xi$ at the sweet spot. Simulation parameters $E_C=0.292$ GHz, $E_J=108.9$ GHz, $\alpha=0.43$ and $f=0.5$.}
	\label{sfig1}
	\vspace{-0.1in}
\end{figure}

\begin{figure}
	\centering
	\includegraphics[width=.48\textwidth]{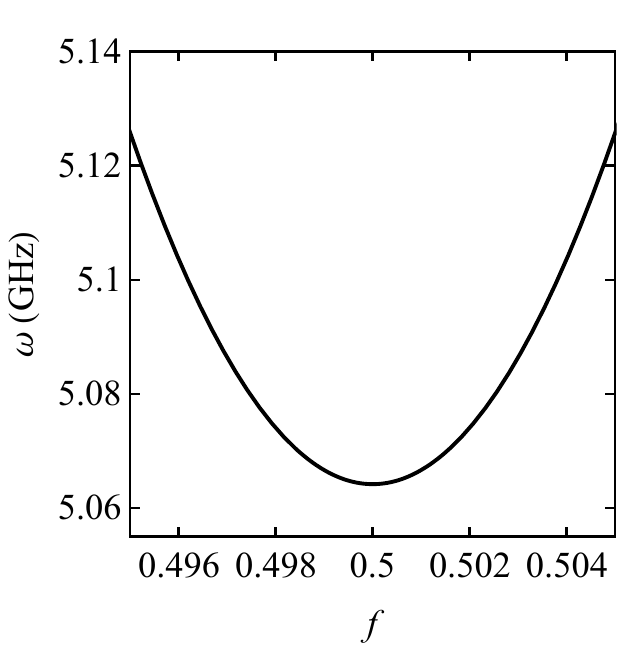}\quad
	\includegraphics[width=.48\textwidth]{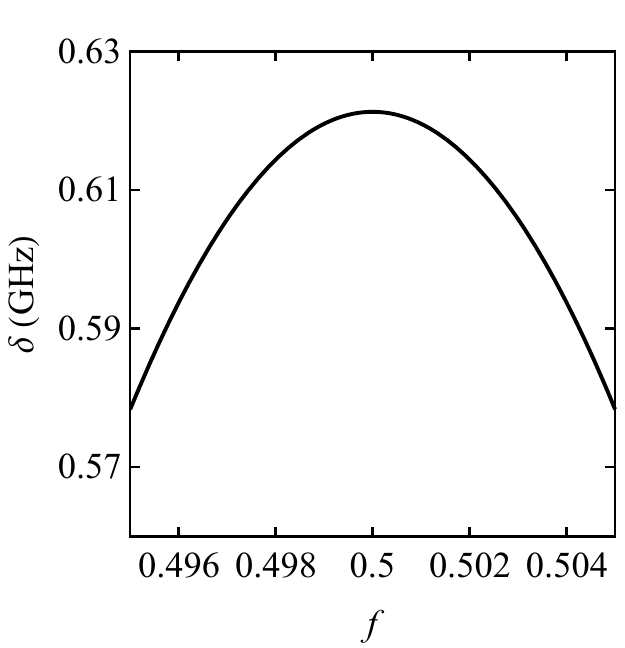}
		\put(-353,170){(a)}
	\put(-183,170){(b)}
	\caption{Frequency (a) and anharmonicity (b) as a function of external flux $f$ using perturbation theory.}
	\label{sfig2}
\end{figure}

In contrast to a transmon, higher order terms (>4) in the expansion of the CSFQ potential also contributes to the eigenvalues, which will change the zero point fluctuation to an unknown number. To be more precise, the Hamiltonian can be quantized in terms of field operators, e.g. $n=i(\hat{a}-\hat{a}^{\dagger})/2\xi$ and  $\phi=\xi(\hat{a}+\hat{a}^{\dagger})$ with $\xi$ being the expansion parameter which will minimize the total energy, the normal ordered Hamiltonian then can be written as 
\beqr \nonumber 
H &=& -\frac{E_c}{\xi^2} \left({a}^{\dagger}-{a}\right)^{2}+ \sum_{u=0}^{\infty}\xi^{2u+2} \sum_{v=0}^{u} \frac{U_{2u+2}(\varphi_0)}{2^{u-v}\left(u-v\right)!} \\ && \times \nonumber  \sum_{w=-\left(v+1\right)}^{v+1}  \frac{\left({a}^{\dagger}\right)^{v+1+w}\left({a}\right)^{v+1-w}}{\left(v+1+w\right)! \left(v+1-w\right)!}  \\ 
&+& \sum_{u=0}^{\infty}\xi^{2u+1} \sum_{v=0}^{u} \frac{U_{2u+1}(\varphi_0)}{2^{u-v}\left(u-v\right)!} \\ && \times \nonumber  \sum_{w=-\left(v+1\right)}^{v} \frac{\left({a}^{\dagger}\right)^{v+1+w}\left({a}\right)^{v-w}}{\left(v+1+w\right)!\left(v-w\right)!}
\eeqr
with
\begin{eqnarray}
U_{n}(\varphi_0)&\equiv&{\partial^{n}U(\varphi_0)}/{\partial\varphi_0^{n}}\\
U(\varphi_0)&\equiv& -2E_{J}\cos\varphi_0/2-\alpha E_{J}\cos\left(2\pi f-\varphi_0\right)
\end{eqnarray}  
where $\varphi_0$ is the phase of minimum $U(\varphi_0)$, i.e. $\varphi_0=-2 \pi \alpha (\delta f)/(1/2-\alpha)$, which vanishes at sweet spot.  
By solving Schr\"{o}dinger equation, the eigenenergies $E_n$ can be obtained using perturbation theory. Unperturbed eigenvalues $E_n^{(0)}$ and first three order corrections are given by
\begin{equation}
\begin{split}
E_{n}^{(0)}&=\left(\sum_{l=1}^{L}\frac{U^{(2l)}\xi^{2l}}{\left(2l-2\right)!!}+\frac{2E_{C}}{\xi^{2}}\right)n\\
&+\sum_{k=2}^{L}\sum_{l=2}^{L}\frac{U^{(2l)}\xi^{2l}}{\left(2l-2k\right)!!}\frac{n!}{\left(n-k\right)!}\\
E_{n}^{(1)}&=0\\
E_{n}^{(2)}&=\sum_{k\neq n}^{n+L}\frac{V_{nk}^{2}}{E_{n}^{(0)}-E_{k}^{(0)}}\\
E_{n}^{(3)}&=\sum_{k\neq n}^{n+L}\sum_{m\neq n}^{n+L}\frac{V_{nm}V_{mk}V_{kn}}{\left(E_{n}^{(0)}-E_{k}^{(0)}\right)\left(E_{n}^{(0)}-E_{m}^{(0)}\right)}
\label{E}
\end{split}
\end{equation}
with
\begin{equation}
\begin{split}
V_{nm}&=x_{|m-n|,|m-n|}\sqrt{\frac{\max(m,n)!}{\min(m,n)!}}\\
&+\sum_{s=0}^{\min(m,n)}x_{\max(m,n)-s,\min(m,n)-s}\sqrt{\frac{n!m!}{s!}}\\
x_{a,b}&=\delta_{ab}\sum_{k=a}^L\sum_{u=0}\Theta(k-a-2u)\frac{U^{(k)}\xi^k}{(k-a)!!a!}\\
&+(1-\delta_{ab})\sum_{k=a}^{L}\sum_{u=0}\Theta(k-a-2u)\frac{U^{(k+b)}\xi^{k+b}}{b(k-a)!!a!}\\
&-\delta_{a2}\delta_{b2}\frac{E_{C}}{\xi^{2}}\nonumber
\end{split}
\end{equation}
Sum up unperturbed energy and all three orders corrections, one can obtain the analytical formula of CSFQ eigenvalues, then frequency $\omega_{01}=(E_1-E_0)/\hbar$ and anharmonicity $\delta=(E_2-2E_1+E_0)/\hbar$ can be evaluated. Here is an example to illustrate how to find these parameters. Consider a CSFQ with $E_C=0.292$ GHz, $E_J=108.9$ GHz, $\alpha=0.43$ and $f=0.5$. Firstly, expand the potential to 20th order (L=10) at the sweet spot and plot the frequency $f_{01}$ as a function of $\xi$. 
Determine $\xi$ by finding the minimum as shown in Fig. \ref{sfig1} (approximately $\xi\approx\phiz/\sqrt{2}$), then substitute $\xi$ back to the Eq. (\ref{E}), the corresponding frequency and anharmonicity spectra are shown in Fig. \ref{sfig2}.
\newpage
\thispagestyle{empty}
\chapter{Hamiltonian of Experimental Circuit}
\label{app:cirH}
\begin{figure} 
	\begin{subfloat}{\label{fig:potential_a}} \end{subfloat}
	\begin{subfloat}{\label{fig:potential_b}} \end{subfloat}
	\begin{subfloat}{\label{fig:potential_c}} \end{subfloat}
	\begin{subfloat}{\label{fig:potential_d}} \end{subfloat}
	\begin{subfloat}{\label{fig:potential_e}} \end{subfloat}
	\begin{subfloat}{\label{fig:potential_f}} \end{subfloat}
	\centering
	\includegraphics[width=1.\textwidth]{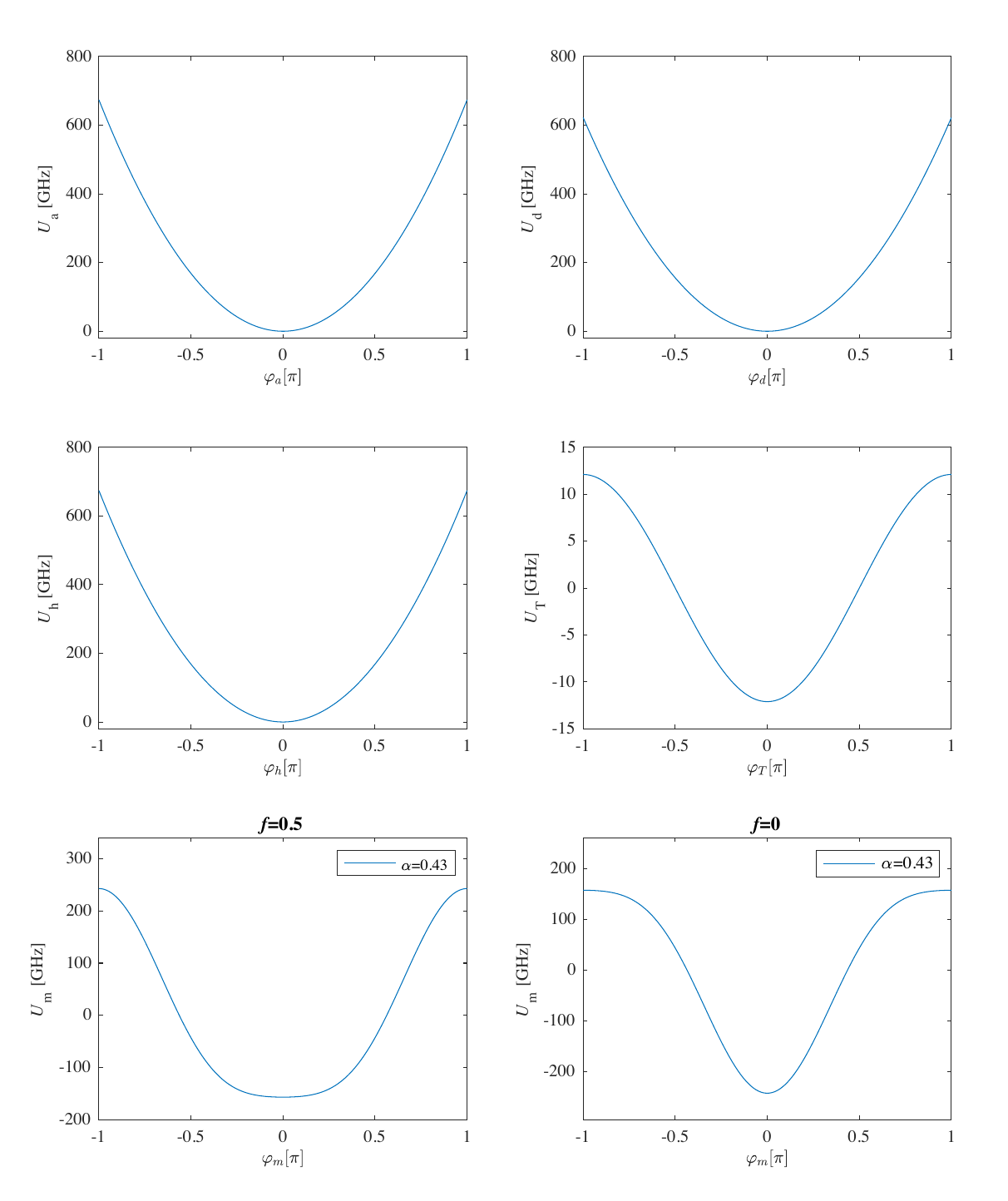}
	\put(-355,410){(a)}
	\put(-185,410){(b)}
	\put(-355,270){(c)}
	\put(-185,270){(d)}
	\put(-355,130){(e)}
	\put(-185,130){(f)}
	\caption{Potential profiles. (a) Readout resonator for transmon. (b) Bus resonator. (c) Readout resonator for CSFQ. (d) Transmon. (e) CSFQ along the $\varphi_m$ direction at $f=0.5$. (f) CSFQ along the $\varphi_m$ direction at $f=0$.}
	\label{fig:potentialtwo}
\end{figure}
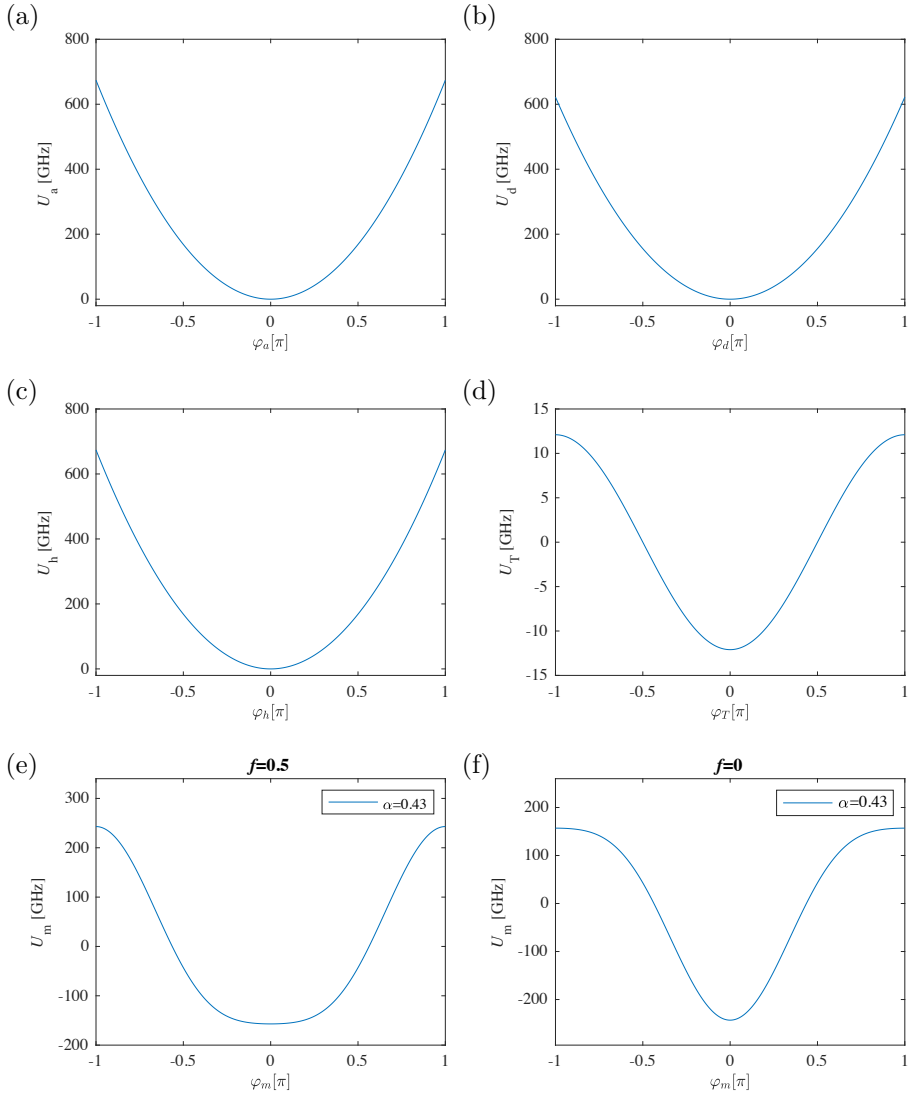

It is worth noting that we put the two redundant phase degrees of freedom in the first place, which will be removed next. In Fig.~\ref{fig:potentialtwo} we plot the potential energies associated with the readout resonator coupled to the transmon (a), the bus resonator (b), the readout resonator coupled to the CSFQ (c), the fixed frequency transmon (d), the CSFQ at the sweet spot (e), and away from the sweet spot (f). 
Since $\varphi_b$ and $\varphi_e$ are not included in the potential, and  $\varphi_p$ can also be neglected as discussed in the section of CFSQ, we can use Cholesky decomposition~\cite{golub1996matrix} to safely remove these three phases, and thus reduce the matrix size of the circuit Hamiltonian from $8\times 8$ to the following $5 \times 5$:
\begin{equation}
H=4\overrightarrow{n}^{T}\frac{e^{2}}{2\mathbf{C'}}\overrightarrow{n}+U,
\end{equation}
where $\overrightarrow{n}=(n_a, n_T, n_d, n_m, n_h)$ is the canonical term of $\overrightarrow{\varphi}$, and  
\begin{equation}
\mathbf{\mathbf{C'}}=\left(\begin{array}{ccccc}
C_{a}' & -\frac{C_{ab}C_{dT}}{C_{T0}}& -\frac{C_{ab}C_{cd}}{C_{T0}}& 0 & 0\\
-\frac{C_{ab}C_{dT}}{C_{T0}} & C_{T}' & \frac{C_{a0}C_{cd}}{C_{T0}} & 0 & 0\\
-\frac{C_{ab}C_{cd}}{C_{T0}} & \frac{C_{a0}C_{cd}}{C_{T0}} & C_{r}' & -\frac{2C_{de}C_{h0}}{C_{m0}} & -\frac{C_{de}C_{gh}}{C_{m0}}\\
0 & 0 & -\frac{2C_{de}C_{h0}}{C_{T0}} & C_{m}' & \frac{2C_{dm}C_{gh}}C_{m0}\\
0 & 0 & -\frac{C_{de}C_{gh}}{C_{m0}} &\frac{ 2C_{dm}C_{gh}}{C_{m0}} & C_{h}'
\end{array}\right).
\end{equation}
where the new notations are defined from existing ones in Fig.~\ref{fig:csfq-transmon} as
\begin{equation}
\begin{split}
C_{T0}	&=C_{ab}+C_{b0}+C_{c0}+C_{cd},C_{h0}	=C_{g0}+C_{gh}\\
C_{m0}	&=C_{de}+C_{e0}+C_{g0}+C_{gh}, C_{dm}=C_{de}+C_{e0}\\
C_{dT}	&=C_{cd}+C_{c0},C_{a0}=C_{ab}+C_{b0}\\
C_{T}'	&=C_{dT}+C_{shT}+C_{T}-{C_{dT}^{2}}/{C_{T0}}\\
C_{m}'	&=2C_J+4(C_{3}+C_{shCSFQ})-4C_{h0}^{2}/C_{m0}+4C_{h0}\\
C_{r}'	&=-C_{cd}^{2}/C_{T0}+C_{cd}+C_{de}+C_{r}-{C_{de}^{2}}/{C_{m0}}\\
C_{a}'	&=-C_{ab}^{2}/C_{T0}+C_{ab}+C_{rT}\\
C_{h}'    &=-C_{gh}^{2}/{C_{m0}}+C_{gh}+C_{rCSFQ}.
\end{split}
\end{equation}

The relationships between the various coupling strengths $g_{ij}$ and the relevant capacitances are given by the following expressions:
\begin{equation}
\begin{split}
g_{hm}	&\propto -\frac{2C_{gh}C_{dm}}{\left(C_{gh}+C_{rCSFQ}\right)\left(C_{gs}C_{m0}-4C_{h0}^{2}\right)} \\
g_{rm}&\propto\frac{2C_{de}C_{h0}}{C_{cder}\left(4C_{h0}^{2}-C_{gs}C_{m0}\right)}\\
g_{aT}&\propto\frac{C_{ab}C_{dT}}{\left(C_{ab}+C_{rT}\right)\left(C_{gT}C_{T0}-C_{a0}^{2}\right)} \\
g_{rT}&\propto -\frac{C_{cd}C_{a0}}{C_{cder}\left(C_{dT}^{2}-C_{gT}C_{T0}\right)}\\
g_{mT}&\propto-\frac{2C_{cd}C_{de}C_{a0}C_{h0}}{C_{cder}\left(C_{gT}C_{T0}-C_{a0}^{2}\right)\left(C_{gs}C_{m0}-4C_{h0}^{2}\right)},
\end{split}
\end{equation}
where $C_{gs}=2C_{0}+4C_{g0}+4C_{gh}+4(C_{3}+C_{shCSFQ})$,  $C_{gT}=C_{cd}+C_{c0}+C_{shT}+C_{T}$, and $C_{cder}=C_{cd}+C_{de}+C_{R}$. 
\newpage
\thispagestyle{empty}

\chapter{CSFQ Coherence Versus Flux}
\label{app:T2}
\begin{figure}[b!]
	\centering
	\makebox[\linewidth]{
		\labellist
		\bfseries
		\pinlabel (a) at 9 410
		\endlabellist
		\includegraphics[width=0.48\linewidth]{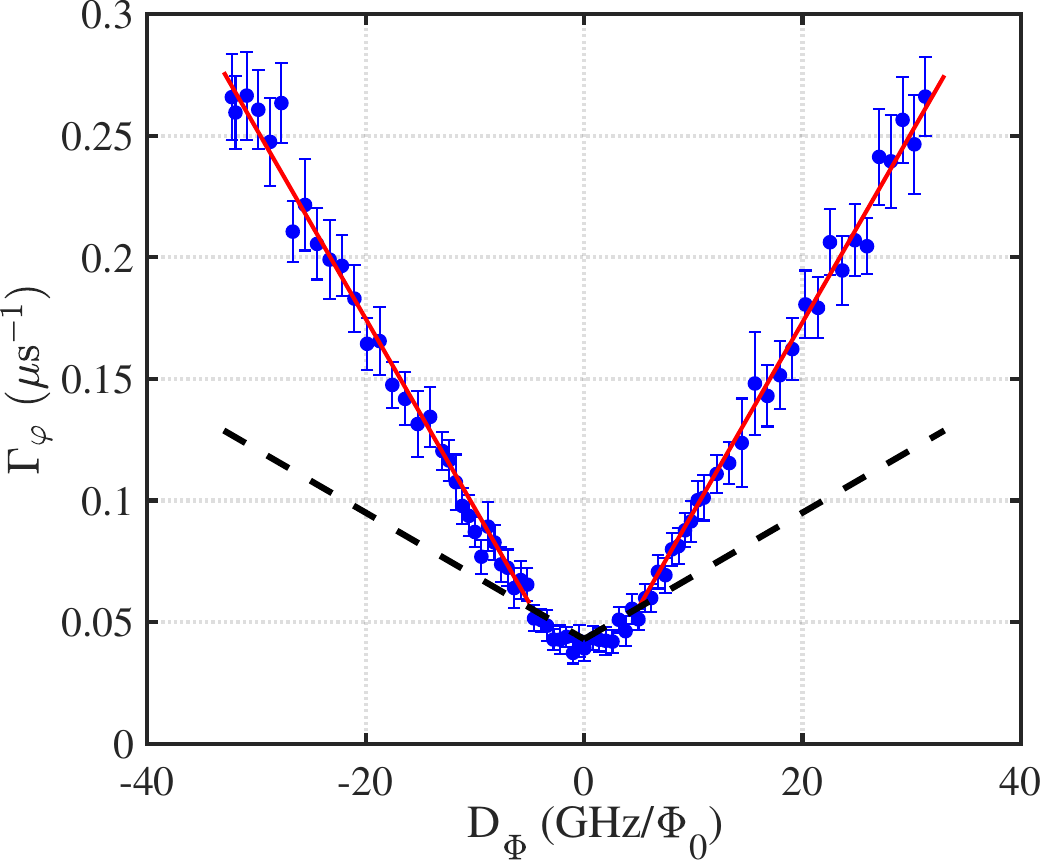}\hspace{0.05in}
		\labellist
		\bfseries
		\pinlabel (b) at 9 410
		\endlabellist
	
		\includegraphics[width=0.455\linewidth]{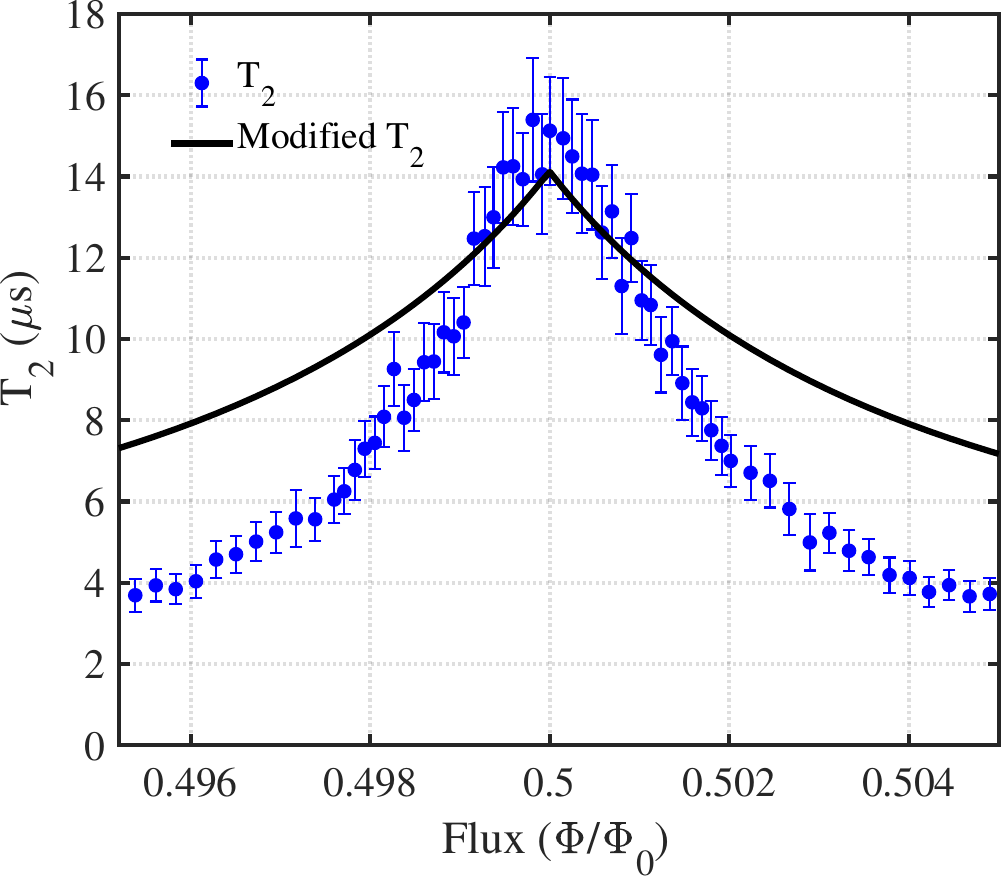}
	}
	\caption{(a) Pure dephasing rate $\Gamma_\varphi$ of the CSFQ vs. qubit frequency gradient $D_\Phi$. Red solid line is a linear fit to the linear portion of data. Black dashed line is the modified pure dephasing rate to account for the reduction in gate errors measured with randomized benchmarking when dephasing is dominated by $1/f$ noise, as discussed in text. (b) Hahn-echo $T_2$ vs. flux and the effective $T_2$ calculated using our modified dephasing model. }
	\label{fig:theoT2}
\end{figure}
For understanding the map of coherence, we must know the coherence times $T_1$ and $T_2$ of each qubit. Transmon frequency is fixed, therefore its $T_1$ and $T_2$ can be extracted  with standard Ramsey sequence. To simulate the gate error for different flux biases, we must model the appropriate flux-dependent dephasing of the CSFQ. Experimental $T_2$ from Ramsey measurement with a single echo refocusing pulse is shown in Fig.~\ref{fig:theoT2}. However, such a dephasing rate overestimates the gate errors away from the sweet spot. It means that the effective $T_2$ must be longer than what is measured with the standard protocol. This behavior may be due to the depolarizing effect from twirling the $1/f$ noise, which is typically the dominant contribution to dephasing in flux tunable CSFQ for bias points away from a sweet spot~\cite{yoshihara2006decoherence, hutchings2017tunable, yan2016theflux}. If we consider Gaussian noise, the random phase accumulated at time $t$ is $\Delta\varphi=D_{\Phi}\int_{0}^{t}dt'\delta\Phi(t')$ with respect to flux $D_{\Phi}=\partial f_{01}/\partial \Phi$, and one can calculate the decay law of the free induction (Ramsey signal) as
\begin{equation}
f_{z,R}(t)=\left\langle \exp(i\Delta\varphi)\right\rangle =\exp\left(-\frac{1}{2}D_{\Phi}^{2}\int_{0}^{t}\int_{0}^{t}dt'dt''\left\langle \delta\Phi(t')\delta\Phi(t'')\right\rangle \right).
\end{equation}
Note that only in the equilibrium, $\left\langle \delta\Phi(t')\delta\Phi(t'')\right\rangle $ reduces to power spectrum $ S_{\Phi}(\left|t'-t''\right|)$  which depends on the time difference, otherwise, time point $t'$ and $t''$ should be considered. In the equilibrium the free induction decay is simplified as
\begin{equation}
f_{z,R}(t)=\exp\left(-\frac{1}{2}D_{\Phi}^{2}\int_{0}^{t}\int_{0}^{t}dt'dt''S_{\Phi}\left(\left|t'-t''\right|\right)\right).
\end{equation}
Fourier transformation of the power spectrum in the frequency domain is written as 
\begin{equation}
S_{\Phi}\left(\left|t'-t''\right|\right)=\frac{1}{2\pi}\int_{-\infty}^{+\infty}S_{\Phi}\left(\omega\right)e^{i\omega\left|t'-t''\right|}d\omega.
\end{equation}
By substituting this Fourier transformation, the induction decay changes to
\begin{equation}
f_{z,R}(t)
=\exp\left[-\frac{t^{2}}{2}D_{\Phi}^{2}\int_{-\infty}^{+\infty}\frac{d\omega}{2\pi}S_{\Phi}\left(\omega\right){\rm {sinc}}^{2}\frac{\omega t}{2}\right].
\end{equation}
here $S_{\Phi}\left(\omega\right)=2\pi A_{\Phi}/|\omega|$ and $\omega_{ir}<|\omega|<\omega_{c}$, the infra-red cutoff $\omega_{ir}$ is usually determined by the measurement protocol and can be set and controlled in experiments~\cite{ithier2005decoherence}. For $1/f$ noise, at times $t\ll1/\omega_{ir}$, the free induction (Ramsey) decay is dominated by the frequencies $\omega<1/t$, so the formula above reduces to
\begin{equation}
f_{z,R}(t)	
=\exp\left[-t^{2}D_{\Phi}^{2}A_{\Phi}\left(-\ln\omega_{ir}t\right)\right]
=\exp\left[-\left(\Gamma_{\varphi}t\right)^{2}\right].
\end{equation}
So pure dephasing rate is
\begin{equation} \Gamma_{\varphi}=2\pi D_{\Phi}\sqrt{A_{\Phi}\left|\ln\omega_{ir}t\right|}.
\end{equation}
The modeled pure dephasing rate can be extracted $\Gamma_\varphi=(0.00288\,{\rm m}\Phi_0)D_\Phi+0.039\ \mu\rm s^{-1}$ for $D_\Phi \ge 0$ shown as the black dashed line in Fig.~\ref{fig:theoT2}(a). Using this modified $\Gamma_\varphi$, we calculate the effective $T_2$ as a function of flux as shown in Fig.~\ref{fig:theoT2}(b). This approach to accounting for gate error measurements in the presence of $1/f$ noise results in calculated coherence-limited gate error vs. flux curves that agree reasonably well with the experimental data.

\chapter{CR Gate Error}
\label{app:CRerror}
\section*{Error of Experimental Circuit}
\begin{figure}[h!]
	\centering
	\includegraphics[width=0.75\linewidth]{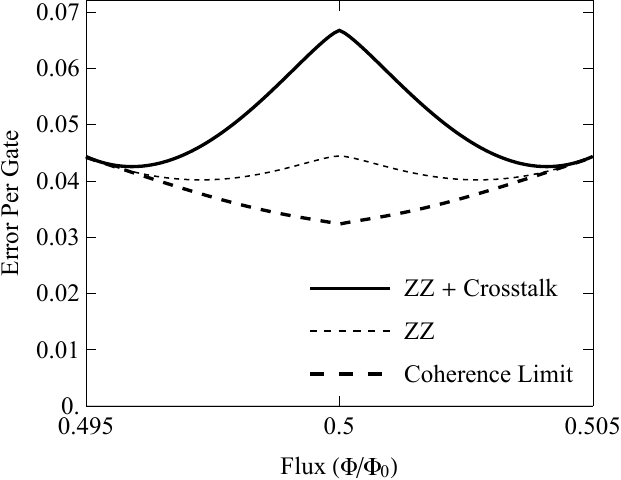}
	\vspace{-0.1in}
	\caption{Error per gate for $t_g=560$ ns with three different cases of error sources. Thick dashed line corresponds to coherence-limited gate error, which sets the lower bound for error per gate. Thin dashed line shows the gate error when ZZ contribution is added in the simulation. Solid line shows the case where both ZZ and classical crosstalk are included.}
	\label{fig:fidelity560}
\end{figure}

To explore the impact of classical crosstalk and ZZ interaction on the gate fidelity, we plot the gate error of echoed CR pulses vs. the flux threading CSFQ at $t_g=560$~ns in Fig.~\ref{fig:fidelity560}. 

\begin{figure}[t!]
	\centering
	\includegraphics[width=0.4\linewidth]{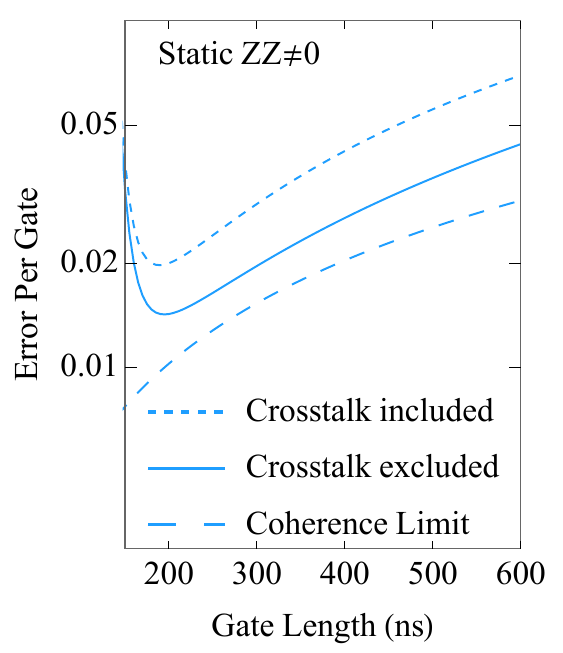}
	\includegraphics[width=0.4\linewidth]{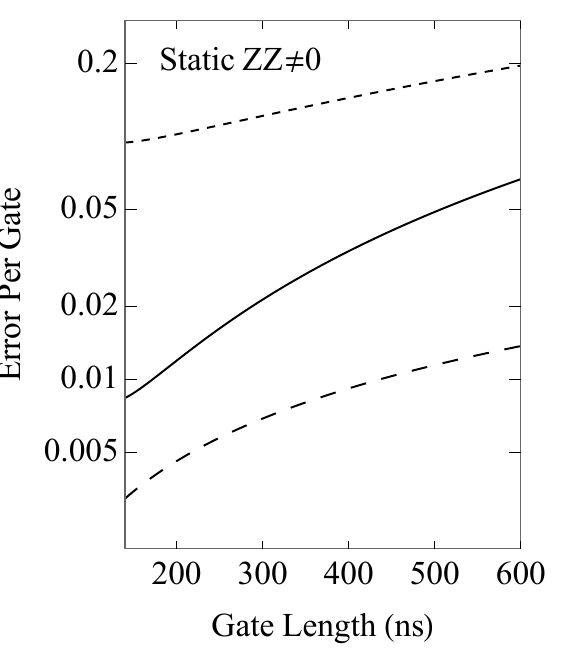}
	\put(-310,160){(a)}
	\put(-155,160){(b)}\\ \vspace{0.15in}
	\includegraphics[width=0.42\linewidth]{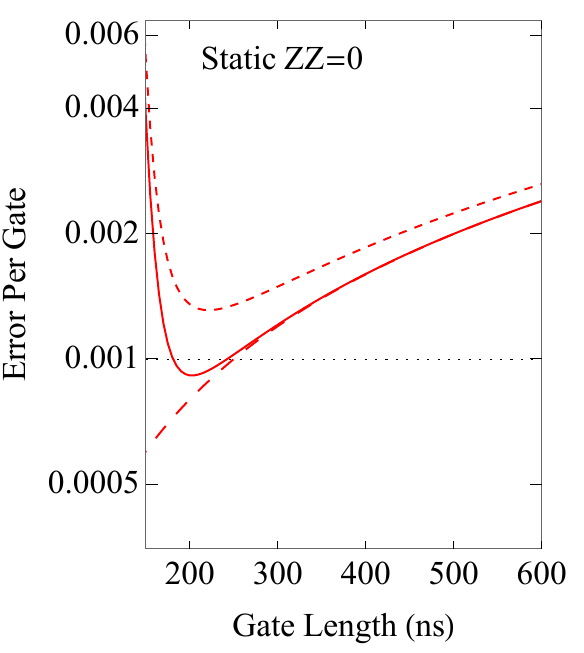}\hspace{3.8in}
	\put(-170,170){(c)}
	\vspace{-0.1in}
	\caption{Two-qubit gate error for three sets of coherence times.  $\left\{T_1^{(1)}, T_2^{(1)}, T_1^{(2)}, T_2^{(2)}\right\}$, where the superscripts indicate qubit label, are (18, 15, 40, 45) for the experimental chip, (40, 54, 43, 67) for the transmon-transmon device in Ref.~\cite{sheldon2016procedure}, and (200, 200, 200, 200) for an ideal CSFQ-transmon device (all times in $\mu$s). (a) Present CSFQ-transmon device. (b) Transmon-transmon device. (c) Ideal CSFQ-transmon device. Note that all three figures share the same legend as in (a). CSFQ is assumed to be at the sweet spot in the simulation. }
	\label{fig:fidelitycoherence}
\end{figure}

The flux-dependent gate error is calculated from the theory model for three cases: only coherence-limited error marked with thick dashed line, ZZ but no classical crosstalk included marked with thin dashed line, and both ZZ and classical crosstalk included marked with solid line. One can see that the gate error due to unwanted terms is the most at the flux sweet spot, despite at which coherence times are the longest and  the ZZ interaction is maximal as shown in Fig.~\ref{fig:ZZ_flux}. Away from the sweet spot, all unwanted terms become suppressed, and therefore the gate fidelity approaches its coherence limit. 

To explore how classical crosstalk, static ZZ interaction and coherence times impact the CR gate both in the CSFQ-transmon circuits with transmon-transmon circuits, we plot the CR gate errors in three samples: experimental CSFQ-transmon device (1), transmon-transmon device (2) and ideal CSFQ-transmon device (3). And plot it in Fig.~\ref{fig:fidelitycoherence}. One can see that eliminating both classical crosstalk and static ZZ interaction while prolonging the coherence times can suppression the CR gate error to be around $0.1\%$. 

\section*{Dynamical Quadratic Factor}
\label{app.eta}
Dynamical quadratic factor is obtained in Eq.~(\ref{eq.eta}) using perturbation theory, and corresponding coefficients are given by
\begin{eqnarray}
A_0&=&2r^3(r+2\gamma+\gamma^2)\nonumber\\
A_1&=&r^2\left[1+4r^2-16\gamma-6\gamma^2+2r(2\gamma^2+4\gamma-5)\right]\nonumber\\
A_2&=&r\left[r^3+22\gamma-2+r(19-32\gamma-12\gamma^2)+2r^2(\gamma^2+2\gamma-10)\right]\nonumber\\
A_3&=&1-5r^3-10\gamma+9\gamma^2+r(44\gamma-15)-2r^2(3\gamma^2+8\gamma-18)\\
A_4&=&4+9r^2-20\gamma+18\gamma^2+r(22\gamma-27)\nonumber\\
A_5&=&7-7r-10\gamma+9\gamma^2\nonumber\\
A_6&=&2\nonumber
\end{eqnarray}

\newpage
\thispagestyle{empty}

\chapter{NONRWA Perturbation Theory}
\label{app:nonrwa}
Pauli coefficients are calculated from two parts: rapid counter rotating terms (CRW) and slow co-rotating terms. RWA neglects the first term while NRWA includes the contribution from fast oscillating terms. Therefore the NRWA gives rise to 
\begin{equation}
\alpha_{\rm UVW}^{\rm tot}=\alpha_{\rm UVW}+\alpha_{\rm UVW}^{\rm coun}
\end{equation}
where U, V and W $\in$$ \{\rm I, X, Y, Z\}$. RWA results have been derived in chapter~\ref{c6}, so here we only discuss contribution from counter-rotating terms. Similarly, these terms can also be defined from perturbation theory as 
\begin{equation}
\begin{aligned}
4\alpha_{\rm ZII}^{\rm coun}&=E_{000}^{\rm coun}+E_{011}^{\rm coun}+E_{001}^{\rm coun}+E_{010}^{\rm coun}-E_{100}^{\rm coun}-E_{101}^{\rm coun}-E_{110}^{\rm coun}-E_{111}^{\rm coun}\\  
4\alpha_{\rm IZI}^{\rm coun}&=E_{000}^{\rm coun}+E_{001}^{\rm coun}+E_{100}^{\rm coun}+E_{101}^{\rm coun}-E_{010}^{\rm coun}-E_{011}^{\rm coun}-E_{110}^{\rm coun}-E_{111}^{\rm coun}\\  
4\alpha_{\rm IIZ}^{\rm coun}&=E_{000}^{\rm coun}+E_{100}^{\rm coun}+E_{010}^{\rm coun}+E_{110}^{\rm coun}-E_{001}^{\rm coun}-E_{011}^{\rm coun}-E_{101}^{\rm coun}-E_{111}^{\rm coun}\\
2\alpha_{\rm ZZI}^{\rm coun}&=E_{000}^{\rm coun}+E_{110}^{\rm coun}-E_{010}^{\rm coun}-E_{100}^{\rm coun}+E_{001}^{\rm coun}+E_{111}^{\rm coun}-E_{101}^{\rm coun}-E_{011}^{\rm coun}\\  
2\alpha_{\rm ZIZ}^{\rm coun}&=E_{000}^{\rm coun}+E_{101}^{\rm coun}-E_{100}^{\rm coun}-E_{001}^{\rm coun}+E_{010}^{\rm coun}+E_{111}^{\rm coun}-E_{011}^{\rm coun}-E_{110}^{\rm coun}\\  
2\alpha_{\rm IZZ}^{\rm coun}&=E_{000}^{\rm coun}+E_{011}^{\rm coun}-E_{001}^{\rm coun}-E_{010}^{\rm coun}+E_{100}^{\rm coun}+E_{111}^{\rm coun}-E_{101}^{\rm coun}-E_{110}^{\rm coun}\\  
\alpha_{\rm ZZZ}^{\rm coun}&=E_{000}^{\rm coun}+E_{011}^{\rm coun}-E_{001}^{\rm coun}-E_{010}^{\rm coun}+E_{101}^{\rm coun}+E_{110}^{\rm coun}-E_{100}^{\rm coun}-E_{111}^{\rm coun}
\end{aligned}
\end{equation}
with $E_{ijk}^{\rm coun}$ being energy shift in state $|ijk\rangle$ $(i,j,k=1,2,3)$ due to CRW terms. For brevity's sake, we define such notations $\Delta_{ij}=\omega_i-\omega_j$ and $\Sigma_{ij}=\omega_{i}+\omega_j$ for qubit frequency detuning and summation, separately. If we consider more than three order perturbation theory, this correction is not purely fast oscillating terms but is a combination of both of the two terms. For instance, the third correction in $|101\rangle$ state contains such a term $\langle101|V|202\rangle\langle202|V|112\rangle\langle112|V|101\rangle$, where the first and third transitions are fast oscillating terms while the second is the slow rotating term. Energy shift from CRW in the computational levels are given by

\begin{align}
E_{000}^{\rm coun}&=-\frac{J_{12}^{2}}{\Sigma_{12}}-\frac{J_{13}^{2}}{\Sigma_{13}}-\frac{J_{23}^{2}}{\Sigma_{23}}+\frac{2J_{12}J_{13}J_{23}}{\Sigma_{12}\Sigma_{13}\Sigma_{23}}(\Sigma_{12}+\Sigma_{13}+\Sigma_{23})\\
E_{001}^{\rm coun}&=-\frac{J_{12}^{2}}{\Sigma_{12}}-\frac{2J_{13}^{2}}{\Sigma_{13}+\delta_3}-\frac{2J_{23}^{2}}{\Sigma_{23}+\delta_3}+2J_{12}J_{13}J_{23}\left[\frac{2}{\Sigma_{12}(\Sigma_{13}+\delta_3)}\right.\nonumber\\
&+\left.\frac{2}{\Sigma_{12}(\Sigma_{23}+\delta_3)}+\frac{2}{(\Sigma_{13}+\delta_3)(\Sigma_{23}+\delta_3)}-\frac{1}{\Delta_{31}\Sigma_{12}}-\frac{1}{\Delta_{32}\Sigma_{12}}\right]\\
E_{010}^{\rm coun}&=-\frac{2J_{12}^{2}}{\Sigma_{12}+\delta_2}-\frac{J_{13}^{2}}{\Sigma_{13}}-\frac{2J_{23}^{2}}{\Sigma_{23}+\delta_2}+2J_{12}J_{13}J_{23}\left[\frac{2}{\Sigma_{13}(\Sigma_{12}+\delta_2)}\right.\nonumber\\
&+\left.\frac{2}{\Sigma_{13}(\Sigma_{23}+\delta_2)}+\frac{2}{(\Sigma_{12}+\delta_2)(\Sigma_{23}+\delta_2)}-\frac{1}{\Delta_{21}\Sigma_{13}}-\frac{1}{\Delta_{23}\Sigma_{13}}\right]\\
E_{100}^{\rm coun}&=-\frac{2J_{12}^{2}}{\Sigma_{12}+\delta_1}-\frac{2J_{13}^{2}}{\Sigma_{13}+\delta_1}-\frac{J_{23}^{2}}{\Sigma_{23}}+2J_{12}J_{13}J_{23}\left[\frac{2}{\Sigma_{23}(\Sigma_{13}+\delta_1)}\right.\nonumber\\
&+\left.\frac{2}{\Sigma_{23}(\Sigma_{12}+\delta_1)}+\frac{2}{(\Sigma_{13}+\delta_1)(\Sigma_{12}+\delta_1)}-\frac{1}{\Delta_{12}\Sigma_{23}}-\frac{1}{\Delta_{13}\Sigma_{23}}\right]\\
E_{011}^{\rm coun}&=-\frac{2J_{12}^{2}}{\Sigma_{12}+\delta_2}-\frac{2J_{13}^{2}}{\Sigma_{13}+\delta_3}+\frac{J_{23}^{2}}{\Sigma_{23}}-\frac{4J_{23}^{2}}{\Sigma_{23}+\delta_2+\delta_3}+2J_{12}J_{13}J_{23}\left[\frac{1}{\Sigma_{23}\Delta_{21}}\right.\nonumber\\
&\left.-\frac{2}{\Sigma_{12}+\delta_2}\left(\frac{1}{\Delta_{31}}+\frac{1}{\Delta_{32}-\delta_2}\right) -\frac{2}{\Sigma_{13}+\delta_3}\left(\frac{1}{\Delta_{21}}+\frac{1}{\Delta_{23}-\delta_3}\right)\right.\nonumber\\
&\left.+\frac{1}{\Sigma_{23}\Delta_{31}}++\frac{4(2\delta_2+2\delta_3+\Sigma_{12}+\Sigma_{13}+\Sigma_{23})}{(\Sigma_{23}+\delta_2+\delta_3)(\Sigma_{12}+\delta_21)(\Sigma_{13}+\delta_3)}\right]\\
E_{101}^{\rm coun}&=-\frac{2J_{12}^{2}}{\Sigma_{12}+\delta_1}+\frac{J_{13}^{2}}{\Sigma_{13}}-\frac{4J_{13}^{2}}{\Sigma_{13}+\delta_1+\delta_3}-\frac{2J_{23}^{2}}{\Sigma_{23}+\delta_3}+2J_{12}J_{13}J_{23}\left[\frac{1}{\Sigma_{13}\Delta_{12}}\right.\nonumber\\
&\left.-\frac{2}{\Sigma_{12}+\delta_1}\left(\frac{1}{\Delta_{32}}+\frac{1}{\Delta_{31}-\delta_1}\right) -\frac{2}{\Sigma_{23}+\delta_3}\left(\frac{1}{\Delta_{12}}+\frac{1}{\Delta_{13}-\delta_3}\right)\right.\nonumber\\
&\left.+\frac{1}{\Sigma_{13}\Delta_{32}}+\frac{4(2\delta_1+2\delta_3+\Sigma_{12}+\Sigma_{13}+\Sigma_{23})}{(\Sigma_{13}+\delta_1+\delta_3)(\Sigma_{12}+\delta_1)(\Sigma_{23}+\delta_3)}\right]\\
E_{110}^{\rm coun}&=\frac{J_{12}^{2}}{\Sigma_{12}}-\frac{4J_{12}^{2}}{\Sigma_{12}+\delta_1+\delta_2}+\frac{2J_{13}^{2}}{\Sigma_{13}+\delta_1}-\frac{2J_{23}^{2}}{\Sigma_{23}+\delta_2}+2J_{12}J_{13}J_{23}\left[\frac{1}{\Sigma_{12}\Delta_{13}}\right.\nonumber\\
&\left.-\frac{2}{\Sigma_{13}+\delta_1}\left(\frac{1}{\Delta_{23}}+\frac{1}{\Delta_{21}-\delta_1}\right) -\frac{2}{\Sigma_{23}+\delta_2}\left(\frac{1}{\Delta_{13}}+\frac{1}{\Delta_{12}-\delta_2}\right)\right.\nonumber\\
&\left.+\frac{1}{\Sigma_{12}\Delta_{23}}+\frac{4(2\delta_1+2\delta_2+\Sigma_{12}+\Sigma_{13}+\Sigma_{23})}{(\Sigma_{12}+\delta_1+\delta_2)(\Sigma_{13}+\delta_1)(\Sigma_{23}+\delta_2)}\right]
\end{align}
\vspace{0.4in}

\begin{align}
E_{111}^{\rm coun}&=\frac{J_{12}^{2}}{\Sigma_{12}}-\frac{4J_{12}^{2}}{\Sigma_{12}+\delta_1+\delta_2}+\frac{J_{13}^{2}}{\Sigma_{13}}-\frac{4J_{13}^{2}}{\Sigma_{13}+\delta_1+\delta_3}+\frac{J_{23}^{2}}{\Sigma_{23}}-\frac{4J_{23}^{2}}{\Sigma_{23}+\delta_2+\delta_3}\nonumber\\
&+2J_{12}J_{13}J_{23}\left[\frac{\Sigma_{12}+\Sigma_{13}+\Sigma_{23}}{\Sigma_{12}\Sigma_{13}\Sigma_{23}}+\frac{2}{\Sigma_{12}}\left(\frac{1}{\Delta_{13}-\delta_3}+\frac{1}{\Delta_{23}-\delta_3}\right)\right.\nonumber\\
&\left.+\frac{2}{\Sigma_{13}}\left(\frac{1}{\Delta_{32}-\delta_2}+\frac{1}{\Delta_{12}-\delta_2}\right)+\frac{2}{\Sigma_{23}}\left(\frac{1}{\Delta_{21}-\delta_1}+\frac{1}{\Delta_{31}-\delta_1}\right)\right.\nonumber\\
&\left.+\frac{4}{\Sigma_{12}+\delta_1+\delta_2}\left(\frac{1}{\Delta_{13}+\delta_1}+\frac{1}{\Delta_{23}+\delta_2} \right) +\frac{4}{\Sigma_{13}+\delta_1+\delta_3}\left(\frac{1}{\Delta_{12}+\delta_1}+\frac{1}{\Delta_{32}+\delta_3} \right)\right.\nonumber\\
&\left.+\frac{4}{\Sigma_{23}+\delta_2+\delta_3}\left(\frac{1}{\Delta_{31}+\delta_3}+\frac{1}{\Delta_{21}+\delta_2} \right)\right.\nonumber\\
&\left.+\frac{8(2\delta_1+2\delta_2+\delta_3+\Sigma_{12}+\Sigma_{13}+\Sigma_{23})}{(\Sigma_{12}+\delta_1+\delta_2)(\Sigma_{13}+\delta_1+\delta_3)(\Sigma_{23}+\delta_2+\delta_3)}\right]
\end{align}

 In fact, these cross terms such as $\Sigma_{ij}\Delta_{jk}$ are dominant compared to pure CRW transitions like $\Sigma_{ij}\Sigma_{jk}$. 
 
\clearpage
\phantom{s}
\thispagestyle{empty}
\clearpage

\end{appendices}

\newpage
\thispagestyle{empty}

\pagenumbering{Roman}
\printbibliography[heading=bibintoc]

\raggedright
\printindex
\newpage
\thispagestyle{empty}

\end{document}